\documentclass[acmsmall,screen,nonacm]{acmart}
\usepackage{flafter}
\usepackage{longtable}
\usepackage{array}

\graphicspath{{figures/}}

\setcopyright{none}
\acmJournal{CSUR}
\begin{document}

\title[Privacy Protection in Self-Evolving Personal Agents]{Privacy Protection in Self-Evolving Personal Agents: Persistent Memory and Cross-Carrier Transfer}

\author{Zibin Lin}
\affiliation{%
  \institution{Shenzhen University}
  \city{Shenzhen}
  \country{China}
}
\email{linaacc9595@gmail.com}

\author{Taotao Wang}
\affiliation{%
  \institution{Shenzhen University}
  \city{Shenzhen}
  \country{China}
}
\email{ttwang@szu.edu.cn}

\author{Shengli Zhang}
\affiliation{%
  \institution{Shenzhen University}
  \city{Shenzhen}
  \country{China}
}
\email{zsl@szu.edu.cn}

\begin{abstract}
Personal agents carry private experience into later decisions through persistent memory. This thematic survey compares memory admission, maintenance, release, and withdrawal by protected change, observer, and validation setting. Direct evidence in the screened corpus comes from controlled memory workflows. Three worked scenarios follow one private fact through booking, feedback-driven learning, and withdrawal. We show when prompt revision and memory reads compose, when skills inherit a private head's bound, and why reader non-use does not establish cleanup of learned state. Diagnostic experiments measure the utility costs of wider label protection and broad memory deletion, and test how missing lineage obscures descendants.
\end{abstract}




\begin{teaserfigure}
\centering
\includegraphics[width=0.88\textwidth,height=0.27\textheight,keepaspectratio]{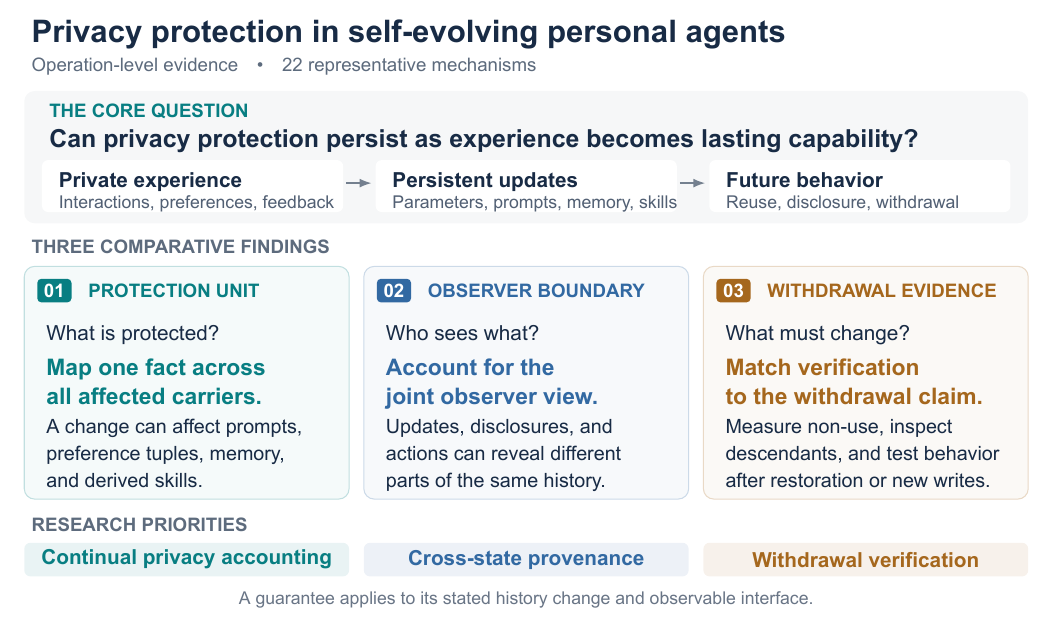}
\caption*{\textbf{Graphical Abstract.} Privacy boundaries across persistent updates, release, and withdrawal.}
\Description{A flow from private interactions, preferences, and feedback through persistent parameters, prompts, memory, and skills to future behavior. Three panels compare protection units, observer boundaries, and withdrawal evidence. The design priorities are continual privacy accounting, cross-state provenance, and withdrawal verification.}
\end{teaserfigure}

\maketitle

\section{Introduction}\label{sec:intro}

A scheduling assistant may retain a user's medical restriction in long-term memory and later use it to propose an appointment. That use has several privacy boundaries. A cloud host can observe a memory access, a booking service can observe the chosen time, and a future session can recover the restriction from a derived summary. If the same restriction also informs a prompt revision or a reusable skill, the pathways multiply. We survey the protection of persistent personal memory and examine what changes when the same private fact reaches another carrier.

Recent benchmarks localize concrete leakage along individual agent interfaces. PrivacyLens~\cite{shao2024privacylens} exposes gaps between normative privacy awareness and actual disclosure during multi-step execution. On long-term memory, MEXTRA~\cite{wang2025mextra} extracts private interactions via black-box prompting, and SPORE~\cite{gao2026spore} leaks stored secrets into external tool-call parameters across sessions. At the skill layer, AntiSkillBench~\cite{xiang2026personaskills} measures content leakage, attribute disclosure, and behavioral impersonation induced by persona skills distilled from personal trajectories. Together, these benchmarks expose interface-level vulnerabilities while leaving open how privacy risk propagates from raw inputs across persistent carriers to external outputs.

We define a \emph{self-evolving personal agent} as an agent that serves one user over time and updates parameters, prompt policies, memories, or skills from experiences, preferences, corrections, or tool feedback~\cite{gao2026selfevolvingsurvey}. PAHF illustrates the feedback loop: pre-action clarification and post-action correction update a per-user memory that can affect the next action~\cite{liang2026pahf}. We center the survey on memory write, maintenance, read, release, and withdrawal. We include results on fixed personal profiles when they evaluate a release surface and study prompt, parameter, preference, and skill methods where they clarify a specific transfer from or interaction with memory.

We also separate two complementary privacy objectives. \emph{Privacy-aware evolution} studies how an agent learns contextual disclosure norms, deciding which attributes a task may consult or share with a recipient. \emph{Privacy-protected evolution} constrains the update process itself, bounding what untrusted endpoints or external observers infer from candidate scores, update logs, or modified state carriers. Personal assistants such as PANOLA~\cite{ulusoy2022panola} and uncertainty-aware decision agents~\cite{ayci2023uncertaintypa} learn disclosure preferences and negotiate release, while their reported release evaluations do not establish privacy guarantees for learning transcripts or persistent state updates. The graphical abstract summarizes our synthesis, and Figure~\ref{fig:1} maps operator interventions across the loop.

Table~\ref{tab:1} situates this review among surveys of adaptation, state governance, and privacy. SAVER tracks where experience acquires influence and which observations reveal a failure~\cite{chen2026safetysurvey}. Always-On Agents links durable state to provenance, recovery, and externally committed effects~\cite{ding2026alwayson}. The RAG accounting survey distinguishes protected units, accounting horizons, and release boundaries~\cite{zhang2026ragaccounting}. We connect these perspectives by tracking a change to the same personal fact through each operation and asking which guarantees survive when an observer sees several outputs. We derive a joint bound for bounded prompt revision and memory reads, and locate its failure at an action that consults raw memory.

\begin{table}[t]
\caption{Analytical Scope of Closest Related Surveys and Comparative Tasks Addressed in This Review}
\label{tab:1}
\footnotesize
\setlength{\tabcolsep}{3.5pt}
\renewcommand{\arraystretch}{1.08}
\begin{tabular}{@{}>{\raggedright\arraybackslash}p{0.23\textwidth}>{\raggedright\arraybackslash}p{0.18\textwidth}>{\raggedright\arraybackslash}p{0.27\textwidth}>{\raggedright\arraybackslash}p{0.26\textwidth}@{}}
\toprule
\textbf{Survey \& Source Locator} & \textbf{Primary Unit of Analysis} & \textbf{Established Foundations} & \textbf{Comparative Task Added Here} \\
\midrule
\emph{Self-Evolving Agents}~\cite{gao2026selfevolvingsurvey}, \S\S2--5; \S5.4, Table~4 & Evolution carriers, timing, and learning paradigms & Separates reusable state carriers and feedback-driven updates horizontally & Differentiates private inputs, observers, and protection units per update operator \\
\addlinespace[1.5pt]
\emph{Safety in Self-Evolving Agents} (SAVER)~\cite{chen2026safetysurvey}, \S\S2--7; App.\ A & Carrier $\times$ adaptation operation, exposure, and response & Models cross-state influence, admission, propagation, recovery, and evidence bounds & Isolates intra-user privacy units, update transcripts, and cross-carrier withdrawal \\
\addlinespace[1.5pt]
Agent security and data-centric privacy~\cite{he2026agentsecuritysurvey,lahjouji2026agentsknowtoomuch}, \S\S3--5 and \S\S3--6, respectively & Agent workflows, risks, and data surfaces & Maps privacy threats, defenses, and cross-interface disclosure & Connects private generation, scoring, and writeback to later observations \\
\addlinespace[1.5pt]
\emph{Privacy Accounting in RAG}~\cite{zhang2026ragaccounting}, \S\S3--5 & Repeated queries and observable releases & Compares granularity, accounting scope, and release boundaries & Holds the history change fixed across prompt, reward, and memory, including fresh writes \\
\addlinespace[1.5pt]
\emph{Long-Term Memory Security}~\cite{lin2026memorysecuritysurvey}, \S2, \S5; App.\ D & Memory lifecycle stages and security objectives & Organizes provenance, access scope, rollback, forgetting, and governance criteria & Contrasts derived memory influence against prompt, parameter, and skill bounds \\
\addlinespace[1.5pt]
\emph{Always-On Agents}~\cite{ding2026alwayson}, \S2.3, \S\S8.5--8.8, \S9 & Durable operational state and its governance & Links authority, provenance, recovery, and committed effects; scores mutations and recovery & Fixes a personal-history change and tests which observer-specific claims survive across carriers \\
\addlinespace[1.5pt]
\emph{Privacy Preserving Prompt Engineering}~\cite{edemacu2025promptsurvey}, Author ver.\ \S\S II--III, \S V & Privacy models, targets, and prompt defenses & Taxonomizes local/central DP and non-DP prompt defenses alongside cost & Compares data access and accounting across generation, scoring, and selection \\
\addlinespace[1.5pt]
\emph{Privacy-Preserving LLM Preference Alignment}~\cite{chen2026preferencesurvey}, TechRxiv v1, \S\S2--5 & DP alignment and federated preference protocols & Organizes central/local DP alignment and model- vs.\ preference-centric FL & Separates tuple, label, and sampling units; tracks reference-model independence \\
\addlinespace[1.5pt]
\emph{Machine Unlearning}~\cite{xu2024unlearnsurvey}, \S\S2--3; Tables~3--7 & Unlearning targets, execution, and verification & Separates deletion targets from empirical and formal verification grounds & Maps behavioral suppression, derived cleanup, version access, and recovery validity \\
\bottomrule
\end{tabular}
\par\smallskip\raggedright\footnotesize\emph{Note:} First-column locators reference the cited texts; publicly archived author versions are cited for prompt privacy~\cite{edemacu2025promptsurvey} and machine unlearning~\cite{xu2024unlearnsurvey}, and TechRxiv v1 for preference alignment~\cite{chen2026preferencesurvey}.\par
\end{table}

Three questions guide the comparison:
\begin{itemize}
\item \textbf{RQ1 (Protection units).} Which changes to persistent personal memory do mechanisms protect, and how do record, attribute, tuple, or user-level units map to the same fact across carriers?
\item \textbf{RQ2 (Observers and endpoints).} Which observations can a host, validator, or recipient link, and under what conditions do protections cover that joint view?
\item \textbf{RQ3 (Persistence and verification).} What evidence supports privacy after a fresh write, artifact revision, or withdrawal, and which conditions govern transfer beyond memory?
\end{itemize}

We map memory protection by operation and follow one fact through booking, feedback-driven learning, and withdrawal. For each scenario we identify which results apply to the specified history change and observer, which fall outside scope, and which lack evidence at the target interface. Sections~\ref{sec:methodology}--\ref{sec:framework} define the scope and coding. Sections~\ref{sec:param-prompt-pref}--\ref{sec:skill} review the mechanisms, Section~\ref{sec:synthesis} compares their boundaries and costs, and Section~\ref{sec:agenda} identifies research priorities.

\section{Review Scope and Methodology}\label{sec:methodology}

\subsection{Research Design and Literature Scope}\label{sec:design}

This thematic survey covers work publicly available through 10 November 2026. Its core is privacy protection for persistent memory in agents that serve one user over time. Prompt, parameter, preference, and skill studies enter when they identify a mechanism or comparator that constrains a specified operation with a plausible path to that setting. Table~\ref{tab:3} illustrates 22 mechanisms and primitives.

The initial search combined thematic queries, archival checks, and one-hop citation tracking. The carrier audit used five fielded arXiv queries and five title and abstract filters over 24,844 ACL Anthology proceedings records. The channels returned 216 and 102 hits, respectively, yielding 191 and 98 unique records within their respective channels. We matched stable identifiers and titles, screened all hits, and checked primary texts for new inclusions. The audit recovered five of nine known direct studies; all nine concern memory, so this check does not calibrate coverage of other carriers. Follow-ups targeted linked observation and withdrawal, and rescreened 40 operation-overlap decisions. Supplementary Sections~S1.14--S1.16 give the windows and dispositions.

\subsection{Inclusion, Exclusion, and Version Selection}\label{sec:inclusion}

Mechanism eligibility requires an accessible primary source by the cutoff, an explicit privacy intervention at an identifiable operation, and enough detail to verify at least two of the protected unit, observer, and trusted endpoints. The operation must act on persistent adaptation, constrain the use or release of retained state, or provide a transferable primitive for one of those steps. Deployment feasibility alone, without a privacy outcome or an applicable analysis, does not count as evidence for protection.

We exclude from mechanism coding static personalization or one-shot inference lacking persistent state or protected read interfaces, ``on-device'' or ``federated'' claims lacking a localizable operator, purely offensive extraction studies, and untraceable summaries. Related surveys and benchmarks supply definitions, threat surfaces, or baselines. A benchmark contributes to mechanism counts when it specifies and supports a protection step with a localized operator, a protected object, and a defined observer or trust boundary. We retain independent studies of the same operation when they contribute a matched comparison, conflicting result, or measured trade-off. Duplicate preprints, OpenReview revisions, and archival publications from one method family collapse to one entry, preferring peer-reviewed archival versions published by 10 November 2026 or a version-pinned preprint available by the cutoff; post-cutoff releases lie outside our scope.

\subsection{Evidence Coding and Synthesis Protocol}\label{sec:coding}

Supplementary Section~S5 records one row per source, protection operation, and evidentiary setting when the source supports distinct claims. Rows identify the carrier, protected change or object, observable interface, and supporting result or specification. Validation settings distinguish formal results, controlled or simulated evaluations, and deployment evidence. The 22 examples in Supplementary Table~S11 also record trusted endpoints, reuse conditions, and uncovered exposures in detail. Evidence distance describes the setting of each operation:
\begin{itemize}
\item \textbf{\texttt{D} (Direct update intervention):} Evidence covers protection of persistent state written, maintained, or withdrawn for one user and carried across decisions or interactions. Controlled loops qualify. Optimizer iterations and held-out tests on a selected candidate do not by themselves establish carryover from a committed update.
\item \textbf{\texttt{S} (Surface or release intervention):} Evaluates reading, minimizing, or releasing retained user state in a personal-agent task.
\item \textbf{\texttt{T} (Transferable mechanism or primitive):} Supports a relevant operation in another setting, including general algorithms and shared federated updates.
\end{itemize}
Guarantee type and evidence distance answer separate questions. An empirical write gate may receive \texttt{D} while a formal composition theorem receives \texttt{T}. A counted operation needs a reported evaluation or an analytical result for the stated protection claim. We record shared pipeline outcomes at the implicated operations and leave the isolated effect of each component unassigned. Design descriptions and worked examples without such support provide background. Supplementary Table~S10 counts each source once at its closest supported operation (\texttt{D}, then \texttt{S}, then \texttt{T}). Section~S5 reports the distribution by operation and distinct sources within each family. Supplementary Section~S3.3 documents an independent recoding of selected boundary cases and shows which conclusions depend on those rulings.

\begin{table}[t]
\caption{Primary Evidence Roles in the 180-Source Corpus}
\label{tab:2}
\footnotesize
\setlength{\tabcolsep}{3.5pt}
\begin{tabular}{@{}>{\raggedright\arraybackslash}p{0.27\textwidth}r>{\raggedright\arraybackslash}p{0.29\textwidth}>{\raggedright\arraybackslash}p{0.28\textwidth}@{}}
\toprule
\textbf{Primary role} & \textbf{Count} & \textbf{Use in this review} & \textbf{Examples} \\
\midrule
Representative mechanisms & 22 & Compare protection operations & DP-OPT, Opal, Reglos \\
Extended or adjacent evidence & 83 & Test mechanisms and transfer conditions & CSLR, PRISM, HomeLLaMA \\
Theoretical and systems primitives & 16 & Supply formal or system foundations & DP-FTRL, Compass, TAPE \\
Risk and evaluation studies & 40 & Characterize exposure and measure residual behavior & MEXTRA, Label-Space Leakage, Claws~\cite{tobaben2026labelspace,dong2026claws} \\
Surveys and state background & 19 & Establish terminology and design context & SAVER, Always-On Agents \\
\midrule
\textbf{Total} & \textbf{180} & \multicolumn{2}{l}{A source has one primary role.} \\
\bottomrule
\end{tabular}
\end{table}

Table~\ref{tab:2} assigns one primary role per source. Risk and evaluation studies can also contribute a tested intervention, as recorded in Supplementary Table~S10. Supplementary Table~S9 gives all source assignments and Section~S5 gives the operation ledger. One contextual graph-model reference lies outside the corpus~\cite{xu2026dynamicgraphcontext}.

\section{From Technical Labels to an Update-Operator Framework}\label{sec:framework}

Differential privacy can protect prompt scores, the training of a reward head, adapter updates, or multi-round memory reads. Each use has its own protected change, observer, trusted endpoints, and composition rule. We trace protection operations along the chain from private inputs through evaluation and updates to persistent state, downstream release, and withdrawal. The four dimensions below connect each guarantee to the part of that chain it covers.

\begin{figure}[tp]
\centering
\includegraphics[width=\textwidth,height=0.36\textheight,keepaspectratio]{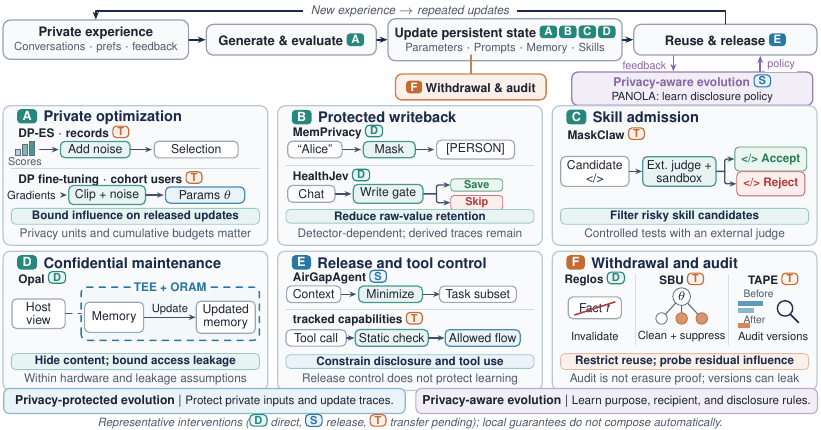}
\caption{Update-operator interventions across the evolution loop: \textbf{(A)} optimization (DP-ES; cohort-user DP fine-tuning, \texttt{T}); \textbf{(B)} writeback protection (MemPrivacy, HealthJev); \textbf{(C)} candidate skill filtering (MaskClaw, \texttt{T}, with an external judge); \textbf{(D)} confidential maintenance (Opal); \textbf{(E)} release and tool gates (AirGapAgent, tracked capabilities) fed by PANOLA disclosure rules; and \textbf{(F)} withdrawal and audit (Reglos, SBU, TAPE). Guarantees do not compose automatically.}\label{fig:1}
\Description{Architectural diagram illustrating privacy mechanisms across the self-evolving personal-agent loop. At the top, a feedback cycle connects Private experience to Generate and evaluate [A], Update persistent state [A, B, C, D] (Parameters, Prompts, Memory, Skills) with an orange branch for F: Withdrawal and audit, and Reuse and release [E], returning to Private experience while feeding learned disclosure preferences via PANOLA privacy-aware evolution back into [E]. Six panels detail each intervention class: Panel A (Private optimization) depicts record-level score noise in DP-ES and cohort-user gradient clipping plus noise in DP fine-tuning tagged [T]; Panel B (Protected writeback) depicts typed entity masking in MemPrivacy and save-or-skip gating in HealthJev; Panel C (Skill admission) depicts judge and sandbox gating in MaskClaw under fixed scenario feedback, with an external judge in the sandbox check; Panel D (Confidential maintenance) depicts an untrusted host view separated from a TEE plus ORAM enclave boundary in Opal; Panel E (Release and tool control) depicts task context minimization in AirGapAgent and static capability flow checks in tracked capabilities; and Panel F (Withdrawal and audit) depicts fact invalidation in Reglos, dependency-tree cleanup and parameter suppression in SBU, and before-and-after version auditing in TAPE. Bottom banners contrast privacy-protected evolution with privacy-aware evolution.}
\end{figure}

\paragraph{Dimension 1: State Carriers and Intervention Operators.}
The first dimension pairs each persistent carrier with the exact update or usage operator constrained by a mechanism. Parameter protections act on gradients, adapters, or reward signals; prompt protections modify generation, scoring, or selection; preference protections bound comparison tuples, choice labels, or policy sampling; and memory or skill protections govern writeback, maintenance, withdrawal, release, or repository admission. Localizing the operator determines which flows are constrained. Keeping conversations on-device reduces upload exposure, yet offers no persistent update bound if distilled summaries enter shared skill repositories or if cloud servers observe accepted prompt revisions over time.

\paragraph{Dimension 2: Protection Units and Intra-User Adjacency.}
The second dimension fixes the \emph{protection unit} by specifying which inputs may differ. Participation privacy changes one person's contribution to a cohort. Intra-user privacy can change a fact, episode, attribute, or time window in the same person's history. When a fact changes $k$ records, the generic group conversion sends $\varepsilon$ to $k\varepsilon$ and $\delta$ to $\delta\sum_{j=0}^{k-1}e^{j\varepsilon}$. A mechanism calibrated to the group can establish a different bound. Pufferfish~\cite{kifer2014pufferfish} compares outputs conditioned on alternative secrets under a specified family of history distributions. It can represent a fact whose evidence is dispersed across turns, with the correlation model and observable outputs made explicit.

\paragraph{Dimension 3: Observable Interfaces and Trusted Endpoints.}
The third dimension identifies the adversary's \emph{observable interface} (or \emph{transcript}) alongside required \emph{trusted endpoints}. Releasing a DP prompt or reward head does not conceal raw training records seen by a trusted curator. Secure multi-party computation (MPC) hides secret-shared inputs from non-colluding servers during execution, yet permitted outputs still reflect those inputs. Entity masking relies on detector recall and local mapping isolation, while trusted execution environments (TEEs) depend on hardware integrity, remote attestation, and declared access-pattern leakage functions. Each mechanism therefore answers a distinct observer question and demands explicit trust boundaries.

\paragraph{Dimension 4: Temporal Horizon, Replay, and Derived-State Withdrawal.}
The fourth dimension tracks continual reuse and derived-artifact propagation. Computation on privatized outputs and public inputs inherits the original DP bound by post-processing. Fresh access to private state requires checking the resulting observation and any applicable composition rule. Likewise, deleting a source record leaves its descendants subject to their own withdrawal requirements. Reuse, replay, and withdrawal therefore depend on which artifacts persist and which interfaces expose them.

Figure~\ref{fig:1} locates these interventions. Table~\ref{tab:3} gives 22 examples of mechanisms and theoretical primitives. Supplementary Table~S11 details their observer, trust, reuse, and withdrawal conditions. Supplementary Section~S5 identifies the setting and supported claim for each documented operation.

\begin{table}[tp]
\caption{Intervention Points and Guarantee Types of 22 Representative Mechanisms and Primitives}\label{tab:3}
\footnotesize
\setlength{\tabcolsep}{3.2pt}
\renewcommand{\arraystretch}{1.01}
\begin{tabular}{@{}>{\raggedright\arraybackslash}p{0.14\textwidth}>{\raggedright\arraybackslash}p{0.20\textwidth}>{\centering\arraybackslash}p{0.05\textwidth}>{\raggedright\arraybackslash}p{0.36\textwidth}>{\raggedright\arraybackslash}p{0.17\textwidth}@{}}
\toprule
\textbf{State Carrier} & \textbf{Mechanism} & \textbf{Code} & \textbf{Coded Operation} & \textbf{Guarantee Type} \\
\midrule
Prompt policy & DP-ES~\cite{liu2026dpes} & \texttt{T} & Minibatch scoring on private records with RDP accounting & Approximate DP \\
Prompt policy & DP-OPT~\cite{hong2024dpopt} & \texttt{T} & DP ensemble generation and exponential-mechanism selection on disjoint validation data & Approximate DP \\
Candidate selection & Private Selection~\cite{liu2019privateselection} & \texttt{T} & Randomized-stopping meta-selection over private candidate mechanisms & Pure / approx.\ DP \\
Parameters & User-Level DP Fine-Tuning~\cite{chua2024privacyunit} & \texttt{T} & Per-user clipping and noise (and group-privacy comparison) for LoRA updates & User-level approx.\ DP \\
Parameters & PrivMeSA~\cite{wang2026privmesa} & \texttt{T} & Risk-guided LoRA optimization on a shared clinical policy & Empirical risk reward \\
Evaluator / preference & Decoupled DP Reward Modeling~\cite{cho2026decoupledreward} & \texttt{T} & DP reward-head training followed by post-processing Best-of-$N$ selection & Tuple approx.\ DP \\
Evaluator / preference & DP Reward Estimation~\cite{raychowdhury2024privatereward} & \texttt{T} & Central objective perturbation or local randomized-response perturbation for preference labels & Central / local label-DP \\
Preference output & KL-Regularization DP~\cite{zhang2025klregularization} & \texttt{T} & Single-shot stochastic action sampling for a fixed prompt & Single-action label-DP \\
Soft prompt & SecP-Tuning~\cite{luo2026secptuning} & \texttt{T} & Two-server MPC forward pass paired with client-side soft-prompt updates & MPC confidentiality \\
Continual parameters & Lifelong DP~\cite{lai2022lifelongdp} & \texttt{T} & Task-partitioned objective perturbation across released model versions & Tuple Lifelong DP \\
Cross-cutting primitive & Concurrent Composition~\cite{henzinger2026concurrentcomposition} & \texttt{T} & Adaptive interleaved queries, updates, and full interaction transcripts & DP composition theorem \\
Memory & MemPrivacy~\cite{chen2026memprivacy} & \texttt{D} & Edge-side typed entity substitution prior to cloud memory writeback & Conditional isolation \\
Memory & DP-MemView~\cite{kim2026dpmemview} & \texttt{S} & Multi-round attribute-level reading and release over stored memory & Per-attribute pure DP \\
Memory / context & Galaxy~\cite{bao2025galaxy} & \texttt{S} & Graded sensitivity masking of personal-agent context before cloud release & Empirical release gating \\
Memory & Opal~\cite{kaviani2026opal} & \texttt{D} & Client-counter-checked enclave ingestion, ORAM indexing, expiry, and summaries & TEE / ORAM confidentiality \\
Memory & HealthJev~\cite{li2026healthjev} & \texttt{D} & Selective writeback gating for privacy-sensitive longitudinal interactions & Empirical governance \\
Memory & Reglos~\cite{yang2026reglos} & \texttt{D} & Tombstone update evaluated with a reader non-use rule & Empirical non-use \\
Parameters / memory & SBU~\cite{wang2026agenticunlearning} & \texttt{T} & Dependency-closure cleanup paired with synchronized parameter suppression & Empirical joint unlearning \\
Skill experience & MaskClaw~\cite{zhao2026maskclaw} & \texttt{T} & Scenario-level privacy-skill scoring and sandbox filtering & Empirical behavioral gating \\
Skill experience & FedSkillGit~\cite{lan2026fedskillgit} & \texttt{T} & Pre-commit payload sanitization and auditing for federated shared skills & Empirical payload filtering \\
Privacy policy & PANOLA~\cite{ulusoy2022panola} & \texttt{S} & Pre-release OSN sharing negotiation via feedback-learned bidding policy & Empirical release governance \\
Context release & AirGapAgent~\cite{bagdasarian2024airgapagent} & \texttt{S} & Pre-interaction minimization of task-relevant context for third parties & Empirical context isolation \\
\bottomrule
\end{tabular}
\par\smallskip\raggedright\footnotesize\emph{Note:} Dist. refers to the representative comparison. Supplementary Table~S11 details the endpoints, conditions, and sources. Supplementary Section~S5 codes individual operations and validation settings.\par
\end{table}

\section{Privacy Mechanisms in Parameter, Prompt, and Preference Updates}\label{sec:param-prompt-pref}

\subsection{Candidate Generation, Scoring, and Selection in Prompt Optimization}\label{sec:prompt-opt}

Prompt self-evolution creates persistent exposure whenever private interaction records shape candidate instructions, exemplars, or rankings. Reviewed mechanisms place formal protection at candidate generation, scoring, or selection. DP-OPT~\cite{hong2024dpopt} runs a local differentially private (DP) ensemble using \texttt{LimitedDomain} approximate-DP token generation (accumulating per-token failure $\delta_0$ across $m$ generated tokens) over private training records, then applies the exponential mechanism on a private validation split so an untrusted cloud model receives only a DP prompt. Parallel composition across generation and selection requires disjointness at the target intra-user protection unit rather than merely by record ID. By contrast, DP-ES~\cite{liu2026dpes} withholds private data from the mutation model and focuses its privacy budget on Gaussian-noised evaluation scores for $K$ candidate prompts per round. Each candidate is scored on a random minibatch (without-replacement subsampling $q = b/n$ under replace-one adjacency), after which top-$k$ selection, mutation, and release follow by post-processing. The two methods also induce distinct trust boundaries: DP-OPT trusts a local generator and validator, while DP-ES trusts a scoring endpoint with private evaluation records. Its R\'enyi differential privacy (RDP) accountant tracks privacy loss across scoring calls.

Prompt carriers span discrete text, continuous soft prompts, and reusable demonstrations. Duan et al.~\cite{duan2023stochasticparrots} introduce two standalone mechanisms. PromptDPSGD trains continuous soft prompts via white-box DP-SGD (per-example clipping and Gaussian noise) and deploys the continuous prompt directly; PromptPATE partitions private data across disjoint teacher prompts to label an unlabeled public dataset via PATE noisy class-label voting, producing discrete demonstrations. Unlike DP-OPT, PromptPATE requires unlabeled public inputs. Differentially private few-shot generation~\cite{tang2024dpfewshot} similarly synthesizes reusable DP exemplars from fixed datasets in a single pass. Migrating these primitives to continual user feedback requires tracking release multiplicity, sample reuse, and intra-user correlation across retraining rounds.

Structural assumptions dictate which conditions require re-verification in a personal agent. A single user's history typically feeds both prompt revision and behavioral evaluation; if an intra-user fact recurs across DP-OPT's training and validation splits, parallel composition fails for that fact and reverts to sequential composition with group scaling. Similarly, DP-ES forfeits minibatch amplification ($q = b/n$) if candidates are scored deterministically over a small recent-turn buffer, and $T$-round iterative feedback composes up to $TK$ RDP evaluations. DP-ES reports a matched task comparison with DP-OPT under a common conservative privacy upper bound~\cite{liu2026dpes}. Its accuracy advantage is larger on GSM8K and MedQA, with close results on BANKING77 and Alpaca. The reported task results and resource costs depend on the benchmark settings and the bound is not an exact budget match. When one user contributes correlated interactions, whole-history protection requires an explicit contribution cap and re-accounting, while intra-user fact protection requires defining an intra-user adjacency unit.

Private selection from private candidates~\cite{liu2019privateselection} targets meta-selection over already-DP candidate oracles when quality scores lack low sensitivity. Randomized stopping releases a winning candidate and conceals rejected draws, bounding privacy loss by a constant factor ($2\varepsilon_1$ or $3\varepsilon_1$) for pure $(\varepsilon_1, 0)$-DP base oracles; approximate $(\varepsilon_1, \delta_1)$-DP oracles ($\delta_1 > 0$) incur a stopping-horizon $\delta$ blow-up ($\propto \delta_1/\gamma$) plus additive $\varepsilon$ overhead. Coded \texttt{T}, this single-pass operator still requires extension to repeated sessions and continual writeback. Effective prompt privacy therefore begins by locating where private inputs enter---generation, scoring, or selection---before fixing the adjacency unit and budget schedule.

Hanke et al. compare locally generated DP prompts or demonstrations with local private weight updates on common tasks. A remote model can still receive the raw inference query after adaptation via a privatized artifact~\cite{hanke2024openprivate}.

\subsection{User-Level Units and Empirical Risk Rewards in Parameter Updates}\label{sec:param-update}

Parameter updates embed private data directly into weights or adapters invoked across unbounded future tasks, creating an observable surface distinct from prompt releases. \emph{Mind the Privacy Unit!}~\cite{chua2024privacyunit} evaluates both Group Privacy (Algorithm~1, capping per-user records under example-level DP-SGD) and User-wise DP-SGD (Algorithms~2--3, per-user gradient clipping and Gaussian noise) for LoRA fine-tuning. Black-box $(\varepsilon, \delta)$ group conversion over $k$ correlated records scales $\varepsilon$ by $k$ and inflates $\delta$ exponentially, whereas per-user clipping and RDP accounting avoid that superlinear $\delta$ blow-up. Still, bounding one participant's aggregate contribution in a multi-user corpus differs structurally from hiding a personal fact inside one user's persistent model. Moreover, random-reshuffling epochs diverge from idealized Poisson subsampling assumed by the accountant, and incremental steps, replay, and cross-task allocation alter accounting premises. Unvalidated for online single-user writeback, this work is coded \texttt{T}.

Private parameter adaptation varies optimization cost and deployment scale. DP zeroth-order fine-tuning~\cite{tang2025dpzo} privatizes loss differences under example adjacency, avoiding per-sample gradient storage. Production Gboard~\cite{xu2023gboard} uses tree-based DP-FTRL, per-device clipping, and participation limits. Slokom and Bellog\'in~\cite{slokom2026federatedrecommendations} deploy clipping and Gaussian perturbation before server aggregation. Their 53-day study measures recommendation use and user control, not privacy outcomes, and does not specify adjacency or continual accounting. We retain it as deployment context without a protection code.

Contribution privacy does not ensure that adapted weights preserve contextual disclosure norms. \emph{Privacy Collapse}~\cite{goel2026privacycollapse} tests held-out tool and memory behavior after offline fine-tuning. Norm-based filtering and conservative data mixing partly mitigate the regression (\S5.4, App.~I; \texttt{T}). The result concerns release behavior, not the confidentiality of the training records.

Earlier private online personalization also separates the shared update from the local policy. P2B~\cite{malekzadeh2020p2b} sends sampled, encoded interaction feedback through a trusted shuffler that suppresses rare context codes. The server uses the filtered batch to train a shared warm-start model. The published analysis uses one-context-vector adjacency for this sharing step. Collaborative bandits~\cite{wang2020collaborativebandits} perturb sufficient statistics under reward-step adjacency and analyze the action transcript. Both are transferable evidence (\texttt{T}). Extending either to a private fact repeated over one user's history requires accounting for all affected interactions and releases.

Privacy-preserving BERT~\cite{qu2021privatebert} applies metric $d_\chi$-privacy ($e^{\eta\,d(x,x')}$) before transmitting token or text representations and recovers utility through privacy-adaptive pretraining. The inverse-distance units of $\eta$ make its budget specific to the embedding metric. Repeated releases can carry semantically and temporally related signals.

Underlying these methods is a modeling asymmetry between cohort training and single-user personalization. Federated user-DP defines adjacency by adding or removing one participant's dataset; in an agent serving one user longitudinally, removing that sole user eliminates the personalization target itself. When the same mechanism admits an empty intermediate history and observes the same outputs, add/remove-user $(\varepsilon,\delta)$-DP implies $(2\varepsilon,(1+e^\varepsilon)\delta)$-DP for replacing that user's history by a remove-then-add argument. A particular fact change is a special case. The cohort result still needs separate checks of its input domain, output scope, and utility in a continuously adapting personal agent.

The isolation boundary also differs from the noise mechanism. PRISP compares private per-user adaptation using a shared task-conditioned anchor, with the learned personal module kept separate\cite{park2026prisp}. Its few-shot LaMP experiments use fixed histories. A provider that receives neither the user history nor the trained module lacks those direct inputs. When a recipient can query the adapted model, extraction and membership tests assess a different boundary. Du et al. compare those outcomes across DP fine-tuning methods and show that the empirical privacy--utility trade-off varies with the adaptation method\cite{du2025privatefinetune}.

Parameter audits must also test sensitive details confined to training inputs rather than target completions. Extraction probing of fine-tuned language models~\cite{szep2026memorization} compares DP, machine unlearning, regularization, and alignment against input-side PII probes, showing that audits must verify extractability of input-only facts, cross-lingual identifiers, and rare personal values.

PrivMeSA~\cite{wang2026privmesa} uses empirical disclosure risk to optimize a shared clinical LoRA with GRPO across 1{,}700 training patients (\S\S3.3--4.1). A separate frozen distiller writes general lessons from completed consultations. During test-time adaptation the policy is fixed and the initially empty lesson store grows across 897 encounters from 200 held-out patients. The exposure metrics concern outbound consultations and registry-based linkage across visits. The training, writeback, and release results all receive \texttt{T}: the evaluated state is shared across patients, and linking visits measures disclosure about a patient. An intra-user parameter claim would require an adapter updated from that user's own history and a test of the interfaces that expose its changes.

Formal DP and empirical risk rewards constrain distinct objects. User-level DP bounds single-participant influence on released parameters while trusting a central trainer and exact clipping, sampling, and accounting; empirical risk rewards steer optimization away from evaluator-detected disclosures. Combining them inside a personal agent requires joint data-dependency analysis across risk-reward updates, lesson distillation, and replay under a shared privacy budget.

PeCL~\cite{zhan2026pecl} adds a task-incremental comparison (\texttt{T}). It perturbs token embeddings and uses sensitivity scores to guide retention as parameters change. Appendix~G calibrates Gaussian noise at a specified scale for a token-level unit. Section~3.2 perturbs only tokens with nonzero sensitivity scores, so a neighboring input can change whether noise is added. In a personal history, one fact may change multiple tokens, contextual features, and the scale or retention decision itself. Transferring the intervention requires an adjacency and a bound that covers these changes and the full set of releases.

A fixed-compute comparison shows how example and user sampling change utility under user adjacency~\cite{charles2025fixedcompute}. PriME uses membership tests on donor adapters in an empirically optimized merge. Its target user and protected donors are distinct~\cite{kim2025prime}.

\subsection{Preference Protection: Complete Tuples, Choice Labels, and Single-Shot Sampling}\label{sec:pref-learning}

Preference alignment must distinguish its protection unit and remain separate from disclosure-preference gates (Section~\ref{sec:memory-maint-release}). Decoupled DP reward modeling~\cite{cho2026decoupledreward} protects the complete comparative tuple \texttt{(prompt, response\_1, response\_2, preference)}: a trusted trainer fits a lightweight DP reward head, and Best-of-$N$ selection scores candidates from a public reference model. Post-processing immunity (Proposition~6) holds only when both the candidate generator $\pi_0$ and the reward head's frozen representation backbone $\phi$ are strictly data-independent and raw tuples are never re-accessed; unseen private user queries and retraining lie outside this static bound (\texttt{T}).

DP reward estimation~\cite{raychowdhury2024privatereward} adopts narrower label-DP (central objective perturbation by a trusted learner or user-side local randomized response), treating prompts and candidate actions as public (\texttt{T}). When sensitive facts reside in prompts or candidate responses, hiding choice labels leaves the primary leakage surface open; conversely, full-tuple mechanisms impose broader trust and noise costs when only choices require protection. Nominal $\varepsilon$ values across label-DP and tuple-DP are therefore incomparable unless protection units match.

PrivDPO~\cite{jiang2026privdpo} encodes label-DP directly into DPO policy updates by adding calibrated randomness along a text-determined 1D preference axis. This boundary assumes public prompts and responses plus an in-distribution reference policy $\pi_{\text{ref}}$ that is data-independent or privately pre-trained (e.g., via PrivSFT); non-private SFT leaves prompt and response memorization exposed. Sparse reward estimation~\cite{ding2026sparsereward} jointly analyzes Sample-DP and randomized-response label-DP in sparse Bradley--Terry--Luce models, characterizing estimation error across granularities while treating prompts and candidates as public.

Recent offline results enlarge the preference comparison. Square $\chi$PO analyzes offline policy fitting under either central sample privacy or local label privacy~\cite{zhou2025squarepo}. A rank-aware release samples a binary top subset under local privacy for a single item relocation in a fixed candidate ranking~\cite{chen2026rankaware}. A private fact can change multiple positions, several lists, or the candidate set itself. Each change requires matching the ranking mechanism's domain and charging the applicable group and repeated releases.

KL-regularized bandits and RLHF~\cite{zhang2025klregularization} distinguish a private action from a private policy. Their pure label-DP result covers one draw from an offline KL-regularized policy under reward, coverage, and boundedness conditions (Theorem~3). Repeated draws require accounting, and the parameters lie outside that bound. Wu et al.~\cite{wu2025onlineprivaterlhf} instead privatize each choice label by randomized response at the client before adaptive reward fitting and policy updates (Section~5, Algorithm~2). Their online result is a regret bound under label LDP. The experiments evaluate the offline algorithm. For a remote learner that must never see raw choices, local privatization protects the label interface even during online learning. Repeated privatizations of a fact-linked label, private contexts and candidates, and other recipient-visible releases require a separate analysis.

DPPrefSyn generates synthetic comparisons from public prompts after private reward fitting~\cite{gao2026dpprefsyn}. PrivAct evaluates contextual disclosure during agent tasks after shared offline preference training~\cite{cheng2026privact}.

\subsection{Boundaries of Confidential Execution and Continual Composition}\label{sec:secure-exec-comp}

\subsubsection{Secure Execution, Federated Sharing, and Permitted Outputs}

Outside differential privacy, SecP-Tuning~\cite{luo2026secptuning} executes secret-shared forward passes of a frozen backbone across two semi-honest, non-colluding online servers while the data owner keeps labels, losses, and CMA-ES soft-prompt updates local (\texttt{T}). Offline preprocessing relies on a trusted third-party dealer $T$ (collusion with either online server breaks secrecy), MPC security covers an approximated fixed-point Random Feature Attention circuit rather than exact Softmax, and forward logit predictions are reconstructed in plaintext to the client across tuning iterations. Its three-A100 LAN/WAN benchmarks quantify cryptographic overhead, leaving allowed-output leakage, user-level adjacency, and consumer-device feasibility unaddressed.

Federated mechanisms separate upload secrecy from output leakage. Personalized federated prompt learning (DP-FPL)~\cite{tran2025dpfpl} decomposes local soft prompts into client-local personalized residuals retained on-device, uploaded local low-rank factors protected by local DP (LDP), and an aggregated global prompt protected by global DP (GDP)---still requiring continual accounting across participation schedules (\texttt{T}). Practical Secure Aggregation~\cite{bonawitz2017secureaggregation} hides individual updates prior to summation, subject to threshold non-collusion ($< t$ corrupted parties), authenticated PKI binding against active Sybil attacks, and cross-round cohort-differencing leakage when participating sets shift by one client; aggregate updates and derived models remain permitted outputs.

The choice of federated prompt mechanism changes what the server learns. SecFPP prevents direct access to prompt vectors during clustering but discloses distances, whose information content the authors analyze under a Gaussian model~\cite{hou2025secfpp}. FedSEPT accounts for noisy prompt updates across communication rounds under its sampler and clipping conditions~\cite{wang2026fedsept}. Both evaluate vision--language settings. For a personal prompt, the comparison turns on whether the server observes a private update, a clustering statistic, or both. A distance based on a private prompt belongs in that observer's view even when the underlying vectors stay hidden.

\begin{figure}[tp]
\centering
\includegraphics[width=\textwidth,height=0.38\textheight,keepaspectratio]{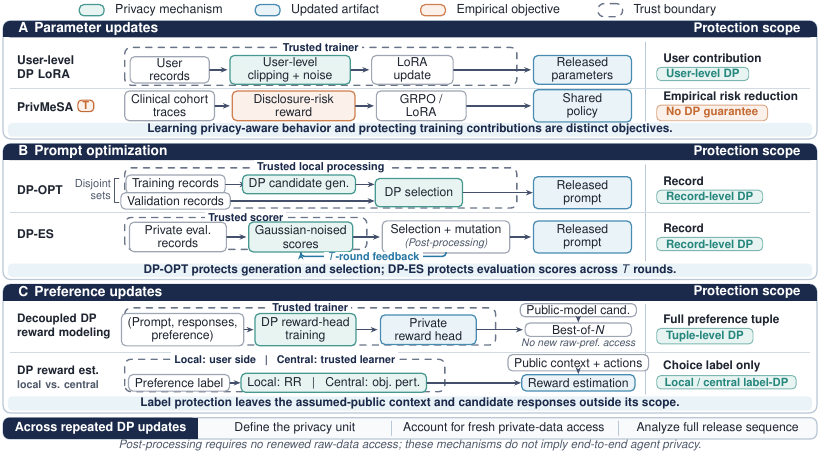}
\caption{Privacy interventions across parameter, prompt, and preference updates. Panel~A contrasts user-level DP LoRA with PrivMeSA's empirical reward for offline shared-policy training; Panel~B contrasts DP-OPT against DP-ES's $T$-round iterative minibatch scoring loop; Panel~C contrasts decoupled tuple-DP reward heads against label-DP estimation. Repeated updates require specifying adjacency units, fresh data access, and release composition.}
\label{fig:2}
\Description{Three-panel horizontal diagram comparing parameter, prompt, and preference update pathways. Panel A (Parameter updates) contrasts User-level DP LoRA, where user records undergo user-level clipping and Gaussian noise inside a trusted trainer to release parameters under user-level DP, against PrivMeSA, which maps cohort clinical traces through a disclosure-risk reward and offline GRPO to a shared LoRA policy, with empirical disclosure-reduction evidence. Panel B (Prompt optimization) contrasts DP-OPT, where disjoint training and validation records inside a trusted local boundary drive DP candidate generation and DP selection to release a prompt under record-level DP, against DP-ES, where private evaluation records inside a trusted scorer produce Gaussian-noised scores feeding selection and mutation post-processing with a T-round feedback loop back to the scorer. Panel C (Preference updates) contrasts Decoupled DP reward modeling, where a trusted trainer produces a private reward head over full tuples for Best-of-N selection with public-model candidates under tuple-level DP, against central or local randomized-response label-DP reward estimation combined with public context and actions. A bottom checklist notes three requirements across repeated DP updates: define the privacy unit, account for fresh private-data access, and analyze the full release sequence.}
\end{figure}

\subsubsection{Continual Release, Adaptive Accounting, and Stopping Conditions}

Continual updates require separating single-mechanism structure from cross-mechanism composition. Lifelong DP~\cite{lai2022lifelongdp} uses objective perturbation and prescribed reuse of fixed, disjoint episodic task datasets or task batches to prove a pure $\varepsilon$-Lifelong-DP bound ($\delta = 0$) across released checkpoints (\texttt{T}). Correlation among distinct task tuples leaves the stated tuple-level guarantee intact. A fact-level change spanning several tuples or task partitions requires separate accounting under that adjacency. Replay outside the protocol can violate its partition or data-access assumptions, and a separately released raw-data summary lies outside the protected checkpoint view.

Concurrent composition~\cite{henzinger2026concurrentcomposition} models adaptively interleaved queries and dataset updates over full transcripts: while approximate DP ($\delta > 0$) forfeits naive parallel composition under adaptive updates, pure $(\varepsilon, 0)$-DP preserves concurrent parallel composition (Corollary~5.5), previewing the pure-DP memory design in Section~\ref{sec:memory-maint-release}. Neither framework confers formal privacy onto empirical gates or skill writebacks. For an explicitly modeled correlated history, inferential privacy offers a different guarantee: compare release distributions conditioned on alternative secret values. Song and Chaudhuri~\cite{song2017timecomposition} give composition results for Markov Quilt under specified time-series conditions. Bai et al.~\cite{bai2026pufferfishcomposition} characterize additional conditions for linear adaptive Pufferfish composition. These results require assumptions about both the mechanism and the history model. Their applicability to feedback-dependent personal-agent traces needs a separate argument.

Recent accountants better match adaptive workflows under strict preconditions. Single-pass tree DP-FTRL~\cite{kairouz2021dpftrl} avoids sampling or shuffling amplification but fails under unprescribed multi-epoch replay without multi-participation matrix or tree accounting. Fully adaptive composition~\cite{whitehouse2023fullyadaptive} selects subsequent mechanisms and budgets from prior outputs, pairing privacy filters (enforcing a pre-committed global $(\varepsilon, \delta)$ cap) with odometers (paying an $O(\sqrt{\log\log(\cdot)})$ time-uniformity penalty over fixed schedules). Continual mean estimation~\cite{george2024continualmean} proves user-level pure DP across multi-user cohorts ($M_t \ll t$); in a single-user agent ($u = 1$, $M_t = t$), whole-user replacement alters 100\% of data and forces $\Omega(1)$ error unless re-instantiated at bounded intra-user event, window, or attribute granularity.

PRISM privatizes client adapter updates, whereas CSLR privatizes replay-list release. Their guarantees concern separate outputs within multi-client learning~\cite{wang2026prism,shihab2026cslr}.

Figure~\ref{fig:2} compares representative mechanisms. Applying their bounds to personal trajectories requires matching each operand to the same history change and tracking all resulting releases.

\section{Memory Writeback, Dynamic Maintenance, Withdrawal, and Read Release}\label{sec:memory}

Long-term memory in a personal agent spans raw interaction logs, user profiles, vector indices, episodic summaries, reflections, and retrieval caches. Ingestion, compression, merging, expiration, retrieval, and subsequent writeback continually alter both state representation and propagation scope. Coverage across these lifecycle stages remains uneven: pre-write isolation, enclave-backed maintenance, and empirical writeback gates supply direct personal-agent update evidence (\texttt{D}), and read-time release admits formal differential privacy (\texttt{S}), while physical erasure, derived-state cleanup, and synchronized parameter withdrawal remain uncertified in long-horizon loops.

Offensive benchmarks confirm that storage isolation and retrieval accuracy do not establish confidentiality. MEXTRA~\cite{wang2025mextra} extracts private interactions from two agent memory architectures under black-box prompting, and SPORE~\cite{gao2026spore} exploits malicious tool responses to exfiltrate user-isolated memories via tool-call arguments and persist trigger payloads across sessions. Memory protection must therefore govern model outputs, tool parameters, cross-session short-term state, and external execution transcripts alongside storage permissions.

\subsection{Writeback Boundaries: Representation Isolation versus Selective Admission}\label{sec:memory-write}

Writeback control balances personal-context utility against persistent retention through two strategies: replacing raw values with isolated surrogates or gating whether an item enters persistent storage. Representation isolation relies on separation between raw values and placeholders, while selective admission relies on sensitivity classification; both require specifying which trusted endpoint reads raw inputs during the decision.

MemPrivacy~\cite{chen2026memprivacy} detects sensitive text spans on a trusted edge device and substitutes typed placeholders prior to cloud memory upload while keeping the recovery map local. SP-Mem~\cite{wang2026spmem} separates sanitized default retrieval views from exact private values, restoring exact values only under task necessity and explicit consent. Both writeback interventions are coded \texttt{D}. The retrieval and restoration path in SP-Mem has a separate scope, with release conditional on consent enforcement. Placeholder replacement still leaves attribute inference from co-occurring placeholders or agent actions open, and SP-Mem's validation relies on synthetic personas and simulated consent. As an adjacent RAG primitive (\texttt{T}), SAGE~\cite{zeng2025sage} replaces private documents with synthetic corpora before indexing, yet requires a trusted synthesizer and re-identification audits over generated text.

HealthJev~\cite{li2026healthjev} tests whether simulated longitudinal health interactions admit restricted attributes, procedures, or events to per-user memory (\texttt{D}). ConMem~\cite{zhang2026conmem} probes the downstream effects of adding a candidate to reconstructed conversations. Its casewise storage judgments quantify over-retention at admission, with repeated writes and confidentiality outside that evaluation (\texttt{T}). The two interventions help distinguish which facts may be written from how a stored fact should be used or disclosed.

Longitudinal evidence changes the comparison between entry and release controls. Kaneko et al.~\cite{kaneko2026disclosureloop} compare anonymizing summaries at write time with instructing a support agent to avoid stating or presupposing attributes unless the user raises them. In their simulated multi-session loop, both interventions lowered measured attribute inference from the accumulated record; the response instruction preserved more coherence. It changes later elicitation and thus affects what reaches memory, while write-time anonymization directly changes what is stored. These are empirical results under the tested user models and memory formats. Dash et al.~\cite{dash2026selfportrait} audit 2,050 memory entries donated by 80 ChatGPT users and separately evaluate pre-ingest query reformulation using a memory-generation proxy. This motivates testing the write boundary on deployed systems and leaves the reformulation's longitudinal effect open. Agent-Memory Protocol~\cite{wu2026amp} provides design context through worked examples of at-rest redaction, purpose-scoped release, and local restoration. These examples specify the interfaces at which retention and disclosure would be tested.

AIM~\cite{johnson2026aim} assigns visibility before writing and enforces per-user scope at a shared vector index (\texttt{T}). Audience-Bound Persistent Memory~\cite{liu2026audiencememory} binds a user's records to their original audiences and assigns an intersection or suppresses the derivative when source audiences differ. Controlled multi-party histories exercise these ingest and derivation operations (\texttt{D}). Context admission has its own authorization evidence (\texttt{S}). The guarantees depend on trusted identity resolution, provenance, and complete mediation. An authorized context may still produce revealing actions, and byte erasure lies outside the claim.

Cross-session linkage gives a second reason to evaluate the read boundary. MemJoin~\cite{yang2026memjoin} follows 288 synthetic cases through six persistent memory systems with a common reader. A sensitive relation may be materialized during writing or inferred only after retrieval joins records. On 287 paired A-Mem cases, masking values in the retrieved context reduced case disclosure from 13.9\% to 0\% and task completion from 20.2\% to 5.2\%. Removing shared cues before the reader reduced disclosure to 5.6\% with 19.2\% completion and increased refusal from 7.2\% to 14.1\% (Table~3). Both interventions act on the read view. These results motivate testing whether admission and write-time masking leave linkable traces across records, alongside whether the read view can preserve task-relevant values while suppressing unnecessary links.

Retention policies act on already-admitted state. MaRS/FiFA couples typed provenance and sensitivity-aware eviction in multi-agent simulations. The study found no significant privacy-score differences across policies despite different utility and cost, and its DP tie-break fired infrequently~\cite{alqithami2025mars}. FSFM measures residual sensitive-record retention in a storage benchmark and leaves recovery through answers and descendants untested~\cite{gu2026fsfm}. Both are \texttt{T} for maintenance. CAMS combines write screening, ongoing re-audit, and extraction interception against a shared test harness (\texttt{T}); its extraction results cannot be attributed to write screening alone~\cite{dhivyasree2026cams}.

HomeLLaMA constrains the update boundary by keeping user profiles and home configuration local (\texttt{D})~\cite{huang2026homellama}. Approved plans and commands become interaction records. A local update inserts or merges profile entries and later retrieves them for action planning (\S IV-C--D). In the 25-day lab study, ten participants provided feedback and rated personalization, and the study tracked optional cloud calls (\S VII-B, Fig.~16). The cloud path receives the current request after obfuscation and user approval. The profile boundary limits direct transmission. The separate fallback experiment measures inference about current commands under its own adversary (\S VI-B.2, \texttt{T}). The local model's offline training uses generic device tasks. The observed longitudinal personalization comes from the evolving local profile path.

\subsection{Dynamic Maintenance and Read Release: Confidential Execution Is Not Update Privacy}\label{sec:memory-maint-release}

\subsubsection{Confidential Maintenance and Oblivious Indexing}

Dynamic maintenance can expose access patterns, update frequencies, and state rollbacks. Opal~\cite{kaviani2026opal} uses a trusted logical client holding a root key and durable monotonic counter $(k,\text{ctr})$ (\texttt{D}). An attested controller authenticates two oblivious RAM (ORAM) trees via Merkle roots and counter-bound MACs. It coordinates trusted embedding and LLM enclaves and performs ingestion, index maintenance, expiration, and summarization within fixed access budgets. Theorem~2.1 bounds host observations for request sequences with the same declared leakage, including public operation types and checkpoint schedules. Intra-enclave side channels and denial of service fall outside that model. The freshness result for sealed state ends at the latest authenticated checkpoint. Expiry reclaims records, while retained summaries have their own lifecycle.

Compass~\cite{zhu2025compass} co-designs HNSW search with a Ring ORAM variant and requires no server-side TEE (\texttt{T}). Opal implements its own Ring ORAM variant with an IVF-PQ index and uses Compass's batched-access ORAM results for trace indistinguishability and authenticated-storage integrity (\S5 and App.~A of Opal~\cite{kaviani2026opal}). Oblix~\cite{mishra2018oblix} uses an SGX enclave with doubly oblivious indexing to address page and cache side channels (\texttt{T}). Each guarantee applies to its storage or search interface. Model and tool disclosures require separate controls.

Post-deletion index confidentiality requires separate specification. Symmetric puncturable encryption~\cite{sun2018backwardsearch} realizes Type-III backward security (\texttt{T}): it hides deleted document identifiers only for keyword--document pairs never queried while active, while still leaking update timestamps and exact \texttt{Delete}-to-\texttt{Insert} cancellation pairs on later searches. Client-cached copies, model contexts, vector summaries, and parameter traces remain outside this storage-interface bound.

\subsubsection{Private Retrieval, Context Release, and Tool Information Flow}

DP-MemView~\cite{kim2026dpmemview} selects from public views by the exponential mechanism, charges all attribute groups intersecting the private read set, and falls back to a generic view when a cap would be exceeded. Under its interface contract and fixed layout, Theorem~1 gives pure $B_a$-DP for changes confined to the indexed group $I_a$. A changed slot outside $I_a$ falls outside that bound. When groups cover all changed slots, Corollary~1 sums the budgets of a minimum-cost cover. An uncovered change receives no finite bound from that corollary. A dynamically rewritten layout needs its own coverage and transcript analysis (\texttt{S} for the evaluated read interface).

Profile-DP~\cite{li2026profiledp} uses a fixed profile and sharing policy (\texttt{T}). DP-SLOT samples fields independently, and DP-BLOCK jointly samples related fields on public support. Theorem~1 and Appendix~A give a pure-DP bound for withheld values with fixed public context and field sets. Reusing a cached sanitized profile with public inputs preserves the bound. Fresh sanitizations compose. MemGov~\cite{christou2026memgov} checks lifecycle status and requester eligibility, then repairs pairwise conflicts before context assembly. Its certificate verifies the supplied constraints on the returned working set (\S\S3--4 and App.~D).

Private RAG designs decouple query, retrieval, and generation endpoints (\texttt{T}). RemoteRAG~\cite{cheng2025remoterag} applies metric \texttt{DistanceDP} ($e^{\eta\,d(\cdot,\cdot)}$) to outbound query embeddings; because $\eta$ carries inverse-distance units scaling distinguishability exponentially with embedding distance and additively across queries, nominal budgets are non-comparable to dimensionless $\varepsilon$-DP. PRAG~\cite{li2026prag} computes homomorphic similarity scores under passive, semi-honest IND-CPA assumptions in non-interactive (\texttt{PRAG-I}) or client-assisted interactive (\texttt{PRAG-II}) modes, leaving active probing, server access patterns, and decrypted generation contexts to separate protection.

Galaxy~\cite{bao2025galaxy} integrates a Persona cognitive tree, a self-evolving Kernel, and a Privacy Gate that applies graded L1--L4 masking before cloud calls and selective restoration upon return. Because its PrivacyLens ablation removes the entire Kernel rather than isolating the Privacy Gate---and never proves that all internal Persona or Kernel updates pass through that gate---its evidence supports empirical release gating (\texttt{S}) rather than protected state evolution (\texttt{D}).

AirGapAgent~\cite{bagdasarian2024airgapagent} separates a trusted minimizer from the third-party-facing agent (\texttt{S}). Its single-turn context-hijacking benchmark on synthetic profiles reports a 97\% protection rate with Gemini Ultra. Contextual integrity studies~\cite{abdi2021privacynorms} identify data type, recipient, purpose, and transmission principle as factors in release decisions. For personalized decisions, PANOLA~\cite{ulusoy2022panola} negotiates sharing through learned bids, and the Uncertainty-Aware Personal Assistant defers under uncertainty~\cite{ayci2023uncertaintypa}. ARIEL~\cite{flemings2026ariel} uses logical entailment over prior user judgments to assess a new request, with an undetermined output when entailed judgments have no majority. Its benchmarks measure decision agreement under non-adversarial requests (\texttt{S}). Evaluating compliance at release and protecting the judgment history during preference updates require distinct tests.

CIMemories varies tasks and recipients to test disclosure appropriateness, and HUSH-Bench distinguishes retrieval of a target from its unsolicited use even when the exact record is absent~\cite{mireshghallah2026cimemories,xu2026hushbench}. MemGate uses a query-conditioned gate before the reader in scripted memory-agent tests~\cite{zhang2026memgate}. In a fixed-pool study, Liu~\cite{liu2026memoryadmission} isolates stricter context admission from presentation of already admitted content. Stricter admission reduces measured cross-domain leakage in one matched configuration. Utility losses and the lack of a robust difference between the two tested renderings limit extrapolation. Even a response that omits sensitive text can expose preferences through the choices it makes. In UMPeek, the evaluated gate compares personalized and non-personalized choices and withholds output as evidence accumulates. Across repeated requests it lowers measured semantic recovery, with residual disclosure and task-score costs (Section~VII, Tables~IV--V)~\cite{li2026umpeek}. The result supports evaluating recipient-visible decisions alongside retrieved evidence and task utility.

Execution-time gates operate outside the model context window. FLOWSEAL~\cite{shim2026controltheflow} enforces lattice provenance with controlled declassification over tool disclosures, and PrivacyFlow~\cite{staab2026privacyflow} weighs purpose, recipient prior knowledge, and inferential consequences before execution to minimize, block, or confirm actions. Securing Agents with Tracked Capabilities~\cite{odersky2026trackedcapabilities} constrains the agent to generating statically typed Scala~3 code in which confidential values use local pure functions and preserve \texttt{Classified} wrappers. Static capability typing blocks explicit plaintext exfiltration over tracked channels under compiler, runtime, and local-model trust. However, it excludes timing and termination covert channels, implicit control-flow branching or declassification (selecting tools from sanitized predicates), and compiler or runtime exception messages returned to an untrusted cloud model across retry turns. All three mechanisms are coded \texttt{T} rather than \texttt{S} because they enforce general per-task tool or program information-flow policies over transient execution inputs rather than gating reads or releases from a persistent longitudinal personal-agent memory store.

Delegation and audit can create additional recipients. Task-Graph CI evaluates propagation gates across several agents; PiSAs measures cross-user visibility as well as inappropriate transfers in shared workflows~\cite{wu2026tgci,gupta2026pisas}. Field-tier minimization limits which raw action parameters enter the authorization and audit views~\cite{oswal2026fieldtier}. These results concern different views and do not bound what the final tool recipient learns.

Client-side draft modification uses retained local profiles and audits inference from query and response. RAG-CT instead tests blocking against PII extraction from fixed retrieval corpora~\cite{salemi2026p3,lyu2026ragct}.

\subsection{Withdrawal and Derived States: From Behavioral Invalidation to Synchronized Unlearning}\label{sec:memory-withdrawal}

Withdrawal mechanisms target distinct outcomes. In Reglos~\cite{yang2026reglos}, the design masks selected identifiers before ingest. A conversational forget request appends a tombstone, and the reader policy instructs the model to avoid the retracted fact. The PersonaMem-v2 outcome observes non-use after the combined intervention. The UI can also delete selected entries. Claims about summaries, backups, embeddings, or weights need separate evidence. Shen et al.~\cite{shen2026revokedmemory} independently test a retrieval guard for revoked organizational policies. Native writeback can leave descendants that continue to affect decisions even when the revoked entry is blocked (Appendix~C.1). This is evidence of behavioral persistence under policy revocation (\texttt{T}), with sensitive-fact erasure outside its scope.

Two benchmarks expose residual risks after source deletion. Deployment-time memorization analysis~\cite{chen2026deploymentmemorization} introduces the Forgetting Residue Score across summarization strength, retrieval breadth, and deletion modes, while MemCut~\cite{zhang2026memcut} separates native retrieval recovery from direct reconstruction out of surviving memory objects. Both confirm that deleting explicit records leaves summaries and dispersed fragments sufficient for secret reconstruction. AfterDeleteBench~\cite{saleh2026afterdeletebench} adds a matched never-exposed control and tests retirement or rebuild of descendants under complete recorded lineage. Its fixed-weight synthetic harness isolates a supported memory path and leaves learned dependencies and incomplete traces open. Memora~\cite{uddin2026memora} detects reliance on superseded facts during later tasks. Its forgetting-aware accuracy measures behavioral use of obsolete state. Bemula~\cite{xie2026bemula} adds no-request controls and measures preservation of unrelated facts separately from target forgetting. AUBench~\cite{lin2026aubenchcontext} expands the outcome to residual knowledge in parameters, memory, and tool-mediated reacquisition.

Synchronized Backflow Unlearning (SBU)~\cite{wang2026agenticunlearning} coordinates memory and parameter cleanup. It blocks retrieval via a target blocklist, prunes exclusive provenance closures over dependency graphs while logically invalidating shared multi-source artifacts, and suppresses parametric residue on $\mathcal{D}_F$ via KL divergence alignment toward a high-entropy reference distribution---a behavioral logit-flattening objective rather than weight-space erasure. In its Store--Query--Delete--Probe medical QA loop, forget-set retrieval hit rate drops from 100\% to 0\%, supplemented by backflow diagnostics and ablations. Because evaluation uses public medical QA without single-user longitudinal histories, shared artifacts remain logically invalidated, multi-agent provenance graphs can miss implicit flows, and retrieval, MIA, and QA scores remain empirical indicators, SBU is coded \texttt{T}.

Secure Forgetting~\cite{ye2026secureforgetting} targets state, trajectory, or environment knowledge and tests target behavior against retained tasks and a never-exposed comparator on public tasks (\texttt{T}). This complements SBU's joint memory and parameter intervention: the comparator can expose residual behavior, while cleanup requires inspecting affected persistent state. External RAG knowledge-base editing~\cite{wang2026ragunlearning} likewise yields behavioral forgetting while leaving base weights intact. Formal baselines define sharper targets (\texttt{T}). Certified data removal~\cite{guo2020certifiedremoval} requires $L_2$-regularized strong convexity, Lipschitz-Hessian smoothness, secret linear objective perturbation at initial training, curator gradient and Hessian access, and bounded cumulative deletions. Crucially, its $(\varepsilon,\delta)$ guarantee bounds only the marginal post-removal distribution $\theta^-$ and leaves the joint pair $(\theta^*, \theta^-)$ exposed if $\theta^*$ was queried without DP. Verifiable machine unlearning~\cite{eisenhofer2025verifiableunlearning} uses initial Proof-of-Training state commitments, SNARKs, and hash chains to prove computational integrity of execution, yet does not establish statistical non-memorization when wrapping approximate rules. Deletion-as-control~\cite{cohen2023deletionascontrol} rejects confidentiality-based deletion in favor of counterfactual post-deletion control and adaptive history-independence simulation conditioned on pre-deletion transcripts. Gupta et al.~\cite{gupta2021adaptiveunlearning} treat requests chosen in response to previously released models and reduce adaptive unlearning to suitable non-adaptive guarantees. Chourasia and Shah~\cite{chourasia2023trueprivacy} show that equivalence to retraining alone can still leave deleted values inferable under such requests. Their post-deletion privacy target covers a sequence of releases. Applying these guarantees across an agent's coupled summaries, indices, skills, backups, and weights remains open.

\begin{figure}[tp]
\centering
\includegraphics[width=\textwidth,height=0.38\textheight,keepaspectratio]{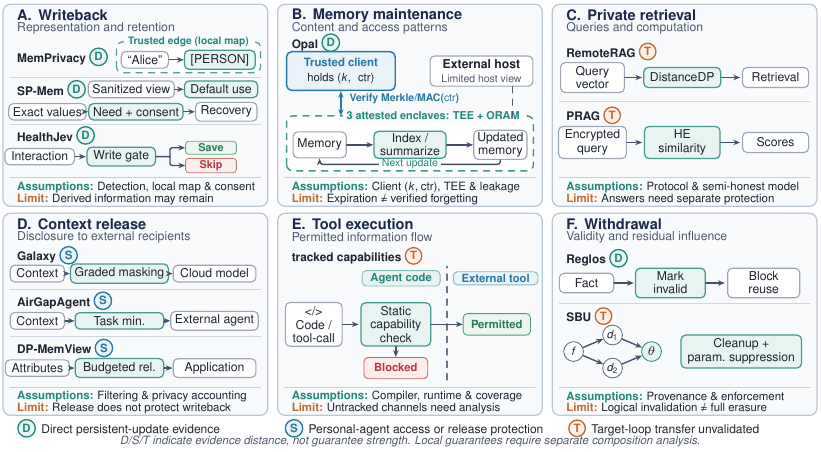}
\caption{Twelve memory and information-flow mechanisms across writeback (A), maintenance (B), private retrieval (C), release (D), tool execution (E), and withdrawal (F). Panel~B separates Opal's trusted client and attested TEE/ORAM updates from the limited host view; Panel~F contrasts Reglos against SBU ($f \to d_1, d_2 \to \theta$). Badges \texttt{D}/\texttt{S}/\texttt{T} indicate evidence distance.}
\label{fig:3}
\Description{Six panels compare mechanisms and their evidence distance. A shows MemPrivacy masking, SP-Mem sanitized views and consent-gated recovery, and HealthJev save-or-skip gating. B separates Opal's limited host view from the trusted client with key and counter, the Merkle/MAC check, and the index and summary update cycle inside the attested TEE/ORAM boundary. C shows RemoteRAG DistanceDP and PRAG encrypted similarity scoring. D compares Galaxy masking, AirGapAgent task minimization, and DP-MemView budgeted release. E shows static capability checking: permitted calls cross the tool boundary, blocked calls stay inside. F contrasts Reglos invalidation with SBU dependency cleanup and parameter suppression.}
\end{figure}

Unlearning also exposes an interface conflict between executing deletion and releasing post-deletion models. Differential leakage analysis~\cite{chen2021unlearningprivacy} shows that querying pre- and post-unlearning versions amplifies membership inference over querying the original model alone, evaluating formal differential privacy separately from empirical output coarseners (label-only release and temperature scaling). Conversely, TAPE~\cite{wang2025tape} uses pre-/post-unlearning posterior differences, shadow models, and reconstructors to audit unlearning without backdoors (\texttt{T}). Coarsening outputs to label-only predictions suppresses the posterior vectors needed by black-box TAPE auditing, requiring withdrawal interfaces to specify version access, query budgets, and output granularity jointly.

EFU~\cite{mohammadi2025efu} addresses request privacy (\texttt{T}). Applying functional encryption to round-level aggregations requires a trusted key-generation authority (KGA) or interactive distributed key generation (DKG) plus at least two non-colluding clients; EFU cryptographically hides pre-aggregation ciphertexts and prevents selective decryption by a compliant server. Hiding learning-versus-unlearning intent in decrypted aggregates, however, relies on heuristic parameter-importance and adversarial-example losses rather than a cryptographic output-indistinguishability proof, leaving intra-agent dependency cleanup uncovered; Xu et al.~\cite{xu2024unlearnsurvey} situate these targets and verification modes within machine unlearning broadly.

TimlMemBench supplies revocation and preservation probes after native ingestion~\cite{zhou2026timlmembench}.

Figure~\ref{fig:3} aligns twelve representative mechanisms across the six memory-lifecycle interfaces. Because observers span cloud stores, untrusted hosts, task models, deletion executors, third-party agents, applications, and external tools---each assuming distinct trusted endpoints---every state transition requires re-specifying adjacency units, permitted outputs, and observable transcripts.

\section{Admission Protection in Skill and Experience Evolution}\label{sec:skill}

Skills convert episodic experience into reusable procedural policies invoked across future tasks. ExpeL~\cite{zhao2024expel} shows how distilling natural-language insights and trajectories drives persistent capability gains, while AntiSkillBench~\cite{xiang2026personaskills} demonstrates that persona skills distilled from private trajectories concentrate sensitive personal facts, enable attribute inference, and facilitate behavioral impersonation. We organize skill-evolution protections across candidate generation, evaluation, persistent admission, and execution, tracing where raw private inputs are read, which artifacts leave the device, and whether admitted skills remain compliant over time.

\subsection{Candidate Generation and Evaluation: The Evaluator Is Also a Recipient}\label{sec:skill-eval}

Skill privacy involves both the input to distillation and the admission of its output. AntiSkillBench~\cite{xiang2026personaskills} tests sanitizing dialogue before distillation and perturbing the synthesized skill (\texttt{T}). CounterPersona~\cite{wang2026counterpersona} intervenes on newly authored records to limit fidelity of unauthorized distillation over a retained archive (\texttt{T}); the original records remain available to the distiller. P2Skill~\cite{ryu2026p2skill} refines reusable cloud-local routing and redaction instructions on benchmark queries (\texttt{T}). Its runtime leakage measurements concern cloud-bound subtasks, while its offline cloud evaluator can see the original refinement prompt. These cases expose different private inputs at the distiller, evaluator, and runtime interfaces.

MaskClaw~\cite{zhao2026maskclaw} provides transferable evidence (\texttt{T}) for candidate skill filtering in five designed scenarios. Section~4.5 specifies a local audit-and-merge update. E3 instead reports fixed correction inputs, text-scorer and sandbox outcomes, and held-out behavior tests with the arbitration store fixed (\S\S5.5, 6.3--6.4, Apps.~E--F). No reported E3 trace links a committed skill version to retrieval and use in a subsequent user interaction. The external multimodal judge is a separate recipient (App.~K, Table~28); the 832-sample static GUI arbitration evaluation does not test persistent updates.

Federated frameworks enforce payload filters at the sharing boundary. FedSkillGit~\cite{lan2026fedskillgit} applies a client-side sanitizer and private-value auditor before committing skill patches or summaries to a version-controlled repository, while FederatedSkill~\cite{yang2026federatedskill} distills local trajectories into compact semantic diffs audited for entity leakage, task reconstruction, and PII canaries before aggregation. Both rely on empirical content filtering for multi-client shared skills and leave a single user's longitudinal library unaddressed, earning code \texttt{T}. Moreover, passing payload audits leaves submission metadata---committer identity, timestamps, and accepted revisions---visible to the coordinating server. Personal admission and federated commits therefore share sanitization primitives under distinct observer and retention models.

Zep Agent Skills documents a trace-to-version pipeline that scans candidates for instance data, limits compilation to the namespace, and retires dependent skills from search after source deletion~\cite{chalef2026zep}. The evaluation runs the admitted snapshot and reports task outcomes. Its privacy controls supply design evidence at the admission and withdrawal boundaries.

SkillLadder evaluates entity and lineage scrubbing before synchronizing skills across simulated users. Covered entity retention reaches zero at its highest scrub level, while source attribution remains above chance~\cite{chen2026skillladder}.

\subsection{Persistent Admission and Execution: Permission Constraints Beyond Content Audits}\label{sec:skill-exec}

Even sanitized skills may harvest fresh personal data or invoke unauthorized tools during execution. SKIPPER~\cite{xie2022skipper} exposes widespread mismatches between declared privacy policies and black-box runtime collection across third-party personal assistant apps, while static tracked capabilities~\cite{odersky2026trackedcapabilities} (Section~\ref{sec:memory-maint-release}) enforce lexical capability permissions and \texttt{Classified} propagation constraints in generated code. SKIPPER supplies risk evidence and tracked capabilities provide an execution primitive, yet neither protects candidate distillation from private failure traces; self-generated personal skills require jointly verifying skill text, declared recipients, and runtime tool invocations.

SkillJack tests both candidate review and runtime monitoring on reused benchmark skills~\cite{ying2026skilljack}. Its exploratory results show that adding a review gate can change which skill is selected and which actions the runtime monitor must handle. Trigger rates are policy-violation proxies rather than disclosure rates, and no private longitudinal user state is protected in the experiment (\texttt{T}).

SkillGuard tests runtime permission mediation on reusable benchmark skills~\cite{pan2026skillguard}. Its effects depend on manifest coverage and the agent scaffold, and the experiments record both blocked violations and task completion. SkillLifeBench separates the selected version from the implementation actually invoked and links that evidence to recovery~\cite{zhang2026skilllife}. Its paired recovery experiments show improved joint task and authorization outcomes when the diagnoser sees execution provenance. Both supply \texttt{T} evidence for verifying reuse. They motivate checking the loaded implementation as well as the version stored in the registry.

Privacy rules and skill contracts also evolve longitudinally. PrivAgentFlow~\cite{tang2026privagentflow} updates workflow privacy policies from historical failures (\texttt{T}), Skill-V~\cite{ma2026skillv} pairs versioned skill libraries with falsifiable contracts and replay-based non-regression checks, and experience-faithfulness evaluations~\cite{zhao2026faithfulness} verify whether stored experiences causally govern downstream decisions or are bypassed at inference time. Integrating these mechanisms requires logging provenance links and admission rationales per revision, enforcing permission invariants at invocation, and re-testing affected paths after fact withdrawal or rollback (Section~\ref{sec:memory-withdrawal}). For the evaluated admission gate, runtime permission enforcement and post-withdrawal skill invalidation remain separate outcomes.

\section{Cross-Carrier Comparison of Privacy Boundaries}\label{sec:synthesis}

\subsection{Matching a Fact, an Observer, and a Claim}\label{sec:synth-triad}

Follow one private constraint---evening appointments are unsuitable---through booking, learning, and withdrawal. It is present in $H_1$ and absent in $H_0$. In each scenario, fix public tasks, options, and permissions; map the change to inputs and derived artifacts; and specify the observer's view. These are worked comparisons, not runs of an integrated agent. Measurements come from component studies or Supplementary Section~S4.

Table~\ref{tab:4} marks results as \emph{transferable} when the input and view match, \emph{out of scope} when they do not, and \emph{insufficient} when the match is unverified. This complements authority and recovery checks: even an authorized action can distinguish private histories~\cite{chen2026safetysurvey,ding2026alwayson,lin2026memorysecuritysurvey}.

\paragraph{Sensitivity to the evidence rule.}
The bounded primary-source rescreen (Supplementary Section~S6.5) retained nine scoped anchors and upheld three boundary exclusions. Requiring measured outcomes after state carryover leaves just two direct comparators in this set: reader non-use and tracked-memory withdrawal. Booking and preference guarantees remain conditional transfers, and skill admission remains a proxy. No checked source evaluates the linked lifecycle view. The rescreen does not reconstruct Stage~2 or estimate coverage beyond these records.

\begin{table}[t]
\caption{Transfer decisions for one private fact. Each decision is conditional on the named observer and claim.}
\label{tab:4}
\footnotesize
\setlength{\tabcolsep}{3.5pt}
\renewcommand{\arraystretch}{1.06}
\begin{tabular}{@{}>{\raggedright\arraybackslash}p{0.15\linewidth}>{\raggedright\arraybackslash}p{0.25\linewidth}>{\raggedright\arraybackslash}p{0.25\linewidth}>{\raggedright\arraybackslash}p{0.29\linewidth}@{}}
\toprule
\textbf{Scenario} & \textbf{Transferable} & \textbf{Out of scope} & \textbf{Insufficient evidence} \\
\midrule
Book: prompt + memory &
Provider, app, and service: bounded prompt revision and fixed-layout reads compose for the declared view. &
Service: raw-dependent slot selection. Host: an extra visible ingest changes public leakage. &
Utility of private booking with all three slots available; linked host privacy after autonomous writes. \\
\addlinespace
Learn: preference head + skill &
Head recipient: label or tuple DP for the matched change. Skill and invocation inherit only as post-processing. &
Provider: raw examples at training. Judge: raw-derived skill text outside the head's bound. &
Skill privacy after commitment and later invocation; permission checks alone do not measure it. \\
\addlinespace
Withdraw: memory + learned state &
Reader: scoped non-use test. Recovery auditor: complete-lineage cleanup as a comparator for tracked memory. &
Version auditor: a marginal removal certificate does not cover paired model outputs. Service: prior receipts remain. &
Joint withdrawal from restored state; privacy of the linked transcript in Section~\ref{sec:synth-linked}. \\
\bottomrule
\end{tabular}
\end{table}

\subsection{Scenario 1: Book an Appointment}\label{sec:synth-compose}

\paragraph{Transfer to linked recipients.}
$H_1$ adds one training and one validation example and changes a preallocated memory value for attribute $a$. The linked observer sees the selected prompt, app transcript, and booking exchange; raw inputs, candidates, and scores stay local. The public calendar has two daytime slots and one evening slot. Stable voting groups with full sampling in a DP-OPT variant limit each fixed-prefix change to one replaced ballot~\cite{hong2024dpopt}. Calibrating for a removal followed by an insertion gives per-call $(\varepsilon_0,\delta_0)$ (Supplementary Section~S6.1). Count all $K$ allowed private calls, including retries and rejections. Conditional on the candidate list, validation-count selection costs $\varepsilon_v$; conditional on the selected prompt, DP-MemView gives pure $B_a$-DP for reads on the common scaffold~\cite{kim2026dpmemview}. The same fact reaches both prompt stages, so sequential composition gives
\[
  (K\varepsilon_0+\varepsilon_v+B_a,\ K\delta_0).
\]
This bound extends to requests, responses, timing, and errors generated only from those protected outputs and public inputs. The worked policy chooses between the two daytime slots in both histories.

\subsection{Scenario 1 Continued: Change the Action or Observer}\label{sec:synth-failure}
If no valid request is for evening in $H_1$, DP implies $\Pr[\text{evening}\mid H_0]\leq\delta$. A raw-data controller that always requests evening in $H_0$ therefore needs $\delta\geq1$~\cite{fu2026privaterecs}. This is an analytical limit on usable slots, not a measured rate. An extra host-visible ingest only in $H_1$ likewise forces $\delta\geq1$ for a linked host view. Opal compares traces with equal declared leakage. Matching operation counts removes one signal; the full host view still needs a conditional simulation or compatible separate bound~\cite{kaviani2026opal}.

\paragraph{Measured proxy trade-off.}
On synthetic PairedMem with the same Qwen2.5-7B-Instruct backbone and 16 fixed queries, DP-MemView Table~2 reports (transcript AUC, target-required utility) of $(0.832,0.524)$ for TaskMin, $(0.500,0.386)$ for TypedMask, and $(0.490,0.506)$ for online DP-MemView ($B_a=2$). TypedMask gives chance-level AUC by construction. Relative to TaskMin, the online interface lowers the auditor's measured distinguishability with a smaller utility loss than typed masking~\cite{kim2026dpmemview}. This is a fixed memory-release comparison, not a booking test. Benchmarks that detect leakage missed by individual tool or answer checks motivate recording the complete linked trace, alongside valid bookings, lost evening options, and refusals~\cite{qiao2026topbench,huang2026cipl,elyagoubi2026agentleak}.

\subsection{Scenario 2: Learn from Feedback and Reuse a Skill}\label{sec:synth-regression}

\paragraph{From a memory fact to a preference head.}
With the same two offered plans, a flipped preference fits label adjacency; changed candidate text requires tuple accounting, including all affected tuples. Reranking inherits the bound only if the generator and online inputs use no unaccounted history~\cite{raychowdhury2024privatereward,cho2026decoupledreward}. Neither local adapter isolation nor cohort DP for released weights covers raw examples supplied to a training provider~\cite{tan2024oppu,chua2024privacyunit}. Section~\ref{sec:synth-diagnostics} tests the utility cost of protecting more labels.

\paragraph{From feedback to an executable skill.}
Skill content and actions inherit the private head's bound only if derived from that head and public inputs. Raw booking feedback can instead reach the judge or shape a service-visible action. MaskClaw's strict text and sandbox pass rates are $43.39\%$ and $16.93\%$ over 189 calendar candidates (Appendix~F, Table~20): gate selectivity, not disclosure or live task completion. SkillLifeBench checks which version actually runs; SkillGuard tests permissions and completion. Neither tests privacy across review, activation, and recipient release~\cite{zhao2026maskclaw,zhang2026skilllife,pan2026skillguard}.

\subsection{Scenario 3: Withdraw the Fact and Recover State}\label{sec:synth-withdrawal}

Withdraw the fact locally in $H_1$ and keep never-exposed $H_0$ as the reference, with public tasks and recovery schedule fixed. Reglos tests reader non-use after a tombstone; MemGov certifies constraints on its returned working set. Neither verifies cleanup of other carriers~\cite{yang2026reglos,christou2026memgov}.

\paragraph{Tracked memory versus new writes.}
In AfterDeleteBench Table~6 (fixed Qwen weights, 200 IID paired profiles), exact source deletion yields reconstruction, new commitment, and later activation rates of $0.735/0.735/0.510$. Adding a sealed literal write gate yields $0.735/0/0.680$; recorded-lineage closure yields zero observed for all three, as does the clean reference. The gate stops the measured new commitment but leaves other routes to later activation. The closure result transfers as a comparator when lineage is complete in a fixed-weight memory harness, not as evidence that prompts, adapters, or skills were repaired~\cite{saleh2026afterdeletebench}.

\begin{figure}[htbp]
\centering
\includegraphics[width=\textwidth,height=0.37\textheight,keepaspectratio]{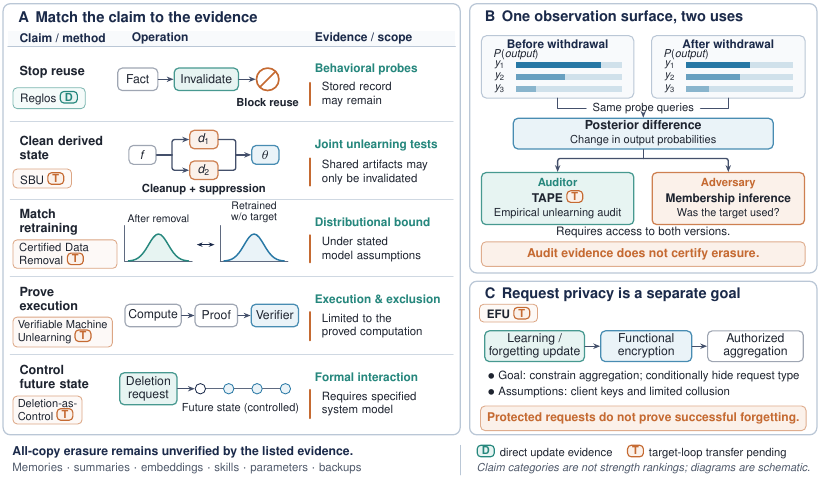}
\caption{Withdrawal claims, verification evidence, and privacy risks. \textbf{A} maps five claims---stopping reuse, cleaning derived state (derived state nodes $d_1, d_2$), matching retraining, proving execution, and controlling interaction---to mechanisms and limits. \textbf{B} contrasts dual-use posteriors across TAPE auditing~\cite{wang2025tape} and version-diff membership inference~\cite{chen2021unlearningprivacy}. \textbf{C} depicts EFU~\cite{mohammadi2025efu} hiding request types without certifying erasure.}
\label{fig:4}
\Description{Three-panel vector diagram. Panel A maps five claims across Reglos [D], SBU [T] (derived state nodes d1, d2 from fact f to theta), Certified Data Removal [T], Verifiable Machine Unlearning [T], and Deletion-as-Control [T]. Panel B compares pre- and post-withdrawal posteriors for an Auditor (TAPE [T]) versus an Adversary (membership inference [Chen et al.]), separating empirical audits from deletion certificates. Panel C shows EFU [T] functional encryption under KGA/DKG assumptions; footer notes unverified all-copy erasure.}
\end{figure}

\paragraph{Why the result does not cover other observers.}
K-Bench finds that answer-only suppression can leave recovery unchanged through tool observations. SBU tests joint memory and weight suppression on public medical QA, not personal recovery~\cite{yu2026kbench,wang2026agenticunlearning}. Probe the restored memory, prompt, weights, and loaded skill against the never-exposed reference. Table~\ref{tab:5} and Figure~\ref{fig:4} distinguish verification targets; Section~\ref{sec:synth-diagnostics} tests cleanup costs and missing lineage.

\begin{table}[htbp]
\centering
\caption{Withdrawal Claims, Verification Evidence, and Observable Boundaries}\label{tab:5}
\footnotesize
\setlength{\tabcolsep}{3.5pt}
\renewcommand{\arraystretch}{1.06}
\begin{tabular}{@{}>{\raggedright\arraybackslash}p{0.18\linewidth}>{\raggedright\arraybackslash}p{0.23\linewidth}>{\raggedright\arraybackslash}p{0.26\linewidth}>{\raggedright\arraybackslash}p{0.27\linewidth}@{}}
\toprule
\textbf{Mechanism or Study} & \textbf{Claim \& Evidence Type} & \textbf{Execution Preconditions} & \textbf{Evidence Boundary} \\
\midrule
Reglos~\cite{yang2026reglos} & Stop future reuse. & Tombstone and reader enforcement. & Reader non-use does not verify stored copies. \\
\addlinespace[2pt]
AfterDeleteBench~\cite{saleh2026afterdeletebench} & Retire or rebuild tracked descendants. & Complete lineage and fixed weights in a synthetic loop. & Measures recovery, commitment, and activation, not all copies. \\
\addlinespace[2pt]
SBU~\cite{wang2026agenticunlearning} & Joint closure cleanup and logit flattening. & Traceable records, vectors, and exclusive derivatives. & Public medical QA; no personal-loop erasure test. \\
\addlinespace[2pt]
TAPE~\cite{wang2025tape} & Empirical audit. & Auditor sees both versions' posteriors. & No certificate; version-diff leakage possible~\cite{chen2021unlearningprivacy}. \\
\addlinespace[2pt]
EFU~\cite{mohammadi2025efu} & Authorized aggregation and request-type masking. & Trusted KGA/DKG and bounded collusion. & Aggregate masking is heuristic; no forgetting proof. \\
\addlinespace[2pt]
Certified Data Removal~\cite{guo2020certifiedremoval} & Distributional closeness to a retrained model. & Initial secret objective perturbation, $L_2$ strong convexity, smooth Hessian, bounded removals. & Bounds marginal $\theta^-$ (not joint $(\theta^*, \theta^-)$); memories, skills, backups uncovered. \\
\addlinespace[2pt]
Verifiable Machine Unlearning~\cite{eisenhofer2025verifiableunlearning} & Proof of prescribed removal and exclusion. & Proof-of-Training commitments and SNARK/hash-chain checks. & Execution integrity, not non-memorization or erasure of copies. \\
\addlinespace[2pt]
Deletion-as-Control~\cite{cohen2023deletionascontrol} & Counterfactual post-deletion interaction control. & Transitions and deletion handlers satisfy adaptive simulation model. & Bounds formal transitions; concrete physical erasure is separate. \\
\bottomrule
\end{tabular}
\end{table}

\subsection{A Linked View Across Booking, Skill Reuse, and Withdrawal}\label{sec:synth-linked}\label{sec:synth-cost}

One observer links three stages by appointment identifier: booking outputs $L_b$ (prompt, app transcript, receipt), skill reuse $L_s$ (judge exchange, invoked version, service request), and withdrawal $L_w$ (notices, restored outputs, tool exchanges). At a fixed checkpoint, withdrawal in $H_1$ and a no-op in $H_0$ stay local, including their branch and timing.

\paragraph{Inherit or account again.}
Charge $L_b$ as in Section~\ref{sec:synth-compose}. Reuse is joint post-processing only if it uses accounted outputs and public inputs. A fresh private fit, read outside the existing cap, or release from an unaccounted checkpoint needs its own bound. For a bounded list of new releases with fixed bounds $(\varepsilon_j,\delta_j)$ for the same history pair at every fixed prefix, including accounted intermediates, the joint bound is $(K\varepsilon_0+\varepsilon_v+B_a+\sum_j\varepsilon_j,\ K\delta_0+\sum_j\delta_j)$. Raw-derived judge inputs and fact-dependent notices have no inherited bound; a notice certain in $H_1$ and absent in $H_0$ forces $\delta\geq1$.

\paragraph{What withdrawal cannot undo.}
Local repair cannot recover earlier receipts or judge copies or reset the ledger. A marginal certificate does not cover paired versions, and execution-state replay assumes no committed effects in the affected suffix~\cite{guo2020certifiedremoval,chen2021unlearningprivacy,yao2026executionunlearning}. This is an analytical trace. Compare utility and cost on matched tasks and privacy units~\cite{hanke2024openprivate,salemi2025ragpeft,das2026finetuning,ma2024privatepeft}.

\subsection{Diagnostic Experiments: Protection Units and Withdrawal}\label{sec:synth-diagnostics}

\begingroup
\setlength{\intextsep}{7pt plus 2pt minus 2pt}
These component diagnostics test Table~\ref{tab:4}, not the full booking trace. Supplementary Section~S4 reports protocols and full results.

\begin{figure}[htbp]
\centering
\includegraphics[width=\linewidth,height=0.47\textheight,keepaspectratio]{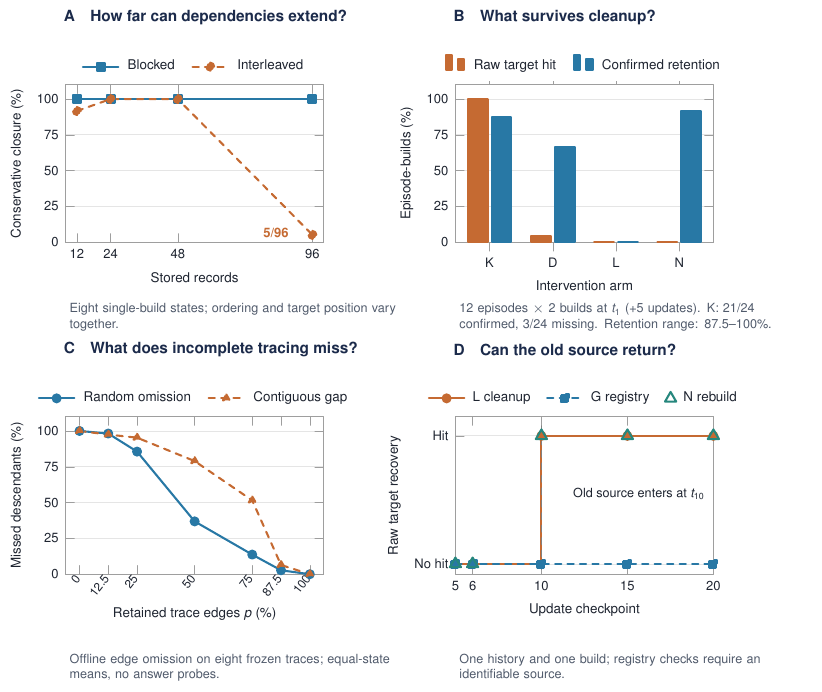}
\caption{Withdrawal diagnostics. \textbf{A}: closure share in eight single-build states. \textbf{B}: raw hits and confirmed retention after five writes (24 paired builds; three keep-arm retention readings missing). \textbf{C}: offline missed descendants. \textbf{D}: exact-source reintroduction at $t_{10}$ in one history.}
\label{fig:5}
\Description{Four panels show closure size, the trade-off between target recovery and non-target retention, missed descendants under two trace-omission patterns, and raw marker recovery after reintroducing a deleted source.}
\end{figure}

\paragraph{Cost of a wider label unit (Stage A).}
We fit a logistic head on 24 synthetic profiles and 24 MovieLens~100K users with fixed item and feature slots. At $\varepsilon=16$, widening protection from one to 16 labels lowers mean balanced accuracy from $79.1\%$ to $57.2\%$ on synthetic profiles and $55.6\%$ to $50.2\%$ on MovieLens (clipped non-private baseline: $56.3\%$; Figure~\ref{fig:6}). This does not test item-history privacy or skill execution.

\begin{figure}[htbp]
\centering
\includegraphics[width=\linewidth,height=0.28\textheight,keepaspectratio]{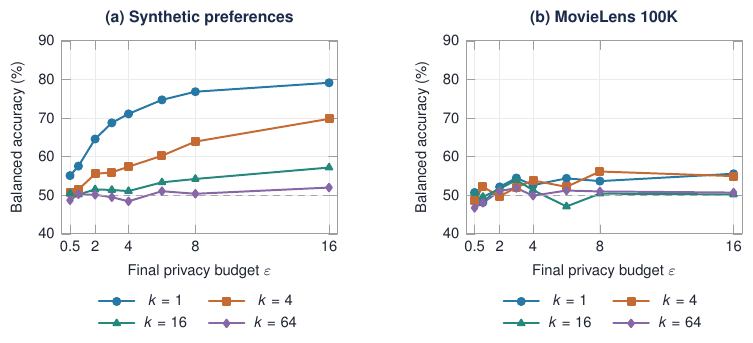}
\caption{Stage A: mean balanced accuracy versus $\varepsilon$ after 16 updates, with $\delta=10^{-6}$ and up to $k$ changed labels in 64 fixed slots. \textbf{A}: synthetic profiles. \textbf{B}: MovieLens 100K.}
\label{fig:6}
\Description{Two line plots compare balanced accuracy across privacy budgets for one, four, sixteen, or sixty-four protected labels in synthetic and MovieLens settings.}
\end{figure}

\paragraph{Deletion scope versus retention.}
After five target-free writes, source-only deletion yields $1/24$ raw marker hits and $16/24$ confirmed non-target recoveries in A-MEM. Both are $0/24$ after closure deletes all five initial records, versus $0/24$ and $22/24$ for never-exposed rebuild (Figure~\ref{fig:5}B). In Stage~B, closure and equal-count random deletion both erase the initial store (Figure~\ref{fig:7}). To isolate targeting, compare equal deletion sizes without saturation and test candidate reach separately~\cite{zhang2026memcut}.

\begin{figure}[htbp]
\centering
\includegraphics[width=\linewidth,height=0.19\textheight,keepaspectratio]{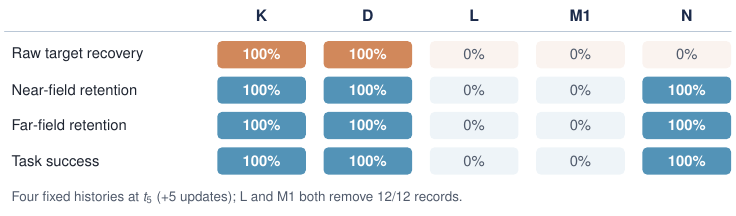}
\caption{Stage B: raw target hits and utility after five updates, averaged over four histories. K: keep; D: root-only deletion; L: closure; M1: equal-count random deletion; N: never-exposed rebuild. L and M1 both remove all 12 initial records.}
\label{fig:7}
\Description{Heatmap comparing raw target hits and near-field, far-field, and task retention. Closure and equal-count random deletion both yield zero raw hits and zero retention.}
\end{figure}

\paragraph{Incomplete lineage (Stage C).}
Across eight single-build states, closure spans $5/96$ records to the full store; write order and target position co-vary. At half trace retention, contiguous gaps miss $79.0\%$ of recorded descendants versus $36.9\%$ under random omission (Figure~\ref{fig:5}A,C). These are reachability losses, not answer leakage.
\par
\endgroup

\section{Research Gaps and Design Agenda}\label{sec:agenda}

\subsection{Provenance-Annotated State Lifecycle Models}\label{sec:agenda-lifecycle}

Lifecycle surveys organize mutation, provenance, and recovery~\cite{lin2026memorysecuritysurvey,ding2026alwayson}; graph rewrites and retention logs model evolving state~\cite{xu2026dynamicgraphcontext,alqithami2025mars}. To connect a history change to its joint observations, annotate edges with which inputs vary together, where they propagate, and which endpoints and outputs are exposed.

\subsection{Verifiable Withdrawal Across Derived States}\label{sec:agenda-withdrawal}

Trace summaries, vectors, skills, and weights, including shared artifacts, through rebuilds and rollbacks. Residue, lineage, and non-use tests distinguish which descendants survive from which affect later decisions~\cite{zhang2026memcut,chen2026deploymentmemorization,saleh2026afterdeletebench,uddin2026memora}. For learned state, specify which observations can influence deletion requests and which version releases the guarantee covers~\cite{gupta2021adaptiveunlearning,chourasia2023trueprivacy}. Empirical probes can expose residue but do not replace the proofs in Table~\ref{tab:5}.

\subsection{Intra-User Protection Units and Continual Privacy Accounting}\label{sec:agenda-accounting}

Map each intra-user fact change to the records and tuples it affects. Adaptive composition and continual mean estimation offer budget primitives, but the cohort participation model must be remapped to a single history~\cite{whitehouse2023fullyadaptive,george2024continualmean}. Group DP bounds a specified set of edits; inferential privacy instead models alternative secrets and correlated histories, with separate composition conditions~\cite{kifer2014pufferfish,song2017timecomposition,bai2026pufferfishcomposition}. The linked view (Section~\ref{sec:synth-linked}) motivates accounting across releases, replay, concurrency, and checkpoints. Renewed consent does not reset past privacy loss.

\subsection{Constrained Evolution of Privacy Mechanisms}\label{sec:agenda-meta}

Feedback-driven updates to skills, harnesses, and policies introduce evaluator drift and exposure to external judges~\cite{zhao2026maskclaw,yang2026reglos,tang2026privagentflow}. Combine immutable safety invariants and budget gates with versioned contracts, replay, and causal probes of whether the stored policy actually governs behavior~\cite{ma2026skillv,zhao2026faithfulness}.

\subsection{Lifecycle Evaluation and User-Visible Control}\label{sec:agenda-eval}

Carry actual state forward through consolidation, withdrawal, recovery, and re-ingestion. Combine answer, tool, and skill probes with tests of recipient appropriateness, cross-carrier reacquisition, and retained utility~\cite{shao2024privacylens,wang2025mextra,gao2026spore,xiang2026personaskills,mireshghallah2026cimemories,xu2026hushbench,xie2026bemula,lin2026aubenchcontext}. Donated memories inform sensitive-field selection~\cite{dash2026selfportrait}; vary linking cues and action-based exposure~\cite{yang2026memjoin,li2026umpeek}. Record consent, service-visible actions, completion, and refusal at each checkpoint~\cite{liu2026sovereignpa,guo2026agentciscope,goel2026agentcibench}.

\begin{table}[htbp]
\centering
\caption{Core Research Gaps, Minimum Mechanism Requirements, and Minimum Verification Evidence}\label{tab:8}
\footnotesize
\setlength{\tabcolsep}{3.5pt}
\renewcommand{\arraystretch}{1.06}
\begin{tabular}{@{}>{\raggedright\arraybackslash}p{0.25\linewidth}>{\raggedright\arraybackslash}p{0.36\linewidth}>{\raggedright\arraybackslash}p{0.34\linewidth}@{}}
\toprule
\textbf{Research Gap} & \textbf{Minimum Mechanism Requirements} & \textbf{Minimum Verification Evidence} \\
\midrule
Continual intra-user privacy (Section~\ref{sec:param-prompt-pref}, Table~\ref{tab:4}) & Define fact/attribute/window adjacency, shared sources, and joint transcript. & Report multi-round composition, exhaustion fallback, and replay accounting. \\
\addlinespace[2pt]
Derived-state withdrawal (Section~\ref{sec:memory-withdrawal}, Table~\ref{tab:5}) & Maintain provenance graphs, dependency closures, shared policies, and weight sync. & Test write--query--delete--probe--recovery--re-ingestion across carriers. \\
\addlinespace[2pt]
Certified deletion (Table~\ref{tab:5}) & Separate logical invalidation, physical erasure, and formal non-reconstructability. & Specify adversary access, certificate semantics, and post-backup/rollback checks. \\
\addlinespace[2pt]
Privacy-mechanism evolution (Section~\ref{sec:skill}) & Enforce immutable invariants, budget gates, regression suites, and rollback admission. & Report pre-/post-mutation surfaces, invariants, and causal failure cases. \\
\addlinespace[2pt]
Personal-device deployment (Section~\ref{sec:synth-cost}) & Decompose local compute, cloud network, and crash-recovery overhead. & Report per-update cost, memory footprint, traffic, latency, and energy draw. \\
\bottomrule
\end{tabular}
\end{table}

Finally, user control requires an inspectable state machine letting users query derived artifacts, track withdrawal propagation, and inspect shared-artifact or retention bounds (Table~\ref{tab:8}).

\section{Limitations of the Review}\label{sec:limitations}

The evidence distribution describes the screened corpus through 10 November 2026. The search windows provide different coverage across carriers and observer interfaces; zero direct studies in a carrier cell does not establish absence in the literature. The historical Stage~2 exclusion trail is incomplete. Agreement counts cover selected recoded cases, not the full corpus. Direct observations come from controlled memory workflows, with source-specific assumptions about state carryover, tracing, and the measured outcome.

The source studies span formal analyses, fixed benchmarks, simulated histories, and prototypes. HomeLLaMA adds a controlled study with real participants and an architectural bound on direct profile transmission. Section~\ref{sec:synthesis} compares worked scenarios, not an integrated implementation. The diagnostic experiments use fixed-slot heads, synthetic memory histories, or offline graphs; they do not estimate end-to-end privacy or utility after deployment.

\section{Conclusion}\label{sec:conclusion}

To transfer a privacy result, fix the history change, trace its derived artifacts, and account for the observer's full view. The booking case exposes lost action options; the feedback case separates reuse of an accounted head from raw-data distillation; withdrawal separates reader non-use from cleanup and prior disclosure. The next test should measure these trade-offs on one linked transcript spanning a fresh update, skill reuse, and withdrawal.

\section*{Data and Reproducibility Availability}

The Supplementary Material PDF (Sections~S1--S6) gives the search, coding, diagnostic, and comparison protocols. The anonymized archive includes search and screening ledgers through Stage~15, frozen diagnostic results, and CPU-runnable integrity checks. Supplementary Section~S4.5 specifies the separate setup needed for a model rerun.

\section*{Ethics Statement}

This review synthesizes publicly available scholarly literature and runs bounded diagnostics on synthetic markers and public MovieLens~100K data without redistributing raw ratings. No human participants, user studies, or animal experiments were conducted, and no new personally identifiable information was collected. Ethics exemption applicability will be confirmed under venue policies prior to publication.

\section*{Conflict of Interest}

The authors declare no competing financial or non-financial interests related to this work.

% =============================================================================
% CAMERA-READY TEMPLATE (Uncomment and populate upon acceptance)
% =============================================================================
% \begin{acks}
% This work was supported by [Funding Agency / Grant Number, or state: ``This
% research received no specific grant from any funding agency in the public,
% commercial, or not-for-profit sectors.'']. The authors thank the anonymous
% reviewers and associate editor for their constructive feedback.
% \end{acks}

% =============================================================================
% CAMERA-READY CRediT TEMPLATE (Uncomment and populate upon acceptance)
% =============================================================================
% \section*{Author Contributions}
% [Author Initials]: Conceptualization, Methodology, Investigation, Data
% Curation, Validation, Formal Analysis, Visualization, Writing -- Original
% Draft, Writing -- Review \& Editing, Supervision.

\section*{Generative AI Usage Disclosure}

Assistive generative AI tools were used for literature discovery, source rechecks, independent recoding of selected operations, evidence-ledger preparation, drafting, and editing. The authors take responsibility for the scope, screening, coding, diagnostic protocols, and manuscript accuracy. Figures are vector diagrams compiled from TikZ/PGFPlots sources and frozen experimental ledgers.

\bibliographystyle{ACM-Reference-Format}
\bibliography{references}


\begin{thebibliography}{181}

%%% the Web address.

\ifx \showCODEN    \undefined \def \showCODEN     #1{\unskip}     \fi
\ifx \showISBNx    \undefined \def \showISBNx     #1{\unskip}     \fi
\ifx \showISBNxiii \undefined \def \showISBNxiii  #1{\unskip}     \fi
\ifx \showISSN     \undefined \def \showISSN      #1{\unskip}     \fi
\ifx \showLCCN     \undefined \def \showLCCN      #1{\unskip}     \fi
\ifx \shownote     \undefined \def \shownote      #1{#1}          \fi
\ifx \showarticletitle \undefined \def \showarticletitle #1{#1}   \fi
\ifx \showURL      \undefined \def \showURL       {\relax}        \fi
% invisible to TeX
\providecommand\bibfield[2]{#2}
\providecommand\bibinfo[2]{#2}
\providecommand\natexlab[1]{#1}
\providecommand\showeprint[2][]{arXiv:#2}

\bibitem[Abdi et~al\mbox{.}(2021)]%
        {abdi2021privacynorms}
\bibfield{author}{\bibinfo{person}{Noura Abdi}, \bibinfo{person}{Xiao Zhan},
  \bibinfo{person}{Kopo~M. Ramokapane}, {and} \bibinfo{person}{Jose Such}.}
  \bibinfo{year}{2021}\natexlab{}.
\newblock \showarticletitle{Privacy Norms for Smart Home Personal Assistants}.
  In \bibinfo{booktitle}{\emph{Proceedings of the 2021 {CHI} Conference on
  Human Factors in Computing Systems}} (Yokohama, Japan).
  \bibinfo{publisher}{Association for Computing Machinery},
  \bibinfo{address}{New York, NY, USA}, Article \bibinfo{articleno}{558},
  \bibinfo{numpages}{14}~pages.
\newblock
\href{https://doi.org/10.1145/3411764.3445122}{doi:\nolinkurl{10.1145/3411764.3445122}}


\bibitem[Alqithami(2025)]%
        {alqithami2025mars}
\bibfield{author}{\bibinfo{person}{Saad Alqithami}.}
  \bibinfo{year}{2025}\natexlab{}.
\newblock \bibinfo{title}{Forgetful but Faithful: A Cognitive Memory
  Architecture and Benchmark for Privacy-Aware Generative Agents}.
\newblock \bibinfo{howpublished}{arXiv preprint arXiv:2512.12856v1}.
\newblock
\urldef\tempurl%
\url{https://arxiv.org/abs/2512.12856v1}
\showURL{%
\tempurl}


\bibitem[Ayci et~al\mbox{.}(2023)]%
        {ayci2023uncertaintypa}
\bibfield{author}{\bibinfo{person}{Gonul Ayci}, \bibinfo{person}{Murat Sensoy},
  \bibinfo{person}{Arzucan {\"{O}}zg{\"{u}}r}, {and} \bibinfo{person}{Pinar
  Yolum}.} \bibinfo{year}{2023}\natexlab{}.
\newblock \showarticletitle{Uncertainty-Aware Personal Assistant for Making
  Personalized Privacy Decisions}.
\newblock \bibinfo{journal}{\emph{ACM Transactions on Internet Technology}}
  \bibinfo{volume}{23}, \bibinfo{number}{1}, Article \bibinfo{articleno}{13}
  (\bibinfo{year}{2023}), \bibinfo{numpages}{24}~pages.
\newblock
\href{https://doi.org/10.1145/3561820}{doi:\nolinkurl{10.1145/3561820}}


\bibitem[Bagdasarian et~al\mbox{.}(2024)]%
        {bagdasarian2024airgapagent}
\bibfield{author}{\bibinfo{person}{Eugene Bagdasarian}, \bibinfo{person}{Ren
  Yi}, \bibinfo{person}{Sahra Ghalebikesabi}, \bibinfo{person}{Peter Kairouz},
  \bibinfo{person}{Marco Gruteser}, \bibinfo{person}{Sewoong Oh},
  \bibinfo{person}{Borja Balle}, {and} \bibinfo{person}{Daniel Ramage}.}
  \bibinfo{year}{2024}\natexlab{}.
\newblock \showarticletitle{{AirGapAgent}: Protecting Privacy-Conscious
  Conversational Agents}. In \bibinfo{booktitle}{\emph{Proceedings of the 2024
  {ACM} {SIGSAC} Conference on Computer and Communications Security}} (Salt
  Lake City, UT, USA). \bibinfo{publisher}{Association for Computing
  Machinery}, \bibinfo{address}{New York, NY, USA},
  \bibinfo{pages}{3868--3882}.
\newblock
\href{https://doi.org/10.1145/3658644.3690350}{doi:\nolinkurl{10.1145/3658644.3690350}}


\bibitem[Bai et~al\mbox{.}(2026)]%
        {bai2026pufferfishcomposition}
\bibfield{author}{\bibinfo{person}{Jiamu Bai}, \bibinfo{person}{Guanlin He},
  \bibinfo{person}{Xin Gu}, \bibinfo{person}{Daniel Kifer}, {and}
  \bibinfo{person}{Kiwan Maeng}.} \bibinfo{year}{2026}\natexlab{}.
\newblock \showarticletitle{Composition for {Pufferfish} Privacy}.
\newblock \bibinfo{journal}{\emph{Proceedings of the VLDB Endowment}}
  \bibinfo{volume}{19}, \bibinfo{number}{10} (\bibinfo{year}{2026}),
  \bibinfo{pages}{2790--2802}.
\newblock
\href{https://doi.org/10.14778/3828612.3828632}{doi:\nolinkurl{10.14778/3828612.3828632}}


\bibitem[Bao et~al\mbox{.}(2025)]%
        {bao2025galaxy}
\bibfield{author}{\bibinfo{person}{Chongyu Bao}, \bibinfo{person}{Ruimin Dai},
  \bibinfo{person}{Yangbo Shen}, \bibinfo{person}{Runyang Jian},
  \bibinfo{person}{Jinghan Zhang}, \bibinfo{person}{Xiaolan Liu}, {and}
  \bibinfo{person}{Kunpeng Liu}.} \bibinfo{year}{2025}\natexlab{}.
\newblock \bibinfo{title}{{Galaxy}: A Cognition-Centered Framework for
  Proactive, Privacy-Preserving, and Self-Evolving {LLM} Agents}.
\newblock \bibinfo{howpublished}{arXiv preprint arXiv:2508.03991v1 [cs.AI]}.
\newblock
\urldef\tempurl%
\url{https://arxiv.org/abs/2508.03991v1}
\showURL{%
\tempurl}


\bibitem[Bonawitz et~al\mbox{.}(2017)]%
        {bonawitz2017secureaggregation}
\bibfield{author}{\bibinfo{person}{Keith Bonawitz}, \bibinfo{person}{Vladimir
  Ivanov}, \bibinfo{person}{Ben Kreuter}, \bibinfo{person}{Antonio Marcedone},
  \bibinfo{person}{H.~Brendan McMahan}, \bibinfo{person}{Sarvar Patel},
  \bibinfo{person}{Daniel Ramage}, \bibinfo{person}{Aaron Segal}, {and}
  \bibinfo{person}{Karn Seth}.} \bibinfo{year}{2017}\natexlab{}.
\newblock \showarticletitle{Practical Secure Aggregation for Privacy-Preserving
  Machine Learning}. In \bibinfo{booktitle}{\emph{Proceedings of the 2017 {ACM}
  {SIGSAC} Conference on Computer and Communications Security}} (Dallas, TX,
  USA). \bibinfo{publisher}{Association for Computing Machinery},
  \bibinfo{address}{New York, NY, USA}, \bibinfo{pages}{1175--1191}.
\newblock
\href{https://doi.org/10.1145/3133956.3133982}{doi:\nolinkurl{10.1145/3133956.3133982}}


\bibitem[Cai et~al\mbox{.}(2026)]%
        {cai2026pairzero}
\bibfield{author}{\bibinfo{person}{Zhijie Cai}, \bibinfo{person}{Yuhao Zheng},
  \bibinfo{person}{Haolong Chen}, \bibinfo{person}{Dongzhu Liu},
  \bibinfo{person}{Bin Wang}, {and} \bibinfo{person}{Guangxu Zhu}.}
  \bibinfo{year}{2026}\natexlab{}.
\newblock \bibinfo{title}{Three Birds, One Stone: Solving the
  Communication-Memory-Privacy Trilemma in {LLM} Fine-tuning Over Wireless
  Networks with Zeroth-Order Optimization}.
\newblock \bibinfo{howpublished}{arXiv preprint arXiv:2604.12401v1}.
\newblock
\urldef\tempurl%
\url{https://arxiv.org/abs/2604.12401v1}
\showURL{%
\tempurl}


\bibitem[Chalef et~al\mbox{.}(2026)]%
        {chalef2026zep}
\bibfield{author}{\bibinfo{person}{Daniel Chalef}, \bibinfo{person}{Peter
  Evans}, \bibinfo{person}{Pavlo Paliychuk}, \bibinfo{person}{Mehul Patel},
  \bibinfo{person}{Preston Rasmussen}, {and} \bibinfo{person}{Jack Ryan}.}
  \bibinfo{year}{2026}\natexlab{}.
\newblock \bibinfo{title}{Learning by Doing: Autonomously Compiling and
  Curating Agent Skills from Verified Traces}.
\newblock \bibinfo{howpublished}{arXiv preprint arXiv:2611.01386v1}.
\newblock
\urldef\tempurl%
\url{https://arxiv.org/abs/2611.01386v1}
\showURL{%
\tempurl}


\bibitem[Chan et~al\mbox{.}(2026)]%
        {chan2026participation}
\bibfield{author}{\bibinfo{person}{T-H.~Hubert Chan}, \bibinfo{person}{Elaine
  Shi}, \bibinfo{person}{Mengshi Zhao}, {and} \bibinfo{person}{Mingxun Zhou}.}
  \bibinfo{year}{2026}\natexlab{}.
\newblock \bibinfo{title}{Continual Learning With Participation Privacy: An
  Auditable Buffering-Aggregation Recipe}.
\newblock \bibinfo{howpublished}{arXiv:2607.07209v2 [cs.CR]}.
\newblock
\urldef\tempurl%
\url{https://arxiv.org/abs/2607.07209v2}
\showURL{%
\tempurl}


\bibitem[Charles et~al\mbox{.}(2025)]%
        {charles2025fixedcompute}
\bibfield{author}{\bibinfo{person}{Zachary Charles}, \bibinfo{person}{Arun
  Ganesh}, \bibinfo{person}{Ryan McKenna}, \bibinfo{person}{H.~Brendan
  McMahan}, \bibinfo{person}{Nicole Mitchell}, \bibinfo{person}{Krishna
  Pillutla}, {and} \bibinfo{person}{Keith Rush}.}
  \bibinfo{year}{2025}\natexlab{}.
\newblock \showarticletitle{Learning with User-Level Differential Privacy Under
  Fixed Compute Budgets}. In \bibinfo{booktitle}{\emph{2025 IEEE Conference on
  Secure and Trustworthy Machine Learning (SaTML)}}. \bibinfo{publisher}{IEEE},
  \bibinfo{address}{Copenhagen, Denmark}, \bibinfo{pages}{901--920}.
\newblock
\href{https://doi.org/10.1109/SATML64287.2025.00055}{doi:\nolinkurl{10.1109/SATML64287.2025.00055}}


\bibitem[Chen et~al\mbox{.}(2026a)]%
        {chen2026safetysurvey}
\bibfield{author}{\bibinfo{person}{Jiahao Chen}, \bibinfo{person}{Zhou Feng},
  \bibinfo{person}{Oubo Ma}, \bibinfo{person}{Yichen Yan},
  \bibinfo{person}{Ruixiao Lin}, \bibinfo{person}{Hangtao Zhang},
  \bibinfo{person}{Linkang Du}, \bibinfo{person}{Hengyu An},
  \bibinfo{person}{Yong Yang}, \bibinfo{person}{Jun Liu},
  \bibinfo{person}{Junhao Li}, \bibinfo{person}{Naen Xu},
  \bibinfo{person}{Chunyi Zhou}, \bibinfo{person}{Yuan Su},
  \bibinfo{person}{Zehao Jin}, \bibinfo{person}{Qianli Ma},
  \bibinfo{person}{Leyi Qi}, \bibinfo{person}{Yiming Wang},
  \bibinfo{person}{Zhe Ma}, \bibinfo{person}{Yuwen Pu},
  \bibinfo{person}{Mengyao Du}, \bibinfo{person}{Yuanyi Song},
  \bibinfo{person}{Enhao Huang}, \bibinfo{person}{Zhihui Fu},
  \bibinfo{person}{Jun Wang}, \bibinfo{person}{Jinfeng Li},
  \bibinfo{person}{Yuefeng Chen}, \bibinfo{person}{Hui Xue},
  \bibinfo{person}{Yiming Li}, \bibinfo{person}{Tianyu Du}, {and}
  \bibinfo{person}{Shouling Ji}.} \bibinfo{year}{2026}\natexlab{a}.
\newblock \bibinfo{title}{Safety in Self-Evolving Agents: A Survey}.
\newblock \bibinfo{howpublished}{arXiv preprint arXiv:2610.00093v1 [cs.CR]}.
\newblock
\urldef\tempurl%
\url{https://arxiv.org/abs/2610.00093v1}
\showURL{%
\tempurl}


\bibitem[Chen et~al\mbox{.}(2026b)]%
        {chen2026preferencesurvey}
\bibfield{author}{\bibinfo{person}{Junwei Chen}, \bibinfo{person}{Pengpeng
  Qiao}, \bibinfo{person}{Wei Yang~Bryan Lim}, \bibinfo{person}{Manjiang Yu},
  {and} \bibinfo{person}{Yang Cao}.} \bibinfo{year}{2026}\natexlab{b}.
\newblock \bibinfo{title}{Privacy-Preserving {LLM} Preference Alignment: A
  Systematic Survey}.
\newblock \bibinfo{howpublished}{{TechRxiv} Preprint}.
\newblock
\href{https://doi.org/10.36227/techrxiv.176978434.43544529/v1}{doi:\nolinkurl{10.36227/techrxiv.176978434.43544529/v1}}


\bibitem[Chen et~al\mbox{.}(2026d)]%
        {chen2026rankaware}
\bibfield{author}{\bibinfo{person}{Junwei Chen}, \bibinfo{person}{Manjiang Yu},
  \bibinfo{person}{Pengpeng Qiao}, {and} \bibinfo{person}{Yang Cao}.}
  \bibinfo{year}{2026}\natexlab{d}.
\newblock \bibinfo{title}{Rank-Aware Differentially Private Release of Listwise
  Preferences for {LLM} Alignment}.
\newblock \bibinfo{howpublished}{arXiv preprint arXiv:2610.37682v2}.
\newblock
\urldef\tempurl%
\url{https://arxiv.org/abs/2610.37682v2}
\showURL{%
\tempurl}


\bibitem[Chen et~al\mbox{.}(2026e)]%
        {chen2026deploymentmemorization}
\bibfield{author}{\bibinfo{person}{Lei~(Rachel) Chen}, \bibinfo{person}{Guilin
  Zhang}, \bibinfo{person}{Kai Zhao}, \bibinfo{person}{Dalmo Cirne},
  \bibinfo{person}{Andy Olsen}, \bibinfo{person}{Xu Chu}, \bibinfo{person}{Zeke
  Miller}, \bibinfo{person}{Alet Blanken}, \bibinfo{person}{Amine Anoun}, {and}
  \bibinfo{person}{Jerry Ting}.} \bibinfo{year}{2026}\natexlab{e}.
\newblock \bibinfo{title}{Deployment-Time Memorization in Foundation-Model
  Agents}.
\newblock \bibinfo{howpublished}{arXiv preprint arXiv:2606.10062v2 [cs.AI]}.
\newblock
\urldef\tempurl%
\url{https://arxiv.org/abs/2606.10062v2}
\showURL{%
\tempurl}
\newblock
\shownote{Presented at the non-archival {ICML} 2026 {MemFM} Workshop}.


\bibitem[Chen et~al\mbox{.}(2021)]%
        {chen2021unlearningprivacy}
\bibfield{author}{\bibinfo{person}{Min Chen}, \bibinfo{person}{Zhikun Zhang},
  \bibinfo{person}{Tianhao Wang}, \bibinfo{person}{Michael Backes},
  \bibinfo{person}{Mathias Humbert}, {and} \bibinfo{person}{Yang Zhang}.}
  \bibinfo{year}{2021}\natexlab{}.
\newblock \showarticletitle{When Machine Unlearning Jeopardizes Privacy}. In
  \bibinfo{booktitle}{\emph{Proceedings of the 2021 {ACM} {SIGSAC} Conference
  on Computer and Communications Security}} (Virtual Event, Republic of Korea).
  \bibinfo{publisher}{Association for Computing Machinery},
  \bibinfo{address}{New York, NY, USA}, \bibinfo{pages}{896--911}.
\newblock
\href{https://doi.org/10.1145/3460120.3484756}{doi:\nolinkurl{10.1145/3460120.3484756}}


\bibitem[Chen et~al\mbox{.}(2026c)]%
        {chen2026skillladder}
\bibfield{author}{\bibinfo{person}{Nan Chen}, \bibinfo{person}{Yi Wang},
  \bibinfo{person}{Bin~Benjamin Zhu}, {and} \bibinfo{person}{Haoyi Xiong}.}
  \bibinfo{year}{2026}\natexlab{c}.
\newblock \showarticletitle{{SkillLadder}: Multi-User Skill Evolution with
  Compliance-Oriented Synchronization and Quality-Gated Synthesis}. In
  \bibinfo{booktitle}{\emph{Findings of the Association for Computational
  Linguistics: EMNLP 2026}}. \bibinfo{publisher}{Association for Computational
  Linguistics}, \bibinfo{address}{Budapest, Hungary},
  \bibinfo{pages}{4553--4573}.
\newblock
\href{https://doi.org/10.18653/2026.findings-emnlp.218}{doi:\nolinkurl{10.18653/2026.findings-emnlp.218}}


\bibitem[Chen et~al\mbox{.}(2026f)]%
        {chen2026memprivacy}
\bibfield{author}{\bibinfo{person}{Yining Chen}, \bibinfo{person}{Jihao Zhao},
  \bibinfo{person}{Bo Tang}, \bibinfo{person}{Haofen Wang},
  \bibinfo{person}{Yue Zhang}, \bibinfo{person}{Fei Huang},
  \bibinfo{person}{Feiyu Xiong}, {and} \bibinfo{person}{Zhiyu Li}.}
  \bibinfo{year}{2026}\natexlab{f}.
\newblock \bibinfo{title}{{MemPrivacy}: Privacy-Preserving Personalized Memory
  Management for Edge-Cloud Agents}.
\newblock \bibinfo{howpublished}{arXiv preprint arXiv:2605.09530v3 [cs.CR]}.
\newblock
\urldef\tempurl%
\url{https://arxiv.org/abs/2605.09530v3}
\showURL{%
\tempurl}


\bibitem[Cheng et~al\mbox{.}(2026)]%
        {cheng2026privact}
\bibfield{author}{\bibinfo{person}{Yuhan Cheng}, \bibinfo{person}{Hancheng Ye},
  \bibinfo{person}{Hai~Helen Li}, \bibinfo{person}{Jingwei Sun}, {and}
  \bibinfo{person}{Yiran Chen}.} \bibinfo{year}{2026}\natexlab{}.
\newblock \showarticletitle{{PrivAct}: Internalizing Contextual Privacy
  Preservation via Multi-Agent Preference Training}. In
  \bibinfo{booktitle}{\emph{Proceedings of the 43rd International Conference on
  Machine Learning}} \emph{(\bibinfo{series}{Proceedings of Machine Learning
  Research}, Vol.~\bibinfo{volume}{306})}. \bibinfo{publisher}{PMLR},
  \bibinfo{address}{Seoul, South Korea}, \bibinfo{pages}{19079--19098}.
\newblock
\urldef\tempurl%
\url{https://proceedings.mlr.press/v306/cheng26o.html}
\showURL{%
\tempurl}


\bibitem[Cheng et~al\mbox{.}(2025)]%
        {cheng2025remoterag}
\bibfield{author}{\bibinfo{person}{Yihang Cheng}, \bibinfo{person}{Lan Zhang},
  \bibinfo{person}{Junyang Wang}, \bibinfo{person}{Mu Yuan}, {and}
  \bibinfo{person}{Yunhao Yao}.} \bibinfo{year}{2025}\natexlab{}.
\newblock \showarticletitle{{RemoteRAG}: A Privacy-Preserving {LLM} Cloud {RAG}
  Service}. In \bibinfo{booktitle}{\emph{Findings of the Association for
  Computational Linguistics: {ACL} 2025}},
  \bibfield{editor}{\bibinfo{person}{Wanxiang Che}, \bibinfo{person}{Joyce
  Nabende}, \bibinfo{person}{Ekaterina Shutova}, {and}
  \bibinfo{person}{Mohammad~Taher Pilehvar}} (Eds.).
  \bibinfo{publisher}{Association for Computational Linguistics},
  \bibinfo{address}{Vienna, Austria}, \bibinfo{pages}{3820--3837}.
\newblock
\showISBNx{979-8-89176-256-5}
\href{https://doi.org/10.18653/v1/2025.findings-acl.197}{doi:\nolinkurl{10.18653/v1/2025.findings-acl.197}}


\bibitem[Cho and Sun(2026)]%
        {cho2026decoupledreward}
\bibfield{author}{\bibinfo{person}{Young~Hyun Cho} {and}
  \bibinfo{person}{Will~Wei Sun}.} \bibinfo{year}{2026}\natexlab{}.
\newblock \bibinfo{title}{Privacy-Preserving Reinforcement Learning from Human
  Feedback via Decoupled Reward Modeling}.
\newblock \bibinfo{howpublished}{arXiv preprint arXiv:2603.22563v1 [stat.ML]}.
\newblock
\urldef\tempurl%
\url{https://arxiv.org/abs/2603.22563v1}
\showURL{%
\tempurl}


\bibitem[Chourasia and Shah(2023)]%
        {chourasia2023trueprivacy}
\bibfield{author}{\bibinfo{person}{Rishav Chourasia} {and}
  \bibinfo{person}{Neil Shah}.} \bibinfo{year}{2023}\natexlab{}.
\newblock \showarticletitle{Forget Unlearning: Towards True Data-Deletion in
  Machine Learning}. In \bibinfo{booktitle}{\emph{Proceedings of the 40th
  International Conference on Machine Learning}}
  \emph{(\bibinfo{series}{Proceedings of Machine Learning Research},
  Vol.~\bibinfo{volume}{202})}, \bibfield{editor}{\bibinfo{person}{Andreas
  Krause}, \bibinfo{person}{Emma Brunskill}, \bibinfo{person}{Kyunghyun Cho},
  \bibinfo{person}{Barbara Engelhardt}, \bibinfo{person}{Sivan Sabato}, {and}
  \bibinfo{person}{Jonathan Scarlett}} (Eds.). \bibinfo{publisher}{PMLR},
  \bibinfo{address}{Honolulu, HI, USA}, \bibinfo{pages}{6028--6073}.
\newblock
\urldef\tempurl%
\url{https://proceedings.mlr.press/v202/chourasia23a.html}
\showURL{%
\tempurl}


\bibitem[Christou et~al\mbox{.}(2026)]%
        {christou2026memgov}
\bibfield{author}{\bibinfo{person}{Panayiotis Christou},
  \bibinfo{person}{Miguel Palma}, \bibinfo{person}{Zefan Du},
  \bibinfo{person}{Kaixun Hua}, {and} \bibinfo{person}{Ying Mao}.}
  \bibinfo{year}{2026}\natexlab{}.
\newblock \showarticletitle{Beyond Retrieval: Deterministic Memory Governance
  with Pairwise Constraints and Certified Repair for Long-Term Language
  Agents}. In \bibinfo{booktitle}{\emph{Proceedings of IJCNLP-AACL 2026 (Volume
  1: Long Papers)}}. \bibinfo{publisher}{Association for Computational
  Linguistics}, \bibinfo{address}{Hengqin, China}, \bibinfo{pages}{5501--5518}.
\newblock
\href{https://doi.org/10.18653/2026.aacl-long.246}{doi:\nolinkurl{10.18653/2026.aacl-long.246}}


\bibitem[Chua et~al\mbox{.}(2024)]%
        {chua2024privacyunit}
\bibfield{author}{\bibinfo{person}{Lynn Chua}, \bibinfo{person}{Badih Ghazi},
  \bibinfo{person}{Yangsibo Huang}, \bibinfo{person}{Pritish Kamath},
  \bibinfo{person}{Ravi Kumar}, \bibinfo{person}{Daogao Liu},
  \bibinfo{person}{Pasin Manurangsi}, \bibinfo{person}{Amer Sinha}, {and}
  \bibinfo{person}{Chiyuan Zhang}.} \bibinfo{year}{2024}\natexlab{}.
\newblock \showarticletitle{Mind the Privacy Unit! User-Level Differential
  Privacy for Language Model Fine-Tuning}. In \bibinfo{booktitle}{\emph{First
  Conference on Language Modeling}}. \bibinfo{publisher}{Conference on Language
  Modeling}, \bibinfo{address}{Philadelphia, PA, USA},
  \bibinfo{numpages}{21}~pages.
\newblock
\urldef\tempurl%
\url{https://proceedings.colm.cc/paper_files/paper/2024/hash/99b419554537c66bf27e5eb7a74c7de4-Abstract-Conference.html}
\showURL{%
\tempurl}


\bibitem[Cohen et~al\mbox{.}(2023)]%
        {cohen2023deletionascontrol}
\bibfield{author}{\bibinfo{person}{Aloni Cohen}, \bibinfo{person}{Adam Smith},
  \bibinfo{person}{Marika Swanberg}, {and} \bibinfo{person}{Prashant~Nalini
  Vasudevan}.} \bibinfo{year}{2023}\natexlab{}.
\newblock \showarticletitle{Control, Confidentiality, and the Right to be
  Forgotten}. In \bibinfo{booktitle}{\emph{Proceedings of the 2023 {ACM}
  {SIGSAC} Conference on Computer and Communications Security}} (Copenhagen,
  Denmark). \bibinfo{publisher}{Association for Computing Machinery},
  \bibinfo{address}{New York, NY, USA}, \bibinfo{pages}{3358--3372}.
\newblock
\href{https://doi.org/10.1145/3576915.3616585}{doi:\nolinkurl{10.1145/3576915.3616585}}


\bibitem[Das et~al\mbox{.}(2026)]%
        {das2026finetuning}
\bibfield{author}{\bibinfo{person}{Soumi Das}, \bibinfo{person}{Camila
  Kolling}, \bibinfo{person}{Mohammad~Aflah Khan}, \bibinfo{person}{Mahsa
  Amani}, \bibinfo{person}{Bishwamittra Ghosh}, \bibinfo{person}{Qinyuan Wu},
  \bibinfo{person}{Till Speicher}, {and} \bibinfo{person}{Krishna~P. Gummadi}.}
  \bibinfo{year}{2026}\natexlab{}.
\newblock \bibinfo{title}{Revisiting Privacy, Utility, and Efficiency
  Trade-offs when Fine-Tuning Large Language Models}.
\newblock \bibinfo{howpublished}{arXiv preprint arXiv:2502.13313v2}.
\newblock
\urldef\tempurl%
\url{https://arxiv.org/abs/2502.13313v2}
\showURL{%
\tempurl}


\bibitem[Dash et~al\mbox{.}(2026)]%
        {dash2026selfportrait}
\bibfield{author}{\bibinfo{person}{Abhisek Dash}, \bibinfo{person}{Soumi Das},
  \bibinfo{person}{Elisabeth Kirsten}, \bibinfo{person}{Qinyuan Wu},
  \bibinfo{person}{Sai~Keerthana Karnam}, \bibinfo{person}{Krishna~P. Gummadi},
  \bibinfo{person}{Thorsten Holz}, \bibinfo{person}{Muhammad~Bilal Zafar},
  {and} \bibinfo{person}{Savvas Zannettou}.} \bibinfo{year}{2026}\natexlab{}.
\newblock \showarticletitle{The Algorithmic Self-Portrait: Deconstructing
  Memory in {ChatGPT}}. In \bibinfo{booktitle}{\emph{Proceedings of the {ACM}
  Web Conference 2026}}. \bibinfo{publisher}{Association for Computing
  Machinery}, \bibinfo{address}{New York, NY, USA},
  \bibinfo{pages}{3471--3482}.
\newblock
\href{https://doi.org/10.1145/3774904.3792671}{doi:\nolinkurl{10.1145/3774904.3792671}}


\bibitem[Ding et~al\mbox{.}(2026a)]%
        {ding2026sparsereward}
\bibfield{author}{\bibinfo{person}{Meng Ding}, \bibinfo{person}{Mingxi Lei},
  \bibinfo{person}{Jie Zhang}, \bibinfo{person}{Jinyan Liu}, {and}
  \bibinfo{person}{Di Wang}.} \bibinfo{year}{2026}\natexlab{a}.
\newblock \bibinfo{title}{Differentially Private Sparse Reward Estimation with
  Preference Feedback}.
\newblock \bibinfo{howpublished}{arXiv preprint arXiv:2610.30099v1 [cs.LG]}.
\newblock
\urldef\tempurl%
\url{https://arxiv.org/abs/2610.30099v1}
\showURL{%
\tempurl}


\bibitem[Ding et~al\mbox{.}(2026b)]%
        {ding2026alwayson}
\bibfield{author}{\bibinfo{person}{Tianyu Ding}, \bibinfo{person}{Aditya
  Nannapaneni}, \bibinfo{person}{Bingfan Liu}, {and} \bibinfo{person}{Ling
  Zhang}.} \bibinfo{year}{2026}\natexlab{b}.
\newblock \bibinfo{title}{Always-On Agents: A Survey of Persistent Memory,
  State, and Governance in {LLM} Agents}.
\newblock \bibinfo{howpublished}{arXiv preprint arXiv:2606.30306v1 [cs.MA]}.
\newblock
\urldef\tempurl%
\url{https://arxiv.org/abs/2606.30306v1}
\showURL{%
\tempurl}


\bibitem[Dong et~al\mbox{.}(2026)]%
        {dong2026claws}
\bibfield{author}{\bibinfo{person}{Ben Dong}, \bibinfo{person}{Zhonghao Guo},
  \bibinfo{person}{Tianyi Lu}, {and} \bibinfo{person}{Qian Wang}.}
  \bibinfo{year}{2026}\natexlab{}.
\newblock \bibinfo{title}{The Claws in Plain Sight: Unauthorized Context
  Disclosure through {LLM} Agent Tool Calls}.
\newblock \bibinfo{howpublished}{arXiv preprint arXiv:2608.20658v1}.
\newblock
\urldef\tempurl%
\url{https://arxiv.org/abs/2608.20658v1}
\showURL{%
\tempurl}


\bibitem[Du et~al\mbox{.}(2025)]%
        {du2025privatefinetune}
\bibfield{author}{\bibinfo{person}{Hao Du}, \bibinfo{person}{Shang Liu}, {and}
  \bibinfo{person}{Yang Cao}.} \bibinfo{year}{2025}\natexlab{}.
\newblock \showarticletitle{Can Differentially Private Fine-Tuning {LLMs}
  Protect Against Privacy Attacks?}. In \bibinfo{booktitle}{\emph{Data and
  Applications Security and Privacy XXXIX}} \emph{(\bibinfo{series}{Lecture
  Notes in Computer Science}, Vol.~\bibinfo{volume}{15722})}.
  \bibinfo{publisher}{Springer Nature Switzerland}, \bibinfo{address}{Cham,
  Switzerland}, \bibinfo{pages}{311--329}.
\newblock
\href{https://doi.org/10.1007/978-3-031-96590-6_17}{doi:\nolinkurl{10.1007/978-3-031-96590-6_17}}


\bibitem[Du et~al\mbox{.}(2023)]%
        {du2023dpforward}
\bibfield{author}{\bibinfo{person}{Minxin Du}, \bibinfo{person}{Xiang Yue},
  \bibinfo{person}{Sherman S.~M. Chow}, \bibinfo{person}{Tianhao Wang},
  \bibinfo{person}{Chenyu Huang}, {and} \bibinfo{person}{Huan Sun}.}
  \bibinfo{year}{2023}\natexlab{}.
\newblock \showarticletitle{{DP-Forward}: Fine-tuning and Inference on Language
  Models with Differential Privacy in Forward Pass}. In
  \bibinfo{booktitle}{\emph{Proceedings of the 2023 ACM SIGSAC Conference on
  Computer and Communications Security}}. \bibinfo{publisher}{ACM},
  \bibinfo{address}{Copenhagen, Denmark}, \bibinfo{pages}{2665--2679}.
\newblock
\href{https://doi.org/10.1145/3576915.3616592}{doi:\nolinkurl{10.1145/3576915.3616592}}


\bibitem[Duan et~al\mbox{.}(2023)]%
        {duan2023stochasticparrots}
\bibfield{author}{\bibinfo{person}{Haonan Duan}, \bibinfo{person}{Adam
  Dziedzic}, \bibinfo{person}{Nicolas Papernot}, {and}
  \bibinfo{person}{Franziska Boenisch}.} \bibinfo{year}{2023}\natexlab{}.
\newblock \showarticletitle{Flocks of Stochastic Parrots: Differentially
  Private Prompt Learning for Large Language Models}. In
  \bibinfo{booktitle}{\emph{Advances in Neural Information Processing Systems}}
  (New Orleans, Louisiana, USA), Vol.~\bibinfo{volume}{36}.
  \bibinfo{publisher}{Neural Information Processing Systems Foundation, Inc.
  (NeurIPS)}, \bibinfo{address}{San Diego, California, USA},
  \bibinfo{pages}{76852--76871}.
\newblock
\href{https://doi.org/10.52202/075280-3358}{doi:\nolinkurl{10.52202/075280-3358}}


\bibitem[Edemacu and Wu(2025)]%
        {edemacu2025promptsurvey}
\bibfield{author}{\bibinfo{person}{Kennedy Edemacu} {and}
  \bibinfo{person}{Xintao Wu}.} \bibinfo{year}{2025}\natexlab{}.
\newblock \showarticletitle{Privacy Preserving Prompt Engineering: A Survey}.
\newblock \bibinfo{journal}{\emph{Comput. Surveys}} \bibinfo{volume}{57},
  \bibinfo{number}{10}, Article \bibinfo{articleno}{255}
  (\bibinfo{year}{2025}), \bibinfo{numpages}{36}~pages.
\newblock
\href{https://doi.org/10.1145/3729219}{doi:\nolinkurl{10.1145/3729219}}


\bibitem[Eisenhofer et~al\mbox{.}(2025)]%
        {eisenhofer2025verifiableunlearning}
\bibfield{author}{\bibinfo{person}{Thorsten Eisenhofer},
  \bibinfo{person}{Doreen Riepel}, \bibinfo{person}{Varun Chandrasekaran},
  \bibinfo{person}{Esha Ghosh}, \bibinfo{person}{Olga Ohrimenko}, {and}
  \bibinfo{person}{Nicolas Papernot}.} \bibinfo{year}{2025}\natexlab{}.
\newblock \showarticletitle{Verifiable and Provably Secure Machine Unlearning}.
  In \bibinfo{booktitle}{\emph{2025 {IEEE} Conference on Secure and Trustworthy
  Machine Learning ({SaTML})}} (Copenhagen, Denmark).
  \bibinfo{publisher}{IEEE}, \bibinfo{address}{Piscataway, NJ, USA},
  \bibinfo{pages}{479--496}.
\newblock
\href{https://doi.org/10.1109/satml64287.2025.00033}{doi:\nolinkurl{10.1109/satml64287.2025.00033}}


\bibitem[El~Yagoubi et~al\mbox{.}(2026)]%
        {elyagoubi2026agentleak}
\bibfield{author}{\bibinfo{person}{Faouzi El~Yagoubi}, \bibinfo{person}{Godwin
  Badu-Marfo}, {and} \bibinfo{person}{Ranwa Al~Mallah}.}
  \bibinfo{year}{2026}\natexlab{}.
\newblock \showarticletitle{{AgentLeak}: A Benchmark for Internal-Channel
  Privacy Leakage in Multi-Agent {LLM} Systems}.
\newblock \bibinfo{journal}{\emph{IEEE Access}}  \bibinfo{volume}{14}
  (\bibinfo{year}{2026}), \bibinfo{pages}{94960--94978}.
\newblock
\href{https://doi.org/10.1109/ACCESS.2026.3704541}{doi:\nolinkurl{10.1109/ACCESS.2026.3704541}}


\bibitem[Ertan et~al\mbox{.}(2026)]%
        {ertan2026paczero}
\bibfield{author}{\bibinfo{person}{Murat~Bilgehan Ertan},
  \bibinfo{person}{Xiaochen Zhu}, \bibinfo{person}{Phuong~Ha Nguyen},
  \bibinfo{person}{Marten van Dijk}, {and} \bibinfo{person}{Srinivas Devadas}.}
  \bibinfo{year}{2026}\natexlab{}.
\newblock \bibinfo{title}{{PACZero}: {PAC-Private} Fine-Tuning of Language
  Models via Sign Quantization}.
\newblock \bibinfo{howpublished}{arXiv preprint arXiv:2605.06505v3}.
\newblock
\urldef\tempurl%
\url{https://arxiv.org/abs/2605.06505v3}
\showURL{%
\tempurl}


\bibitem[Feng et~al\mbox{.}(2025)]%
        {feng2025premia}
\bibfield{author}{\bibinfo{person}{Qizhang Feng}, \bibinfo{person}{Siva~Rajesh
  Kasa}, \bibinfo{person}{Santhosh~Kumar Kasa}, \bibinfo{person}{Hyokun Yun},
  \bibinfo{person}{Choon~Hui Teo}, {and} \bibinfo{person}{Sravan~Babu
  Bodapati}.} \bibinfo{year}{2025}\natexlab{}.
\newblock \showarticletitle{Exposing Privacy Gaps: Membership Inference Attack
  on Preference Data for {LLM} Alignment}. In
  \bibinfo{booktitle}{\emph{Proceedings of the 28th International Conference on
  Artificial Intelligence and Statistics}} \emph{(\bibinfo{series}{Proceedings
  of Machine Learning Research}, Vol.~\bibinfo{volume}{258})}.
  \bibinfo{publisher}{PMLR}, \bibinfo{address}{Mai Khao, Thailand},
  \bibinfo{pages}{5221--5229}.
\newblock
\urldef\tempurl%
\url{https://proceedings.mlr.press/v258/feng25a.html}
\showURL{%
\tempurl}


\bibitem[Flemings et~al\mbox{.}(2026)]%
        {flemings2026ariel}
\bibfield{author}{\bibinfo{person}{James Flemings}, \bibinfo{person}{Ren Yi},
  \bibinfo{person}{Octavian Suciu}, \bibinfo{person}{Kassem Fawaz},
  \bibinfo{person}{Murali Annavaram}, {and} \bibinfo{person}{Marco Gruteser}.}
  \bibinfo{year}{2026}\natexlab{}.
\newblock \showarticletitle{Personalizing Agent Privacy Decisions via Logical
  Entailment}.
\newblock \bibinfo{journal}{\emph{Proceedings on Privacy Enhancing
  Technologies}} \bibinfo{volume}{2026}, \bibinfo{number}{3}
  (\bibinfo{year}{2026}), \bibinfo{pages}{378--407}.
\newblock
\href{https://doi.org/10.56553/popets-2026-0087}{doi:\nolinkurl{10.56553/popets-2026-0087}}


\bibitem[Fr{\'e}ry et~al\mbox{.}(2025)]%
        {frery2025helora}
\bibfield{author}{\bibinfo{person}{Jordan Fr{\'e}ry}, \bibinfo{person}{Roman
  Bredeh{\"o}ft}, \bibinfo{person}{Jakub Klemsa}, \bibinfo{person}{Arthur
  Meyre}, {and} \bibinfo{person}{Andrei Stoian}.}
  \bibinfo{year}{2025}\natexlab{}.
\newblock \bibinfo{title}{Private {LoRA} Fine-tuning of Open-Source {LLMs} with
  Homomorphic Encryption}.
\newblock \bibinfo{howpublished}{arXiv:2505.07329v1 [cs.CR]}.
\newblock
\urldef\tempurl%
\url{https://arxiv.org/abs/2505.07329v1}
\showURL{%
\tempurl}


\bibitem[Fu et~al\mbox{.}(2026)]%
        {fu2026privaterecs}
\bibfield{author}{\bibinfo{person}{Xingyu Fu}, \bibinfo{person}{Ningyuan Chen},
  \bibinfo{person}{Pin Gao}, {and} \bibinfo{person}{Yang Li}.}
  \bibinfo{year}{2026}\natexlab{}.
\newblock \showarticletitle{Privacy-Preserving Personalized Recommender
  Systems}.
\newblock \bibinfo{journal}{\emph{Manufacturing \& Service Operations
  Management}} \bibinfo{volume}{28}, \bibinfo{number}{1}
  (\bibinfo{year}{2026}), \bibinfo{pages}{271--289}.
\newblock
\href{https://doi.org/10.1287/msom.2023.0271}{doi:\nolinkurl{10.1287/msom.2023.0271}}


\bibitem[Gao and Yang(2026)]%
        {gao2026dpprefsyn}
\bibfield{author}{\bibinfo{person}{Fengyu Gao} {and} \bibinfo{person}{Jing
  Yang}.} \bibinfo{year}{2026}\natexlab{}.
\newblock \showarticletitle{Differentially Private Preference Data Synthesis
  for Large Language Model Alignment}. In \bibinfo{booktitle}{\emph{Proceedings
  of the 43rd International Conference on Machine Learning}}
  \emph{(\bibinfo{series}{Proceedings of Machine Learning Research},
  Vol.~\bibinfo{volume}{306})}. \bibinfo{publisher}{PMLR},
  \bibinfo{address}{Seoul, South Korea}, \bibinfo{pages}{33724--33749}.
\newblock
\urldef\tempurl%
\url{https://proceedings.mlr.press/v306/gao26ae.html}
\showURL{%
\tempurl}


\bibitem[Gao et~al\mbox{.}(2026b)]%
        {gao2026selfevolvingsurvey}
\bibfield{author}{\bibinfo{person}{Huan-ang Gao}, \bibinfo{person}{Jiayi Geng},
  \bibinfo{person}{Wenyue Hua}, \bibinfo{person}{Mengkang Hu},
  \bibinfo{person}{Xinzhe Juan}, \bibinfo{person}{Hongzhang Liu},
  \bibinfo{person}{Shilong Liu}, \bibinfo{person}{Jiahao Qiu},
  \bibinfo{person}{Xuan Qi}, \bibinfo{person}{Qihan Ren},
  \bibinfo{person}{Yiran Wu}, \bibinfo{person}{Hongru Wang},
  \bibinfo{person}{Han Xiao}, \bibinfo{person}{Yuhang Zhou},
  \bibinfo{person}{Shaokun Zhang}, \bibinfo{person}{Jiayi Zhang},
  \bibinfo{person}{Jinyu Xiang}, \bibinfo{person}{Yixiong Fang},
  \bibinfo{person}{Qiwen Zhao}, \bibinfo{person}{Dongrui Liu},
  \bibinfo{person}{Cheng Qian}, \bibinfo{person}{Zhenhailong Wang},
  \bibinfo{person}{Minda Hu}, \bibinfo{person}{Huazheng Wang},
  \bibinfo{person}{Qingyun Wu}, \bibinfo{person}{Heng Ji}, {and}
  \bibinfo{person}{Mengdi Wang}.} \bibinfo{year}{2026}\natexlab{b}.
\newblock \showarticletitle{A Survey of Self-Evolving Agents: What, When, How,
  and Where to Evolve on the Path to Artificial Super Intelligence}.
\newblock \bibinfo{journal}{\emph{Transactions on Machine Learning Research}}
  \bibinfo{volume}{2026}, \bibinfo{number}{1} (\bibinfo{year}{2026}),
  \bibinfo{numpages}{77}~pages.
\newblock
\urldef\tempurl%
\url{https://openreview.net/forum?id=CTr3bovS5F}
\showURL{%
\tempurl}


\bibitem[Gao et~al\mbox{.}(2026a)]%
        {gao2026spore}
\bibfield{author}{\bibinfo{person}{Xinyu Gao}, \bibinfo{person}{Wenyu Chen},
  \bibinfo{person}{Xiangtao Meng}, \bibinfo{person}{Li Wang},
  \bibinfo{person}{Chuanchao Zang}, \bibinfo{person}{Jianing Wang},
  \bibinfo{person}{Zheng Li}, {and} \bibinfo{person}{Shanqing Guo}.}
  \bibinfo{year}{2026}\natexlab{a}.
\newblock \bibinfo{title}{Isolated but Exposed: Persistence-Based Memory
  Extraction Attack on {LLM} Agents}.
\newblock \bibinfo{howpublished}{arXiv preprint arXiv:2607.23444v1 [cs.CR]}.
\newblock
\urldef\tempurl%
\url{https://arxiv.org/abs/2607.23444v1}
\showURL{%
\tempurl}


\bibitem[George et~al\mbox{.}(2024)]%
        {george2024continualmean}
\bibfield{author}{\bibinfo{person}{Anand~Jerry George},
  \bibinfo{person}{Lekshmi Ramesh}, \bibinfo{person}{Aditya~Vikram Singh},
  {and} \bibinfo{person}{Himanshu Tyagi}.} \bibinfo{year}{2024}\natexlab{}.
\newblock \showarticletitle{Continual Mean Estimation Under User-Level
  Privacy}.
\newblock \bibinfo{journal}{\emph{IEEE Journal on Selected Areas in Information
  Theory}}  \bibinfo{volume}{5} (\bibinfo{year}{2024}),
  \bibinfo{pages}{28--43}.
\newblock
\href{https://doi.org/10.1109/jsait.2024.3366086}{doi:\nolinkurl{10.1109/jsait.2024.3366086}}


\bibitem[Goel et~al\mbox{.}(2026)]%
        {goel2026privacycollapse}
\bibfield{author}{\bibinfo{person}{Anmol Goel}, \bibinfo{person}{Cornelius
  Emde}, \bibinfo{person}{Sangdoo Yun}, \bibinfo{person}{Seong~Joon Oh}, {and}
  \bibinfo{person}{Martin Gubri}.} \bibinfo{year}{2026}\natexlab{}.
\newblock \showarticletitle{Privacy Collapse: Benign Fine-Tuning Can Break
  Contextual Privacy in Language Models}. In
  \bibinfo{booktitle}{\emph{Proceedings of the 64th Annual Meeting of the
  Association for Computational Linguistics (Volume 1: Long Papers)}},
  \bibfield{editor}{\bibinfo{person}{Maria Liakata},
  \bibinfo{person}{Viviane~P. Moreira}, \bibinfo{person}{Jiajun Zhang}, {and}
  \bibinfo{person}{David Jurgens}} (Eds.). \bibinfo{publisher}{Association for
  Computational Linguistics}, \bibinfo{address}{San Diego, California, United
  States}, \bibinfo{pages}{8870--8892}.
\newblock
\href{https://doi.org/10.18653/v1/2026.acl-long.400}{doi:\nolinkurl{10.18653/v1/2026.acl-long.400}}


\bibitem[Goel and Gurevych(2026)]%
        {goel2026agentcibench}
\bibfield{author}{\bibinfo{person}{Anmol Goel} {and} \bibinfo{person}{Iryna
  Gurevych}.} \bibinfo{year}{2026}\natexlab{}.
\newblock \showarticletitle{Capable but Careless: Do Computer-Use Agents Follow
  Contextual Integrity?}. In \bibinfo{booktitle}{\emph{Proceedings of the 2026
  Conference on Empirical Methods in Natural Language Processing}},
  \bibfield{editor}{\bibinfo{person}{Hung-Yi Lee}, \bibinfo{person}{Sunipa
  Dev}, \bibinfo{person}{Junyi~Jessy Li}, {and} \bibinfo{person}{Desmond
  Elliott}} (Eds.). \bibinfo{publisher}{Association for Computational
  Linguistics}, \bibinfo{address}{Budapest, Hungary},
  \bibinfo{pages}{1235--1266}.
\newblock
\href{https://doi.org/10.18653/2026.emnlp-main.55}{doi:\nolinkurl{10.18653/2026.emnlp-main.55}}


\bibitem[Gu et~al\mbox{.}(2026)]%
        {gu2026fsfm}
\bibfield{author}{\bibinfo{person}{Yingjie Gu}, \bibinfo{person}{Wenjian
  Xiong}, \bibinfo{person}{Liqiang Wang}, \bibinfo{person}{Pengcheng Ren},
  \bibinfo{person}{Chao Li}, \bibinfo{person}{Xiaojing Zhang},
  \bibinfo{person}{Yijuan Guo}, \bibinfo{person}{Qi Sun},
  \bibinfo{person}{Jingyao Ma}, {and} \bibinfo{person}{Shidang Shi}.}
  \bibinfo{year}{2026}\natexlab{}.
\newblock \bibinfo{title}{{FSFM}: A Biologically-Inspired Framework for
  Selective Forgetting of Agent Memory}.
\newblock \bibinfo{howpublished}{arXiv preprint arXiv:2604.20300v2}.
\newblock
\urldef\tempurl%
\url{https://arxiv.org/abs/2604.20300v2}
\showURL{%
\tempurl}


\bibitem[Guo et~al\mbox{.}(2026b)]%
        {guo2026cider}
\bibfield{author}{\bibinfo{person}{Bingcan Guo}, \bibinfo{person}{Eryue Xu},
  \bibinfo{person}{Jijie Zhou}, \bibinfo{person}{Zhiping Zhang}, {and}
  \bibinfo{person}{Tianshi Li}.} \bibinfo{year}{2026}\natexlab{b}.
\newblock \bibinfo{title}{{CIDER}: A Dataset of Contextual Disclosure
  Boundaries for Privacy Preference Alignment}.
\newblock \bibinfo{howpublished}{arXiv:2608.09164v2 [cs.AI]}.
\newblock
\urldef\tempurl%
\url{https://arxiv.org/abs/2608.09164v2}
\showURL{%
\tempurl}


\bibitem[Guo et~al\mbox{.}(2026a)]%
        {guo2026agentciscope}
\bibfield{author}{\bibinfo{person}{Chengquan Guo}, \bibinfo{person}{Zhaorun
  Chen}, \bibinfo{person}{Zinan Lin}, \bibinfo{person}{Xiaodong Liu}, {and}
  \bibinfo{person}{Bo Li}.} \bibinfo{year}{2026}\natexlab{a}.
\newblock \bibinfo{title}{{AgentCIScope}: Grounded Contextual Integrity
  Evaluation for {LLM} Agents}.
\newblock \bibinfo{howpublished}{arXiv preprint arXiv:2610.19158v1}.
\newblock
\urldef\tempurl%
\url{https://arxiv.org/abs/2610.19158v1}
\showURL{%
\tempurl}


\bibitem[Guo et~al\mbox{.}(2020)]%
        {guo2020certifiedremoval}
\bibfield{author}{\bibinfo{person}{Chuan Guo}, \bibinfo{person}{Tom Goldstein},
  \bibinfo{person}{Awni Hannun}, {and} \bibinfo{person}{Laurens van~der
  Maaten}.} \bibinfo{year}{2020}\natexlab{}.
\newblock \showarticletitle{Certified Data Removal from Machine Learning
  Models}. In \bibinfo{booktitle}{\emph{Proceedings of the 37th International
  Conference on Machine Learning}} \emph{(\bibinfo{series}{Proceedings of
  Machine Learning Research}, Vol.~\bibinfo{volume}{119})},
  \bibfield{editor}{\bibinfo{person}{Hal Daum{\'{e}}, III} {and}
  \bibinfo{person}{Aarti Singh}} (Eds.). \bibinfo{publisher}{PMLR},
  \bibinfo{address}{Virtual Event}, \bibinfo{pages}{3832--3842}.
\newblock
\urldef\tempurl%
\url{https://proceedings.mlr.press/v119/guo20c.html}
\showURL{%
\tempurl}


\bibitem[Gupta et~al\mbox{.}(2026)]%
        {gupta2026pisas}
\bibfield{author}{\bibinfo{person}{Shubham Gupta}, \bibinfo{person}{Nazanin
  Mohammadi~Sepahvand}, \bibinfo{person}{Abhinav Kumar}, \bibinfo{person}{Cem
  Subakan}, \bibinfo{person}{Spandana Gella}, \bibinfo{person}{Pierre-Andr{\'e}
  No{\"e}l}, \bibinfo{person}{Perouz Taslakian}, \bibinfo{person}{Eugene
  Bagdasarian}, {and} \bibinfo{person}{Valentina Zantedeschi}.}
  \bibinfo{year}{2026}\natexlab{}.
\newblock \bibinfo{title}{{PiSAs}: Benchmarking Contextual Integrity in
  Multi-User Agentic Systems}.
\newblock \bibinfo{howpublished}{arXiv preprint arXiv:2607.05318v2}.
\newblock
\urldef\tempurl%
\url{https://arxiv.org/abs/2607.05318v2}
\showURL{%
\tempurl}


\bibitem[Gupta et~al\mbox{.}(2021)]%
        {gupta2021adaptiveunlearning}
\bibfield{author}{\bibinfo{person}{Varun Gupta}, \bibinfo{person}{Christopher
  Jung}, \bibinfo{person}{Seth Neel}, \bibinfo{person}{Aaron Roth},
  \bibinfo{person}{Saeed Sharifi-Malvajerdi}, {and} \bibinfo{person}{Chris
  Waites}.} \bibinfo{year}{2021}\natexlab{}.
\newblock \showarticletitle{Adaptive Machine Unlearning}. In
  \bibinfo{booktitle}{\emph{Advances in Neural Information Processing
  Systems}}, \bibfield{editor}{\bibinfo{person}{M.~Ranzato},
  \bibinfo{person}{A.~Beygelzimer}, \bibinfo{person}{Y.~Dauphin},
  \bibinfo{person}{P.S. Liang}, {and} \bibinfo{person}{J.~Wortman Vaughan}}
  (Eds.), Vol.~\bibinfo{volume}{34}. \bibinfo{publisher}{Curran Associates,
  Inc.}, \bibinfo{address}{Red Hook, NY, USA}, \bibinfo{pages}{16319--16330}.
\newblock
\urldef\tempurl%
\url{https://proceedings.neurips.cc/paper_files/paper/2021/file/87f7ee4fdb57bdfd52179947211b7ebb-Paper.pdf}
\showURL{%
\tempurl}


\bibitem[Hanke et~al\mbox{.}(2024)]%
        {hanke2024openprivate}
\bibfield{author}{\bibinfo{person}{Vincent Hanke}, \bibinfo{person}{Tom
  Blanchard}, \bibinfo{person}{Franziska Boenisch}, \bibinfo{person}{Iyiola~E.
  Olatunji}, \bibinfo{person}{Michael Backes}, {and} \bibinfo{person}{Adam
  Dziedzic}.} \bibinfo{year}{2024}\natexlab{}.
\newblock \showarticletitle{Open {LLMs} are Necessary for Current Private
  Adaptations and Outperform their Closed Alternatives}. In
  \bibinfo{booktitle}{\emph{Advances in Neural Information Processing
  Systems}}, Vol.~\bibinfo{volume}{37}. \bibinfo{publisher}{Neural Information
  Processing Systems Foundation, Inc. (NeurIPS)}, \bibinfo{address}{Vancouver,
  BC, Canada}, \bibinfo{pages}{1220--1250}.
\newblock
\href{https://doi.org/10.52202/079017-0038}{doi:\nolinkurl{10.52202/079017-0038}}


\bibitem[He et~al\mbox{.}(2026)]%
        {he2026agentsecuritysurvey}
\bibfield{author}{\bibinfo{person}{Feng He}, \bibinfo{person}{Tianqing Zhu},
  \bibinfo{person}{Dayong Ye}, \bibinfo{person}{Bo Liu},
  \bibinfo{person}{Wanlei Zhou}, {and} \bibinfo{person}{Philip~S. Yu}.}
  \bibinfo{year}{2026}\natexlab{}.
\newblock \showarticletitle{The Emerged Security and Privacy of {LLM} Agent: A
  Survey with Case Studies}.
\newblock \bibinfo{journal}{\emph{Comput. Surveys}} \bibinfo{volume}{58},
  \bibinfo{number}{6}, Article \bibinfo{articleno}{162} (\bibinfo{year}{2026}),
  \bibinfo{numpages}{36}~pages.
\newblock
\href{https://doi.org/10.1145/3773080}{doi:\nolinkurl{10.1145/3773080}}


\bibitem[Henzinger et~al\mbox{.}(2026)]%
        {henzinger2026concurrentcomposition}
\bibfield{author}{\bibinfo{person}{Monika Henzinger}, \bibinfo{person}{Roodabeh
  Safavi}, {and} \bibinfo{person}{Salil Vadhan}.}
  \bibinfo{year}{2026}\natexlab{}.
\newblock \showarticletitle{Concurrent Composition for Differentially Private
  Continual Mechanisms}.
\newblock \bibinfo{journal}{\emph{Proceedings of the ACM on Management of
  Data}} \bibinfo{volume}{4}, \bibinfo{number}{2}, Article
  \bibinfo{articleno}{99} (\bibinfo{year}{2026}), \bibinfo{numpages}{26}~pages.
\newblock
\href{https://doi.org/10.1145/3801895}{doi:\nolinkurl{10.1145/3801895}}


\bibitem[Hong et~al\mbox{.}(2024)]%
        {hong2024dpopt}
\bibfield{author}{\bibinfo{person}{Junyuan Hong},
  \bibinfo{person}{Jiachen~(Tianhao) Wang}, \bibinfo{person}{Chenhui Zhang},
  \bibinfo{person}{Zhangheng Li}, \bibinfo{person}{Bo Li}, {and}
  \bibinfo{person}{Zhangyang Wang}.} \bibinfo{year}{2024}\natexlab{}.
\newblock \showarticletitle{{DP}-{OPT}: Make Large Language Model Your
  Privacy-Preserving Prompt Engineer}. In \bibinfo{booktitle}{\emph{Proceedings
  of the Twelfth International Conference on Learning Representations}},
  Vol.~\bibinfo{volume}{2024}. \bibinfo{publisher}{OpenReview.net},
  \bibinfo{address}{Vienna, Austria}, \bibinfo{pages}{58003--58026}.
\newblock
\urldef\tempurl%
\url{https://proceedings.iclr.cc/paper_files/paper/2024/hash/fffe5a7804c40465ef2432386850c2c7-Abstract-Conference.html}
\showURL{%
\tempurl}


\bibitem[Hou et~al\mbox{.}(2025)]%
        {hou2025secfpp}
\bibfield{author}{\bibinfo{person}{Sizai Hou}, \bibinfo{person}{Songze Li},
  {and} \bibinfo{person}{Baturalp Buyukates}.} \bibinfo{year}{2025}\natexlab{}.
\newblock \bibinfo{title}{Privacy-preserving Prompt Personalization in
  Federated Learning for Multimodal Large Language Models}.
\newblock \bibinfo{howpublished}{arXiv preprint arXiv:2505.22447v1}.
\newblock
\urldef\tempurl%
\url{https://arxiv.org/abs/2505.22447v1}
\showURL{%
\tempurl}


\bibitem[Huang et~al\mbox{.}(2026b)]%
        {huang2026cipl}
\bibfield{author}{\bibinfo{person}{Tao Huang}, \bibinfo{person}{Guosen Wu},
  \bibinfo{person}{Guolong Zheng}, \bibinfo{person}{Jiayang Meng},
  \bibinfo{person}{Chen Hou}, \bibinfo{person}{Xu Yang},
  \bibinfo{person}{Xuechao Yang}, {and} \bibinfo{person}{Feng Xia}.}
  \bibinfo{year}{2026}\natexlab{b}.
\newblock \bibinfo{title}{{CIPL}: A Channel-Aware Framework for Recoverable
  Privacy Leakage in {LLM} Agents}.
\newblock \bibinfo{howpublished}{arXiv preprint arXiv:2609.21686v1}.
\newblock
\urldef\tempurl%
\url{https://arxiv.org/abs/2609.21686v1}
\showURL{%
\tempurl}


\bibitem[Huang et~al\mbox{.}(2026a)]%
        {huang2026homellama}
\bibfield{author}{\bibinfo{person}{Xinyu Huang}, \bibinfo{person}{Leming Shen},
  \bibinfo{person}{Zijing Ma}, {and} \bibinfo{person}{Yuanqing Zheng}.}
  \bibinfo{year}{2026}\natexlab{a}.
\newblock \showarticletitle{Towards Privacy-Preserving and Personalized Smart
  Homes via Tailored Small Language Models}.
\newblock \bibinfo{journal}{\emph{IEEE Transactions on Mobile Computing}}
  \bibinfo{volume}{25}, \bibinfo{number}{7} (\bibinfo{year}{2026}),
  \bibinfo{pages}{9505--9520}.
\newblock
\href{https://doi.org/10.1109/TMC.2026.3654976}{doi:\nolinkurl{10.1109/TMC.2026.3654976}}


\bibitem[Jiang et~al\mbox{.}(2026)]%
        {jiang2026privdpo}
\bibfield{author}{\bibinfo{person}{Yangfan Jiang}, \bibinfo{person}{Fei Wei},
  \bibinfo{person}{Ergute Bao}, \bibinfo{person}{Xiaokui Xiao},
  \bibinfo{person}{Yaliang Li}, {and} \bibinfo{person}{Bolin Ding}.}
  \bibinfo{year}{2026}\natexlab{}.
\newblock \bibinfo{title}{Private Direct Preference Optimization for {LLM}
  Alignment}.
\newblock \bibinfo{howpublished}{arXiv preprint arXiv:2608.05040v1 [cs.CR]}.
\newblock
\urldef\tempurl%
\url{https://arxiv.org/abs/2608.05040v1}
\showURL{%
\tempurl}
\newblock
\shownote{Accepted at the 2026 {ACM} {SIGSAC} Conference on Computer and
  Communications Security ({CCS} 2026)}.


\bibitem[Jin et~al\mbox{.}(2026)]%
        {jin2026privpair}
\bibfield{author}{\bibinfo{person}{Heng Jin}, \bibinfo{person}{Chaoyu Zhang},
  \bibinfo{person}{Hexuan Yu}, \bibinfo{person}{Wenjing Lou}, {and}
  \bibinfo{person}{Y.~Thomas Hou}.} \bibinfo{year}{2026}\natexlab{}.
\newblock \bibinfo{title}{Privacy-Preserving Split Learning for Federated {LLM}
  Fine-Tuning}.
\newblock \bibinfo{howpublished}{arXiv preprint arXiv:2609.09794v1}.
\newblock
\urldef\tempurl%
\url{https://arxiv.org/abs/2609.09794v1}
\showURL{%
\tempurl}


\bibitem[Johnson et~al\mbox{.}(2026)]%
        {johnson2026aim}
\bibfield{author}{\bibinfo{person}{Zachary Johnson},
  \bibinfo{person}{Nigel~Boachie Kumankumah}, \bibinfo{person}{Somya
  Chatterjee}, \bibinfo{person}{Tejas Sathyamurthi}, \bibinfo{person}{Min
  Chen}, \bibinfo{person}{Xinyi~Alice Li}, \bibinfo{person}{Xiao Wang},
  \bibinfo{person}{Emily~Morgan Gelchie}, \bibinfo{person}{Jessica Lin},
  \bibinfo{person}{Sadid~A. Hasan}, {and} \bibinfo{person}{Sulaiman Vesal}.}
  \bibinfo{year}{2026}\natexlab{}.
\newblock \bibinfo{title}{{AIM}: A Privacy-Aware Interoperable Memory Framework
  for Multi-Agent Multi-User {LLM} Systems}.
\newblock \bibinfo{howpublished}{arXiv preprint arXiv:2609.12320v1}.
\newblock
\href{https://doi.org/10.48550/arXiv.2609.12320}{doi:\nolinkurl{10.48550/arXiv.2609.12320}}


\bibitem[Kairouz et~al\mbox{.}(2021)]%
        {kairouz2021dpftrl}
\bibfield{author}{\bibinfo{person}{Peter Kairouz}, \bibinfo{person}{Brendan
  McMahan}, \bibinfo{person}{Shuang Song}, \bibinfo{person}{Om Thakkar},
  \bibinfo{person}{Abhradeep Thakurta}, {and} \bibinfo{person}{Zheng Xu}.}
  \bibinfo{year}{2021}\natexlab{}.
\newblock \showarticletitle{Practical and Private (Deep) Learning Without
  Sampling or Shuffling}. In \bibinfo{booktitle}{\emph{Proceedings of the 38th
  International Conference on Machine Learning}}
  \emph{(\bibinfo{series}{Proceedings of Machine Learning Research},
  Vol.~\bibinfo{volume}{139})}, \bibfield{editor}{\bibinfo{person}{Marina
  Meila} {and} \bibinfo{person}{Tong Zhang}} (Eds.). \bibinfo{publisher}{PMLR},
  \bibinfo{address}{Virtual Event}, \bibinfo{pages}{5213--5225}.
\newblock
\urldef\tempurl%
\url{https://proceedings.mlr.press/v139/kairouz21b.html}
\showURL{%
\tempurl}


\bibitem[Kaneko et~al\mbox{.}(2026)]%
        {kaneko2026disclosureloop}
\bibfield{author}{\bibinfo{person}{Masahiro Kaneko}, \bibinfo{person}{Hiromi
  Arai}, {and} \bibinfo{person}{Timothy Baldwin}.}
  \bibinfo{year}{2026}\natexlab{}.
\newblock \bibinfo{title}{Disclosure Begets Disclosure: A Self-Reinforcing
  Privacy Loop in {LLM} Agents with Persistent Memory}.
\newblock \bibinfo{howpublished}{arXiv preprint arXiv:2610.28315v2 [cs.CR]}.
\newblock
\urldef\tempurl%
\url{https://arxiv.org/abs/2610.28315v2}
\showURL{%
\tempurl}


\bibitem[Kanpak et~al\mbox{.}(2026)]%
        {kanpak2026heoft}
\bibfield{author}{\bibinfo{person}{Halil~{\.{I}}brahim Kanpak},
  \bibinfo{person}{Sinem Sav}, {and} \bibinfo{person}{Alptekin
  K{\"u}p{\c{c}}{\"u}}.} \bibinfo{year}{2026}\natexlab{}.
\newblock \bibinfo{title}{{HE-OFT}: Privacy-Preserving One-Shot Federated
  Fine-Tuning under Homomorphic Encryption}.
\newblock \bibinfo{howpublished}{arXiv preprint arXiv:2610.08255v1}.
\newblock
\urldef\tempurl%
\url{https://arxiv.org/abs/2610.08255v1}
\showURL{%
\tempurl}


\bibitem[Kaviani et~al\mbox{.}(2026)]%
        {kaviani2026opal}
\bibfield{author}{\bibinfo{person}{Darya Kaviani}, \bibinfo{person}{Alp~Eren
  Ozdarendeli}, \bibinfo{person}{Jinhao Zhu}, \bibinfo{person}{Yu Ding}, {and}
  \bibinfo{person}{Raluca~Ada Popa}.} \bibinfo{year}{2026}\natexlab{}.
\newblock \bibinfo{title}{{Opal}: Private Memory for Personal {AI}}.
\newblock \bibinfo{howpublished}{arXiv preprint arXiv:2604.02522v1 [cs.CR]}.
\newblock
\urldef\tempurl%
\url{https://arxiv.org/abs/2604.02522v1}
\showURL{%
\tempurl}


\bibitem[Kifer and Machanavajjhala(2014)]%
        {kifer2014pufferfish}
\bibfield{author}{\bibinfo{person}{Daniel Kifer} {and} \bibinfo{person}{Ashwin
  Machanavajjhala}.} \bibinfo{year}{2014}\natexlab{}.
\newblock \showarticletitle{{Pufferfish}: A Framework for Mathematical Privacy
  Definitions}.
\newblock \bibinfo{journal}{\emph{ACM Transactions on Database Systems}}
  \bibinfo{volume}{39}, \bibinfo{number}{1}, Article \bibinfo{articleno}{3}
  (\bibinfo{year}{2014}), \bibinfo{numpages}{36}~pages.
\newblock
\href{https://doi.org/10.1145/2514689}{doi:\nolinkurl{10.1145/2514689}}


\bibitem[Kim et~al\mbox{.}(2026)]%
        {kim2026dpmemview}
\bibfield{author}{\bibinfo{person}{Jong~Wook Kim}, \bibinfo{person}{Byoungjae
  Min}, \bibinfo{person}{Kennedy Edemacu}, \bibinfo{person}{Yoonhyuk Choi},
  \bibinfo{person}{Sae-Hong Cho}, {and} \bibinfo{person}{Beakcheol Jang}.}
  \bibinfo{year}{2026}\natexlab{}.
\newblock \bibinfo{title}{{DP}-{MemView}: A Memory Interface for
  Attribute-Level Transcript Privacy in Long-Term {LLM} Agents}.
\newblock \bibinfo{howpublished}{arXiv preprint arXiv:2608.03130v1 [cs.CR]}.
\newblock
\urldef\tempurl%
\url{https://arxiv.org/abs/2608.03130v1}
\showURL{%
\tempurl}


\bibitem[Kim et~al\mbox{.}(2025)]%
        {kim2025prime}
\bibfield{author}{\bibinfo{person}{Kyuyoung Kim}, \bibinfo{person}{Jinwoo
  Shin}, {and} \bibinfo{person}{Jaehyung Kim}.}
  \bibinfo{year}{2025}\natexlab{}.
\newblock \showarticletitle{Personalized Language Models via Privacy-Preserving
  Evolutionary Model Merging}. In \bibinfo{booktitle}{\emph{Proceedings of the
  2025 Conference on Empirical Methods in Natural Language Processing}}.
  \bibinfo{publisher}{Association for Computational Linguistics},
  \bibinfo{address}{Suzhou, China}, \bibinfo{pages}{34453--34468}.
\newblock
\href{https://doi.org/10.18653/v1/2025.emnlp-main.1747}{doi:\nolinkurl{10.18653/v1/2025.emnlp-main.1747}}


\bibitem[Lahjouji and Colaco(2026)]%
        {lahjouji2026agentsknowtoomuch}
\bibfield{author}{\bibinfo{person}{Nada Lahjouji} {and}
  \bibinfo{person}{Ashwin~Gerard Colaco}.} \bibinfo{year}{2026}\natexlab{}.
\newblock \bibinfo{title}{Agents That Know Too Much: A Data-Centric Survey of
  Privacy in {LLM} Agents}.
\newblock \bibinfo{howpublished}{arXiv preprint arXiv:2606.26627v1 [cs.CR]}.
\newblock
\urldef\tempurl%
\url{https://arxiv.org/abs/2606.26627v1}
\showURL{%
\tempurl}


\bibitem[Lai et~al\mbox{.}(2022)]%
        {lai2022lifelongdp}
\bibfield{author}{\bibinfo{person}{Phung Lai}, \bibinfo{person}{Han Hu},
  \bibinfo{person}{Hai Phan}, \bibinfo{person}{Ruoming Jin},
  \bibinfo{person}{My Thai}, {and} \bibinfo{person}{An Chen}.}
  \bibinfo{year}{2022}\natexlab{}.
\newblock \showarticletitle{Lifelong {DP}: Consistently Bounded Differential
  Privacy in Lifelong Machine Learning}. In
  \bibinfo{booktitle}{\emph{Proceedings of The 1st Conference on Lifelong
  Learning Agents}} \emph{(\bibinfo{series}{Proceedings of Machine Learning
  Research}, Vol.~\bibinfo{volume}{199})},
  \bibfield{editor}{\bibinfo{person}{Sarath Chandar}, \bibinfo{person}{Razvan
  Pascanu}, {and} \bibinfo{person}{Doina Precup}} (Eds.).
  \bibinfo{publisher}{PMLR}, \bibinfo{address}{Montreal, QC, Canada},
  \bibinfo{pages}{778--797}.
\newblock
\urldef\tempurl%
\url{https://proceedings.mlr.press/v199/lai22a.html}
\showURL{%
\tempurl}


\bibitem[Lan et~al\mbox{.}(2026)]%
        {lan2026fedskillgit}
\bibfield{author}{\bibinfo{person}{Boxun Lan}, \bibinfo{person}{Jiamei Li},
  \bibinfo{person}{Weijian Li}, \bibinfo{person}{Chengyu Yang},
  \bibinfo{person}{Zhuang Qi}, \bibinfo{person}{Chengcheng Wan},
  \bibinfo{person}{Xiaohong Chen}, {and} \bibinfo{person}{Ming Hu}.}
  \bibinfo{year}{2026}\natexlab{}.
\newblock \bibinfo{title}{{FedSkillGit}: Federated Skill Evolution via Version
  Control}.
\newblock \bibinfo{howpublished}{arXiv preprint arXiv:2610.21348v1 [cs.AI]}.
\newblock
\urldef\tempurl%
\url{https://arxiv.org/abs/2610.21348v1}
\showURL{%
\tempurl}


\bibitem[Li et~al\mbox{.}(2026a)]%
        {li2026healthjev}
\bibfield{author}{\bibinfo{person}{Haoran Li}, \bibinfo{person}{Jiebi Deng},
  \bibinfo{person}{Tong Jin}, \bibinfo{person}{Jinghong Han},
  \bibinfo{person}{Yuxin Wang}, \bibinfo{person}{Zexin Wang},
  \bibinfo{person}{Jinyu Sun}, \bibinfo{person}{Qingyi Si},
  \bibinfo{person}{Weikang Gong}, \bibinfo{person}{Xiahai Zhuang},
  \bibinfo{person}{Jia You}, \bibinfo{person}{Wei Cheng},
  \bibinfo{person}{Jianfeng Feng}, {and} \bibinfo{person}{Hongcheng Guo}.}
  \bibinfo{year}{2026}\natexlab{a}.
\newblock \bibinfo{title}{A Self-Evolving Agent for Longitudinal Personal
  Health Management}.
\newblock \bibinfo{howpublished}{arXiv preprint arXiv:2607.13940v2 [cs.AI]}.
\newblock
\urldef\tempurl%
\url{https://arxiv.org/abs/2607.13940v2}
\showURL{%
\tempurl}


\bibitem[Li et~al\mbox{.}(2026c)]%
        {li2026umpeek}
\bibfield{author}{\bibinfo{person}{Haoyang Li}, \bibinfo{person}{Yaxin Xiao},
  \bibinfo{person}{Qingqing Ye}, \bibinfo{person}{Huadi Zheng}, {and}
  \bibinfo{person}{Haibo Hu}.} \bibinfo{year}{2026}\natexlab{c}.
\newblock \bibinfo{title}{Inferring Hidden User Models from the Behavior of
  Personalized {LLM} Agents}.
\newblock \bibinfo{howpublished}{arXiv preprint arXiv:2609.03815v1}.
\newblock
\urldef\tempurl%
\url{https://arxiv.org/abs/2609.03815v1}
\showURL{%
\tempurl}


\bibitem[Li et~al\mbox{.}(2026b)]%
        {li2026profiledp}
\bibfield{author}{\bibinfo{person}{Mingchen Li}, \bibinfo{person}{Zifan Peng},
  \bibinfo{person}{Kewei Sha}, \bibinfo{person}{Lavinia~F. Pieptea},
  \bibinfo{person}{Heng Fan}, \bibinfo{person}{Junhua Ding},
  \bibinfo{person}{Meikang Qiu}, {and} \bibinfo{person}{Yunhe Feng}.}
  \bibinfo{year}{2026}\natexlab{b}.
\newblock \bibinfo{title}{{Profile-DP}: Differential Privacy for
  User-Controlled Profile Release in Personalized {LLM} Inference}.
\newblock \bibinfo{howpublished}{arXiv preprint arXiv:2610.27213v1}.
\newblock
\urldef\tempurl%
\url{https://arxiv.org/abs/2610.27213v1}
\showURL{%
\tempurl}


\bibitem[Li et~al\mbox{.}(2026d)]%
        {li2026prag}
\bibfield{author}{\bibinfo{person}{Zhijun Li}, \bibinfo{person}{Minghui Xu},
  \bibinfo{person}{Huayi Qi}, \bibinfo{person}{Wenxuan Yu},
  \bibinfo{person}{Tingchuang Zhang}, \bibinfo{person}{Qiao Zhang},
  \bibinfo{person}{GuangYong Shang}, \bibinfo{person}{Zhen Ma}, {and}
  \bibinfo{person}{Xiuzhen Cheng}.} \bibinfo{year}{2026}\natexlab{d}.
\newblock \bibinfo{title}{{PRAG}: End-to-End Privacy-Preserving
  Retrieval-Augmented Generation}.
\newblock \bibinfo{howpublished}{arXiv preprint arXiv:2604.26525v2 [cs.CR]}.
\newblock
\urldef\tempurl%
\url{https://arxiv.org/abs/2604.26525v2}
\showURL{%
\tempurl}


\bibitem[Liang et~al\mbox{.}(2026)]%
        {liang2026pahf}
\bibfield{author}{\bibinfo{person}{Kaiqu Liang}, \bibinfo{person}{Julia Kruk},
  \bibinfo{person}{Shengyi Qian}, \bibinfo{person}{Xianjun Yang},
  \bibinfo{person}{Shengjie Bi}, \bibinfo{person}{Yuanshun Yao},
  \bibinfo{person}{Shaoliang Nie}, \bibinfo{person}{Mingyang Zhang},
  \bibinfo{person}{Lijuan Liu}, \bibinfo{person}{Jaime Fern{\'a}ndez~Fisac},
  \bibinfo{person}{Shuyan Zhou}, {and} \bibinfo{person}{Saghar Hosseini}.}
  \bibinfo{year}{2026}\natexlab{}.
\newblock \bibinfo{title}{Learning Personalized Agents from Human Feedback}.
\newblock \bibinfo{howpublished}{arXiv preprint arXiv:2602.16173v1}.
\newblock
\href{https://doi.org/10.48550/arXiv.2602.16173}{doi:\nolinkurl{10.48550/arXiv.2602.16173}}


\bibitem[Lin et~al\mbox{.}(2026c)]%
        {lin2026aubenchcontext}
\bibfield{author}{\bibinfo{person}{Junkai Lin}, \bibinfo{person}{Ge Xu},
  \bibinfo{person}{Zheng Xu}, \bibinfo{person}{Junkai Chen}, {and}
  \bibinfo{person}{Xiangyuan Lan}.} \bibinfo{year}{2026}\natexlab{c}.
\newblock \bibinfo{title}{Can Agents Truly Forget? Explore Machine Unlearning
  for {LLM} Agents}.
\newblock \bibinfo{howpublished}{arXiv preprint arXiv:2610.17659v1 [cs.AI]}.
\newblock
\href{https://doi.org/10.48550/arXiv.2610.17659}{doi:\nolinkurl{10.48550/arXiv.2610.17659}}


\bibitem[Lin et~al\mbox{.}(2026a)]%
        {lin2026memorysecuritysurvey}
\bibfield{author}{\bibinfo{person}{Zehao Lin}, \bibinfo{person}{Xixuan Hao},
  \bibinfo{person}{Renyu Fu}, \bibinfo{person}{Shaobo Cui},
  \bibinfo{person}{Kai Chen}, \bibinfo{person}{Chunyu Li},
  \bibinfo{person}{Zhiyu Li}, {and} \bibinfo{person}{Feiyu Xiong}.}
  \bibinfo{year}{2026}\natexlab{a}.
\newblock \showarticletitle{A Survey on Long-Term Memory Security in {LLM}
  Agents: Attacks, Defenses, and Governance Across the Memory Lifecycle}. In
  \bibinfo{booktitle}{\emph{Proceedings of the 2026 Conference on Empirical
  Methods in Natural Language Processing}},
  \bibfield{editor}{\bibinfo{person}{Hung-Yi Lee}, \bibinfo{person}{Sunipa
  Dev}, \bibinfo{person}{Junyi~Jessy Li}, {and} \bibinfo{person}{Desmond
  Elliott}} (Eds.). \bibinfo{publisher}{Association for Computational
  Linguistics}, \bibinfo{address}{Budapest, Hungary},
  \bibinfo{pages}{52221--52235}.
\newblock
\showISBNx{979-8-89176-459-0}
\href{https://doi.org/10.18653/2026.emnlp-main.2296}{doi:\nolinkurl{10.18653/2026.emnlp-main.2296}}


\bibitem[Lin et~al\mbox{.}(2026b)]%
        {lin2026tipo}
\bibfield{author}{\bibinfo{person}{Zhixin Lin}, \bibinfo{person}{Jungang Li},
  \bibinfo{person}{Dongliang Xu}, \bibinfo{person}{Shidong Pan},
  \bibinfo{person}{Yibo Shi}, \bibinfo{person}{Yuchi Liu},
  \bibinfo{person}{Yuecong Min}, {and} \bibinfo{person}{Yue Yao}.}
  \bibinfo{year}{2026}\natexlab{b}.
\newblock \bibinfo{title}{Mobile {GUI} Agent Privacy Personalization with
  Trajectory Induced Preference Optimization}.
\newblock \bibinfo{howpublished}{arXiv preprint arXiv:2604.11259v1}.
\newblock
\urldef\tempurl%
\url{https://arxiv.org/abs/2604.11259v1}
\showURL{%
\tempurl}


\bibitem[Liu and Ding(2026)]%
        {liu2026memoryadmission}
\bibfield{author}{\bibinfo{person}{Chang Liu} {and} \bibinfo{person}{Deliang
  Ding}.} \bibinfo{year}{2026}\natexlab{}.
\newblock \bibinfo{title}{What to Admit and How to Present: Governing
  Persistent Memory in {LLM} Agents}.
\newblock \bibinfo{howpublished}{arXiv preprint arXiv:2610.11188v1}.
\newblock
\urldef\tempurl%
\url{https://arxiv.org/abs/2610.11188v1}
\showURL{%
\tempurl}


\bibitem[Liu(2026a)]%
        {liu2026sovereignpa}
\bibfield{author}{\bibinfo{person}{Dylan~Zongmin Liu}.}
  \bibinfo{year}{2026}\natexlab{a}.
\newblock \bibinfo{title}{{SovereignPA-Bench}: Evaluating User-Owned Personal
  Agents under Evolving Intent, Platform Mediation, and Consent Constraints}.
\newblock \bibinfo{howpublished}{arXiv preprint arXiv:2607.05363v1}.
\newblock
\urldef\tempurl%
\url{https://arxiv.org/abs/2607.05363v1}
\showURL{%
\tempurl}


\bibitem[Liu and Talwar(2019)]%
        {liu2019privateselection}
\bibfield{author}{\bibinfo{person}{Jingcheng Liu} {and} \bibinfo{person}{Kunal
  Talwar}.} \bibinfo{year}{2019}\natexlab{}.
\newblock \showarticletitle{Private Selection from Private Candidates}. In
  \bibinfo{booktitle}{\emph{Proceedings of the 51st Annual {ACM} {SIGACT}
  Symposium on Theory of Computing}} (Phoenix, AZ, USA).
  \bibinfo{publisher}{Association for Computing Machinery},
  \bibinfo{address}{New York, NY, USA}, \bibinfo{pages}{298--309}.
\newblock
\href{https://doi.org/10.1145/3313276.3316377}{doi:\nolinkurl{10.1145/3313276.3316377}}


\bibitem[Liu(2026b)]%
        {liu2026audiencememory}
\bibfield{author}{\bibinfo{person}{Sibo Liu}.}
  \bibinfo{year}{2026}\natexlab{b}.
\newblock \bibinfo{title}{Audience-Bound Persistent Memory: Authorization
  Across the Memory Lifecycle}.
\newblock \bibinfo{howpublished}{arXiv preprint arXiv:2609.36373v1}.
\newblock
\urldef\tempurl%
\url{https://arxiv.org/abs/2609.36373v1}
\showURL{%
\tempurl}


\bibitem[Liu et~al\mbox{.}(2026a)]%
        {liu2026privtune}
\bibfield{author}{\bibinfo{person}{Yi Liu}, \bibinfo{person}{Weixiang Han},
  \bibinfo{person}{Chengjun Cai}, \bibinfo{person}{Xingliang Yuan}, {and}
  \bibinfo{person}{Cong Wang}.} \bibinfo{year}{2026}\natexlab{a}.
\newblock \showarticletitle{{PrivTune}: Efficient and Privacy-Preserving
  Fine-Tuning of Large Language Models via Device-Cloud Collaboration}. In
  \bibinfo{booktitle}{\emph{2026 {IEEE} Conference on Computer Communications
  ({INFOCOM})}} (Tokyo, Japan). \bibinfo{publisher}{IEEE},
  \bibinfo{address}{Piscataway, NJ, USA}, \bibinfo{pages}{1--10}.
\newblock
\href{https://doi.org/10.1109/INFOCOM59046.2026.11571688}{doi:\nolinkurl{10.1109/INFOCOM59046.2026.11571688}}


\bibitem[Liu et~al\mbox{.}(2026b)]%
        {liu2026dpes}
\bibfield{author}{\bibinfo{person}{Ziniu Liu}, \bibinfo{person}{Aiping Li},
  \bibinfo{person}{Yue Han}, \bibinfo{person}{Han Yu}, \bibinfo{person}{Junjian
  Zhang}, \bibinfo{person}{Dong Zhu}, \bibinfo{person}{Changjian Li}, {and}
  \bibinfo{person}{Shiqiang Zhang}.} \bibinfo{year}{2026}\natexlab{b}.
\newblock \showarticletitle{{DP}-{ES}: Differentially Private Evolution
  Strategies for Prompt Optimization}. In \bibinfo{booktitle}{\emph{Proceedings
  of the 2026 Conference on Empirical Methods in Natural Language Processing}},
  \bibfield{editor}{\bibinfo{person}{Hung-Yi Lee}, \bibinfo{person}{Sunipa
  Dev}, \bibinfo{person}{Junyi~Jessy Li}, {and} \bibinfo{person}{Desmond
  Elliott}} (Eds.). \bibinfo{publisher}{Association for Computational
  Linguistics}, \bibinfo{address}{Budapest, Hungary},
  \bibinfo{pages}{5529--5545}.
\newblock
\showISBNx{979-8-89176-459-0}
\href{https://doi.org/10.18653/2026.emnlp-main.242}{doi:\nolinkurl{10.18653/2026.emnlp-main.242}}


\bibitem[Luo et~al\mbox{.}(2026)]%
        {luo2026secptuning}
\bibfield{author}{\bibinfo{person}{Jinglong Luo}, \bibinfo{person}{Zhuo Zhang},
  \bibinfo{person}{Yehong Zhang}, \bibinfo{person}{Shiyu Liu},
  \bibinfo{person}{Ye Dong}, \bibinfo{person}{Hui Wang}, \bibinfo{person}{Yue
  Yu}, \bibinfo{person}{Xun Zhou}, {and} \bibinfo{person}{Zenglin Xu}.}
  \bibinfo{year}{2026}\natexlab{}.
\newblock \showarticletitle{{SecP-Tuning}: Efficient Privacy-Preserving Prompt
  Tuning for Large Language Models via {MPC}}. In
  \bibinfo{booktitle}{\emph{Proceedings of the Fourteenth International
  Conference on Learning Representations}}, Vol.~\bibinfo{volume}{2026}.
  \bibinfo{publisher}{OpenReview.net}, \bibinfo{address}{Rio de Janeiro,
  Brazil}, \bibinfo{pages}{78760--78780}.
\newblock
\urldef\tempurl%
\url{https://proceedings.iclr.cc/paper_files/paper/2026/hash/7f2e5d16c2c5ac161ce7aa44af35c0fe-Abstract-Conference.html}
\showURL{%
\tempurl}


\bibitem[Lyu et~al\mbox{.}(2026)]%
        {lyu2026ragct}
\bibfield{author}{\bibinfo{person}{Xingyu Lyu}, \bibinfo{person}{Jiayimei
  Wang}, \bibinfo{person}{Jianfeng He}, \bibinfo{person}{Ning Wang},
  \bibinfo{person}{Yidan Hu}, {and} \bibinfo{person}{Yimin Chen}.}
  \bibinfo{year}{2026}\natexlab{}.
\newblock \bibinfo{title}{{RAG-CT}: Mitigating Privacy Risks on
  Retrieval-Augmented Generation Systems via Scanning Prompt Distribution}.
\newblock \bibinfo{howpublished}{arXiv:2609.16095v1 [cs.CL]}.
\newblock
\urldef\tempurl%
\url{https://arxiv.org/abs/2609.16095v1}
\showURL{%
\tempurl}


\bibitem[Ma et~al\mbox{.}(2026)]%
        {ma2026skillv}
\bibfield{author}{\bibinfo{person}{Jie Ma}, \bibinfo{person}{Zhipeng Qian},
  \bibinfo{person}{Yufei Ma}, \bibinfo{person}{Zihan Liang},
  \bibinfo{person}{Jiayi Ji}, \bibinfo{person}{Qingpeng Cai},
  \bibinfo{person}{Ben Chen}, \bibinfo{person}{Peng Jiang}, {and}
  \bibinfo{person}{Xiaoshuai Sun}.} \bibinfo{year}{2026}\natexlab{}.
\newblock \bibinfo{title}{{Skill-V}: Verifiable Self-Evolving Skill Library for
  Interactive Agents}.
\newblock \bibinfo{howpublished}{arXiv preprint arXiv:2610.11781v1 [cs.CV]}.
\newblock
\urldef\tempurl%
\url{https://arxiv.org/abs/2610.11781v1}
\showURL{%
\tempurl}


\bibitem[Ma et~al\mbox{.}(2024)]%
        {ma2024privatepeft}
\bibfield{author}{\bibinfo{person}{Olivia Ma}, \bibinfo{person}{Jonathan
  Passerat-Palmbach}, {and} \bibinfo{person}{Dmitrii Usynin}.}
  \bibinfo{year}{2024}\natexlab{}.
\newblock \bibinfo{title}{Efficient and Private: Memorisation under
  differentially private parameter-efficient fine-tuning in language models}.
\newblock \bibinfo{howpublished}{arXiv preprint arXiv:2411.15831v1}.
\newblock
\urldef\tempurl%
\url{https://arxiv.org/abs/2411.15831v1}
\showURL{%
\tempurl}


\bibitem[Malekzadeh et~al\mbox{.}(2020)]%
        {malekzadeh2020p2b}
\bibfield{author}{\bibinfo{person}{Mohammad Malekzadeh},
  \bibinfo{person}{Dimitrios Athanasakis}, \bibinfo{person}{Hamed Haddadi},
  {and} \bibinfo{person}{Ben Livshits}.} \bibinfo{year}{2020}\natexlab{}.
\newblock \showarticletitle{Privacy-Preserving Bandits}. In
  \bibinfo{booktitle}{\emph{Proceedings of Machine Learning and Systems}},
  \bibfield{editor}{\bibinfo{person}{Inderjit~S. Dhillon},
  \bibinfo{person}{Dimitris~S. Papailiopoulos}, {and} \bibinfo{person}{Vivienne
  Sze}} (Eds.), Vol.~\bibinfo{volume}{2}. \bibinfo{publisher}{mlsys.org},
  \bibinfo{address}{Austin, TX, USA}, \bibinfo{pages}{350--362}.
\newblock
\urldef\tempurl%
\url{https://proceedings.mlsys.org/paper_files/paper/2020/hash/6ba1453022da1faf642d4de27241d53f-Abstract.html}
\showURL{%
\tempurl}


\bibitem[Mireshghallah et~al\mbox{.}(2026)]%
        {mireshghallah2026cimemories}
\bibfield{author}{\bibinfo{person}{Niloofar Mireshghallah},
  \bibinfo{person}{Neal Mangaokar}, \bibinfo{person}{Narine Kokhlikyan},
  \bibinfo{person}{Arman Zharmagambetov}, \bibinfo{person}{Manzil Zaheer},
  \bibinfo{person}{Saeed Mahloujifar}, {and} \bibinfo{person}{Kamalika
  Chaudhuri}.} \bibinfo{year}{2026}\natexlab{}.
\newblock \showarticletitle{{CIMemories}: A Compositional Benchmark for
  Contextual Integrity in {LLMs}}. In \bibinfo{booktitle}{\emph{The Fourteenth
  International Conference on Learning Representations}}.
  \bibinfo{publisher}{OpenReview.net}, \bibinfo{address}{Rio de Janeiro,
  Brazil}, \bibinfo{numpages}{20}~pages.
\newblock
\urldef\tempurl%
\url{https://openreview.net/forum?id=YnNIp38v1M}
\showURL{%
\tempurl}


\bibitem[Mishra et~al\mbox{.}(2018)]%
        {mishra2018oblix}
\bibfield{author}{\bibinfo{person}{Pratyush Mishra}, \bibinfo{person}{Rishabh
  Poddar}, \bibinfo{person}{Jerry Chen}, \bibinfo{person}{Alessandro Chiesa},
  {and} \bibinfo{person}{Raluca~Ada Popa}.} \bibinfo{year}{2018}\natexlab{}.
\newblock \showarticletitle{{Oblix}: An Efficient Oblivious Search Index}. In
  \bibinfo{booktitle}{\emph{2018 {IEEE} Symposium on Security and Privacy
  ({SP})}} (San Francisco, CA, USA). \bibinfo{publisher}{IEEE},
  \bibinfo{address}{Piscataway, NJ, USA}, \bibinfo{pages}{279--296}.
\newblock
\href{https://doi.org/10.1109/sp.2018.00045}{doi:\nolinkurl{10.1109/sp.2018.00045}}


\bibitem[Mohammadi et~al\mbox{.}(2025)]%
        {mohammadi2025efu}
\bibfield{author}{\bibinfo{person}{Samaneh Mohammadi},
  \bibinfo{person}{Vasileios Tsouvalas}, \bibinfo{person}{Iraklis Symeonidis},
  \bibinfo{person}{Ali Balador}, \bibinfo{person}{Tanir Ozcelebi},
  \bibinfo{person}{Francesco Flammini}, {and} \bibinfo{person}{Nirvana
  Meratnia}.} \bibinfo{year}{2025}\natexlab{}.
\newblock \showarticletitle{{EFU}: Enforcing Federated Unlearning via
  Functional Encryption}. In \bibinfo{booktitle}{\emph{Proceedings of the 34th
  {ACM} International Conference on Information and Knowledge Management}}
  (Seoul, Republic of Korea). \bibinfo{publisher}{Association for Computing
  Machinery}, \bibinfo{address}{New York, NY, USA},
  \bibinfo{pages}{2148--2158}.
\newblock
\href{https://doi.org/10.1145/3746252.3761091}{doi:\nolinkurl{10.1145/3746252.3761091}}


\bibitem[Odersky et~al\mbox{.}(2026)]%
        {odersky2026trackedcapabilities}
\bibfield{author}{\bibinfo{person}{Martin Odersky}, \bibinfo{person}{Yaoyu
  Zhao}, \bibinfo{person}{Yichen Xu}, \bibinfo{person}{Oliver Bra{\v{c}}evac},
  {and} \bibinfo{person}{Cao~Nguyen Pham}.} \bibinfo{year}{2026}\natexlab{}.
\newblock \showarticletitle{Securing Agents with Tracked Capabilities}. In
  \bibinfo{booktitle}{\emph{Proceedings of the {ACM} Conference on {AI} and
  Agentic Systems}} (San Jose, CA, USA). \bibinfo{publisher}{Association for
  Computing Machinery}, \bibinfo{address}{New York, NY, USA},
  \bibinfo{pages}{812--838}.
\newblock
\href{https://doi.org/10.1145/3786335.3813127}{doi:\nolinkurl{10.1145/3786335.3813127}}


\bibitem[Oswal and Cadeddu(2026)]%
        {oswal2026fieldtier}
\bibfield{author}{\bibinfo{person}{Jiten Oswal} {and} \bibinfo{person}{John
  Cadeddu}.} \bibinfo{year}{2026}\natexlab{}.
\newblock \bibinfo{title}{Separating Disclosure from Authorization: Field-Tier
  Minimization for Agent Action Mediation}.
\newblock \bibinfo{howpublished}{arXiv preprint arXiv:2608.25474v1}.
\newblock
\urldef\tempurl%
\url{https://arxiv.org/abs/2608.25474v1}
\showURL{%
\tempurl}


\bibitem[Pan et~al\mbox{.}(2026)]%
        {pan2026skillguard}
\bibfield{author}{\bibinfo{person}{Shidong Pan}, \bibinfo{person}{Xiaoyu Sun},
  \bibinfo{person}{Tianyi Zhang}, \bibinfo{person}{Dianshu Liao},
  \bibinfo{person}{Kaiwen Yang}, {and} \bibinfo{person}{Zhenchang Xing}.}
  \bibinfo{year}{2026}\natexlab{}.
\newblock \bibinfo{title}{{SkillGuard}: A Permission-Centric Framework for
  Agent Skill Security}.
\newblock \bibinfo{howpublished}{arXiv preprint arXiv:2606.03024v2}.
\newblock
\urldef\tempurl%
\url{https://arxiv.org/abs/2606.03024v2}
\showURL{%
\tempurl}


\bibitem[Park et~al\mbox{.}(2026)]%
        {park2026prisp}
\bibfield{author}{\bibinfo{person}{Junho Park}, \bibinfo{person}{Dohoon Kim},
  {and} \bibinfo{person}{Taesup Moon}.} \bibinfo{year}{2026}\natexlab{}.
\newblock \showarticletitle{{PRISP}: Privacy-Safe Few-Shot Personalization via
  Lightweight Adaptation}. In \bibinfo{booktitle}{\emph{Proceedings of the 64th
  Annual Meeting of the Association for Computational Linguistics (Volume 1:
  Long Papers)}}. \bibinfo{publisher}{Association for Computational
  Linguistics}, \bibinfo{address}{San Diego, CA, USA},
  \bibinfo{pages}{24986--25003}.
\newblock
\href{https://doi.org/10.18653/v1/2026.acl-long.1146}{doi:\nolinkurl{10.18653/v1/2026.acl-long.1146}}


\bibitem[Qi et~al\mbox{.}(2026)]%
        {qi2026surveyalignment}
\bibfield{author}{\bibinfo{person}{Xiuyuan Qi}, \bibinfo{person}{Yi Wang},
  {and} \bibinfo{person}{Wenjie Wang}.} \bibinfo{year}{2026}\natexlab{}.
\newblock \bibinfo{title}{Differential Privacy Protected Preference Alignment
  on Survey Derived Data}.
\newblock \bibinfo{howpublished}{arXiv:2610.16458v1 [cs.CR]}.
\newblock
\urldef\tempurl%
\url{https://arxiv.org/abs/2610.16458v1}
\showURL{%
\tempurl}


\bibitem[Qiao et~al\mbox{.}(2026)]%
        {qiao2026topbench}
\bibfield{author}{\bibinfo{person}{Yuxuan Qiao}, \bibinfo{person}{Dongqin Liu},
  \bibinfo{person}{Hongchang Yang}, \bibinfo{person}{Wei Zhou}, {and}
  \bibinfo{person}{Songlin Hu}.} \bibinfo{year}{2026}\natexlab{}.
\newblock \showarticletitle{Agent Tools Orchestration Leaks More: Dataset,
  Benchmark, and Mitigation}. In \bibinfo{booktitle}{\emph{Findings of the
  Association for Computational Linguistics: {EMNLP} 2026}},
  \bibfield{editor}{\bibinfo{person}{Hung-Yi Lee}, \bibinfo{person}{Sunipa
  Dev}, \bibinfo{person}{Jessy Li}, {and} \bibinfo{person}{Desmond Elliott}}
  (Eds.). \bibinfo{publisher}{Association for Computational Linguistics},
  \bibinfo{address}{Budapest, Hungary}, \bibinfo{pages}{27344--27363}.
\newblock
\href{https://doi.org/10.18653/2026.findings-emnlp.1293}{doi:\nolinkurl{10.18653/2026.findings-emnlp.1293}}


\bibitem[Qu et~al\mbox{.}(2021)]%
        {qu2021privatebert}
\bibfield{author}{\bibinfo{person}{Chen Qu}, \bibinfo{person}{Weize Kong},
  \bibinfo{person}{Liu Yang}, \bibinfo{person}{Mingyang Zhang},
  \bibinfo{person}{Michael Bendersky}, {and} \bibinfo{person}{Marc Najork}.}
  \bibinfo{year}{2021}\natexlab{}.
\newblock \showarticletitle{Natural Language Understanding with
  Privacy-Preserving {BERT}}. In \bibinfo{booktitle}{\emph{Proceedings of the
  30th {ACM} International Conference on Information and Knowledge Management}}
  (Virtual Event, QLD, Australia). \bibinfo{publisher}{Association for
  Computing Machinery}, \bibinfo{address}{New York, NY, USA},
  \bibinfo{pages}{1488--1497}.
\newblock
\href{https://doi.org/10.1145/3459637.3482281}{doi:\nolinkurl{10.1145/3459637.3482281}}


\bibitem[Ram{\'i}rez et~al\mbox{.}(2026)]%
        {ramirez2026peep}
\bibfield{author}{\bibinfo{person}{Guillem Ram{\'i}rez},
  \bibinfo{person}{Alexandra Birch}, {and} \bibinfo{person}{Ivan Titov}.}
  \bibinfo{year}{2026}\natexlab{}.
\newblock \showarticletitle{Controlling What You Share: Assessing Language
  Model Adherence to Privacy Preferences}. In
  \bibinfo{booktitle}{\emph{Findings of the Association for Computational
  Linguistics: ACL 2026}}. \bibinfo{publisher}{Association for Computational
  Linguistics}, \bibinfo{address}{San Diego, CA, USA},
  \bibinfo{pages}{40024--40049}.
\newblock
\href{https://doi.org/10.18653/v1/2026.findings-acl.1990}{doi:\nolinkurl{10.18653/v1/2026.findings-acl.1990}}


\bibitem[Ray~Chowdhury et~al\mbox{.}(2024)]%
        {raychowdhury2024privatereward}
\bibfield{author}{\bibinfo{person}{Sayak Ray~Chowdhury},
  \bibinfo{person}{Xingyu Zhou}, {and} \bibinfo{person}{Nagarajan Natarajan}.}
  \bibinfo{year}{2024}\natexlab{}.
\newblock \showarticletitle{Differentially Private Reward Estimation with
  Preference Feedback}. In \bibinfo{booktitle}{\emph{Proceedings of The 27th
  International Conference on Artificial Intelligence and Statistics}}
  \emph{(\bibinfo{series}{Proceedings of Machine Learning Research},
  Vol.~\bibinfo{volume}{238})}, \bibfield{editor}{\bibinfo{person}{Sanjoy
  Dasgupta}, \bibinfo{person}{Stephan Mandt}, {and} \bibinfo{person}{Yingzhen
  Li}} (Eds.). \bibinfo{publisher}{PMLR}, \bibinfo{address}{Valencia, Spain},
  \bibinfo{pages}{4843--4851}.
\newblock
\urldef\tempurl%
\url{https://proceedings.mlr.press/v238/ray-chowdhury24a.html}
\showURL{%
\tempurl}


\bibitem[Ryu et~al\mbox{.}(2026)]%
        {ryu2026p2skill}
\bibfield{author}{\bibinfo{person}{Myunghoon Ryu}, \bibinfo{person}{Geunpyo
  Park}, \bibinfo{person}{Sungjoon Lee}, \bibinfo{person}{XinYu Piao}, {and}
  \bibinfo{person}{Jong-Kook Kim}.} \bibinfo{year}{2026}\natexlab{}.
\newblock \bibinfo{title}{{P2Skill}: Privacy Preserving Skill Distillation for
  Cloud-Local {LLM} Inference Systems}.
\newblock \bibinfo{howpublished}{arXiv preprint arXiv:2608.14094v2}.
\newblock
\urldef\tempurl%
\url{https://arxiv.org/abs/2608.14094v2}
\showURL{%
\tempurl}


\bibitem[Saleh et~al\mbox{.}(2026)]%
        {saleh2026afterdeletebench}
\bibfield{author}{\bibinfo{person}{Ehsan Saleh}, \bibinfo{person}{Roozbeh
  Nahavandi}, \bibinfo{person}{Sama Salarian}, \bibinfo{person}{Raef Bassily},
  {and} \bibinfo{person}{Andrew Perrault}.} \bibinfo{year}{2026}\natexlab{}.
\newblock \bibinfo{title}{{AfterDeleteBench}: Auditing Post-Deletion
  Information Persistence in {LLM} Agents}.
\newblock \bibinfo{howpublished}{arXiv preprint arXiv:2610.31366v1 [cs.CR]}.
\newblock
\urldef\tempurl%
\url{https://arxiv.org/abs/2610.31366v1}
\showURL{%
\tempurl}


\bibitem[Salemi and Zamani(2025)]%
        {salemi2025ragpeft}
\bibfield{author}{\bibinfo{person}{Alireza Salemi} {and} \bibinfo{person}{Hamed
  Zamani}.} \bibinfo{year}{2025}\natexlab{}.
\newblock \showarticletitle{Comparing Retrieval-Augmentation and
  Parameter-Efficient Fine-Tuning for Privacy-Preserving Personalization of
  Large Language Models}. In \bibinfo{booktitle}{\emph{Proceedings of the 2025
  International ACM SIGIR Conference on Innovative Concepts and Theories in
  Information Retrieval (ICTIR)}}. \bibinfo{publisher}{ACM},
  \bibinfo{address}{Padua, Italy}, \bibinfo{pages}{286--296}.
\newblock
\href{https://doi.org/10.1145/3731120.3744595}{doi:\nolinkurl{10.1145/3731120.3744595}}


\bibitem[Salemi and Zamani(2026)]%
        {salemi2026p3}
\bibfield{author}{\bibinfo{person}{Alireza Salemi} {and} \bibinfo{person}{Hamed
  Zamani}.} \bibinfo{year}{2026}\natexlab{}.
\newblock \bibinfo{title}{Improving User Privacy in Personalized Generation:
  Client-Side Retrieval-Augmented Modification of Server-Side Generated
  Speculations}.
\newblock \bibinfo{howpublished}{arXiv:2601.17569v1 [cs.CL]}.
\newblock
\urldef\tempurl%
\url{https://arxiv.org/abs/2601.17569v1}
\showURL{%
\tempurl}


\bibitem[Schneider et~al\mbox{.}(2026)]%
        {schneider2026sea}
\bibfield{author}{\bibinfo{person}{Chris Schneider}, \bibinfo{person}{Philipp
  Schoenegger}, {and} \bibinfo{person}{Ben Bariach}.}
  \bibinfo{year}{2026}\natexlab{}.
\newblock \bibinfo{title}{Separable Expert Architecture: Toward
  Privacy-Preserving {LLM} Personalization via Composable Adapters and
  Deletable User Proxies}.
\newblock \bibinfo{howpublished}{arXiv preprint arXiv:2604.21571v1}.
\newblock
\urldef\tempurl%
\url{https://arxiv.org/abs/2604.21571v1}
\showURL{%
\tempurl}


\bibitem[Shaaban and Elmahallawy(2026)]%
        {shaaban2026jepaprune}
\bibfield{author}{\bibinfo{person}{Mohamed Shaaban} {and}
  \bibinfo{person}{Mohamed Elmahallawy}.} \bibinfo{year}{2026}\natexlab{}.
\newblock \bibinfo{title}{Prune the Leaky Neurons: Mitigating Private Data
  Leakage in Fine-Tuned {LLMs} via Joint-Embedding-Guided Pruning}.
\newblock \bibinfo{howpublished}{arXiv preprint arXiv:2610.22792v1}.
\newblock
\urldef\tempurl%
\url{https://arxiv.org/abs/2610.22792v1}
\showURL{%
\tempurl}


\bibitem[Shao et~al\mbox{.}(2024)]%
        {shao2024privacylens}
\bibfield{author}{\bibinfo{person}{Yijia Shao}, \bibinfo{person}{Tianshi Li},
  \bibinfo{person}{Weiyan Shi}, \bibinfo{person}{Yanchen Liu}, {and}
  \bibinfo{person}{Diyi Yang}.} \bibinfo{year}{2024}\natexlab{}.
\newblock \showarticletitle{{PrivacyLens}: Evaluating Privacy Norm Awareness of
  Language Models in Action}. In \bibinfo{booktitle}{\emph{Advances in Neural
  Information Processing Systems}} (Vancouver, BC, Canada),
  Vol.~\bibinfo{volume}{37}. \bibinfo{publisher}{Neural Information Processing
  Systems Foundation, Inc. (NeurIPS)}, \bibinfo{address}{San Diego, California,
  USA}, \bibinfo{pages}{89373--89407}.
\newblock
\href{https://doi.org/10.52202/079017-2837}{doi:\nolinkurl{10.52202/079017-2837}}


\bibitem[Shen et~al\mbox{.}(2026)]%
        {shen2026revokedmemory}
\bibfield{author}{\bibinfo{person}{Yi~Ting Shen}, \bibinfo{person}{Kentaroh
  Toyoda}, {and} \bibinfo{person}{Alex Leung}.}
  \bibinfo{year}{2026}\natexlab{}.
\newblock \bibinfo{title}{Revoked but Still Authoritative: An Empirical Study
  of Revocation Enforcement in Agent-Memory Systems}.
\newblock \bibinfo{howpublished}{arXiv preprint arXiv:2609.08258v1}.
\newblock
\urldef\tempurl%
\url{https://arxiv.org/abs/2609.08258v1}
\showURL{%
\tempurl}


\bibitem[Shi et~al\mbox{.}(2025)]%
        {shi2025guardedtuning}
\bibfield{author}{\bibinfo{person}{Haonan Shi}, \bibinfo{person}{Tu Ouyang},
  {and} \bibinfo{person}{An Wang}.} \bibinfo{year}{2025}\natexlab{}.
\newblock \bibinfo{title}{Navigating the Designs of Privacy-Preserving
  Fine-tuning for Large Language Models}.
\newblock \bibinfo{howpublished}{arXiv:2501.04323v3 [cs.LG]}.
\newblock
\urldef\tempurl%
\url{https://arxiv.org/abs/2501.04323v3}
\showURL{%
\tempurl}


\bibitem[Shi et~al\mbox{.}(2022)]%
        {shi2022justfinetune}
\bibfield{author}{\bibinfo{person}{Weiyan Shi}, \bibinfo{person}{Ryan Shea},
  \bibinfo{person}{Si Chen}, \bibinfo{person}{Chiyuan Zhang},
  \bibinfo{person}{Ruoxi Jia}, {and} \bibinfo{person}{Zhou Yu}.}
  \bibinfo{year}{2022}\natexlab{}.
\newblock \showarticletitle{Just Fine-tune Twice: Selective Differential
  Privacy for Large Language Models}. In \bibinfo{booktitle}{\emph{Proceedings
  of the 2022 Conference on Empirical Methods in Natural Language Processing}}.
  \bibinfo{publisher}{Association for Computational Linguistics},
  \bibinfo{address}{Abu Dhabi, United Arab Emirates},
  \bibinfo{pages}{6327--6340}.
\newblock
\href{https://doi.org/10.18653/v1/2022.emnlp-main.425}{doi:\nolinkurl{10.18653/v1/2022.emnlp-main.425}}


\bibitem[Shihab et~al\mbox{.}(2026)]%
        {shihab2026cslr}
\bibfield{author}{\bibinfo{person}{Ibne~Farabi Shihab}, \bibinfo{person}{Abu
  Sa-Adat Mohamed Moon-Im~Al Ahsan}, {and} \bibinfo{person}{Anuj Sharma}.}
  \bibinfo{year}{2026}\natexlab{}.
\newblock \bibinfo{title}{Canonicalized Stable-List Replay for Private
  Federated Continual Learning over Language-Model Embeddings}.
\newblock \bibinfo{howpublished}{arXiv:2606.00426v1 [cs.LG]}.
\newblock
\urldef\tempurl%
\url{https://arxiv.org/abs/2606.00426v1}
\showURL{%
\tempurl}


\bibitem[Shim et~al\mbox{.}(2026)]%
        {shim2026controltheflow}
\bibfield{author}{\bibinfo{person}{Minsun Shim}, \bibinfo{person}{Ramisha~Raida
  Karim}, \bibinfo{person}{Ruthwik Jakkula}, \bibinfo{person}{Kaiwen Zhou},
  \bibinfo{person}{Xin Liu}, \bibinfo{person}{Xin~Eric Wang}, {and}
  \bibinfo{person}{Zhou Li}.} \bibinfo{year}{2026}\natexlab{}.
\newblock \bibinfo{title}{Confuse the Model, Control the Flow: Understanding
  and Mitigating Privacy Leakage from {LLM} Agents with Information Flow
  Control}.
\newblock \bibinfo{howpublished}{arXiv preprint arXiv:2609.14003v1 [cs.CR]}.
\newblock
\urldef\tempurl%
\url{https://arxiv.org/abs/2609.14003v1}
\showURL{%
\tempurl}


\bibitem[Slokom and Bellog{\'i}n(2026)]%
        {slokom2026federatedrecommendations}
\bibfield{author}{\bibinfo{person}{Manel Slokom} {and}
  \bibinfo{person}{Alejandro Bellog{\'i}n}.} \bibinfo{year}{2026}\natexlab{}.
\newblock \showarticletitle{Beyond Centralization: User-Controlled Federated
  Recommendations in Practice}. In \bibinfo{booktitle}{\emph{Proceedings of the
  34th ACM Conference on User Modeling, Adaptation and Personalization}}.
  \bibinfo{publisher}{Association for Computing Machinery},
  \bibinfo{address}{New York, NY, USA}, \bibinfo{pages}{462--465}.
\newblock
\href{https://doi.org/10.1145/3774935.3812710}{doi:\nolinkurl{10.1145/3774935.3812710}}


\bibitem[Song and Chaudhuri(2017)]%
        {song2017timecomposition}
\bibfield{author}{\bibinfo{person}{Shuang Song} {and} \bibinfo{person}{Kamalika
  Chaudhuri}.} \bibinfo{year}{2017}\natexlab{}.
\newblock \showarticletitle{Composition Properties of Inferential Privacy for
  Time-Series Data}. In \bibinfo{booktitle}{\emph{2017 55th Annual Allerton
  Conference on Communication, Control, and Computing (Allerton)}} (Monticello,
  IL, USA). \bibinfo{publisher}{IEEE}, \bibinfo{address}{Piscataway, NJ, USA},
  \bibinfo{pages}{814--821}.
\newblock
\href{https://doi.org/10.1109/allerton.2017.8262823}{doi:\nolinkurl{10.1109/allerton.2017.8262823}}


\bibitem[Staab et~al\mbox{.}(2026)]%
        {staab2026privacyflow}
\bibfield{author}{\bibinfo{person}{Robin Staab}, \bibinfo{person}{Tom
  Wollschl{\"{a}}ger}, \bibinfo{person}{Vahab Mirrokni}, {and}
  \bibinfo{person}{Reza Shokri}.} \bibinfo{year}{2026}\natexlab{}.
\newblock \bibinfo{title}{{PrivacyFlow}: Privacy in {AI} Agents through
  Information Flow Analysis and Consequence-Aware Risk Management}.
\newblock \bibinfo{howpublished}{arXiv preprint arXiv:2610.40714v1 [cs.CR]}.
\newblock
\urldef\tempurl%
\url{https://arxiv.org/abs/2610.40714v1}
\showURL{%
\tempurl}


\bibitem[Sun et~al\mbox{.}(2018)]%
        {sun2018backwardsearch}
\bibfield{author}{\bibinfo{person}{Shi-Feng Sun}, \bibinfo{person}{Xingliang
  Yuan}, \bibinfo{person}{Joseph~K. Liu}, \bibinfo{person}{Ron Steinfeld},
  \bibinfo{person}{Amin Sakzad}, \bibinfo{person}{Viet Vo}, {and}
  \bibinfo{person}{Surya Nepal}.} \bibinfo{year}{2018}\natexlab{}.
\newblock \showarticletitle{Practical Backward-Secure Searchable Encryption
  from Symmetric Puncturable Encryption}. In
  \bibinfo{booktitle}{\emph{Proceedings of the 2018 {ACM} {SIGSAC} Conference
  on Computer and Communications Security}} (Toronto, ON, Canada).
  \bibinfo{publisher}{Association for Computing Machinery},
  \bibinfo{address}{New York, NY, USA}, \bibinfo{pages}{763--780}.
\newblock
\href{https://doi.org/10.1145/3243734.3243782}{doi:\nolinkurl{10.1145/3243734.3243782}}


\bibitem[Szep et~al\mbox{.}(2026)]%
        {szep2026memorization}
\bibfield{author}{\bibinfo{person}{Marton Szep}, \bibinfo{person}{Jorge
  Marin~Ruiz}, \bibinfo{person}{Georgios Kaissis}, \bibinfo{person}{Paulina
  Seidl}, \bibinfo{person}{R{\"u}diger von Eisenhart-Rothe},
  \bibinfo{person}{Florian Hinterwimmer}, {and} \bibinfo{person}{Daniel
  Rueckert}.} \bibinfo{year}{2026}\natexlab{}.
\newblock \showarticletitle{Unintended Memorization of Sensitive Information in
  Fine-Tuned Language Models}. In \bibinfo{booktitle}{\emph{Proceedings of the
  19th Conference of the European Chapter of the Association for Computational
  Linguistics (Volume 1: Long Papers)}},
  \bibfield{editor}{\bibinfo{person}{Vera Demberg}, \bibinfo{person}{Kentaro
  Inui}, {and} \bibinfo{person}{Llu{\'i}s Marquez}} (Eds.).
  \bibinfo{publisher}{Association for Computational Linguistics},
  \bibinfo{address}{Rabat, Morocco}, \bibinfo{pages}{6461--6480}.
\newblock
\showISBNx{979-8-89176-380-7}
\href{https://doi.org/10.18653/v1/2026.eacl-long.304}{doi:\nolinkurl{10.18653/v1/2026.eacl-long.304}}


\bibitem[T. et~al\mbox{.}(2026)]%
        {dhivyasree2026cams}
\bibfield{author}{\bibinfo{person}{Dhivyasree T.}, \bibinfo{person}{Saravanan
  S.}, \bibinfo{person}{Arun~Kumar Ramamoorthy}, {and}
  \bibinfo{person}{Uma~Maheswari Balasubramanian}.}
  \bibinfo{year}{2026}\natexlab{}.
\newblock \showarticletitle{Cognitive Autonomous Memory Security ({CAMS})
  against injection and extraction attacks in long-term memory of {AI} agents}.
\newblock \bibinfo{journal}{\emph{Egyptian Informatics Journal}}
  \bibinfo{volume}{34}, Article \bibinfo{articleno}{100983}
  (\bibinfo{year}{2026}), \bibinfo{numpages}{15}~pages.
\newblock
\href{https://doi.org/10.1016/j.eij.2026.100983}{doi:\nolinkurl{10.1016/j.eij.2026.100983}}


\bibitem[Tan et~al\mbox{.}(2024)]%
        {tan2024oppu}
\bibfield{author}{\bibinfo{person}{Zhaoxuan Tan}, \bibinfo{person}{Qingkai
  Zeng}, \bibinfo{person}{Yijun Tian}, \bibinfo{person}{Zheyuan Liu},
  \bibinfo{person}{Bing Yin}, {and} \bibinfo{person}{Meng Jiang}.}
  \bibinfo{year}{2024}\natexlab{}.
\newblock \showarticletitle{Democratizing Large Language Models via
  Personalized Parameter-Efficient Fine-tuning}. In
  \bibinfo{booktitle}{\emph{Proceedings of the 2024 Conference on Empirical
  Methods in Natural Language Processing}}. \bibinfo{publisher}{Association for
  Computational Linguistics}, \bibinfo{address}{Miami, FL, USA},
  \bibinfo{pages}{6476--6491}.
\newblock
\href{https://doi.org/10.18653/v1/2024.emnlp-main.372}{doi:\nolinkurl{10.18653/v1/2024.emnlp-main.372}}


\bibitem[Tang et~al\mbox{.}(2026)]%
        {tang2026privagentflow}
\bibfield{author}{\bibinfo{person}{Tianyi Tang}, \bibinfo{person}{Tianyi Ma},
  \bibinfo{person}{Yueming Lyu}, \bibinfo{person}{Yew-Soon Ong},
  \bibinfo{person}{Ivor Tsang}, {and} \bibinfo{person}{Haiyan Yin}.}
  \bibinfo{year}{2026}\natexlab{}.
\newblock \showarticletitle{{PrivAgentFlow}: Agentic Workflow for Automating
  Privacy Preservation in Web Agents}. In \bibinfo{booktitle}{\emph{Findings of
  the Association for Computational Linguistics: {EMNLP} 2026}},
  \bibfield{editor}{\bibinfo{person}{Hung-Yi Lee}, \bibinfo{person}{Sunipa
  Dev}, \bibinfo{person}{Jessy Li}, {and} \bibinfo{person}{Desmond Elliott}}
  (Eds.). \bibinfo{publisher}{Association for Computational Linguistics},
  \bibinfo{address}{Budapest, Hungary}, \bibinfo{pages}{34471--34491}.
\newblock
\showISBNx{979-8-89176-462-0}
\href{https://doi.org/10.18653/2026.findings-emnlp.1620}{doi:\nolinkurl{10.18653/2026.findings-emnlp.1620}}


\bibitem[Tang et~al\mbox{.}(2025)]%
        {tang2025dpzo}
\bibfield{author}{\bibinfo{person}{Xinyu Tang}, \bibinfo{person}{Ashwinee
  Panda}, \bibinfo{person}{Milad Nasr}, \bibinfo{person}{Saeed Mahloujifar},
  {and} \bibinfo{person}{Prateek Mittal}.} \bibinfo{year}{2025}\natexlab{}.
\newblock \showarticletitle{Private Fine-Tuning of Large Language Models with
  Zeroth-Order Optimization}.
\newblock \bibinfo{journal}{\emph{Transactions on Machine Learning Research}}
  \bibinfo{volume}{2025}, \bibinfo{number}{1} (\bibinfo{year}{2025}),
  \bibinfo{pages}{1--27}.
\newblock
\urldef\tempurl%
\url{https://openreview.net/forum?id=3Y3o0yFZfu}
\showURL{%
\tempurl}


\bibitem[Tang et~al\mbox{.}(2024)]%
        {tang2024dpfewshot}
\bibfield{author}{\bibinfo{person}{Xinyu Tang}, \bibinfo{person}{Richard Shin},
  \bibinfo{person}{Huseyin Inan}, \bibinfo{person}{Andre Manoel},
  \bibinfo{person}{Niloofar Mireshghallah}, \bibinfo{person}{Zinan Lin},
  \bibinfo{person}{Sivakanth Gopi}, \bibinfo{person}{Janardhan Kulkarni}, {and}
  \bibinfo{person}{Robert Sim}.} \bibinfo{year}{2024}\natexlab{}.
\newblock \showarticletitle{Privacy-Preserving In-Context Learning with
  Differentially Private Few-Shot Generation}. In
  \bibinfo{booktitle}{\emph{Proceedings of the Twelfth International Conference
  on Learning Representations}}, Vol.~\bibinfo{volume}{2024}.
  \bibinfo{publisher}{OpenReview.net}, \bibinfo{address}{Vienna, Austria},
  \bibinfo{pages}{33058--33077}.
\newblock
\urldef\tempurl%
\url{https://proceedings.iclr.cc/paper_files/paper/2024/hash/8cb564df771e9eacbfe9d72bd46a24a9-Abstract-Conference.html}
\showURL{%
\tempurl}


\bibitem[Tao et~al\mbox{.}(2026)]%
        {tao2026aegis}
\bibfield{author}{\bibinfo{person}{Ye Tao}, \bibinfo{person}{Hong Shen},
  \bibinfo{person}{Hui Tian}, \bibinfo{person}{Xin Wang}, {and}
  \bibinfo{person}{Can Wang}.} \bibinfo{year}{2026}\natexlab{}.
\newblock \bibinfo{title}{{AEGIS}: Attention-Embedding Gradient Isolation
  Shield - Triple-Channel Gradient Masking for Privacy-Preserving Federated
  {LLM} Fine-Tuning}.
\newblock \bibinfo{howpublished}{arXiv preprint arXiv:2608.19534v1}.
\newblock
\urldef\tempurl%
\url{https://arxiv.org/abs/2608.19534v1}
\showURL{%
\tempurl}


\bibitem[Tobaben et~al\mbox{.}(2026)]%
        {tobaben2026labelspace}
\bibfield{author}{\bibinfo{person}{Marlon Tobaben}, \bibinfo{person}{Talal
  Alrawajfeh}, \bibinfo{person}{Marcus Klasson}, \bibinfo{person}{Mikko
  Heikkil{\"{a}}}, \bibinfo{person}{Arno Solin}, {and} \bibinfo{person}{Antti
  Honkela}.} \bibinfo{year}{2026}\natexlab{}.
\newblock \showarticletitle{Privacy Leakage via Output Label Space and
  Differentially Private Continual Learning}.
\newblock \bibinfo{journal}{\emph{Transactions on Machine Learning Research}}
  \bibinfo{volume}{2026}, \bibinfo{number}{8} (\bibinfo{year}{2026}),
  \bibinfo{numpages}{46}~pages.
\newblock
\urldef\tempurl%
\url{https://openreview.net/forum?id=ZshFgRQWrm}
\showURL{%
\tempurl}


\bibitem[Tran et~al\mbox{.}(2025)]%
        {tran2025dpfpl}
\bibfield{author}{\bibinfo{person}{Linh Tran}, \bibinfo{person}{Wei Sun},
  \bibinfo{person}{Stacy Patterson}, {and} \bibinfo{person}{Ana Milanova}.}
  \bibinfo{year}{2025}\natexlab{}.
\newblock \showarticletitle{Privacy-Preserving Personalized Federated Prompt
  Learning for Multimodal Large Language Models}. In
  \bibinfo{booktitle}{\emph{Proceedings of the Thirteenth International
  Conference on Learning Representations}}, Vol.~\bibinfo{volume}{2025}.
  \bibinfo{publisher}{OpenReview.net}, \bibinfo{address}{Singapore},
  \bibinfo{pages}{27362--27380}.
\newblock
\urldef\tempurl%
\url{https://proceedings.iclr.cc/paper_files/paper/2025/hash/4431224d3762aa655f0aee4eaf04ff16-Abstract-Conference.html}
\showURL{%
\tempurl}


\bibitem[Uddin et~al\mbox{.}(2026)]%
        {uddin2026memora}
\bibfield{author}{\bibinfo{person}{Md~Nayem Uddin}, \bibinfo{person}{Kumar
  Shubham}, \bibinfo{person}{Eduardo Blanco}, \bibinfo{person}{Chitta Baral},
  {and} \bibinfo{person}{Gengyu Wang}.} \bibinfo{year}{2026}\natexlab{}.
\newblock \showarticletitle{From Recall to Forgetting: Benchmarking Long-Term
  Memory for Personalized Agents}. In \bibinfo{booktitle}{\emph{Findings of the
  Association for Computational Linguistics: ACL 2026}}.
  \bibinfo{publisher}{Association for Computational Linguistics},
  \bibinfo{address}{San Diego, California, USA}, \bibinfo{pages}{26814--26841}.
\newblock
\href{https://doi.org/10.18653/v1/2026.findings-acl.1337}{doi:\nolinkurl{10.18653/v1/2026.findings-acl.1337}}


\bibitem[Ulusoy and Yolum(2022)]%
        {ulusoy2022panola}
\bibfield{author}{\bibinfo{person}{Onuralp Ulusoy} {and}
  \bibinfo{person}{P{\i}nar Yolum}.} \bibinfo{year}{2022}\natexlab{}.
\newblock \showarticletitle{{PANOLA}: A Personal Assistant for Supporting Users
  in Preserving Privacy}.
\newblock \bibinfo{journal}{\emph{ACM Transactions on Internet Technology}}
  \bibinfo{volume}{22}, \bibinfo{number}{1}, Article \bibinfo{articleno}{27}
  (\bibinfo{year}{2022}), \bibinfo{numpages}{32}~pages.
\newblock
\href{https://doi.org/10.1145/3471187}{doi:\nolinkurl{10.1145/3471187}}


\bibitem[Wang et~al\mbox{.}(2025a)]%
        {wang2025mextra}
\bibfield{author}{\bibinfo{person}{Bo Wang}, \bibinfo{person}{Weiyi He},
  \bibinfo{person}{Shenglai Zeng}, \bibinfo{person}{Zhen Xiang},
  \bibinfo{person}{Yue Xing}, \bibinfo{person}{Jiliang Tang}, {and}
  \bibinfo{person}{Pengfei He}.} \bibinfo{year}{2025}\natexlab{a}.
\newblock \showarticletitle{Unveiling Privacy Risks in {LLM} Agent Memory}. In
  \bibinfo{booktitle}{\emph{Proceedings of the 63rd Annual Meeting of the
  Association for Computational Linguistics (Volume 1: Long Papers)}},
  \bibfield{editor}{\bibinfo{person}{Wanxiang Che}, \bibinfo{person}{Joyce
  Nabende}, \bibinfo{person}{Ekaterina Shutova}, {and}
  \bibinfo{person}{Mohammad~Taher Pilehvar}} (Eds.).
  \bibinfo{publisher}{Association for Computational Linguistics},
  \bibinfo{address}{Vienna, Austria}, \bibinfo{pages}{25241--25260}.
\newblock
\showISBNx{979-8-89176-251-0}
\href{https://doi.org/10.18653/v1/2025.acl-long.1227}{doi:\nolinkurl{10.18653/v1/2025.acl-long.1227}}


\bibitem[Wang et~al\mbox{.}(2026e)]%
        {wang2026agenticunlearning}
\bibfield{author}{\bibinfo{person}{Bin Wang}, \bibinfo{person}{Fan Wang},
  \bibinfo{person}{Pingping Wang}, \bibinfo{person}{Jinyu Cong},
  \bibinfo{person}{Yang Yu}, \bibinfo{person}{Yilong Yin},
  \bibinfo{person}{Zhongyi Han}, {and} \bibinfo{person}{Benzheng Wei}.}
  \bibinfo{year}{2026}\natexlab{e}.
\newblock \bibinfo{title}{Agentic Unlearning: When {LLM} Agent Meets Machine
  Unlearning}.
\newblock \bibinfo{howpublished}{arXiv preprint arXiv:2602.17692v2 [cs.LG]}.
\newblock
\urldef\tempurl%
\url{https://arxiv.org/abs/2602.17692v2}
\showURL{%
\tempurl}


\bibitem[Wang et~al\mbox{.}(2026f)]%
        {wang2026privmesa}
\bibfield{author}{\bibinfo{person}{Dannong Wang}, \bibinfo{person}{Yuran
  Zhang}, \bibinfo{person}{Bian Sun}, \bibinfo{person}{Alex Stinard},
  \bibinfo{person}{Yuzhang Shang}, \bibinfo{person}{Song Wang}, {and}
  \bibinfo{person}{Yu Tian}.} \bibinfo{year}{2026}\natexlab{f}.
\newblock \bibinfo{title}{{PrivMeSA}: Privacy-Aware Self-Evolving Multi-Agent
  System for Medicine via Local-Remote {LLM} Collaboration}.
\newblock \bibinfo{howpublished}{arXiv preprint arXiv:2609.38458v1 [cs.AI]}.
\newblock
\urldef\tempurl%
\url{https://arxiv.org/abs/2609.38458v1}
\showURL{%
\tempurl}


\bibitem[Wang et~al\mbox{.}(2020)]%
        {wang2020collaborativebandits}
\bibfield{author}{\bibinfo{person}{Huazheng Wang}, \bibinfo{person}{Qian Zhao},
  \bibinfo{person}{Qingyun Wu}, \bibinfo{person}{Shubham Chopra},
  \bibinfo{person}{Abhinav Khaitan}, {and} \bibinfo{person}{Hongning Wang}.}
  \bibinfo{year}{2020}\natexlab{}.
\newblock \showarticletitle{Global and Local Differential Privacy for
  Collaborative Bandits}. In \bibinfo{booktitle}{\emph{Fourteenth {ACM}
  Conference on Recommender Systems}} (Virtual Event, Brazil).
  \bibinfo{publisher}{Association for Computing Machinery},
  \bibinfo{address}{New York, NY, USA}, \bibinfo{pages}{150--159}.
\newblock
\href{https://doi.org/10.1145/3383313.3412254}{doi:\nolinkurl{10.1145/3383313.3412254}}


\bibitem[Wang et~al\mbox{.}(2026d)]%
        {wang2026counterpersona}
\bibfield{author}{\bibinfo{person}{Pengwei Wang}, \bibinfo{person}{Zihan Wang},
  \bibinfo{person}{Hangcheng Cao}, \bibinfo{person}{Qingchuan Zhao},
  \bibinfo{person}{Hongwei Li}, {and} \bibinfo{person}{Guowen Xu}.}
  \bibinfo{year}{2026}\natexlab{d}.
\newblock \bibinfo{title}{{CounterPersona}: Append-Only Defense Against
  Unauthorized Persona Skill Distillation}.
\newblock \bibinfo{howpublished}{arXiv preprint arXiv:2609.15097v1}.
\newblock
\urldef\tempurl%
\url{https://arxiv.org/abs/2609.15097v1}
\showURL{%
\tempurl}


\bibitem[Wang et~al\mbox{.}(2025c)]%
        {wang2025privacychecker}
\bibfield{author}{\bibinfo{person}{Shouju Wang}, \bibinfo{person}{Fenglin Yu},
  \bibinfo{person}{Xirui Liu}, \bibinfo{person}{Xiaoting Qin},
  \bibinfo{person}{Jue Zhang}, \bibinfo{person}{Qingwei Lin},
  \bibinfo{person}{Dongmei Zhang}, {and} \bibinfo{person}{Saravan Rajmohan}.}
  \bibinfo{year}{2025}\natexlab{c}.
\newblock \showarticletitle{Privacy in Action: Towards Realistic Privacy
  Mitigation and Evaluation for {LLM}-Powered Agents}. In
  \bibinfo{booktitle}{\emph{Findings of the Association for Computational
  Linguistics: EMNLP 2025}}, \bibfield{editor}{\bibinfo{person}{Christos
  Christodoulopoulos}, \bibinfo{person}{Tanmoy Chakraborty},
  \bibinfo{person}{Carolyn Rose}, {and} \bibinfo{person}{Violet Peng}} (Eds.).
  \bibinfo{publisher}{Association for Computational Linguistics},
  \bibinfo{address}{Suzhou, China}, \bibinfo{pages}{17055--17074}.
\newblock
\href{https://doi.org/10.18653/v1/2025.findings-emnlp.925}{doi:\nolinkurl{10.18653/v1/2025.findings-emnlp.925}}


\bibitem[Wang et~al\mbox{.}(2026g)]%
        {wang2026ragunlearning}
\bibfield{author}{\bibinfo{person}{Shang Wang}, \bibinfo{person}{Tianqing Zhu},
  \bibinfo{person}{Dayong Ye}, {and} \bibinfo{person}{Wanlei Zhou}.}
  \bibinfo{year}{2026}\natexlab{g}.
\newblock \showarticletitle{When Machine Unlearning Meets Retrieval-Augmented
  Generation ({RAG}): Keep Secret or Forget Knowledge?}
\newblock \bibinfo{journal}{\emph{IEEE Transactions on Dependable and Secure
  Computing}} \bibinfo{volume}{23}, \bibinfo{number}{2} (\bibinfo{year}{2026}),
  \bibinfo{pages}{2030--2045}.
\newblock
\href{https://doi.org/10.1109/tdsc.2025.3620832}{doi:\nolinkurl{10.1109/tdsc.2025.3620832}}


\bibitem[Wang et~al\mbox{.}(2026c)]%
        {wang2026spmem}
\bibfield{author}{\bibinfo{person}{Wenjie Wang}, \bibinfo{person}{Wenhe Si},
  \bibinfo{person}{Xinyue Xu}, {and} \bibinfo{person}{Yue Xu}.}
  \bibinfo{year}{2026}\natexlab{c}.
\newblock \bibinfo{title}{What to Remember, What to Reveal: Privacy-Aware
  Memory for Conversational Agents}.
\newblock \bibinfo{howpublished}{arXiv preprint arXiv:2608.16551v1 [cs.CR]}.
\newblock
\urldef\tempurl%
\url{https://arxiv.org/abs/2608.16551v1}
\showURL{%
\tempurl}


\bibitem[Wang et~al\mbox{.}(2025b)]%
        {wang2025tape}
\bibfield{author}{\bibinfo{person}{Weiqi Wang}, \bibinfo{person}{Zhiyi Tian},
  \bibinfo{person}{An Liu}, {and} \bibinfo{person}{Shui Yu}.}
  \bibinfo{year}{2025}\natexlab{b}.
\newblock \showarticletitle{{TAPE}: Tailored Posterior Difference for Auditing
  of Machine Unlearning}. In \bibinfo{booktitle}{\emph{Proceedings of the {ACM}
  on Web Conference 2025}} (Sydney, NSW, Australia).
  \bibinfo{publisher}{Association for Computing Machinery},
  \bibinfo{address}{New York, NY, USA}, \bibinfo{pages}{3061--3072}.
\newblock
\href{https://doi.org/10.1145/3696410.3714875}{doi:\nolinkurl{10.1145/3696410.3714875}}


\bibitem[Wang et~al\mbox{.}(2026a)]%
        {wang2026fedsept}
\bibfield{author}{\bibinfo{person}{Yuhua Wang}, \bibinfo{person}{Xiaodong Li},
  \bibinfo{person}{Yihao Guo}, \bibinfo{person}{Yuxiang Jia},
  \bibinfo{person}{Qinnan Zhang}, \bibinfo{person}{Yifan Sun},
  \bibinfo{person}{Hainan Zhang}, \bibinfo{person}{Yongxin Tong}, {and}
  \bibinfo{person}{Zhiming Zheng}.} \bibinfo{year}{2026}\natexlab{a}.
\newblock \bibinfo{title}{Towards Privacy-Preserving Federated Prompt Tuning
  under Data Heterogeneity: A Subspace-Decomposed Expert Approach}.
\newblock \bibinfo{howpublished}{arXiv preprint arXiv:2607.21417v1}.
\newblock
\urldef\tempurl%
\url{https://arxiv.org/abs/2607.21417v1}
\showURL{%
\tempurl}


\bibitem[Wang et~al\mbox{.}(2026b)]%
        {wang2026prism}
\bibfield{author}{\bibinfo{person}{Ziyi Wang}, \bibinfo{person}{Liang Liu},
  \bibinfo{person}{Zhu Meng}, {and} \bibinfo{person}{Zhicheng Zhao}.}
  \bibinfo{year}{2026}\natexlab{b}.
\newblock \showarticletitle{{PRISM}: Private Rank-Invariant Subspace Memory for
  Federated Continual Fine-Tuning under Local Differential Privacy}. In
  \bibinfo{booktitle}{\emph{Proceedings of the 2026 Conference on Empirical
  Methods in Natural Language Processing}}. \bibinfo{publisher}{Association for
  Computational Linguistics}, \bibinfo{address}{Budapest, Hungary},
  \bibinfo{pages}{7204--7223}.
\newblock
\href{https://doi.org/10.18653/2026.emnlp-main.313}{doi:\nolinkurl{10.18653/2026.emnlp-main.313}}


\bibitem[Watanabe and Uchikoshi(2025)]%
        {watanabe2025zkadvice}
\bibfield{author}{\bibinfo{person}{Hiroki Watanabe} {and}
  \bibinfo{person}{Motonobu Uchikoshi}.} \bibinfo{year}{2025}\natexlab{}.
\newblock \showarticletitle{Generating Privacy-Preserving Personalized Advice
  with Zero-Knowledge Proofs and {LLMs}}. In
  \bibinfo{booktitle}{\emph{Companion Proceedings of the ACM on Web Conference
  2025}}. \bibinfo{publisher}{ACM}, \bibinfo{address}{Sydney, NSW, Australia},
  \bibinfo{pages}{1385--1389}.
\newblock
\href{https://doi.org/10.1145/3701716.3715597}{doi:\nolinkurl{10.1145/3701716.3715597}}


\bibitem[Wei(2026)]%
        {wei2026trishield}
\bibfield{author}{\bibinfo{person}{Cheng Wei}.}
  \bibinfo{year}{2026}\natexlab{}.
\newblock \bibinfo{title}{{TriShield}: Zero-Utility-Loss Defense Against
  Privacy Backdoors in Federated Language Model Fine-Tuning via Orthogonal
  Gradient Projection and Optimizer State Entanglement}.
\newblock \bibinfo{howpublished}{arXiv:2607.27940v2 [cs.LG]}.
\newblock
\urldef\tempurl%
\url{https://arxiv.org/abs/2607.27940v2}
\showURL{%
\tempurl}


\bibitem[Whitehouse et~al\mbox{.}(2023)]%
        {whitehouse2023fullyadaptive}
\bibfield{author}{\bibinfo{person}{Justin Whitehouse}, \bibinfo{person}{Aaditya
  Ramdas}, \bibinfo{person}{Ryan Rogers}, {and} \bibinfo{person}{Steven Wu}.}
  \bibinfo{year}{2023}\natexlab{}.
\newblock \showarticletitle{Fully-Adaptive Composition in Differential
  Privacy}. In \bibinfo{booktitle}{\emph{Proceedings of the 40th International
  Conference on Machine Learning}} \emph{(\bibinfo{series}{Proceedings of
  Machine Learning Research}, Vol.~\bibinfo{volume}{202})},
  \bibfield{editor}{\bibinfo{person}{Andreas Krause}, \bibinfo{person}{Emma
  Brunskill}, \bibinfo{person}{Kyunghyun Cho}, \bibinfo{person}{Barbara
  Engelhardt}, \bibinfo{person}{Sivan Sabato}, {and} \bibinfo{person}{Jonathan
  Scarlett}} (Eds.). \bibinfo{publisher}{PMLR}, \bibinfo{address}{Honolulu, HI,
  USA}, \bibinfo{pages}{36990--37007}.
\newblock
\urldef\tempurl%
\url{https://proceedings.mlr.press/v202/whitehouse23a.html}
\showURL{%
\tempurl}


\bibitem[Wu et~al\mbox{.}(2026a)]%
        {wu2026tgci}
\bibfield{author}{\bibinfo{person}{Hengyu Wu}, \bibinfo{person}{Blaise
  Delattre}, {and} \bibinfo{person}{Yang Cao}.}
  \bibinfo{year}{2026}\natexlab{a}.
\newblock \bibinfo{title}{Judging Before Propagation: Task-Graph Contextual
  Integrity for Multi-Agent Privacy}.
\newblock \bibinfo{howpublished}{arXiv preprint arXiv:2610.39532v1}.
\newblock
\urldef\tempurl%
\url{https://arxiv.org/abs/2610.39532v1}
\showURL{%
\tempurl}


\bibitem[Wu et~al\mbox{.}(2026b)]%
        {wu2026amp}
\bibfield{author}{\bibinfo{person}{Junde Wu}, \bibinfo{person}{Minhao Hu},
  \bibinfo{person}{Jiayuan Zhu}, \bibinfo{person}{Jiaye Wang}, {and}
  \bibinfo{person}{Yueming Jin}.} \bibinfo{year}{2026}\natexlab{b}.
\newblock \showarticletitle{Agent-Memory Protocol: A Privacy-Focused Protocol
  for {LLM} Agents and User Memory Interaction}. In
  \bibinfo{booktitle}{\emph{Proceedings of The Second {AAAI} Bridge Program on
  AI for Medicine and Healthcare}} \emph{(\bibinfo{series}{Proceedings of
  Machine Learning Research}, Vol.~\bibinfo{volume}{317})},
  \bibfield{editor}{\bibinfo{person}{Junde Wu}, \bibinfo{person}{Jiazhen Pan},
  \bibinfo{person}{Jiayuan Zhu}, \bibinfo{person}{Luyang Luo},
  \bibinfo{person}{Yitong Li}, \bibinfo{person}{Min Xu},
  \bibinfo{person}{Yueming Jin}, {and} \bibinfo{person}{Daniel Rueckert}}
  (Eds.). \bibinfo{publisher}{PMLR}, \bibinfo{address}{Singapore},
  \bibinfo{pages}{293--301}.
\newblock
\urldef\tempurl%
\url{https://proceedings.mlr.press/v317/wu26a.html}
\showURL{%
\tempurl}


\bibitem[Wu et~al\mbox{.}(2026c)]%
        {wu2026collabmem}
\bibfield{author}{\bibinfo{person}{Shun Wu}, \bibinfo{person}{Di Wu},
  \bibinfo{person}{XueYou Zhang}, {and} \bibinfo{person}{Kang Liu}.}
  \bibinfo{year}{2026}\natexlab{c}.
\newblock \showarticletitle{{COLLABMEM}: A Diagnostic Benchmark for
  Collaborative Memory Operations in Multi-Agent {LLM} Systems}. In
  \bibinfo{booktitle}{\emph{Proceedings of the 2026 Conference on Empirical
  Methods in Natural Language Processing}}. \bibinfo{publisher}{Association for
  Computational Linguistics}, \bibinfo{address}{Budapest, Hungary},
  \bibinfo{pages}{27343--27356}.
\newblock
\href{https://doi.org/10.18653/2026.emnlp-main.1194}{doi:\nolinkurl{10.18653/2026.emnlp-main.1194}}


\bibitem[Wu et~al\mbox{.}(2025)]%
        {wu2025onlineprivaterlhf}
\bibfield{author}{\bibinfo{person}{Yulian Wu}, \bibinfo{person}{Rushil
  Thareja}, \bibinfo{person}{Praneeth Vepakomma}, {and}
  \bibinfo{person}{Francesco Orabona}.} \bibinfo{year}{2025}\natexlab{}.
\newblock \bibinfo{title}{Offline and Online {KL}-Regularized {RLHF} under
  Differential Privacy}.
\newblock \bibinfo{howpublished}{arXiv preprint arXiv:2510.13512v1}.
\newblock
\urldef\tempurl%
\url{https://arxiv.org/abs/2510.13512v1}
\showURL{%
\tempurl}


\bibitem[Xiang et~al\mbox{.}(2026)]%
        {xiang2026personaskills}
\bibfield{author}{\bibinfo{person}{Yongli Xiang}, \bibinfo{person}{Zhifang
  Zhang}, \bibinfo{person}{Bojun Yang}, \bibinfo{person}{Ziming Hong},
  \bibinfo{person}{Lei Feng}, \bibinfo{person}{Miao Xu}, {and}
  \bibinfo{person}{Tongliang Liu}.} \bibinfo{year}{2026}\natexlab{}.
\newblock \bibinfo{title}{When Agents Learn to Be You: Benchmarking Privacy
  Leakage, Impersonation Risk, and Defenses in Persona Skills}.
\newblock \bibinfo{howpublished}{arXiv preprint arXiv:2608.03700v1 [cs.CR]}.
\newblock
\urldef\tempurl%
\url{https://arxiv.org/abs/2608.03700v1}
\showURL{%
\tempurl}


\bibitem[Xie et~al\mbox{.}(2026a)]%
        {xie2026bemula}
\bibfield{author}{\bibinfo{person}{Bofan Xie}, \bibinfo{person}{Chaokun Wang},
  \bibinfo{person}{Guanchen Ge}, \bibinfo{person}{Songyao Wang},
  \bibinfo{person}{Fang Niu}, \bibinfo{person}{Hang Zhang}, {and}
  \bibinfo{person}{Wei Wang}.} \bibinfo{year}{2026}\natexlab{a}.
\newblock \bibinfo{title}{Can Agents Forget When Asked? A Benchmark for Memory
  Unlearning in {LLM} Agents}.
\newblock \bibinfo{howpublished}{arXiv preprint arXiv:2610.18052v1 [cs.AI]}.
\newblock
\urldef\tempurl%
\url{https://arxiv.org/abs/2610.18052v1}
\showURL{%
\tempurl}


\bibitem[Xie et~al\mbox{.}(2022)]%
        {xie2022skipper}
\bibfield{author}{\bibinfo{person}{Fuman Xie}, \bibinfo{person}{Yanjun Zhang},
  \bibinfo{person}{Chuan Yan}, \bibinfo{person}{Suwan Li}, \bibinfo{person}{Lei
  Bu}, \bibinfo{person}{Kai Chen}, \bibinfo{person}{Zi Huang}, {and}
  \bibinfo{person}{Guangdong Bai}.} \bibinfo{year}{2022}\natexlab{}.
\newblock \showarticletitle{Scrutinizing Privacy Policy Compliance of Virtual
  Personal Assistant Apps}. In \bibinfo{booktitle}{\emph{Proceedings of the
  37th {IEEE}/{ACM} International Conference on Automated Software
  Engineering}} (Rochester, MI, USA). \bibinfo{publisher}{Association for
  Computing Machinery}, \bibinfo{address}{New York, NY, USA}, Article
  \bibinfo{articleno}{90}, \bibinfo{numpages}{13}~pages.
\newblock
\href{https://doi.org/10.1145/3551349.3560416}{doi:\nolinkurl{10.1145/3551349.3560416}}


\bibitem[Xie et~al\mbox{.}(2026b)]%
        {xie2026veiltune}
\bibfield{author}{\bibinfo{person}{Huiyu Xie}, \bibinfo{person}{Tanping Zhou},
  \bibinfo{person}{Tao Hu}, \bibinfo{person}{Xinlei Liu}, {and}
  \bibinfo{person}{Hailong Ma}.} \bibinfo{year}{2026}\natexlab{b}.
\newblock \bibinfo{title}{{VeilTune}: Secure Fine-Tuning of Private {LoRA}
  Adapters on Public Pretrained Language Models}.
\newblock \bibinfo{howpublished}{arXiv preprint arXiv:2611.11817v1}.
\newblock
\urldef\tempurl%
\url{https://arxiv.org/abs/2611.11817v1}
\showURL{%
\tempurl}


\bibitem[Xu et~al\mbox{.}(2024)]%
        {xu2024unlearnsurvey}
\bibfield{author}{\bibinfo{person}{Heng Xu}, \bibinfo{person}{Tianqing Zhu},
  \bibinfo{person}{Lefeng Zhang}, \bibinfo{person}{Wanlei Zhou}, {and}
  \bibinfo{person}{Philip~S. Yu}.} \bibinfo{year}{2024}\natexlab{}.
\newblock \showarticletitle{Machine Unlearning: A Survey}.
\newblock \bibinfo{journal}{\emph{Comput. Surveys}} \bibinfo{volume}{56},
  \bibinfo{number}{1}, Article \bibinfo{articleno}{9} (\bibinfo{year}{2024}),
  \bibinfo{numpages}{36}~pages.
\newblock
\href{https://doi.org/10.1145/3603620}{doi:\nolinkurl{10.1145/3603620}}


\bibitem[Xu et~al\mbox{.}(2026a)]%
        {xu2026dpdylora}
\bibfield{author}{\bibinfo{person}{Jie Xu}, \bibinfo{person}{Karthikeyan
  Saravanan}, \bibinfo{person}{Rogier van Dalen}, \bibinfo{person}{Haaris
  Mehmood}, \bibinfo{person}{David Tuckey}, {and} \bibinfo{person}{Mete Ozay}.}
  \bibinfo{year}{2026}\natexlab{a}.
\newblock \showarticletitle{{DP-DyLoRA}: Fine-Tuning Transformer-Based Models
  On-Device under Differentially Private Federated Learning using Dynamic
  Low-Rank Adaptation}. In \bibinfo{booktitle}{\emph{2026 {IEEE} International
  Conference on Communications ({ICC})}} (Glasgow, United Kingdom).
  \bibinfo{publisher}{IEEE}, \bibinfo{address}{Piscataway, NJ, USA},
  \bibinfo{pages}{1--7}.
\newblock
\href{https://doi.org/10.1109/ICC59461.2026.11587764}{doi:\nolinkurl{10.1109/ICC59461.2026.11587764}}


\bibitem[Xu et~al\mbox{.}(2026b)]%
        {xu2026hushbench}
\bibfield{author}{\bibinfo{person}{Lingxiang Xu}, \bibinfo{person}{Jiaoyun
  Yang}, \bibinfo{person}{Min Hu}, {and} \bibinfo{person}{Ning An}.}
  \bibinfo{year}{2026}\natexlab{b}.
\newblock \bibinfo{title}{{HUSH-Bench}: Measuring Memory-Use Boundaries for
  Sensitive History in Conversational Agents}.
\newblock \bibinfo{howpublished}{arXiv preprint arXiv:2606.06055v2}.
\newblock
\href{https://doi.org/10.48550/arXiv.2606.06055}{doi:\nolinkurl{10.48550/arXiv.2606.06055}}


\bibitem[Xu et~al\mbox{.}(2026c)]%
        {xu2026dynamicgraphcontext}
\bibfield{author}{\bibinfo{person}{Yuanyuan Xu}, \bibinfo{person}{Wenjie
  Zhang}, \bibinfo{person}{Yin Chen}, \bibinfo{person}{Xuemin Lin}, {and}
  \bibinfo{person}{Ying Zhang}.} \bibinfo{year}{2026}\natexlab{c}.
\newblock \bibinfo{title}{Self-Evolving Agents as Dynamic Graph Transformation:
  A Survey and New Perspective}.
\newblock \bibinfo{howpublished}{arXiv preprint arXiv:2608.18104v1 [cs.AI]}.
\newblock
\href{https://doi.org/10.48550/arXiv.2608.18104}{doi:\nolinkurl{10.48550/arXiv.2608.18104}}


\bibitem[Xu et~al\mbox{.}(2023)]%
        {xu2023gboard}
\bibfield{author}{\bibinfo{person}{Zheng Xu}, \bibinfo{person}{Yanxiang Zhang},
  \bibinfo{person}{Galen Andrew}, \bibinfo{person}{Christopher Choquette},
  \bibinfo{person}{Peter Kairouz}, \bibinfo{person}{Brendan McMahan},
  \bibinfo{person}{Jesse Rosenstock}, {and} \bibinfo{person}{Yuanbo Zhang}.}
  \bibinfo{year}{2023}\natexlab{}.
\newblock \showarticletitle{Federated Learning of {Gboard} Language Models with
  Differential Privacy}. In \bibinfo{booktitle}{\emph{Proceedings of the 61st
  Annual Meeting of the Association for Computational Linguistics (Volume 5:
  Industry Track)}}, \bibfield{editor}{\bibinfo{person}{Sunayana Sitaram},
  \bibinfo{person}{Beata Beigman~Klebanov}, {and} \bibinfo{person}{Jason~D
  Williams}} (Eds.). \bibinfo{publisher}{Association for Computational
  Linguistics}, \bibinfo{address}{Toronto, Canada}, \bibinfo{pages}{629--639}.
\newblock
\href{https://doi.org/10.18653/v1/2023.acl-industry.60}{doi:\nolinkurl{10.18653/v1/2023.acl-industry.60}}


\bibitem[Yang et~al\mbox{.}(2026b)]%
        {yang2026p3m}
\bibfield{author}{\bibinfo{person}{Dishu Yang}, \bibinfo{person}{Jingjing Liu},
  {and} \bibinfo{person}{Jize Li}.} \bibinfo{year}{2026}\natexlab{b}.
\newblock \bibinfo{title}{Balancing Privacy, Utility, and Safety in {LLM}
  Alignment through Preference Optimization}.
\newblock \bibinfo{howpublished}{arXiv:2608.30141v1 [cs.CR]}.
\newblock
\urldef\tempurl%
\url{https://arxiv.org/abs/2608.30141v1}
\showURL{%
\tempurl}


\bibitem[Yang et~al\mbox{.}(2026d)]%
        {yang2026federatedskill}
\bibfield{author}{\bibinfo{person}{Jingbo Yang}, \bibinfo{person}{Guanyu Yao},
  \bibinfo{person}{Yang Zhang}, \bibinfo{person}{Ramana~Rao Kompella},
  \bibinfo{person}{Gaowen Liu}, {and} \bibinfo{person}{Shiyu Chang}.}
  \bibinfo{year}{2026}\natexlab{d}.
\newblock \bibinfo{title}{{FederatedSkill}: Federated Learning for Agentic
  Skill Evolution}.
\newblock \bibinfo{howpublished}{arXiv preprint arXiv:2606.03143v1 [cs.LG]}.
\newblock
\urldef\tempurl%
\url{https://arxiv.org/abs/2606.03143v1}
\showURL{%
\tempurl}


\bibitem[Yang et~al\mbox{.}(2026c)]%
        {yang2026memjoin}
\bibfield{author}{\bibinfo{person}{Qishun Yang}, \bibinfo{person}{Shu Yang},
  \bibinfo{person}{Difei Xu}, \bibinfo{person}{Di Wang}, {and}
  \bibinfo{person}{Meng Ding}.} \bibinfo{year}{2026}\natexlab{c}.
\newblock \bibinfo{title}{Can Agents Keep Secrets They Were Never Told?
  Benchmarking Latent Privacy Risks Across Sessions with {MemJoin}}.
\newblock \bibinfo{howpublished}{arXiv preprint arXiv:2610.24359v1 [cs.CR]}.
\newblock
\href{https://doi.org/10.48550/arXiv.2610.24359}{doi:\nolinkurl{10.48550/arXiv.2610.24359}}


\bibitem[Yang et~al\mbox{.}(2026a)]%
        {yang2026reglos}
\bibfield{author}{\bibinfo{person}{Sen Yang}, \bibinfo{person}{Jiaduo Lin},
  \bibinfo{person}{Zehong Huang}, {and} \bibinfo{person}{Wen Liu}.}
  \bibinfo{year}{2026}\natexlab{a}.
\newblock \bibinfo{title}{{Reglos}: A Self-Evolving Personal Memory Agent with
  Recursive Self-Improvement}.
\newblock \bibinfo{howpublished}{arXiv preprint arXiv:2611.06049v1 [cs.AI]}.
\newblock
\urldef\tempurl%
\url{https://arxiv.org/abs/2611.06049v1}
\showURL{%
\tempurl}


\bibitem[Yao et~al\mbox{.}(2026)]%
        {yao2026executionunlearning}
\bibfield{author}{\bibinfo{person}{Chao Yao}, \bibinfo{person}{Yangbo Wei},
  \bibinfo{person}{Zhen Huang}, \bibinfo{person}{Junhong Qian},
  \bibinfo{person}{Chenle Chen}, \bibinfo{person}{Shaoqiang Lu},
  \bibinfo{person}{Chen Wu}, {and} \bibinfo{person}{Lei He}.}
  \bibinfo{year}{2026}\natexlab{}.
\newblock \bibinfo{title}{Forgetting Without Restarting: Execution-State
  Unlearning for Stateful {LLM} Agents}.
\newblock \bibinfo{howpublished}{arXiv preprint arXiv:2609.04875v1 [cs.CR]}.
\newblock
\urldef\tempurl%
\url{https://arxiv.org/abs/2609.04875v1}
\showURL{%
\tempurl}


\bibitem[Ye et~al\mbox{.}(2026)]%
        {ye2026secureforgetting}
\bibfield{author}{\bibinfo{person}{Dayong Ye}, \bibinfo{person}{Tianqing Zhu},
  \bibinfo{person}{Congcong Zhu}, \bibinfo{person}{Feng He},
  \bibinfo{person}{Qi He}, \bibinfo{person}{Wang Shang}, \bibinfo{person}{Bo
  Liu}, {and} \bibinfo{person}{Wanlei Zhou}.} \bibinfo{year}{2026}\natexlab{}.
\newblock \bibinfo{title}{Secure Forgetting: A Framework for Privacy-Driven
  Unlearning in Large Language Model ({LLM})-Based Agents}.
\newblock \bibinfo{howpublished}{arXiv preprint arXiv:2604.00430v1}.
\newblock
\urldef\tempurl%
\url{https://arxiv.org/abs/2604.00430v1}
\showURL{%
\tempurl}


\bibitem[Ying et~al\mbox{.}(2026)]%
        {ying2026skilljack}
\bibfield{author}{\bibinfo{person}{Zonghao Ying}, \bibinfo{person}{Xiangfan
  Wu}, \bibinfo{person}{Huiyu Wu}, \bibinfo{person}{Xing Zheng},
  \bibinfo{person}{Huangsheng Cheng}, \bibinfo{person}{Xiaorong Shi}, {and}
  \bibinfo{person}{Jing Guo}.} \bibinfo{year}{2026}\natexlab{}.
\newblock \bibinfo{title}{{SkillJack}: Persistent Skill Backdoors in
  Self-Evolving Agents}.
\newblock \bibinfo{howpublished}{arXiv preprint arXiv:2608.03509v2}.
\newblock
\urldef\tempurl%
\url{https://arxiv.org/abs/2608.03509v2}
\showURL{%
\tempurl}


\bibitem[Yu et~al\mbox{.}(2026)]%
        {yu2026kbench}
\bibfield{author}{\bibinfo{person}{Guangsheng Yu}, \bibinfo{person}{Yanna
  Jiang}, \bibinfo{person}{Qin Wang}, \bibinfo{person}{Baihe Ma}, {and}
  \bibinfo{person}{Xu Wang}.} \bibinfo{year}{2026}\natexlab{}.
\newblock \bibinfo{title}{{K-Bench}: A Benchmark for {LLM} Unlearning in
  Agentic Deployments}.
\newblock \bibinfo{howpublished}{arXiv preprint arXiv:2609.12808v1}.
\newblock
\urldef\tempurl%
\url{https://arxiv.org/abs/2609.12808v1}
\showURL{%
\tempurl}


\bibitem[Zeng et~al\mbox{.}(2025)]%
        {zeng2025sage}
\bibfield{author}{\bibinfo{person}{Shenglai Zeng}, \bibinfo{person}{Jiankun
  Zhang}, \bibinfo{person}{Pengfei He}, \bibinfo{person}{Jie Ren},
  \bibinfo{person}{Tianqi Zheng}, \bibinfo{person}{Hanqing Lu},
  \bibinfo{person}{Han Xu}, \bibinfo{person}{Hui Liu}, \bibinfo{person}{Yue
  Xing}, {and} \bibinfo{person}{Jiliang Tang}.}
  \bibinfo{year}{2025}\natexlab{}.
\newblock \showarticletitle{Mitigating the Privacy Issues in
  Retrieval-Augmented Generation ({RAG}) via Pure Synthetic Data}. In
  \bibinfo{booktitle}{\emph{Proceedings of the 2025 Conference on Empirical
  Methods in Natural Language Processing}},
  \bibfield{editor}{\bibinfo{person}{Christos Christodoulopoulos},
  \bibinfo{person}{Tanmoy Chakraborty}, \bibinfo{person}{Carolyn Rose}, {and}
  \bibinfo{person}{Violet Peng}} (Eds.). \bibinfo{publisher}{Association for
  Computational Linguistics}, \bibinfo{address}{Suzhou, China},
  \bibinfo{pages}{24527--24558}.
\newblock
\showISBNx{979-8-89176-332-6}
\href{https://doi.org/10.18653/v1/2025.emnlp-main.1247}{doi:\nolinkurl{10.18653/v1/2025.emnlp-main.1247}}


\bibitem[Zhan et~al\mbox{.}(2026)]%
        {zhan2026pecl}
\bibfield{author}{\bibinfo{person}{Bihao Zhan}, \bibinfo{person}{Jie Zhou},
  \bibinfo{person}{Junsong Li}, \bibinfo{person}{Yutao Yang},
  \bibinfo{person}{Shilian Chen}, \bibinfo{person}{Qianjun Pan},
  \bibinfo{person}{Xin Li}, \bibinfo{person}{Wen Wu}, \bibinfo{person}{Xingjiao
  Wu}, \bibinfo{person}{Qin Chen}, \bibinfo{person}{Hang Yan},
  \bibinfo{person}{Liang Dou}, {and} \bibinfo{person}{Liang He}.}
  \bibinfo{year}{2026}\natexlab{}.
\newblock \showarticletitle{Forget What's Sensitive, Remember What Matters:
  Token-Level Differential Privacy in Memory Sculpting for Continual Learning}.
  In \bibinfo{booktitle}{\emph{Proceedings of the 2026 Conference on Empirical
  Methods in Natural Language Processing}}. \bibinfo{publisher}{Association for
  Computational Linguistics}, \bibinfo{address}{Budapest, Hungary},
  \bibinfo{pages}{847--861}.
\newblock
\href{https://doi.org/10.18653/2026.emnlp-main.39}{doi:\nolinkurl{10.18653/2026.emnlp-main.39}}


\bibitem[Zhang et~al\mbox{.}(2026e)]%
        {zhang2026routingdp}
\bibfield{author}{\bibinfo{person}{Annan Zhang}, \bibinfo{person}{Shengjie
  Song}, \bibinfo{person}{Wenshuai Yao}, \bibinfo{person}{Lihao Zhang}, {and}
  \bibinfo{person}{Tianmiao}.} \bibinfo{year}{2026}\natexlab{e}.
\newblock \bibinfo{title}{Sensitivity Inflation from Cross-Record Precision
  Routing in Private Quantized Fine-Tuning}.
\newblock \bibinfo{howpublished}{arXiv:2611.01411v1 [cs.LG]}.
\newblock
\urldef\tempurl%
\url{https://arxiv.org/abs/2611.01411v1}
\showURL{%
\tempurl}


\bibitem[Zhang et~al\mbox{.}(2026a)]%
        {zhang2026memcut}
\bibfield{author}{\bibinfo{person}{Cheng Zhang}, \bibinfo{person}{Xuanning
  Cao}, \bibinfo{person}{Zhicheng Zheng}, {and} \bibinfo{person}{Hui Xiong}.}
  \bibinfo{year}{2026}\natexlab{a}.
\newblock \bibinfo{title}{{MemCut}: Every Secret Leaves a Trail in the
  Long-Term Memory of {LLM} Agents}.
\newblock \bibinfo{howpublished}{arXiv preprint arXiv:2610.21405v1 [cs.AI]}.
\newblock
\urldef\tempurl%
\url{https://arxiv.org/abs/2610.21405v1}
\showURL{%
\tempurl}


\bibitem[Zhang et~al\mbox{.}(2026d)]%
        {zhang2026skilllife}
\bibfield{author}{\bibinfo{person}{Chenxu Zhang}, \bibinfo{person}{Zhou Liu},
  \bibinfo{person}{Wu Xin}, {and} \bibinfo{person}{Wentao Zhang}.}
  \bibinfo{year}{2026}\natexlab{d}.
\newblock \bibinfo{title}{{SkillLifeBench}: Benchmarking Security Across the
  Lifecycle of Agent Skills}.
\newblock \bibinfo{howpublished}{arXiv preprint arXiv:2610.35215v1}.
\newblock
\urldef\tempurl%
\url{https://arxiv.org/abs/2610.35215v1}
\showURL{%
\tempurl}


\bibitem[Zhang et~al\mbox{.}(2026b)]%
        {zhang2026memgate}
\bibfield{author}{\bibinfo{person}{Jiawen Zhang}, \bibinfo{person}{Kejia Chen},
  \bibinfo{person}{Jiachen Ma}, \bibinfo{person}{Yangfan Hu},
  \bibinfo{person}{Lipeng He}, \bibinfo{person}{Yechao Zhang},
  \bibinfo{person}{Jian Liu}, \bibinfo{person}{Xiaohu Yang},
  \bibinfo{person}{Tianwei Zhang}, {and} \bibinfo{person}{Ruoxi Jia}.}
  \bibinfo{year}{2026}\natexlab{b}.
\newblock \bibinfo{title}{Beyond Similarity: Trustworthy Memory Search for
  Personal {AI} Agents}.
\newblock \bibinfo{howpublished}{arXiv preprint arXiv:2606.06054v1}.
\newblock
\urldef\tempurl%
\url{https://arxiv.org/abs/2606.06054v1}
\showURL{%
\tempurl}


\bibitem[Zhang et~al\mbox{.}(2026c)]%
        {zhang2026conmem}
\bibfield{author}{\bibinfo{person}{Shuning Zhang}, \bibinfo{person}{Tenghao
  Ji}, \bibinfo{person}{Jingruo Chen}, \bibinfo{person}{Eryue Xu},
  \bibinfo{person}{Bohao Chu}, {and} \bibinfo{person}{Xin Yi}.}
  \bibinfo{year}{2026}\natexlab{c}.
\newblock \bibinfo{title}{{ConMem}: Counterfactual Multi-Scenario Probing for
  Resolving {LLM} Memory Misalignment}.
\newblock \bibinfo{howpublished}{arXiv preprint arXiv:2610.33432v1}.
\newblock
\urldef\tempurl%
\url{https://arxiv.org/abs/2610.33432v1}
\showURL{%
\tempurl}


\bibitem[Zhang et~al\mbox{.}(2025)]%
        {zhang2025klregularization}
\bibfield{author}{\bibinfo{person}{Yizhou Zhang}, \bibinfo{person}{Kishan
  Panaganti}, \bibinfo{person}{Laixi Shi}, \bibinfo{person}{Juba Ziani}, {and}
  \bibinfo{person}{Adam Wierman}.} \bibinfo{year}{2025}\natexlab{}.
\newblock \bibinfo{title}{{KL}-Regularization Itself Is Differentially Private
  in Bandits and {RLHF}}.
\newblock \bibinfo{howpublished}{arXiv preprint arXiv:2505.18407v2 [cs.LG]}.
\newblock
\urldef\tempurl%
\url{https://arxiv.org/abs/2505.18407v2}
\showURL{%
\tempurl}


\bibitem[Zhang(2026)]%
        {zhang2026ragaccounting}
\bibfield{author}{\bibinfo{person}{Zhiyuan Zhang}.}
  \bibinfo{year}{2026}\natexlab{}.
\newblock \bibinfo{title}{From One Query to Many: A Critical Survey of Privacy
  Accounting in {RAG}}.
\newblock \bibinfo{howpublished}{arXiv preprint arXiv:2610.21245v2}.
\newblock
\urldef\tempurl%
\url{https://arxiv.org/abs/2610.21245v2}
\showURL{%
\tempurl}


\bibitem[Zhao et~al\mbox{.}(2024)]%
        {zhao2024expel}
\bibfield{author}{\bibinfo{person}{Andrew Zhao}, \bibinfo{person}{Daniel
  Huang}, \bibinfo{person}{Quentin Xu}, \bibinfo{person}{Matthieu Lin},
  \bibinfo{person}{Yong-Jin Liu}, {and} \bibinfo{person}{Gao Huang}.}
  \bibinfo{year}{2024}\natexlab{}.
\newblock \showarticletitle{{ExpeL}: {LLM} Agents Are Experiential Learners}.
\newblock \bibinfo{journal}{\emph{Proceedings of the AAAI Conference on
  Artificial Intelligence}} \bibinfo{volume}{38}, \bibinfo{number}{17}
  (\bibinfo{year}{2024}), \bibinfo{pages}{19632--19642}.
\newblock
\href{https://doi.org/10.1609/aaai.v38i17.29936}{doi:\nolinkurl{10.1609/aaai.v38i17.29936}}


\bibitem[Zhao et~al\mbox{.}(2026a)]%
        {zhao2026faithfulness}
\bibfield{author}{\bibinfo{person}{Weixiang Zhao}, \bibinfo{person}{Yingshuo
  Wang}, \bibinfo{person}{Yichen Zhang}, \bibinfo{person}{Yang Deng},
  \bibinfo{person}{Yanyan Zhao}, \bibinfo{person}{Wanxiang Che},
  \bibinfo{person}{Bing Qin}, {and} \bibinfo{person}{Ting Liu}.}
  \bibinfo{year}{2026}\natexlab{a}.
\newblock \showarticletitle{Large Language Model Agents Are Not Always Faithful
  Self-Evolvers}. In \bibinfo{booktitle}{\emph{Proceedings of the 43rd
  International Conference on Machine Learning}}
  \emph{(\bibinfo{series}{Proceedings of Machine Learning Research},
  Vol.~\bibinfo{volume}{306})}, \bibfield{editor}{\bibinfo{person}{Tong Zhang},
  \bibinfo{person}{Miroslav Dudik}, \bibinfo{person}{Martin Jaggi},
  \bibinfo{person}{Alekh Agarwal}, \bibinfo{person}{Sharon Li},
  \bibinfo{person}{Dale Schuurmans}, \bibinfo{person}{Jerry Zhu},
  \bibinfo{person}{Felix Berkenkamp}, \bibinfo{person}{Hanze Dong}, {and}
  \bibinfo{person}{Alberto Bietti}} (Eds.). \bibinfo{publisher}{PMLR},
  \bibinfo{address}{Seoul, Republic of Korea}, \bibinfo{pages}{162243--162268}.
\newblock
\urldef\tempurl%
\url{https://proceedings.mlr.press/v306/zhao26aw.html}
\showURL{%
\tempurl}


\bibitem[Zhao et~al\mbox{.}(2026b)]%
        {zhao2026maskclaw}
\bibfield{author}{\bibinfo{person}{Yanqiu Zhao}, \bibinfo{person}{Dongying
  Zheng}, \bibinfo{person}{Kaibo Huang}, \bibinfo{person}{Yukun Wei},
  \bibinfo{person}{Zhongliang Yang}, {and} \bibinfo{person}{Linna Zhou}.}
  \bibinfo{year}{2026}\natexlab{b}.
\newblock \bibinfo{title}{{MaskClaw}: Edge-Side Personalized Privacy
  Arbitration for {GUI} Agents with Behavior-Driven Skill Evolution}.
\newblock \bibinfo{howpublished}{arXiv preprint arXiv:2605.28646v2 [cs.CR]}.
\newblock
\urldef\tempurl%
\url{https://arxiv.org/abs/2605.28646v2}
\showURL{%
\tempurl}


\bibitem[Zhou et~al\mbox{.}(2026)]%
        {zhou2026timlmembench}
\bibfield{author}{\bibinfo{person}{Fangxing Zhou}, \bibinfo{person}{Yifei Sun},
  \bibinfo{person}{Jiachen Li}, \bibinfo{person}{Jinyu Xu},
  \bibinfo{person}{Yuhan Wang}, \bibinfo{person}{Ruixin Li}, {and}
  \bibinfo{person}{Qing Xie}.} \bibinfo{year}{2026}\natexlab{}.
\newblock \showarticletitle{{TimlMemBench}: A Diagnostic Benchmark for Governed
  Use of Long-Term Memory in Dialogue Agents}. In
  \bibinfo{booktitle}{\emph{Findings of the Association for Computational
  Linguistics: EMNLP 2026}}. \bibinfo{publisher}{Association for Computational
  Linguistics}, \bibinfo{address}{Budapest, Hungary},
  \bibinfo{pages}{10624--10638}.
\newblock
\href{https://doi.org/10.18653/2026.findings-emnlp.507}{doi:\nolinkurl{10.18653/2026.findings-emnlp.507}}


\bibitem[Zhou et~al\mbox{.}(2025)]%
        {zhou2025squarepo}
\bibfield{author}{\bibinfo{person}{Xingyu Zhou}, \bibinfo{person}{Yulian Wu},
  \bibinfo{person}{Wenqian Weng}, {and} \bibinfo{person}{Francesco Orabona}.}
  \bibinfo{year}{2025}\natexlab{}.
\newblock \showarticletitle{Square{$\chi$}{PO}: Differentially Private and
  Robust {$\chi^2$}-Preference Optimization in Offline Direct Alignment}. In
  \bibinfo{booktitle}{\emph{Proceedings of the 42nd International Conference on
  Machine Learning}} \emph{(\bibinfo{series}{Proceedings of Machine Learning
  Research}, Vol.~\bibinfo{volume}{267})}. \bibinfo{publisher}{PMLR},
  \bibinfo{address}{Vancouver, Canada}, \bibinfo{pages}{79389--79413}.
\newblock
\urldef\tempurl%
\url{https://proceedings.mlr.press/v267/zhou25ad.html}
\showURL{%
\tempurl}


\bibitem[Zhu et~al\mbox{.}(2025)]%
        {zhu2025compass}
\bibfield{author}{\bibinfo{person}{Jinhao Zhu}, \bibinfo{person}{Liana Patel},
  \bibinfo{person}{Matei Zaharia}, {and} \bibinfo{person}{Raluca~Ada Popa}.}
  \bibinfo{year}{2025}\natexlab{}.
\newblock \showarticletitle{{Compass}: Encrypted Semantic Search with High
  Accuracy}. In \bibinfo{booktitle}{\emph{19th {USENIX} Symposium on Operating
  Systems Design and Implementation ({OSDI} 25)}}. \bibinfo{publisher}{USENIX
  Association}, \bibinfo{address}{Boston, MA}, \bibinfo{pages}{915--938}.
\newblock
\showISBNx{978-1-939133-47-2}
\urldef\tempurl%
\url{https://www.usenix.org/conference/osdi25/presentation/zhu-jinhao}
\showURL{%
\tempurl}


\end{thebibliography}

\end{document}

\title[Supplementary Material: Privacy Protection in Self-Evolving Personal Agents]{Supplementary Material for: Privacy Protection in Self-Evolving Personal Agents: Persistent Memory and Cross-Carrier Transfer}

\author{Zibin Lin}
\affiliation{%
  \institution{Shenzhen University}
  \city{Shenzhen}
  \country{China}
}
\email{linaacc9595@gmail.com}

\author{Taotao Wang}
\affiliation{%
  \institution{Shenzhen University}
  \city{Shenzhen}
  \country{China}
}
\email{ttwang@szu.edu.cn}

\author{Shengli Zhang}
\affiliation{%
  \institution{Shenzhen University}
  \city{Shenzhen}
  \country{China}
}
\email{zsl@szu.edu.cn}

\begin{abstract}
This supplement documents searches and candidate decisions for literature available through 10 November 2026. It includes a deduplicated source ledger, primary-source locators for 22 examples, and diagnostic protocols and results. Sections~\ref{sec:supp-operation-evidence} and~\ref{sec:supp-deployment-cases} give the operation matrix and source-level checks for the three deployment comparisons.
\end{abstract}

\begin{CCSXML}
<ccs2012>
   <concept>
       <concept_id>10002978.10003029.10011150</concept_id>
       <concept_desc>Security and privacy~Privacy protections</concept_desc>
       <concept_significance>500</concept_significance>
   </concept>
   <concept>
       <concept_id>10010147.10010178.10010219.10010221</concept_id>
       <concept_desc>Computing methodologies~Intelligent agents</concept_desc>
       <concept_significance>500</concept_significance>
   </concept>
   <concept>
       <concept_id>10010147.10010257.10010282.10010284</concept_id>
       <concept_desc>Computing methodologies~Continual learning</concept_desc>
       <concept_significance>300</concept_significance>
   </concept>
   <concept>
       <concept_id>10002951.10003260.10003261.10003271</concept_id>
       <concept_desc>Information systems~Personalization</concept_desc>
       <concept_significance>300</concept_significance>
   </concept>
</ccs2012>
\end{CCSXML}

\ccsdesc[500]{Security and privacy~Privacy protections}
\ccsdesc[500]{Computing methodologies~Intelligent agents}
\ccsdesc[300]{Computing methodologies~Continual learning}
\ccsdesc[300]{Information systems~Personalization}

\keywords{personal agents, self-evolution, privacy protection, continual learning, long-term memory, machine unlearning}

\maketitle

\section{Literature Search Protocol, Version-Pinning Policy, and Screening Records}
\label{sec:supp-search}

\subsection{Review Scope, Cutoff Date, and Multi-Stage Search Architecture}
\label{sec:supp-search-scope}

The review uses a bounded thematic search for privacy mechanisms, their prerequisites, and evidence about their interfaces. The literature cutoff is \textbf{10 November 2026}. Each retained work had a publicly accessible primary manuscript or authoritative archival record by that date. The initial search ran on 10--11 November and the revision checks on 18--19 November.

To ensure sufficient depth across both direct personal-agent prototypes and transferable formal primitives without conflating general agent capabilities with privacy guarantees, the original 75-source corpus was assembled in three stages and then expanded:
\begin{enumerate}
  \item \textbf{Stage 1: Core Direct-Intersection Baseline and Bounded Gap Audit (Sources \#1--\#25; executed 10--11 November 2026).} Starting from an initial seed analysis of update-operator mechanisms, we executed a structured gap audit targeting four sparse areas of the mechanism map: (i) the direct intersection of personal agents, self-evolution, and privacy; (ii) persistent memory forgetting and derived-state deletion; (iii) user-level, continual, and cross-round formal guarantees; and (iv) private feedback, preference, experience, and skill updates. This stage established the 25-source core baseline comprising 20 representative mechanisms (including newly identified \texttt{SBU}~\cite{wang2026agenticunlearning} coded as \texttt{T} and \texttt{Galaxy}~\cite{bao2025galaxy} coded as \texttt{S}), four adjacent safety and privacy surveys~\cite{chen2026safetysurvey,lahjouji2026agentsknowtoomuch,edemacu2025promptsurvey,chen2026preferencesurvey}, and one operational definition survey on self-evolving agents~\cite{gao2026selfevolvingsurvey}.
  \item \textbf{Stage 2: Operator-Adjacent Thematic Expansion (Sources \#26--\#62; $+37$ unique sources; executed 11 November 2026).} While holding the long-term single-user personal-agent scope fixed, we expanded retrieval across seven adjacent mechanism, primitive, and evaluation themes: prompt and parameter differential privacy (DP), preference protection, continual learning and adaptive composition, private retrieval and oblivious dynamic indexing, machine unlearning and deletion verification, information-flow control and skill evolution, and lifecycle attack and residue benchmarks. All 25 core sources were retained and 37 non-duplicate sources were added, yielding 62 unique sources.
  \item \textbf{Stage 3: Targeted ACM Venue Supplementation (Sources \#63--\#75; $+13$ unique sources; executed 11 November 2026).} To address under-representation of ACM archival venues in the 62-source corpus, we conducted a targeted search over the \texttt{10.1145} DOI prefix (publication window \texttt{2018-01-01} to \texttt{2026-11-10}) across four focal areas: preference learning in personal privacy assistants, agent context and tool information flow, machine unlearning verification and version-differential leakage, and backward-secure dynamic searchable encryption. This stage added 13 peer-reviewed ACM sources (bringing the initial corpus to \textbf{75 unique sources}, of which 22 representative mechanisms are compared at operator level in Main Table~3) and documented six substantively related excluded ACM candidates.
  \item \textbf{Stage 4: Targeted coverage and boundary check (Sources \#76--\#86; $+11$ net; 18 November 2026).} We held the literature cutoff fixed and screened the works listed in Section~\ref{sec:supp-targeted-search}. Ten publications were newly added and AUBench was moved from a contextual citation into the evidence corpus. This stage brought the corpus to 86 sources.
  \item \textbf{Stage 5: Cross-session relation follow-up (Source \#87; $+1$; 18 November 2026).} An exact-title and arXiv identifier check added MemJoin as a risk and evaluation source with a scoped read-intervention code. This stage brought the corpus to 87 sources.
  \item \textbf{Stage 6: Core-coverage recheck (Sources \#88--\#93; $+6$; 18 November 2026).} We reran focused discovery across the four carrier families, verified the two benchmark omissions, and rechecked the coding of a skill-defense paper. Section~\ref{sec:supp-core-recheck} gives candidate decisions. This stage brought the corpus to 93 sources.
  \item \textbf{Stage 7: Fixed-cutoff retrieval and coding audit (Sources \#94--\#100; $+7$; 18 November 2026).} Four field queries and two referrals provide a complete candidate record for this pass (Section~\ref{sec:supp-cutoff-rerun}). We recoded PrivMeSA and checked representative-set substitutions in Sections~\ref{sec:supp-boundary-review} and~\ref{sec:supp-representative-sensitivity}. The corpus reached 100 sources.
  \item \textbf{Stage 8: Broader discovery and boundary re-adjudication (Sources \#101--\#110; $+10$; 18 November 2026).} Ranked web discovery, Crossref proceedings queries, and targeted primary-source checks cover private online feedback, prompt optimization, memory admission and reads, release behavior, skill distillation, withdrawal, and overlapping surveys (Section~\ref{sec:supp-broader-search}). The corpus reached 110 sources.
  \item \textbf{Stage 9: Targeted source and evidence check (Sources \#111--\#112; $+2$; 18 November 2026).} We screened two additional publication records and rechecked protocol examples (Section~\ref{sec:supp-final-screening}).
  \item \textbf{Stage 10: Fixed-window retrieval and evidence audit (Sources \#113--\#115; $+3$; 19 November 2026).} Section~\ref{sec:supp-stage10} lists the queries, candidates, and coding changes.
  \item \textbf{Stage 11: Targeted disclosure, skill, and withdrawal screening (Sources \#116--\#124; $+9$; 19 November 2026).} Section~\ref{sec:supp-stage11} records the complete returned windows for three fielded queries and two referral checks. This pass brought the corpus to 124 sources.
  \item \textbf{Stage 12: Carrier-stratified recheck (Sources \#125--\#138; $+14$; 19 November 2026).} Section~\ref{sec:supp-stage12} records the returned windows, candidate dispositions, and additional referrals. Section~\ref{sec:supp-independent-recoding} reports the separate coding check.
  \item \textbf{Stage 13: Full-window carrier audit and archival check (Sources \#139--\#165; $+27$; 19 November 2026).} Section~\ref{sec:supp-stage13} records the complete arXiv and anthology windows and the targeted NeurIPS check.
  \item \textbf{Stage 14: Observer and withdrawal follow-up (Sources \#166--\#169; $+4$; 19 November 2026).} Section~\ref{sec:supp-stage14} checks retrieval of known direct studies and adds matched evidence on joint observation, withdrawal, and tuning costs.
  \item \textbf{Stage 15: Observer and withdrawal windows and E6 rescreening (Sources \#170--\#180; $+11$; 19 November 2026).} Section~\ref{sec:supp-stage15} records both complete candidate windows, an AgentLeak referral and the revised dispositions for all 40 previous operation-overlap exclusions.
\end{enumerate}

\subsection{Queried Information Sources, Access Boundaries, and Stopping Criteria}
\label{sec:supp-search-sources}

Candidate discovery and primary-source verification combined scholarly indexes, DOI metadata registries, open-access preprint APIs, and official conference/journal proceedings. Crucially, anti-bot gates, human-verification challenges, and session timeouts encountered during interactive web queries were recorded strictly as incomplete channel access rather than zero hits or negative domain evidence.

In \textbf{IEEE Xplore} (queried 10 November 2026), a strict three-way intersection over personal agents, privacy, and continual/self-evolving updates returned 0 hits. A broader query (\texttt{personalized AND agent AND privacy}) returned 335 records; however, narrowing those records with \texttt{self-evolving}, \texttt{memory}, or \texttt{continual learning} yielded no candidates with a localizable privacy mechanism intervening in persistent state updates. These counts apply strictly to the recorded query syntax and date rather than the entire IEEE Xplore database.

In the \textbf{ACM Digital Library} (queried 10--11 November 2026), submitting the interactive advanced-search query \texttt{"personal agent" AND privacy AND ("self-evolving" OR "continual learning" OR memory)} triggered a Cloudflare verification block, and subsequent automated browser retries timed out without returning sequential result pages. Rather than recording zero hits, Stage~3 pivoted to Crossref metadata queries scoped to \texttt{prefix:10.1145} and exact-title lookups, followed by manual verification against official ACM DOI pages or author-archived camera-ready PDFs. Similarly, \textbf{Google Scholar} exact-title and forward-citation queries (10 November 2026) triggered human-verification challenges; this channel was logged as incomplete and neither interpreted as zero citations nor used to satisfy stopping criteria. Table~\ref{tab:supp-sources} summarizes all queried information sources and their interpretive boundaries.

\begin{table}[htbp]
\centering
\small
\setlength{\tabcolsep}{3pt}
\renewcommand{\arraystretch}{1.18}
\caption{Queried Scholarly Indexes and Repositories, Execution Status, Primary Role, and Interpretive Boundaries}
\label{tab:supp-sources}
\begin{tabular}{@{}>{\raggedright\arraybackslash}p{0.20\linewidth}>{\raggedright\arraybackslash}p{0.21\linewidth}>{\raggedright\arraybackslash}p{0.28\linewidth}>{\raggedright\arraybackslash}p{0.25\linewidth}@{}}
\toprule
\textbf{Information Source} & \textbf{Execution Status} & \textbf{Primary Role in Review} & \textbf{Interpretive Boundary} \\
\midrule
IEEE Xplore Advanced Search & Completed bounded queries (\texttt{2026-11-10}) & Check intersection of personalization, self-evolution, privacy, memory, and continual learning & Strict query returned 0; broad query returned 335 (0 qualifying after operator filtering); counts apply only to recorded syntax and date \\
ACM Digital Library / Crossref (\texttt{10.1145}) & Interactive search blocked by Cloudflare; Crossref DOI metadata and exact-title checks completed (\texttt{2026-11-10/11}) & Targeted supplementation on personal privacy assistants, agent release control, unlearning auditing, and encrypted indexing & Access block not recorded as 0 hits; not claimed as exhaustive ACM DL sweep; all 13 added items verified via ACM DOI or camera-ready PDF \\
Google Scholar & Blocked by human-verification gate (\texttt{2026-11-10}) & Exact-title lookup and forward-citation tracking & Not recorded as zero citations; excluded from stopping-rule evaluation \\
Bing Web Search & Completed bounded exact-title and semantic queries (\texttt{2026-11-10}) & Discover primary-manuscript entry points (e.g., surfacing \texttt{Reglos}~\cite{yang2026reglos} prior to arXiv verification) & Secondary summaries, blogs, and mirrors never used as technical evidence \\
arXiv API and Primary HTML/PDF & Completed 13 field-level API queries and full-text screening (\texttt{2026-11-11}) & Bounded gap audit, exact version pinning (\texttt{vN}), and section/algorithm/theorem localization & Raw query hit totals reflect exact arXiv field syntax at execution and are not extrapolated field-wide \\
Archival Proceedings (\texttt{ACL Anthology}, \texttt{OpenReview}, \texttt{PMLR}, \texttt{NeurIPS}, \texttt{ICLR}, \texttt{COLM}, \texttt{TMLR}, \texttt{USENIX}, \texttt{IEEE}, \texttt{AAAI}) & Verified per candidate (\texttt{2026-11-10/11}) & Confirm canonical peer-reviewed versions, bibliographic metadata, formal theorems, and experimental setups & Used for candidate-level primary verification; not claimed as independent exhaustive sweeps of each venue \\
\bottomrule
\end{tabular}
\end{table}

\paragraph{Two-Tier Stopping Rule and Methodological Boundaries.}
Search termination followed a two-tier protocol. At the \emph{direct-intersection tier} (Stage~1), retrieval halted once all newly surfaced candidates were verified against primary texts and two consecutive query families plus accessible one-hop exact-title/related-paper checks yielded no new personal-agent mechanism category (producing only duplicates, evaluation benchmarks, or established exclusion categories). In Stage~1, after arXiv queries \texttt{A1} and \texttt{A2} surfaced \texttt{Galaxy}~\cite{bao2025galaxy} and \texttt{SBU}~\cite{wang2026agenticunlearning}, queries \texttt{A3}--\texttt{A13} and accessible \texttt{Q5} checks returned only previously captured methods, background benchmarks, or excluded items. At the \emph{thematic expansion tier} (Stages~2--3), retrieval halted once all seven operator themes possessed mechanism, boundary, and risk/evaluation evidence and successive candidates fell into duplicate versions, identical technical precedents, or established exclusion codes. Because interactive ACM DL result pagination and Google Scholar forward-citation tracking were obstructed, and citation chasing was bounded to one hop, this protocol supports reproducible mechanism comparison within the initial 75-source corpus without claiming field-wide exhaustiveness; absence of evidence in our bounded search is never restated as proof of non-existence across the literature.

\subsection{Multi-Stage Search Queries and Field-Level Execution Counts}
\label{sec:supp-search-queries}

\subsubsection{Stage 1 Core Gap-Audit Query Families (Q1--Q5) and arXiv API Queries (A1--A13)}
Stage~1 defined four database-agnostic Boolean query families (\texttt{Q1}--\texttt{Q4}) and a bounded one-hop citation/related-work protocol (\texttt{Q5}):
\begin{verbatim}
Q1 (Direct intersection: personal agents, self-evolution, and privacy):
   ("personal agent" OR "personal AI" OR "personal assistant"
    OR "personalized agent" OR "long-term LLM agent")
   AND (privacy OR private OR confidentiality OR sensitive OR disclosure)
   AND ("self-evolving" OR "continual learning" OR "online learning"
    OR "memory update" OR writeback OR "skill evolution")

Q2 (Persistent memory, forgetting, and derived-artifact deletion):
   ("agent memory" OR "long-term memory" OR "persistent memory"
    OR "personal memory agent")
   AND (privacy OR redaction OR forgetting OR deletion OR unlearning
    OR "right to be forgotten" OR "certified deletion"
    OR "derived data" OR "derived artifact")
   AND (update OR writeback OR persistent OR continual OR "self-evolving")

Q3 (User-level, continual, and cross-round formal guarantees):
   ("personal agent" OR "LLM agent" OR "agent memory")
   AND ("user-level differential privacy"
    OR "continual differential privacy"
    OR "privacy composition" OR "cross-round" OR "privacy budget"
    OR "machine unlearning" OR "certified deletion"
    OR "derived data deletion")

Q4 (Private feedback, preference, lesson, skill, and experience updates):
   ("self-evolving agent" OR "personalized agent")
   AND (feedback OR preference OR lesson OR skill OR experience)
   AND (privacy OR private OR sanitization OR redaction
    OR "secure aggregation" OR "differential privacy")
\end{verbatim}
For \texttt{Q5} (\emph{one-hop citation and related-paper check}), we executed exact-title lookups for \texttt{PrivMeSA}~\cite{wang2026privmesa}, \texttt{HealthJev}~\cite{li2026healthjev}, \texttt{FedSkillGit}~\cite{lan2026fedskillgit}, \texttt{Reglos}~\cite{yang2026reglos}, \texttt{MemPrivacy}~\cite{chen2026memprivacy}, \texttt{Opal}~\cite{kaviani2026opal}, and \texttt{MaskClaw}~\cite{zhao2026maskclaw}, inspecting strictly one hop of backward references and accessible related-paper listings to prevent unbounded drift into generic agent-memory papers.

To operationalize \texttt{Q1}--\texttt{Q4} reproducibly over the arXiv API on 11 November 2026, the composite expressions were decomposed into 13 exact field queries (\texttt{A1}--\texttt{A13}). Table~\ref{tab:supp-arxiv} reports the exact API query strings, raw all-time return counts prior to cutoff/eligibility filtering, and screening summaries. Zero-hit queries (\texttt{A5}--\texttt{A9}, \texttt{A13}) reflect only exact phrase matching in the arXiv API at execution time.

\begin{table}[htbp]
\centering
\small
\setlength{\tabcolsep}{3.5pt}
\renewcommand{\arraystretch}{1.16}
\caption{Stage 1 Exact arXiv API Field Queries (\texttt{A1}--\texttt{A13}), Raw Return Counts, and Screening Outcomes}
\label{tab:supp-arxiv}
\begin{tabular}{@{}>{\raggedright\arraybackslash}p{0.06\linewidth}>{\raggedright\arraybackslash}p{0.49\linewidth}>{\raggedleft\arraybackslash}p{0.07\linewidth}>{\raggedright\arraybackslash}p{0.31\linewidth}@{}}
\toprule
\textbf{ID} & \textbf{arXiv API Query String} & \textbf{Hits} & \textbf{Screening Summary} \\
\midrule
\texttt{A1} & \texttt{all:"privacy-preserving" AND all:"self-evolving" AND all:agent} & 3 & \texttt{Galaxy}~\cite{bao2025galaxy} (included, \texttt{S}), \texttt{FederatedSkill} (\texttt{R06}), 1 survey \\
\texttt{A2} & \texttt{all:"machine unlearning" AND all:agent AND all:memory} & 4 & \texttt{SBU}~\cite{wang2026agenticunlearning} (included, \texttt{T}), \texttt{AUBench} (\texttt{R03}, benchmark), 2 surveys \\
\texttt{A3} & \texttt{all:"continual privacy" AND all:agent} & 1 & \texttt{FOCAL} (\texttt{R04}, excluded \texttt{E3}) \\
\texttt{A4} & \texttt{all:"privacy-preserving" AND all:"personal agent"} & 3 & \texttt{PPMI} (\texttt{R05}, \texttt{E1}), \texttt{Setoka} benchmark (\texttt{R12}, \texttt{E4}), 1 unrelated FL paper \\
\texttt{A5} & \texttt{all:"user-level differential privacy" AND all:agent} & 0 & No API return \\
\texttt{A6} & \texttt{all:"differential privacy" AND all:"personalized agent"} & 0 & No API return \\
\texttt{A7} & \texttt{all:"privacy composition" AND all:agent} & 0 & No API return \\
\texttt{A8} & \texttt{all:"certified deletion" AND all:agent} & 0 & No API return \\
\texttt{A9} & \texttt{all:"derived data deletion" AND all:agent} & 0 & No API return \\
\texttt{A10} & \texttt{all:privacy AND all:"memory update" AND all:agent} & 1 & Unrelated Rowhammer / federated-learning paper \\
\texttt{A11} & \texttt{all:privacy AND all:"skill evolution" AND all:agent} & 4 & \texttt{FedSkillGit}~\cite{lan2026fedskillgit}, \texttt{MaskClaw}~\cite{zhao2026maskclaw}, \texttt{FederatedSkill}, 1 non-private skill paper (\texttt{R07}) \\
\texttt{A12} & \texttt{all:"private feedback" AND all:agent} & 2 & Both outside target personal-agent update chain \\
\texttt{A13} & \texttt{all:privacy AND all:feedback AND all:"self-evolving agent"} & 0 & No API return \\
\bottomrule
\end{tabular}
\end{table}

\subsubsection{Stage 2 Operator-Adjacent Expansion Queries (EQ1--EQ4)}
Stage~2 expanded the 25-source core baseline to 62 sources using four composite query families (denoted \texttt{E1}--\texttt{E4} in the Stage~2 protocol and written here as \texttt{EQ1}--\texttt{EQ4} to avoid confusion with exclusion codes \texttt{E1}--\texttt{E9}):
\begin{verbatim}
EQ1 (Prompt/parameter DP, preference protection, and continual composition):
    ("differential privacy" OR "secure aggregation" OR MPC)
    AND ("prompt learning" OR "prompt optimization" OR fine-tuning
         OR preference OR RLHF OR "continual learning")

EQ2 (Private retrieval, encrypted semantic search, and dynamic indexing):
    ("LLM agent" OR "personal agent" OR "conversational agent" OR RAG)
    AND (memory OR retrieval OR index)
    AND (privacy OR private OR confidentiality OR ORAM
         OR "homomorphic encryption")

EQ3 (Machine unlearning, certified removal, and deletion verification):
    (agent OR RAG OR memory)
    AND (unlearning OR deletion OR forgetting OR "certified removal"
         OR "verifiable unlearning" OR "right to be forgotten")

EQ4 (Information-flow control, skill evolution, and lifecycle attacks):
    ("LLM agent" OR "personal agent" OR "self-evolving agent")
    AND ("information flow" OR skill OR experience OR memory)
    AND (privacy OR leakage OR extraction OR impersonation OR defense)
\end{verbatim}
Across the seven Stage~2 themes, \texttt{EQ1} targeted prompt/parameter DP and preference protection; \texttt{EQ1} and \texttt{EQ3} jointly covered continual learning and adaptive composition; \texttt{EQ2} covered private retrieval and oblivious dynamic indexing; \texttt{EQ3} covered machine unlearning and deletion verification; \texttt{EQ4} covered information-flow control and skill evolution; and \texttt{EQ2} and \texttt{EQ4} jointly covered extraction attacks and lifecycle evaluation benchmarks.

\subsubsection{Stage 3 Targeted ACM (DOI Prefix 10.1145) Supplementation Queries (ACM-Q1--ACM-Q10)}
In Stage~3, Crossref queries were restricted to \texttt{prefix:10.1145} with publication dates from \texttt{2018-01-01} to \texttt{2026-11-10} across five title/topic probes (\texttt{ACM-Q1}--\texttt{ACM-Q5}, recorded as \texttt{A1}--\texttt{A5} in the Stage~3 log), followed by five targeted semantic/title verification combinations (\texttt{ACM-Q6}--\texttt{ACM-Q10}, recorded as \texttt{A6}--\texttt{A10} in the Stage~3 log) to audit canonical ACM versions and mechanism boundaries:
\begin{verbatim}
Crossref prefix:10.1145 topic probes (2018-01-01 to 2026-11-10):
  ACM-Q1: "machine unlearning"
  ACM-Q2: "privacy preserving language model inference"
  ACM-Q3: "private retrieval augmented generation"
  ACM-Q4: "privacy personal assistant"
  ACM-Q5: "information flow agent"

Follow-up canonical ACM and boundary verification combinations:
  ACM-Q6:  ("personal assistant" AND privacy
            AND (learning OR preference OR uncertainty))
  ACM-Q7:  (agent AND (context OR tool OR capability) AND privacy)
  ACM-Q8:  (unlearning AND (audit OR verification OR leakage
            OR "functional encryption"))
  ACM-Q9:  ("searchable encryption" AND (deletion OR "backward secure"))
  ACM-Q10: ("collaborative bandit" AND ("differential privacy" OR private))
\end{verbatim}

\subsection{Inclusion Criteria, Exclusion Codes, and Version-Pinning Policy}
\label{sec:supp-search-rules}

\paragraph{Mechanism Eligibility and Orthogonal Coding Rules.}
A study had to satisfy six criteria for mechanism eligibility. Table~\ref{tab:supp-75-summary} gives the current mechanism and background counts.
\begin{enumerate}
  \item \textbf{Cutoff and primary availability:} An original research paper or authoritative archival record was publicly accessible on or before 10 November 2026.
  \item \textbf{Persistent state or protected state interface:} The system updates persistent state across parameters, prompts/control policies, memories, or skills/experiences, constrains read/context release over such state, or provides a formal composition, selection, indexing, or unlearning-verification primitive directly mappable to those operators.
  \item \textbf{Localizable operator intervention:} The primary text identifies the specific candidate-generation, scoring, writeback, maintenance, read-release, withdrawal/deletion, or skill-admission step where protection or auditing intervenes.
  \item \textbf{Explicit threat/interface specification:} At least two of the following three elements are confirmable from the primary text: \emph{protection unit (protected object)}, \emph{observer}, and \emph{trusted endpoint / trust assumption}.
  \item \textbf{Verifiable support:} The protection claim has a reported evaluation or an analytical result that can be located in the primary paper. Design descriptions and worked examples alone supply background.
  \item \textbf{Operational evidence-distance assignment:} Evidence distance (\texttt{D}, \texttt{S}, \texttt{T}) is coded by proximity to a long-term single-user personal-agent update loop---never inferred merely from surface labels such as \texttt{private}, \texttt{local}, \texttt{on-device}, \texttt{federated}, or \texttt{self-evolving}---and is recorded orthogonally to guarantee type (DP, cryptographic/system confidentiality, or empirical governance/behavioral gating).
\end{enumerate}

Sources without a credited protection operation establish selected definitions, interface threat surfaces, design context, and evaluation targets. A study with both risk measurements and a specified and evaluated defense can enter the risk stratum and contribute a mechanism code for that defense. When the primary text locates a privacy control and specifies its interface, an evaluation can qualify the operation for coding. We code the guarantee independently and credit a stated DP budget only to the extent that the source supports the corresponding adjacency, accounting, and observable release. Table~\ref{tab:supp-excodes} gives the screening codes.

\begin{table}[htbp]
\centering
\small
\setlength{\tabcolsep}{4pt}
\renewcommand{\arraystretch}{1.15}
\caption{Standardized Mechanism-Screening Exclusion and Disposition Codes (\texttt{E1}--\texttt{E9})}
\label{tab:supp-excodes}
\begin{tabular}{@{}>{\raggedright\arraybackslash}p{0.08\linewidth}>{\raggedright\arraybackslash}p{0.88\linewidth}@{}}
\toprule
\textbf{Code} & \textbf{Exclusion or Non-Mechanism Disposition Criterion} \\
\midrule
\texttt{E1} & Static personalization or single-turn inference without a protected persistent-state interface or a specified primitive that transfers to one. \\
\texttt{E2} & Neither updates a persistent state carrier nor constrains read/context release over a persistent state. \\
\texttt{E3} & No identifiable privacy intervention at an operation; a \texttt{local}, \texttt{on-device}, \texttt{federated}, \texttt{safe}, or \texttt{privacy-first} label cannot establish the operation. \\
\texttt{E4} & Attack, leakage/risk analysis, survey, or evaluation benchmark only, without intervening in a protection workflow (eligible only as non-mechanism background/risk context unless also contributing a transferable mitigation or private release mechanism, as in~\cite{tobaben2026labelspace,chen2021unlearningprivacy}). \\
\texttt{E5} & Secondary commentary, blog post, or promotional page that cannot be verified against a primary research manuscript. \\
\texttt{E6} & Duplicate version, mirror, or renamed release within a method family. Earlier ledgers also used this code for operation overlap. The follow-up retains independent studies that contribute new comparisons or endpoints. \\
\texttt{E7} & Non-personal-agent domain whose mechanism lies outside the target state-update and release chain. \\
\texttt{E8} & Restricted full text or insufficient accessible technical evidence as of the cutoff date. \\
\texttt{E9} & First public release occurred after the 10 November 2026 literature cutoff date. \\
\bottomrule
\end{tabular}
\end{table}

\paragraph{Deduplication and Canonical Version-Pinning Policy.}
Preprints, OpenReview submissions, non-archival workshop papers, and peer-reviewed conference or journal articles representing the same method family were deduplicated into a single corpus entry. Version selection followed three deterministic rules:
\begin{itemize}
  \item \textbf{Archival precedence:} Whenever a peer-reviewed conference or journal version was verifiable on or before 10 November 2026 (via \texttt{ACL Anthology}, \texttt{ACM Digital Library}, \texttt{IEEE Xplore}, \texttt{PMLR}, \texttt{NeurIPS}, \texttt{ICLR}, \texttt{COLM}, \texttt{TMLR}, \texttt{USENIX}, or \texttt{AAAI}), the archival publication was adopted as the canonical citation (e.g., STOC~2019 for~\cite{liu2019privateselection}, PACMMOD/PODS~2026 for~\cite{henzinger2026concurrentcomposition}, and TMLR formal editions for~\cite{gao2026selfevolvingsurvey,tang2025dpzo,tobaben2026labelspace}). For journals whose online-first date precedes volume/issue assignment, the bibliography records the official volume publication year while noting the online-first year in verification logs (specifically \texttt{PANOLA}~\cite{ulusoy2022panola}, ACM TOIT Vol.~22, 2022, first online 2021; and \texttt{Machine Unlearning: A Survey}~\cite{xu2024unlearnsurvey}, ACM CSUR Vol.~56, 2024, first online 2023).
  \item \textbf{Explicit versioned preprint pinning:} Works without formal archival pagination by 10 November 2026 were pinned to an exact versioned preprint (\texttt{arXiv:...vN} with its DataCite record, or \texttt{TechRxiv v1} Crossref DOI for~\cite{chen2026preferencesurvey}), using the submission year of the cited version. In particular, \texttt{PrivDPO}~\cite{jiang2026privdpo} was verified on the official ACM CCS~2026 accepted-papers list, but because proceedings pagination was not yet registered at the cutoff date, it is pinned to \texttt{arXiv:2608.05040v1}. Likewise, \texttt{Deployment-Time Memorization}~\cite{chen2026deploymentmemorization} appeared at the non-archival ICML~2026 MemFM Workshop and is pinned to \texttt{arXiv:2606.10062v2} (with author attribution verified directly against the versioned PDF).
  \item \textbf{One-to-one citation integrity:} Every corpus source appears in the supplementary bibliography and maps to one BibTeX key across the submission.
\end{itemize}

\subsection{Candidate-Level Screening Disposition Logs}
\label{sec:supp-search-dispositions}

\paragraph{Stage 1 Core Gap-Search Dispositions (R01--R12).}
Table~\ref{tab:supp-r01-r12} records the candidate-level screening decisions from the Stage~1 bounded gap audit. Two candidates entered the representative mechanism set: \texttt{SBU}~\cite{wang2026agenticunlearning} (\texttt{R01}, coded \texttt{T} because its synchronized parameter--memory unlearning and dependency-closure cleanup were evaluated on public medical QA rather than a longitudinal single-user loop, and its shared-artifact logical invalidation is empirical rather than certified deletion) and \texttt{Galaxy}~\cite{bao2025galaxy} (\texttt{R02}, coded \texttt{S} because its Kernel Privacy Gate masks outbound cloud LLM calls and selectively demasks responses in a personal self-evolving architecture, without proving that every internal Persona/Kernel self-modification path traverses the gate under cross-round guarantees). Note on cross-stage progression for \texttt{R06}: \texttt{FederatedSkill}~\cite{yang2026federatedskill} was deferred from Main Table~3's 22-mechanism core baseline under representative non-duplication (\texttt{E6}, avoiding duplicate federated skill-sanitization representatives alongside \texttt{FedSkillGit}~\cite{lan2026fedskillgit}, \#17), and retained in the extended mechanism stratum during Stage~2 thematic expansion (Source~\#62, coded \texttt{T}) as an adjacent federated skill-distillation and PII-canary audit reference in Main Section~6.1.

\begingroup
\footnotesize
\setlength{\tabcolsep}{2.6pt}
\renewcommand{\arraystretch}{1.18}
\begin{longtable}{>{\raggedright\arraybackslash}p{0.05\linewidth}>{\raggedright\arraybackslash}p{0.23\linewidth}>{\raggedright\arraybackslash}p{0.08\linewidth}>{\raggedright\arraybackslash}p{0.22\linewidth}>{\raggedright\arraybackslash}p{0.14\linewidth}>{\raggedright\arraybackslash}p{0.20\linewidth}}
\caption{Stage 1 Bounded Core Gap-Search Candidate Screening Dispositions (\texttt{R01}--\texttt{R12})}
\label{tab:supp-r01-r12}\\
\toprule
\textbf{ID} & \textbf{Candidate and Pinned Version} & \textbf{Query} & \textbf{State Carrier and Operator Locus} & \textbf{Decision / Code} & \textbf{Impact on Review Synthesis} \\
\midrule
\endfirsthead
\caption[]{Stage 1 Bounded Core Gap-Search Candidate Screening Dispositions (\texttt{R01}--\texttt{R12}) (Continued)}\\
\toprule
\textbf{ID} & \textbf{Candidate and Pinned Version} & \textbf{Query} & \textbf{State Carrier and Operator Locus} & \textbf{Decision / Code} & \textbf{Impact on Review Synthesis} \\
\midrule
\endhead
\midrule
\multicolumn{6}{r}{\footnotesize\emph{Continued on next page}} \\
\endfoot
\bottomrule
\endlastfoot
\texttt{R01} & \emph{Agentic Unlearning} (\texttt{SBU})~\cite{wang2026agenticunlearning}, \texttt{arXiv:2602.17692v2} & \texttt{A2} & Closed-loop agent updating parameters and persistent memory; \S3.2, Alg.~1, Fig.~6 dependency closure + parameter suppression & \textbf{Include (\texttt{T})} & Adds joint parameter--memory unlearning and derived-artifact cleanup; public medical QA setup (does not add \texttt{D}) \\
\texttt{R02} & \emph{Galaxy}~\cite{bao2025galaxy}, \texttt{arXiv:2508.03991v1} & \texttt{A1} & Personal self-evolving assistant (Persona cognitive tree + Kernel); Fig.~4 pre-cloud tiered masking and selective demasking & \textbf{Include (\texttt{S})} & Adds personal self-evolving outbound context gate; internal update-call coverage unproven (not upgraded to \texttt{D}) \\
\texttt{R03} & \emph{Can Agents Truly Forget?} (\texttt{AUBench})~\cite{lin2026aubenchcontext} & \texttt{A2} & Provenance-tracked benchmark across parameters, persistent memory, and tool re-acquisition & \textbf{Exclude from mech.\ (\texttt{E4})} & Evaluation rather than a new protection mechanism. Initially contextual; included in the risk and evaluation stratum at Stage~4. \\
\texttt{R04} & \emph{FOCAL}, Yin et al., 2026, \texttt{arXiv:2604.19541v2} & \texttt{A3} & Continuous desktop activity logging and task-local memory via on-device filter-plan-log pipeline & \textbf{Exclude (\texttt{E3})} & Efficiency and on-device isolation only, without formal threat model or update-privacy operator \\
\texttt{R05} & \emph{PPMI}, Bae et al., 2025, \texttt{arXiv:2506.17336} & \texttt{A4} & Personal data retrieval via homomorphically encrypted vector search and local generation; no continual writeback & \textbf{Exclude from mech.\ (\texttt{E1})} & Static/one-shot encrypted read protection outside a continual self-evolving writeback loop \\
\texttt{R06} & \emph{FederatedSkill}~\cite{yang2026federatedskill}, \texttt{arXiv:2606.03143v1} & \texttt{A1}, \texttt{A11} & Federated skill evolution with trajectory distillation, semantic diffing, and PII canary auditing & \textbf{Table~3 dedup.\ (\texttt{E6}); Stage~2 ext.\ (\texttt{T})} & Deferred from Main Table~3 for representative non-duplication alongside \texttt{FedSkillGit} (\#17); retained in Stage~2 (\#62, \texttt{T}) as federated skill contrast \\
\texttt{R07} & \emph{Verifiable Self-Evolution for Open-Ended Dialogue Skills}, \texttt{arXiv:2607.18973} & \texttt{A11} & Feedback-driven dialogue skill evolution via future-feedback prediction & \textbf{Exclude (\texttt{E3})} & No privacy mechanism protecting private user feedback or persistent skill writeback \\
\texttt{R08} & \emph{AI-EduAgent} & IEEE Xplore & Decentralized educational personalization; abstract lacks identifiable update-privacy operator & \textbf{Exclude (\texttt{E3})} & No verifiable protection over private update inputs or persistent state \\
\texttt{R09} & \emph{Memory Management Strategies for Long-Term Dialogue Coherence in Chatbots} & IEEE Xplore & Long-term dialogue memory management and personalization without privacy operator & \textbf{Exclude (\texttt{E3})} & Memory management only; no privacy mechanism on writeback or release \\
\texttt{R10} & \emph{Machine Learning in E-Commerce: Trends, Applications, and Challenges} & IEEE Xplore & Broad e-commerce ML survey outside personal-agent state evolution & \textbf{Exclude (\texttt{E7})} & Outside target personal-agent update chain \\
\texttt{R11} & Exact-title mirrors, secondary summaries, and promotional pages & \texttt{Q5} & Secondary/duplicate web pages surfaced during one-hop title checks & \textbf{Exclude / Dedup (\texttt{E5}/\texttt{E6})} & Deduplicated against verified primary papers; zero new methods \\
\texttt{R12} & \emph{Setoka} and analogous risk/capability benchmarks & \texttt{A4} & Evaluates personal-agent risk or task capability without intervening in state protection & \textbf{Exclude (\texttt{E4})} & Evaluation benchmark only; does not alter mechanism taxonomy \\
\end{longtable}
\endgroup

\paragraph{Stage 3 Targeted ACM Supplementation: 13 Included Sources and 6 Excluded Candidates.}
Tables~\ref{tab:supp-acm-included} and~\ref{tab:supp-acm-excluded} document the initial Stage~3 screening. Stage~8 re-adjudicates one exclusion. Stage~3 added 13 ACM sources without changing the direct-update source count at that stage. \texttt{PANOLA}~\cite{ulusoy2022panola} and the Uncertainty-Aware Personal Assistant~\cite{ayci2023uncertaintypa} learn privacy decisions from feedback and evaluate outbound disclosure controls. \texttt{TAPE}~\cite{wang2025tape} and Unlearning Leakage~\cite{chen2021unlearningprivacy} show that differences between earlier and later model versions can supply both an audit signal and membership evidence. \texttt{EFU}~\cite{mohammadi2025efu} examines how to hide the intent of an unlearning request.

\begingroup
\footnotesize
\setlength{\tabcolsep}{2.8pt}
\renewcommand{\arraystretch}{1.18}
\begin{longtable}{>{\raggedleft\arraybackslash}p{0.035\linewidth}>{\raggedright\arraybackslash}p{0.205\linewidth}>{\raggedright\arraybackslash}p{0.250\linewidth}>{\raggedright\arraybackslash}p{0.145\linewidth}>{\centering\arraybackslash}p{0.045\linewidth}>{\raggedright\arraybackslash}p{0.230\linewidth}}
\caption{Stage 3 Targeted ACM Supplementation: 13 Included ACM Sources (\#63--\#75)}
\label{tab:supp-acm-included}\\
\toprule
\textbf{\#} & \textbf{Source and Citation} & \textbf{ACM Venue and DOI} & \textbf{Primary Role} & \textbf{Code} & \textbf{Main Text Placement and Analytical Increment} \\
\midrule
\endfirsthead
\caption[]{Stage 3 Targeted ACM Supplementation: 13 Included ACM Sources (\#63--\#75) (Continued)}\\
\toprule
\textbf{\#} & \textbf{Source and Citation} & \textbf{ACM Venue and DOI} & \textbf{Primary Role} & \textbf{Code} & \textbf{Main Text Placement and Analytical Increment} \\
\midrule
\endhead
\midrule
\multicolumn{6}{r}{\footnotesize\emph{Continued on next page}} \\
\endfoot
\bottomrule
\endlastfoot
63 & \texttt{PANOLA}~\cite{ulusoy2022panola} & ACM TOIT 2022 (online 2021), \nolinkurl{10.1145/3471187} & Representative mechanism & \texttt{S} & \S1, \S2.3, \S3 (Table~3), \S5.2; learns user privacy preferences and negotiates outbound release via auctions without protecting preference updates \\
64 & \texttt{Uncertainty-Aware PA}~\cite{ayci2023uncertaintypa} & ACM TOIT 2023, \nolinkurl{10.1145/3561820} & Extended mechanism & \texttt{S} & \S1, \S2.3, \S5.2; personalizes privacy labels and defers uncertain disclosure decisions back to the user \\
65 & \texttt{AirGapAgent}~\cite{bagdasarian2024airgapagent} & ACM CCS 2024, \nolinkurl{10.1145/3658644.3690350} & Representative mechanism & \texttt{S} & \S2.3, \S3 (Table~3), \S5.2; task-scoped context minimization prior to third-party interaction (protects release, not persistent writeback) \\
66 & \texttt{Privacy-Preserving BERT}~\cite{qu2021privatebert} & ACM CIKM 2021, \nolinkurl{10.1145/3459637.3482281} & Extended mechanism & \texttt{T} & \S4.2; client-side token/representation $d_{\chi}$-privacy with privacy-adaptive pretraining; single-text rather than longitudinal user adjacency \\
67 & \texttt{Unlearning Leakage}~\cite{chen2021unlearningprivacy} & ACM CCS 2021, \nolinkurl{10.1145/3460120.3484756} & Risk \& evaluation study (dual-coded) & \texttt{T} & \S2.3, \S5.3, \S7.1; demonstrates membership leakage from pre-/post-unlearning version differentials alongside label-only, temperature, and DP mitigations \\
68 & \texttt{TAPE}~\cite{wang2025tape} & ACM WebConf 2025, \nolinkurl{10.1145/3696410.3714875} & Theoretical \& systems primitive & \texttt{T} & \S5.3, \S7.1; audits multi-example unlearning via tailored posterior differences (empirical audit signal rather than deletion certificate) \\
69 & \texttt{EFU}~\cite{mohammadi2025efu} & ACM CIKM 2025, \nolinkurl{10.1145/3746252.3761091} & Extended mechanism & \texttt{T} & \S5.3, \S7.1; functional encryption enforces authorized federated unlearning aggregation and conditionally hides unlearning request intent \\
70 & \texttt{Tracked Capabilities}~\cite{odersky2026trackedcapabilities} & ACM CAIS 2026, \nolinkurl{10.1145/3786335.3813127} & Extended mechanism & \texttt{T} & \S5.2, \S6.2; static capabilities, local purity, and derived labels constrain tool-execution information flow \\
71 & \texttt{Backward-Secure SSE}~\cite{sun2018backwardsearch} & ACM CCS 2018, \nolinkurl{10.1145/3243734.3243782} & Extended mechanism & \texttt{T} & \S5.2; post-deletion backward security for dynamic encrypted indexes via symmetric puncturable encryption; excludes prior fetched copies and weights \\
72 & \texttt{Machine Unlearning Survey}~\cite{xu2024unlearnsurvey} & ACM CSUR 2024 (online 2023), \nolinkurl{10.1145/3603620} & Surveys \& state background & --- & \S1 (Table~1), \S5.3; anchors exact vs.\ approximate unlearning, streaming deletion requests, and verification taxonomy \\
73 & \texttt{SKIPPER}~\cite{xie2022skipper} & IEEE/ACM ASE 2022, \nolinkurl{10.1145/3551349.3560416} & Risk \& evaluation study & --- & \S6.2; audits policy-to-behavior inconsistencies in third-party virtual personal assistant apps (risk audit, not a protection algorithm) \\
74 & \texttt{Privacy Norms for SPA}~\cite{abdi2021privacynorms} & ACM CHI 2021, \nolinkurl{10.1145/3411764.3445122} & Surveys \& state background & --- & \S5.2; grounds smart-assistant data types, recipients, purposes, and transmission principles in contextual integrity \\
75 & \texttt{Collaborative Bandits DP}~\cite{wang2020collaborativebandits} & ACM RecSys 2020, \nolinkurl{10.1145/3383313.3412254} & Extended mechanism & \texttt{T} & \S4.2; central and local DP for online personalized bandit feedback under single-reward event-level adjacency \\
\end{longtable}
\endgroup

\begingroup
\small
\setlength{\tabcolsep}{3.5pt}
\renewcommand{\arraystretch}{1.18}
\begin{longtable}{>{\raggedright\arraybackslash}p{0.31\linewidth}>{\raggedright\arraybackslash}p{0.14\linewidth}>{\raggedright\arraybackslash}p{0.48\linewidth}}
\caption{Stage 3 Targeted ACM Supplementation: Six Initially Excluded ACM Candidates and Screening Rationales}
\label{tab:supp-acm-excluded}\\
\toprule
\textbf{Candidate Title} & \textbf{Disposition} & \textbf{Exact Screening Rationale} \\
\midrule
\endfirsthead
\caption[]{Stage 3 Targeted ACM Supplementation: Six Initially Excluded ACM Candidates and Screening Rationales (Continued)}\\
\toprule
\textbf{Candidate Title} & \textbf{Disposition} & \textbf{Exact Screening Rationale} \\
\midrule
\endhead
\midrule
\multicolumn{3}{r}{\footnotesize\emph{Continued on next page}} \\
\endfoot
\bottomrule
\endlastfoot
\emph{EvidenceBot: A Privacy-Preserving, Customizable RAG-Based Tool} & Exclude (\texttt{E3}/\texttt{E6}) & Local execution and relevant-chunk minimization narrow the exposure surface, but introduce no new adversary model or guarantee class beyond included \texttt{RemoteRAG}~\cite{cheng2025remoterag}, \texttt{PRAG}~\cite{li2026prag}, and \texttt{Opal}~\cite{kaviani2026opal}. \\
\emph{Towards Reliable Forgetting: A Survey on Machine Unlearning Verification} & Exclude (\texttt{E4}/\texttt{E6}) & Substantially overlaps in analytical function within this review with the included ACM CSUR machine unlearning survey~\cite{xu2024unlearnsurvey} and \texttt{TAPE}~\cite{wang2025tape}; retained in the screening log without redundant corpus expansion. \\
\emph{THOR: Secure Transformer Inference with Homomorphic Encryption} & Exclude (\texttt{E1}) & Protects static one-shot Transformer inference via homomorphic encryption without addressing persistent personal-agent state carriers, tool boundaries, or continual update loops. \\
\emph{RECALLbot: Designing Agentic Memory and Reciprocal Disclosure} & Exclude (\texttt{E3}) & Evaluates human--agent relationship building and increased reciprocal self-disclosure via agentic memory, without an identifiable privacy-protection guarantee or update-protection verification. \\
\emph{Private Web Search with Tiptoe} & Exclude (\texttt{E6}/\texttt{E7}) & Private querying over a public web corpus duplicates the retrieval-side locus of \texttt{PRAG}~\cite{li2026prag} and \texttt{Compass}~\cite{zhu2025compass} without adding evidence on mutable personal state updates. \\
\emph{Beyond Centralization: User-Controlled Federated Recommendations in Practice} & Initial \texttt{E3}, revised to \texttt{T} & The Stage~3 screening overlooked client-side clipping and Gaussian perturbation in the published Sections~2.1 and~3. Stage~8 includes it as source \#106 for that update-sharing intervention (Section~\ref{sec:supp-broader-search}). \\
\end{longtable}
\endgroup

\subsection{Targeted Revision Search and Candidate Decisions}\label{sec:supp-targeted-search}

On 18 November 2026 we ran focused web searches for evidence that could change the protected unit, the comparison between writing and release, or the interpretation of withdrawal. The literature cutoff remained 10 November 2026. The search strings included \texttt{persistent memory personal agent privacy write-time response intervention}, \texttt{persistent memory agent after deletion benchmark provenance closure preservation}, and \texttt{longitudinal privacy secret pairs Pufferfish Markov Quilt composition adaptive machine unlearning deletion privacy}, with exact-title and identifier follow-up on arXiv, publisher, and proceedings records. The following log records the candidates taken to primary-source review for this check. Ranked search returns do not provide stable channel-wide denominators. The original Stage~2 records establish 37 net inclusions but do not recover a complete per-channel retrieval, deduplication, or exclusion trail.

\begin{description}
\item[\#76, disclosure feedback loop~\cite{kaneko2026disclosureloop}.] Include as extended mechanism, \texttt{D}. The pinned arXiv v2 (22 October 2026), Sections~2.5 and~3.5, Figure~4, and Appendix~I distinguish writing anonymized summaries from modifying the response policy. The former intervenes in persistent personal state. The latter has empirical evidence over simulated sessions and is recorded as a read/release-side comparator, without a second source count.
\item[\#77, Algorithmic Self-Portrait~\cite{dash2026selfportrait}.] Include primarily as risk and evaluation, also \texttt{T}. The Web Conference 2026 study (DOI \nolinkurl{10.1145/3774904.3792671}), Sections~3--4, audits donated ChatGPT memory entries. Section~7.4 tests a separate query-reformulation mitigation using a memory extractor as a proxy, which leaves sustained deployment effects unresolved.
\item[\#78, Agent-Memory Protocol~\cite{wu2026amp}.] Stage~4 assigned \texttt{T} to a protocol specification with two worked use cases. Stage~9 retains the source as design background and supersedes this code (Section~\ref{sec:supp-final-screening}).
\item[\#79, AfterDeleteBench~\cite{saleh2026afterdeletebench}.] Include primarily as risk and evaluation, also \texttt{D}. The arXiv v1 (22 October 2026), Sections~3--4, evaluates recorded-lineage retirement or rebuild in persistent synthetic personal memory. Complete source annotations, a controlled memory harness, exact-string targets, and fixed weights bound the result. The write filter covers a separate reinsertion event.
\item[\#80--\#81, Bemula and AUBench~\cite{xie2026bemula,lin2026aubenchcontext}.] Include as risk and evaluation, without mechanism codes. Bemula (arXiv v1, 13 October 2026, Sections~3.4 and~4) conditions forgetting and preservation on paired no-request probes. AUBench (arXiv v1, 13 October 2026, Sections~4--5) distinguishes residual influence through parameters, memory, and tools. Its \texttt{E4} mechanism disposition remains applicable. The corpus decision changes because that evaluation covers a distinct path relevant to withdrawal.
\item[\#82--\#84, inferential privacy and composition~\cite{kifer2014pufferfish,song2017timecomposition,bai2026pufferfishcomposition}.] Include as theoretical primitives, \texttt{T}. The archival TODS 2014, Allerton 2017, and PVLDB 2026 versions specify secrets and data-generating assumptions, give scoped time-series composition, and state extra conditions for linear adaptive composition, respectively. These sources do not evaluate a personal-agent update loop.
\item[\#85--\#86, adaptive deletion~\cite{gupta2021adaptiveunlearning,chourasia2023trueprivacy}.] Include as theoretical primitives, \texttt{T}. NeurIPS 2021 studies adaptive deletion requests for retraining-based guarantees. ICML 2023 specifies post-deletion privacy and separates it from equivalence to retraining. They provide different targets for an agent that acts on previously visible versions.
\item[Scope exclusions.] \emph{Agent Zero Memory} (arXiv \nolinkurl{2608.29606v1}) evaluates attributed retrieval and answer utility. \emph{Memory Provenance Laundering in LLM Agents} (arXiv \nolinkurl{2607.29167v1}) tests whether untrusted observations acquire action authority. These were screened out of the targeted expansion because their evaluated outcomes do not address the confidentiality or withdrawal comparisons (\texttt{E7}).
\item[Cutoff exclusion.] \emph{Forget the Lineage} (arXiv \nolinkurl{2611.18133v1}) was relevant to derived-state erasure but first appeared on 13 November 2026, after the fixed cutoff (\texttt{E9}).
\end{description}

Stage~4 changed direct-update counts from $7$ to $9$ and transferable counts from $44$ to $51$, while read/release counts stayed at $5$ and non-mechanism counts rose from $19$ to $21$. The two additional direct codes concern persistent memory. The response-side comparison changes the synthesis because a release intervention can also alter what will be written in later turns. The new withdrawal studies separate source recovery, preservation of unrelated state, and reacquisition, while the new formal sources qualify claims about correlated facts and adaptive deletion. The 22 reference mechanisms in Main Table~3 are retained to keep their original operator comparisons traceable. The added studies enter the matched comparisons in Sections~5, 7, and~8.

\paragraph{Stage 5 exact-title follow-up.}
On 18 November 2026 we searched for \texttt{MemJoin cross-session sensitive relation persistent agent memory cross-record cue removal read intervention arxiv 2610.24359}, then checked the arXiv version record and the primary text. The cutoff remains 10 November 2026. MemJoin~\cite{yang2026memjoin} (\#87, arXiv \nolinkurl{2610.24359v1}, posted 17 October 2026) enters the risk and evaluation stratum. Sections~3.1--3.3 define the person-linked relation and score disclosure in ordinary-task answers. Sections~4.3--4.4, Tables~2--3, and Appendix~F.1 test cue removal on retrieved context before a common reader over persistent multi-session memory. The protected object is the sensitive relation and the observed surface is the answer delivered to the task recipient. The operation meets \texttt{S} because it changes the per-query view of already-stored state. The synthetic harness measures empirical mitigation, with no tested persistent write intervention or formal privacy bound. This adds one \texttt{S} to Stage~4's count, for $9$ \texttt{D}, $6$ \texttt{S}, and $51$ \texttt{T} across the 87 sources at the end of Stage~5.

\subsection{Core-Coverage Recheck and Candidate Dispositions}\label{sec:supp-core-recheck}

On 18 November 2026, we ran ranked web searches across the update, release, and withdrawal operators with the 10 November 2026 cutoff held fixed. Search strings paired the carrier or operation with its privacy or evaluation endpoint:
\begin{itemize}
\item \textbf{Memory release:} \texttt{personalized LLM memory privacy sensitive facts contextual integrity federated user history 2026 agent ICLR ACL HUSH CIMemories selective retention}.
\item \textbf{Model and prompt updates:} \texttt{persistent user history personalized LLM agent privacy continual learning privacy DP user fact preference update agent memory cross session before November 2026 arxiv private personalized fine tuning}.
\item \textbf{Skill distillation and admission:} \texttt{privacy preserving LLM agent personal skill distillation user trajectories skill admission privacy audit personal agent skills 2026 MaskClaw AntiSkillBench}.
\item \textbf{Withdrawal:} \texttt{cross-carrier personal agent unlearning memory parameter withdrawal provenance privacy 2026 Memora AUBench afterdeletebench personal agent}.
\end{itemize}
Additional proceedings queries paired \emph{personal agent long-term memory} with \emph{privacy}, \emph{selective privacy protection}, and \emph{cross session memory deletion} in ACL Anthology, and checked exact titles for CIMemories, HUSH-Bench, and PAHF. The SAVER survey's memory discussion supplied terminology for a follow-up search on sensitive-history use boundaries~\cite{chen2026safetysurvey}. We matched titles, DOI records, and arXiv identifiers against Table~\ref{tab:supp-75-ledger}, then verified fixed versions through ACL Anthology, ICLR proceedings/OpenReview, or arXiv. The pass ended after primary-text screening of the targeted candidates below. Counts refer to these audited decisions. Ranked results provide no database-wide denominator, and this recheck leaves the historical Stage~2 exclusion trail incomplete.

\begin{description}
\item[\#88, CIMemories~\cite{mireshghallah2026cimemories}.] Include as risk and evaluation (ICLR 2026, Sections~4--5, especially the violation/completeness, repeated-task, and privacy-prompting evaluations). The synthetic memory is evaluated across recipient--task contexts and runs. Generic privacy instructions serve as comparison conditions, and the outcomes reveal a violation--completeness trade-off. No mechanism code (\texttt{E4}).
\item[\#89, HUSH-Bench~\cite{xu2026hushbench}.] Include as risk and evaluation (arXiv v2, revised 3 August 2026, Sections~3.3--3.4 and~4.4--4.5, Tables~4--5). The benchmark isolates exact-target exposure in retrieved input, unsolicited use of sensitive history in the answer, and use after explicit invitation. Retrieval interfaces and invitation are diagnostic comparison conditions. No mechanism code (\texttt{E4}).
\item[\#90, Memora~\cite{uddin2026memora}.] Include as risk and evaluation (Findings of ACL 2026, benchmark construction and Forgetting-Aware Memory Accuracy). The benchmark penalizes use of superseded memories in longitudinal tasks. It supplies an obsolete-state reuse check for withdrawal evaluation. No mechanism code (\texttt{E4}).
\item[\#91, AIM~\cite{johnson2026aim}.] Include as an extended mechanism, \texttt{T} (arXiv v1, 11 September 2026, Sections~3.4, 3.7, and~5). Pre-storage visibility classification and index filtering restrict which other users can retrieve private entries. Classification and enforcement depend on the trusted platform. The multi-user prototype tests access scope rather than inference about an attribute within one user's history.
\item[\#92, PAHF~\cite{liang2026pahf}.] Include as state background (arXiv v1, 18 February 2026, Sections~3.2 and~4.2). Pre-action clarification and post-action correction show how a natural-language preference can enter persistent memory and shape subsequent actions. The evaluation targets adaptation to preference drift. No mechanism code (\texttt{E4}).
\item[\#93, PeCL~\cite{zhan2026pecl}.] Include as an extended mechanism, \texttt{T} (EMNLP 2026, Sections~3.2--3.3 and Appendix~G). It perturbs clipped token embeddings and uses sensitivity-guided retention during task-incremental training. Its appendix calibrates a Gaussian step for a specified noise scale. A joint adaptation-and-release guarantee would need to control the data-dependent scale, neighboring contextual features, retention decisions, and repeated access. Code \texttt{T} credits the local intervention and task-level evaluation.
\item[Recode \#59, AntiSkillBench~\cite{xiang2026personaskills}.] Keep the primary risk role and add \texttt{T} for its active pre-distillation sanitization (arXiv v1, Sections~4.4 and~5.3, Table~2 and Appendix~C.1). The operation transforms persona dialogue before a distiller sees it. The study tests scripted trace corpora and fixed distillation runs. Its provenance signals address unauthorized reuse detection as a separate claim.
\item[Duplicates.] The CIMemories preprint (\nolinkurl{2511.14937}) and Memora preprint (\nolinkurl{2604.20006}) collapse into their archival records (\texttt{E6}). SP-Mem, DP-ES, AUBench, and MaskClaw already have corpus rows.
\item[Scope and cutoff exclusions.] CAPTURE (arXiv \nolinkurl{2609.02265v1}) focuses on preference drift versus poisoning (\texttt{E7}). TransMem (arXiv \nolinkurl{2611.10683v1}) evaluates transition-aware retrieval for changing preferences (\texttt{E3}); PAHF supplies the feedback-loop comparison. \emph{Privacy Leakage from Recurrent State in LLM Inference} (arXiv \nolinkurl{2611.16048v1}) first appeared on 11 November 2026 (\texttt{E9}).
\end{description}

The recheck changes three comparisons. Memory evaluation now separates contextual appropriateness, exact-target retrieval, and response-level use. Withdrawal evaluation distinguishes superseded-state reuse from source and descendant erasure. The skill analysis now contrasts protecting distillation inputs with screening a candidate at admission. The additional mechanisms and the reclassified study add three \texttt{T} codes. The \texttt{D} and \texttt{S} counts remain at 9 and 6.

\subsection{Fixed-Cutoff Retrieval Rerun and Candidate Dispositions}
\label{sec:supp-cutoff-rerun}

On 18 November 2026 we queried the public arXiv Atom API at \nolinkurl{export.arxiv.org/api/query}. We parenthesized each field expression below and appended the shared date filter. The filter limits this discovery pass to modern agents; earlier foundational sources remain eligible in the main corpus. The API returned 1 parameter/prompt/preference hit (P), 8 memory hits (M), 7 skill hits (S), and 3 withdrawal hits (W). All results appear below in rank order. We matched versions before coding and checked primary texts for newly included or closely related candidates. Two targeted referrals were retrieved separately by identifier. The log records all 21 dispositions for this pass. The historical Stage~2 log remains incomplete.

\begingroup
\small
\begin{verbatim}
Shared filter: AND submittedDate:[202301010000 TO 202611102359]
Options: start=0, max_results=10, sortBy=relevance,
         sortOrder=descending

P: (ti:agent OR ti:personalization)
   AND (abs:"differential privacy" OR abs:"private feedback")
   AND (abs:"prompt optimization" OR abs:"continual learning"
        OR abs:"preference learning" OR abs:"parameter update")
M: (ti:agent OR ti:memory)
   AND (abs:"long-term memory" OR abs:"persistent memory")
   AND (abs:"privacy preservation" OR abs:"sensitive information")
S: (ti:agent OR ti:skill)
   AND (abs:"skill evolution" OR abs:"skill distillation")
   AND (abs:privacy OR abs:confidentiality)
W: ti:agent AND (ti:unlearning OR ti:forgetting OR ti:deletion)
   AND (abs:privacy OR abs:withdrawal)
\end{verbatim}
\endgroup

\begin{description}
\item[P1.] One-Bit Model Aggregation (\nolinkurl{2507.03973v1}): \texttt{E7}. The protocol protects aggregation in generic federated learning. It does not probe personal-history states or an agent-facing output.
\item[M1.] Social Hippocampus Memory Learning (\nolinkurl{2603.25614v1}): \texttt{E7}. Its shared latent memories transfer predictive representations across clients. They are evaluated as federated classifier state, outside the personal memory and release chain.
\item[M2.] When Agents Remember Too Much (\nolinkurl{2607.06595v1}): \texttt{E4} from this pass. Attack-centered memory integrity evidence does not add a confidentiality or withdrawal intervention to the comparison.
\item[M3.] SP-Mem (\nolinkurl{2608.16551v1}): existing source \#39, \texttt{D} for writeback~\cite{wang2026spmem}.
\item[M4.] ConMem (\nolinkurl{2610.33432v1}, 24 October 2026): include as \#99, \texttt{T}. Sections~3--5 define a candidate-storage decision, risk-sensitive probes, and a trusted response evaluator. The evaluation predicts admission for independently reconstructed histories without carrying writes from one case to the next~\cite{zhang2026conmem}.
\item[M5.] According to Me (\nolinkurl{2603.01990v1}): \texttt{E4}. Evaluates personal multimodal recall and referential QA. Its recall outcomes do not test an intervention at the privacy boundary.
\item[M6.] Agents of Chaos (\nolinkurl{2602.20021v1}): \texttt{E4} from this pass. Exploratory red-teaming reports incidents across autonomous agents. It does not evaluate a protective update to personal state.
\item[M7.] CIMemories (\nolinkurl{2511.14937v1}): \texttt{E6}, same method family as the archival record \#88~\cite{mireshghallah2026cimemories}.
\item[M8.] Trustworthy Agentic AI (\nolinkurl{2609.22712v1}): \texttt{E4}. Broad trustworthiness survey and proposed development lifecycle. The operation-level comparison uses the closer surveys in Main Table~1.
\item[S1--S2.] FederatedSkill (\nolinkurl{2606.03143v1}) and AntiSkillBench (\nolinkurl{2608.03700v1}): existing \texttt{T} rows \#62 and \#59. The former filters shared skills and the latter tests pre-distillation sanitization~\cite{yang2026federatedskill,xiang2026personaskills}.
\item[S3.] CounterPersona (\nolinkurl{2609.15097v1}, 14 September 2026): include as \#98, \texttt{T}. Sections~3--5 describe an append-only intervention on the input to unauthorized distillation and test downstream behavioral fidelity on fixed archives. The original archive remains visible to the distiller~\cite{wang2026counterpersona}.
\item[S4.] P2Skill (\nolinkurl{2608.14094v2}, 13 October 2026): include as \#97, \texttt{T}. Sections~3.2--3.3 and 4 specify reusable routing and redaction skills with offline refinement. The cloud evaluator sees original benchmark inputs. Runtime leakage results concern the released subtasks~\cite{ryu2026p2skill}.
\item[S5.] MaskClaw (\nolinkurl{2605.28646v2}, 29 May 2026): existing source \#10. The pre-cutoff v2 replaces v1 without adding a source. Stage~7 provisionally coded E3 as \texttt{D}; Stage~10 changes this to \texttt{T} (Section~\ref{sec:supp-boundary-review})~\cite{zhao2026maskclaw}.
\item[S6.] FedSkillGit (\nolinkurl{2610.21348v1}): existing \texttt{T} row \#17 for pre-commit filtering on shared skills~\cite{lan2026fedskillgit}.
\item[S7.] Verifiable Self-Evolution for Open-Ended Dialogue Skills (\nolinkurl{2607.18973v1}): retain \texttt{E3} from Table~S4. The feedback mechanism evaluates adaptation without testing privacy controls.
\item[W1.] MaRS/FiFA (\nolinkurl{2512.12856v1}, 14 December 2025): include as \#94, \texttt{T}. Sections~3.2.8, 3.3.3, 4.3 and 6.6.4 cover provenance, sensitivity-aware retention, and leakage scoring in a multi-agent simulation. The optional DP tie-break has a limited selection scope. The simulation detects no significant privacy-score difference across retention policies~\cite{alqithami2025mars}.
\item[W2.] Agents Are All You Need for LLM Unlearning (\nolinkurl{2502.00406v2}): \texttt{E7}. The intervention filters inference from a fixed LLM and tests model-knowledge targets. It does not retire user-derived persistent state or evaluate releases from a personal history.
\item[W3.] FSFM (\nolinkurl{2604.20300v2}, 23 April 2026): include as \#100, \texttt{T}. Sections~3.2 and 4.4 test sensitive-record filtering and pruning with a trusted memory manager on a capacity benchmark. Retention metrics do not measure answer disclosure or descendant recovery~\cite{gu2026fsfm}.
\item[Referral 1.] Secure Forgetting (\nolinkurl{2604.00430v1}, 1 April 2026): include as \#95, \texttt{T}. Sections~III--VI and VIII specify a trusted agent-side unlearning intervention and compare behavior over public state, trajectory, and environment tasks, including a trained-without reference at the threat-evaluation stage~\cite{ye2026secureforgetting}.
\item[Referral 2.] CAMS (DOI \nolinkurl{10.1016/j.eij.2026.100983}, June 2026 issue): include as \#96, \texttt{T}. The publisher text describes write checks (\S3.5.4), re-audit (\S3.6), and extraction interception. The joint extraction benchmark does not isolate the effect of write admission or follow an individual adaptive agent~\cite{dhivyasree2026cams}.
\end{description}

Five arXiv hits and two referrals added seven \texttt{T} sources. Six arXiv results matched prior ledger entries, and eight received scope or non-mechanism exclusions. An additional version check updated MemPrivacy to v3 (14 May 2026) after verifying the writeback operator and personal-memory evaluation~\cite{chen2026memprivacy}. PrivMeSA's recode moved one source from \texttt{D} to \texttt{T}. The Stage~7 source counts were $(8,6,62)$ in D/S/T order. They apply to the screened corpus at that stage.

\subsection{Broader Discovery and Boundary Re-adjudication}
\label{sec:supp-broader-search}

On 18 November 2026 we broadened discovery beyond the arXiv title and abstract intersections. The screening unit was a research work, deduplicated by title, DOI, and preprint identifiers. We screened 16 entries in the ranked web windows below, 12 Crossref proceedings records, and seven targeted records. We used index and venue records for discovery, primary texts for operator coding, and the 10 November 2026 cutoff for version selection. The 35 decisions include 10 additions, seven matches to earlier corpus records, and 18 exclusions. These counts describe the screened windows rather than the coverage of either index.

\paragraph{Ranked discovery.}
We submitted each exact query below to the OpenAI web search interface and screened the stated prefix of identifiable research records in the returned order. Mirrors of the same work share one position, and news, repositories without a paper, and promotional pages are outside the windows. Linked publisher, venue, author, and arXiv records supplied verification.
\begin{description}
\item[R1 (first 3).] \texttt{online preference label differential privacy KL regularized RLHF private optimization personal feedback prompt tuning research}. Online and offline KL-regularized RLHF enters as \#103. Decoupled private reward modeling~\cite{cho2026decoupledreward} and sparse reward estimation~\cite{ding2026sparsereward} match the existing corpus.
\item[R2 (first 5).] \texttt{personal LLM agents memory read access contextual leakage memory retrieval gate policy revocation withdrawal research}. \emph{Revoked but Still Authoritative} enters as \#108. \emph{Memory as a Controlled Process} (\nolinkurl{2607.13591v3}) learns utility-oriented memory operations without privacy evaluation (\texttt{E3}). \emph{Your LLM Agent Can Leak Your Data} (\nolinkurl{10.18653/v1/2026.findings-acl.1257}) tests a threat and uses existing retrieval guards as baselines (\texttt{E4}). \emph{What to Admit and How to Present} enters as \#109, and MemGate enters as \#107.
\item[R3 (first 3).] \texttt{personal LLM agent privacy-preserving skill distillation user trajectory profiling redaction admission}. P2Skill~\cite{ryu2026p2skill} and AntiSkillBench~\cite{xiang2026personaskills} match the corpus. An indexed OpenReview distillation manuscript (\nolinkurl{https://openreview.net/forum?id=gvYCdNQ3Wp}) could not be verified behind the access challenge (\texttt{E8}).
\item[R4 (first 2).] \texttt{LLM agent privacy survey release boundaries accounting repeated queries retrieval augmented generation}. \emph{From One Query to Many} enters as \#104. \emph{Agents That Know Too Much}~\cite{lahjouji2026agentsknowtoomuch} matches the corpus.
\item[R5 (first 3).] \texttt{differential privacy discrete prompt optimization language model private example scoring prompts user private adapters continual optimization ICLR ACL}. DP-OPT~\cite{hong2024dpopt} and DP-ES~\cite{liu2026dpes} match the corpus. \emph{Locally Differentially Private Document Generation Using Zero Shot Prompting} (\nolinkurl{10.18653/v1/2023.findings-emnlp.566}) perturbs a document for one-shot release. At this boundary it duplicates the standalone text-privatization comparison and does not target a persistent prompt update (\texttt{E6}).
\end{description}

\paragraph{Proceedings retrieval via Crossref.}
We used \texttt{query.bibliographic} at \nolinkurl{https://api.crossref.org/works} with default relevance ranking. Each request used \texttt{rows=3}. The filters were \nolinkurl{from-pub-date:2023-01-01}, \nolinkurl{until-pub-date:2026-11-10}, and \nolinkurl{type:proceedings-article}. The lists below give every returned DOI and its screening decision.
\begin{description}
\item[V1.] \texttt{private online preference learning differential privacy human feedback}. Online iterative RLHF (\nolinkurl{10.52202/079017-2598}) specifies adaptation without a protective mechanism (\texttt{E3}). Private cyber threat intelligence (\nolinkurl{10.1109/ubmk67458.2025.11206831}) and overlapping grouped learning (\nolinkurl{10.1109/isit57864.2024.10619521}) address organizational collaboration or model sharing across groups outside the personal-agent interfaces (\texttt{E7}).
\item[V2.] \texttt{LLM agent memory privacy information retrieval access control revocation}. Strategic information retrieval (\nolinkurl{10.1145/3726302.3730125}) concerns publisher and retriever incentives (\texttt{E7}). The clinical RAG/multi-agent risk study (\nolinkurl{10.1109/icetci71772.2026.11710536}) and VAIL (\nolinkurl{10.1109/bigdia71496.2026.11684708}) had proceedings records, but we could not verify a relevant personal privacy operation in accessible primary text (\texttt{E8}).
\item[V3.] \texttt{privacy preserving personal agent skill distillation user trajectory}. Risk-scored membership-private knowledge distillation (\nolinkurl{10.5220/0015079200004103}), synthetic mobility generation (\nolinkurl{10.5220/0014634400004061}), and a cancer research platform (\nolinkurl{10.5220/0012718100003767}) address model training or data workflows without a personal-agent skill or protected state interface (\texttt{E7}).
\item[V4.] \texttt{large language model agent privacy and security survey differential privacy}. The DP-enhanced parameter-efficient fine-tuning study (\nolinkurl{10.1145/3709018.3736329}) supplies a membership-risk evaluation of existing fine-tuning methods without a new protection operator (\texttt{E4}). The general LLM ethics and security survey (\nolinkurl{10.1109/vcris63677.2024.10813382}) and the Cohere DP system (\nolinkurl{10.1109/sp54263.2024.00122}) address broad model risks or general DP query management without a specific personal-agent operation (\texttt{E7}).
\end{description}

\paragraph{Targeted decisions.}
We checked three titles or persistent identifiers covering profile release, behavioral disclosure, and the earlier CSUR survey, plus the Stage~3 exclusion. The exact query \texttt{personal agents privacy persistent memory admission privacy policy classifier memory governance 2610.11188} also surfaced three adjacent records. Audience-Bound Persistent Memory enters as \#110. PLACE (\nolinkurl{2610.25179v1}, Sections~2--3) evaluates stale-state correctness and ineligible task-value influence without a confidentiality threat model (\texttt{E7}). MemGov (\nolinkurl{https://jidmis.org/index.php/jidmis/article/view/3221}) targets cross-department access and policy compliance in a synthetic enterprise workload (\texttt{E7}).

\paragraph{New corpus entries.}
\begin{description}
\item[\#101.] Profile-DP, arXiv v1 (19 October 2026), extended \texttt{T}. Theorem~1, Section~4.4, and Appendix~A give a local release bound for withheld values of a fixed personal profile while authorized attributes and the sharing context are public. Reusing a privatized profile and regenerating it have different accounting. The inference experiments hold the source profile fixed~\cite{li2026profiledp}.
\item[\#102.] UMPeek, arXiv v1 (3 September 2026), risk/evaluation with \texttt{S} for the stateful counterfactual release gate. Section~VII and Tables~IV--V evaluate disclosure and utility under repeated requests in four personalized benchmark settings. The external observer sees ordinary choices and responses. A trusted evaluator compares personalized and unpersonalized behavior before release. The real-system study in Section~VI covers the threat assessment~\cite{li2026umpeek}.
\item[\#103.] Online and offline KL-regularized RLHF, arXiv v1 (15 October 2025), extended \texttt{T}. Section~5 and Algorithm~2 localize randomized response to user-side preference labels in the online contextual-bandit model. The policy learns from reported labels, while the theorem's privacy unit is the label. Section~6 tests the offline algorithm. The NeurIPS ResponsibleFM workshop version belongs to the same family~\cite{wu2025onlineprivaterlhf}.
\item[\#104.] \emph{From One Query to Many}, arXiv v2 (23 October 2026), survey/background. Sections~4--5 compare protection granularity, accounting scope, and observable releases in RAG. This is a baseline for the read-time analysis and for distinguishing the cross-carrier contribution~\cite{zhang2026ragaccounting}.
\item[\#105.] He et al., ACM Computing Surveys 58(6), article 162 (published online 9 December 2025, April 2026 issue), survey/background. Sections~3--5 organize agent threats and defenses, including attacker inference and workflow exposure~\cite{he2026agentsecuritysurvey}.
\item[\#106.] \emph{Beyond Centralization}, archival UMAP 2026 demo paper, extended \texttt{T}. Sections~2.1 and~3 identify client-side clipping and Gaussian perturbation on model updates derived from viewing histories before sharing with an aggregator. The 53-day, 22-participant deployment measures recommendation use and user control. The paper reports a nominal privacy parameter without a complete adjacency, failure probability, or repeated-release accounting contract. We record the deployment and intervention, and leave its DP guarantee unclassified~\cite{slokom2026federatedrecommendations}.
\item[\#107.] MemGate, arXiv v1 (4 June 2026), extended \texttt{S}. Sections~IV--V and Figure~3 evaluate a query-conditioned gate on retrieved personal memory before generation. The gate receives candidate memories and reduces measured cross-domain failures in benchmark tests using memory frameworks and the OpenClaw runtime. The tests use scripted tasks~\cite{zhang2026memgate}.
\item[\#108.] \emph{Revoked but Still Authoritative}, arXiv v1 (8 September 2026), risk/evaluation with \texttt{T}. Section~4.2 and Appendix~C.1 (Table~14) locate a read-time validity guard and test later decisions when operating on retained policy records. The store-level filter prevents the measured unsafe action in the single-read condition. Derived journal records can still influence decisions after write-back. The target is non-use of a revoked policy in organizational scenarios~\cite{shen2026revokedmemory}.
\item[\#109.] \emph{What to Admit and How to Present}, arXiv v1 (8 October 2026), extended \texttt{S}. Sections~3.2 and~4.1 and Sections~5.1--5.5 distinguish the admission of recalled personal information into the generator's context from rendering an unchanged admitted set. The adjudicator sees the retrieved candidates. The evaluation uses fixed memory pools and a personal-memory benchmark substrate and measures task-specific failure and beneficial personalization~\cite{liu2026memoryadmission}.
\item[\#110.] Audience-Bound Persistent Memory, arXiv v1 (28 September 2026), extended \texttt{D} for write and derivation controls and \texttt{S} for per-attempt context admission. Sections~3.1--3.5, Section~6.4, and Appendix~E specify audience binding at ingest, restrictions on derived audiences, and mediated admission into model calls. The longitudinal unit is one owner across synthetic multi-party histories. The native checks cover a scoped validation frame. The formal guarantee concerns assembled context under reliable identity, provenance, and complete mediation~\cite{liu2026audiencememory}.
\end{description}

At the end of Stage~8 the corpus contained 110 sources, with 84 mechanism-eligible sources and 26 risk or background sources without a mechanism code. The additions contributed one \texttt{D}, three \texttt{S}, and four \texttt{T} source rollups. The audience-bound study adds evidence about memory writing and derivation, and its retrieval result is recorded independently. The online RLHF and federated recommendation results establish adjacent update evidence under different protection units or deployment scopes. The historical Stage~2 exclusion log remains incomplete. Evidence gaps refer to the screened sources.

\subsection{Targeted Source and Evidence Check}
\label{sec:supp-final-screening}

On 18 November 2026 we checked the two named candidate papers against the 10 November 2026 cutoff and the mechanism criteria, using exact titles and official PoPETs and MLSys proceedings records. Both candidates entered the corpus. We selected the archival versions and treated the matching arXiv records as the same works. We also revisited one existing source using its PMLR version.

\begin{description}
\item[\#111, ARIEL~\cite{flemings2026ariel}.] Include as extended \texttt{S}. Sections~3--4 define a release decision over third-party requests using a history of one user's sharing judgments. Sections~5--6 evaluate judgment agreement and abstention on smart-home and education benchmarks under non-adversarial requests. The evaluated endpoint is a decision about sharing. The study leaves actual transmission, confidentiality of the stored judgments, and protection of feedback updates outside the measured claim.
\item[\#112, P2B~\cite{malekzadeh2020p2b}.] Include as extended \texttt{T}. Sections~3.1--3.3 identify client sampling, context encoding, and trusted shuffling and suppression of feedback before server-side training of a shared warm-start model. Sections~2.1 and~4 analyze one-context-vector adjacency in the population sharing mechanism. Section~5 evaluates recommendation utility. The evaluated and analyzed setting is transferable to private online adaptation, with a separate accounting question for repeated changes in one personal-agent history.
\item[\#78, Agent-Memory Protocol~\cite{wu2026amp}.] Retain as design background. The radiology and finance walkthroughs describe storage and release operations. Their protection claims lack a reported intervention evaluation or analytical result at those interfaces. The two design rows retain their \textbf{I} tags and contribute no evidence-distance counts.
\end{description}

At the end of Stage~9, 85 of 112 sources contributed 121 counted operations. Stage~10 supersedes that snapshot.

\subsection{Fixed-Window Retrieval and Evidence Audit}\label{sec:supp-stage10}

On 19 November 2026 we queried Crossref and arXiv for missed work on privacy regressions, persistent personalization, skill gates, and withdrawal. For Crossref we used \nolinkurl{api.crossref.org/works}, the exact \texttt{query.bibliographic} strings in Table~\ref{tab:supp-stage10-queries}, default relevance, \texttt{rows=5}, and \texttt{type:proceedings-article} with publication dates 1 January 2024 through 10 November 2026. For \nolinkurl{export.arxiv.org/api/query} we used the shown expressions in parentheses followed by \texttt{AND submittedDate:[202401010000 TO 202611102359]}, starting at zero with \texttt{max\_results=5} and \texttt{sortBy=relevance}. The start date applies only to this pass. For retained work, we checked that the cited version was available by the original cutoff.

\begin{table}[htbp]
\centering
\footnotesize
\setlength{\tabcolsep}{3pt}
\caption{Stage 10 Search Requests and Observation Windows}\label{tab:supp-stage10-queries}
\begin{tabular}{@{}ll>{\raggedright\arraybackslash}p{0.58\linewidth}r@{}}
\toprule
ID & Index & Query before date and type filters & Returned \\
\midrule
C1 & Crossref & benign fine tuning contextual privacy memory tools agents & 5 \\
C2 & Crossref & personalized agent long term memory privacy persistent retention release & 5 \\
C3 & Crossref & large language model agent skill evolution privacy distillation audit & HTTP 429 \\
C4 & Crossref & agent memory unlearning deletion provenance derived state & HTTP 429 \\
A1 & arXiv & \texttt{all:"skill evolution" AND all:privacy} & 4 \\
A2 & arXiv & \texttt{all:"personal agent" AND all:privacy AND all:memory} & 4 \\
A3 & arXiv & \texttt{all:"memory" AND (all:unlearning OR all:withdrawal) AND all:provenance AND all:agent} & 5 \\
\bottomrule
\end{tabular}
\end{table}

Crossref reported 47,485 and 63,492 matches for C1 and C2, respectively, of which only the first five per query were screened. C3 and C4 yielded no screenable response after retry and are not zero-hit searches. The narrower arXiv queries returned all 13 results reported by the API for those expressions. We matched DOI or base arXiv ID and title against the corpus, then screened unmatched records against the privacy outcome and target-operation criteria. Table~\ref{tab:supp-stage10-decisions} lists all 23 returned items in rank order. Existing means the same work is already represented; scope exclusion and exemplar selection are separate decisions. Unresolved means there was insufficient accessible evidence for a decision. This pass does not reconstruct Stage~2 or measure independent coding agreement.

\begingroup
\footnotesize
\setlength{\tabcolsep}{3pt}
\begin{longtable}{@{}>{\raggedright\arraybackslash}p{0.08\linewidth}>{\raggedright\arraybackslash}p{0.35\linewidth}>{\raggedright\arraybackslash}p{0.52\linewidth}@{}}
\caption{Stage 10 Candidate Dispositions by Query and Rank}\label{tab:supp-stage10-decisions}\\
\toprule Hit & Persistent identifier & Decision and reason \\
\midrule
\endfirsthead
\toprule Hit & Persistent identifier & Decision and reason \\
\midrule
\endhead
\bottomrule
\endfoot
C1.1 & \nolinkurl{10.18653/v1/2026.acl-long.400} & Include \#113: \emph{Privacy Collapse}. \S4.1 and Fig.~2 document regressions; \S5.4 and App.~I test mitigations (\texttt{T}/B)~\cite{goel2026privacycollapse}. \\
C1.2 & \nolinkurl{10.1364/ofc.2025.w1d.7} & Exclude at title screen: distributed fiber sensing. \\
C1.3 & \nolinkurl{10.1109/csmla70906.2026.11724418} & Exclude at title screen: incremental classification and device memory. \\
C1.4 & \nolinkurl{10.1145/3737899.3768528} & Exclude at title screen: memory-efficient federated graph computation. \\
C1.5 & \nolinkurl{10.18653/2026.emnlp-main.2674} & Exclude at title screen: remote-sensing agent training. \\
C2.1 & \nolinkurl{10.1109/icact70625.2026.11748725} & Unresolved: title focuses on retrieval quality; primary text not accessible for a protection check. Not counted. \\
C2.2 & \nolinkurl{10.1109/icdcece65353.2025.11036013} & Exclude at title screen: English-learning resource recommendation. \\
C2.3 & \nolinkurl{10.1109/icdsis61070.2024.10594603} & Exclude at title screen: library book recommendation. \\
C2.4 & \nolinkurl{10.18653/v1/2026.findings-acl.1337} & Existing \#90: same archival Memora record~\cite{uddin2026memora}. \\
C2.5 & \nolinkurl{10.1117/12.3056646} & Exclude at title screen: glucose prediction. \\
A1.1 & \nolinkurl{2605.28646v2} & Existing \#10: MaskClaw; E3 changes to \texttt{T}/B~\cite{zhao2026maskclaw}. \\
A1.2 & \nolinkurl{2606.03143v1} & Existing \#62: FederatedSkill~\cite{yang2026federatedskill}. \\
A1.3 & \nolinkurl{2610.21348v1} & Existing \#17: FedSkillGit~\cite{lan2026fedskillgit}. \\
A1.4 & \nolinkurl{2607.18973v1} & Exclude at abstract screen: offline skill validation predicts feedback, not privacy protection. \\
A2.1 & \nolinkurl{2607.27056v2} & Exclude at abstract screen: Setoka evaluates user understanding; synthetic benchmark construction is not protection of agent state. \\
A2.2 & \nolinkurl{2608.18638v1} & Exclude at abstract screen: workshop report, not an operator study or a scoped survey. \\
A2.3 & \nolinkurl{2602.13319v2} & Exclude at abstract screen: prediction of perspectives, not protection of persisted perspectives. \\
A2.4 & \nolinkurl{2512.03608v1} & Exclude at title screen: inference memory hardware. \\
A3.1 & \nolinkurl{2610.17659v1} & Existing \#81: AUBench, risk and evaluation~\cite{lin2026aubenchcontext}. \\
A3.2 & \nolinkurl{2609.04875v1} & Include \#114: Theorem~1 and \S5 support a transferable runtime-withdrawal operation (\texttt{T}/F+B)~\cite{yao2026executionunlearning}. \\
A3.3 & \nolinkurl{2608.25553v3} & Exclude after checking \S\S2--4: validates superseded task constraints, not withdrawal of private influence or observer disclosure. \\
A3.4 & \nolinkurl{2606.30306v1} & Include \#115: \emph{Always-On Agents}. \S2.3, \S8, and \S9 cover governance and recovery. No operation credit~\cite{ding2026alwayson}. \\
A3.5 & \nolinkurl{2611.05873v1} & Exclude at abstract screen: scoped experience supports software testing, not protection of personal histories. \\
\end{longtable}
\endgroup

\paragraph{Rollup and scope.}
The window contains three new inclusions, five already represented works, 14 exclusions, and one unresolved record. Rechecking \#106 found support for deployment feasibility but no measured privacy outcome or usable privacy accounting; it remains in the corpus without operation credit~\cite{slokom2026federatedrecommendations}. The MaskClaw change moves an existing row from \texttt{D} to \texttt{T}. The revised total is 115 sources, of which 86 yield 122 credited operations. Closest source codes are $8/10/68$ for \texttt{D}/\texttt{S}/\texttt{T}; all ten \texttt{D} rows concern controlled memory operations.

\subsection{Targeted Screening of Disclosure, Skill Reuse, and Withdrawal}\label{sec:supp-stage11}

On 19 November 2026, a targeted pass used the arXiv Atom endpoint \nolinkurl{export.arxiv.org/api/query} with the expressions below. Each expression was wrapped in parentheses and followed by \texttt{AND submittedDate:[202401010000 TO 202611102359]}. GET requests used \texttt{sortBy=relevance} and \texttt{max\_results=8}, starting at zero. All four requests returned HTTP 200. Q2 used a second page at offset eight. We screened all $6+10+7=23$ reported hits, not a relevance-ranked subset, and checked two referral records (R1 from review and R2 from the references and App.~17.3 of PiSAs). AgentCIScope was also a review lead and already appears in Q1, so it is counted only once. We checked cutoff eligibility and preferred the archival version of AgentCIBench. This pass exhausts these query outputs, not the literature, and does not reconstruct Stage~2.

\begin{description}
\item[Q1. Application-side disclosure (6 hits).] \mbox{}\\
\texttt{(ti:"contextual integrity" OR ti:"sovereignpa-bench" OR ti:"disclosure")}\\
\texttt{AND (ti:agent OR ti:agents)}\\
\texttt{AND (abs:"tool calls" OR abs:consent OR abs:recipient)}
\item[Q2. Skill reuse (10 hits).] \mbox{}\\
\texttt{(ti:skill OR ti:skills) AND (abs:privacy OR abs:sanitization)}\\
\texttt{AND (abs:"skill evolution" OR abs:distillation OR abs:"cross-session")}
\item[Q3. Cross-carrier withdrawal (7 hits).] \mbox{}\\
\texttt{(ti:unlearning OR ti:withdrawal OR ti:deletion OR ti:revocation)}\\
\texttt{AND (ti:agent OR ti:agents OR ti:memory)}\\
\texttt{AND (abs:provenance OR abs:rollback OR abs:replay)}
\end{description}

\paragraph{Candidate decisions in retrieval order.}
\begin{description}
\item[Q1.1 (2610.39532v1).] Include \#116, Task-Graph CI (\texttt{T}/B). Sections~4--5 test propagation-scoped release over recipient histories in multi-agent workflows; no persistent personal update is tested~\cite{wu2026tgci}.
\item[Q1.2 (2610.19158v1).] Include \#117, AgentCIScope, evaluation only. Sections~3--4 measure disclosures in executable cross-application workflows. Independent task repetitions do not establish adaptive state carryover~\cite{guo2026agentciscope}.
\item[Q1.3 (2607.05318v2).] Include \#118, PiSAs (\texttt{T}/B). Section~5.1 and Apps.~16 and~19.1 compare prompts and topologies for cross-user disclosures. Credit the shared-workflow release comparison, not a personal writeback effect~\cite{gupta2026pisas}.
\item[Q1.4 (2606.23189v2).] Include \#119, AgentCIBench, evaluation only; use the EMNLP archival record \nolinkurl{10.18653/2026.emnlp-main.55}. Executable personal-app tasks test disclosure, not an intervention on persisted personal updates~\cite{goel2026agentcibench}.
\item[Q1.5 (2608.25474v1).] Include \#120, Field-Tier Minimization (\texttt{T}/B). Sections~3, 8--9 test and analyze projections at an authorization and audit interface. The pilot itself does not establish privacy outcomes at scale~\cite{oswal2026fieldtier}.
\item[Q1.6 (2608.20658v1).] Include \#121, Claws in Plain Sight (\texttt{T}/B). Section~7.2 tests the monitor on deterministic replays to an inert sink; neither native-call effectiveness nor protected adaptive state is established~\cite{dong2026claws}.
\item[Q2.1--Q2.4.] Existing \#98 CounterPersona (2609.15097v1), \#10 MaskClaw (2605.28646v2), \#17 FedSkillGit (2610.21348v1), and \#62 FederatedSkill (2606.03143v1). Retain existing codes~\cite{wang2026counterpersona,zhao2026maskclaw,lan2026fedskillgit,yang2026federatedskill}.
\item[Q2.5 (2608.03509v2).] Include \#122, SkillJack (\texttt{T}/B). Section~4.9 separates candidate review from runtime monitoring on benchmark skill banks and measures policy-violation proxies. Actual downstream skill reuse is tested, but not protected reuse of one user's private state~\cite{ying2026skilljack}.
\item[Q2.6 (2607.18973v1).] Exclude: prediction of subsequent feedback measures skill utility, not a privacy outcome.
\item[Q2.7 (2605.21810v1).] Exclude: hardware design skill optimization and isolation of a hidden task verifier do not protect personal history.
\item[Q2.8--Q2.9.] Existing \#59 AntiSkillBench (2608.03700v1) and \#97 P2Skill (2608.14094v2); retain existing codes~\cite{xiang2026personaskills,ryu2026p2skill}.
\item[Q2.10 (2603.06902v1).] Exclude: entity-agnostic rules derived from simulated clinical trajectories are evaluated for task completion; no privacy outcome or applicable bound is reported.
\item[Q3.1 (2609.04875v1).] Existing \#114 Execution-State Unlearning. Retain \texttt{T}/F+B for controlled runtime equivalence under the replay theorem's premises~\cite{yao2026executionunlearning}.
\item[Q3.2--Q3.5.] Exclude as authority or action revocation rather than withdrawal of retained personal influence: Root-Scoped Quiescence (2609.21284v1, \S10.1) certifies no late acceptance at manifested sinks; NAQD Env (2609.38460v1) tests selective suspension and resumption of plans; Portico (2606.22504v1, \S8) revokes mediated capabilities while leaving model context unchanged; VERA (2608.30091v1, \S\S III and IX) verifies affected delegation paths, not forgetting or disclosure of a withdrawn fact.
\item[Q3.6--Q3.7.] Existing \#81 AUBench (2610.17659v1) and \#79 AfterDeleteBench (2610.31366v1). The former provides benchmark coverage; the latter retains direct credit for tracked memory descendants~\cite{lin2026aubenchcontext,saleh2026afterdeletebench}.
\item[R1 (2607.05363v1).] Include \#123, SovereignPA-Bench (\texttt{S}/B). Sections~3 and~5, Table~2 compare bundled policies on fixed personal-agent scenarios. Attribute neither the effect to a single control nor the evidence to longitudinal update protection~\cite{liu2026sovereignpa}.
\item[R2 (10.18653/v1/2025.findings-emnlp.925).] Include \#124, PrivacyChecker (\texttt{S}/B). Sections~3--4 evaluate contextual flow filtering on static PrivacyLens and executable tool workloads. PiSAs evaluates the same filter on different data as a comparator, not as a new operator~\cite{wang2025privacychecker,gupta2026pisas}.
\end{description}

\paragraph{Effect on the evidence distribution.}
Of 25 screened records, nine were newly included, nine were already represented, and seven were excluded. Two new sources contribute \texttt{S}, five contribute \texttt{T}, and two are evaluation-only. One new \texttt{T} source (SkillJack) contributes two distinct operation rows. The totals move from 115 to 124 sources, from 86 to 93 sources with counted protection operations, and from 122 to 130 operations. The closest source codes change from $8/10/68$ to $8/12/73$; the operation counts change from $10/19/93$ to $10/21/99$ (\texttt{D}/\texttt{S}/\texttt{T}). The ten direct rows across eight controlled memory sources are unchanged. No new direct skill-reuse or cross-carrier withdrawal result was identified in this pass.

\subsection{Carrier-stratified retrieval and additional referrals}
\label{sec:supp-stage12}

On 19 November 2026 we queried the arXiv Atom API at \nolinkurl{https://export.arxiv.org/api/query}. We wrapped each expression in parentheses and appended \texttt{AND submittedDate:[201801010000 TO 202611102359]}. Requests used \texttt{start=0}, \texttt{max\_results=8}, and \texttt{sortBy=relevance}. All five requests returned HTTP 200. We screened the entire returned window of 37 distinct records using titles and abstracts, then examined the primary text for operationally relevant or ambiguous candidates. Twenty-one records were excluded, ten matched retained sources, and six were admitted. The API date restriction applies to first submission. We checked version dates and used the eligible archival editions where available.

\begin{description}
\item[P (8/16).] {\footnotesize\ttfamily\raggedright (ti:prompt OR ti:prompts)\newline AND (ti:privacy OR ti:private OR abs:\char34{}differential privacy\char34{})\newline AND (abs:personalization OR abs:personalized OR abs:\char34{}user feedback\char34{} OR abs:\char34{}continual learning\char34{})\par}
\item[W (8/149).] {\footnotesize\ttfamily\raggedright (ti:\char34{}continual learning\char34{} OR ti:\char34{}fine-tuning\char34{} OR ti:personalization OR ti:personalized)\newline AND (ti:private OR ti:privacy)\newline AND (abs:\char34{}language model\char34{} OR abs:\char34{}language models\char34{} OR abs:adapter)\par}
\item[R (8/38).] {\footnotesize\ttfamily\raggedright (ti:preference OR ti:personalized OR ti:bandit OR ti:bandits)\newline AND (ti:privacy OR ti:private)\newline AND (abs:\char34{}online learning\char34{} OR abs:\char34{}preference feedback\char34{} OR abs:\char34{}language models\char34{})\par}
\item[M (7/7).] {\footnotesize\ttfamily\raggedright (ti:memory OR ti:forgetting)\newline AND (ti:privacy OR ti:private OR ti:revocation OR ti:withdrawal OR ti:deletion)\newline AND (abs:\char34{}persistent memory\char34{} OR abs:\char34{}long-term memory\char34{} OR abs:\char34{}personal agent\char34{})\newline AND (abs:agent OR abs:agents)\par}
\item[K (6/6).] {\footnotesize\ttfamily\raggedright (ti:skill OR ti:skills)\newline AND (abs:privacy OR abs:sanitization OR abs:confidentiality)\newline AND (abs:agent OR abs:agents)\newline AND (abs:personal OR abs:personalized OR abs:distillation OR abs:commit OR abs:reuse)\par}
\end{description}

\paragraph{Candidate dispositions by query and rank.}
An existing match keeps its earlier source index. An excluded record can be relevant to generic privacy or adaptation while contributing no additional covered operation in this pass.

\noindent\textbf{Query P}\par
\begin{itemize}\setlength\itemsep{0pt}
\item P1 (\nolinkurl{2501.13904v3}). Existing. DP-FPL, source 33; same method family.
\item P2 (\nolinkurl{2505.22447v1}). Include 125. SecFPP; inspect leakage of clustering distances.
\item P3 (\nolinkurl{2603.26032v1}). Exclude. Current-prompt character perturbation, without retained state or an update operator.
\item P4 (\nolinkurl{2510.03662v1}). Exclude. Minimizes a fresh prompt without a persistent read or update.
\item P5 (\nolinkurl{2305.15008v1}). Exclude. Single-inference regurgitation and sanitization tests.
\item P6 (\nolinkurl{2607.21417v1}). Include 126. FedSEPT; assess accounting for repeated federated prompt releases.
\item P7 (\nolinkurl{2603.28972v1}). Exclude. Routes current inputs without a protected persistent update experiment.
\item P8 (\nolinkurl{2607.02932v1}). Exclude. Current-prompt obfuscation and profiling, without retained state.
\end{itemize}
\noindent\textbf{Query W}\par
\begin{itemize}\setlength\itemsep{0pt}
\item W1 (\nolinkurl{2305.13865v3}). Exclude. Generic private fine-tuning initialization; evaluation centers on utility and model size.
\item W2 (\nolinkurl{2312.15603v1}). Exclude. Generic private training architecture; no additional personal-state observation in this pass.
\item W3 (\nolinkurl{2504.21036v2}). Include 127. Use the DBSec archival study for empirical extraction, membership, and utility.
\item W4 (\nolinkurl{2312.17342v1}). Exclude. Encrypted inference on private inputs; no per-user history or retained-state boundary in the evaluation.
\item W5 (\nolinkurl{2611.10578v1}). Exclude. Egocentric video privacy reasoning rather than protection of a personal language agent update.
\item W6 (\nolinkurl{2210.15042v3}). Exclude. Generic private fine-tuning methods on standard benchmarks.
\item W7 (\nolinkurl{2604.27833v1}). Exclude. Federated class-prototype perturbation; the task setting is generic classification.
\item W8 (\nolinkurl{2110.06500v2}). Exclude. Generic parameter-efficient DP training on standard NLP tasks.
\end{itemize}
\noindent\textbf{Query R}\par
\begin{itemize}\setlength\itemsep{0pt}
\item R1 (\nolinkurl{2206.13192v2}). Exclude. Multi-agent combinatorial bandit coordination; lacks a retained personal-data interface.
\item R2 (\nolinkurl{2404.09413v1}). Exclude. General contextual-bandit regret analysis; no personal history or task.
\item R3 (\nolinkurl{2508.07672v1}). Exclude. Conceptual privacy-alignment workflow without a tested operator.
\item R4 (\nolinkurl{2610.30099v1}). Existing. Differentially private sparse reward estimation, source 32.
\item R5 (\nolinkurl{2610.37682v2}). Include 128. Rank-aware release extends protection from a binary label to an item relocation in a list.
\item R6 (\nolinkurl{2604.11259v1}). Include 129. TIPO evaluates contextual action compliance after offline trajectory alignment.
\item R7 (\nolinkurl{2505.21395v1}). Include 130. Use the ICML archival paper for central sample and local label protection.
\item R8 (\nolinkurl{2602.08268v2}). Exclude. User-controlled granularity with personalization and cost results; no privacy outcome for updates or transfer.
\item R1 (\nolinkurl{10.1287/msom.2023.0271}). Include 132. Exact-title check. Formal comparator for a recipient inferring preferences from an observed action.
\item R2 (\nolinkurl{2611.01386v1}). Include 138. Keyword referral. Privacy-related admission and retirement are design context; the benchmark measures task outcomes.
\end{itemize}
\noindent\textbf{Query M}\par
\begin{itemize}\setlength\itemsep{0pt}
\item M1 (\nolinkurl{2609.08258v1}). Existing. Revoked but Still Authoritative, source 108.
\item M2 (\nolinkurl{2608.16551v1}). Existing. SP-Mem, source 39.
\item M3 (\nolinkurl{2603.02240v1}). Exclude. Local-first multi-agent memory; tests trust and poisoning defenses rather than confidentiality or withdrawal.
\item M4 (\nolinkurl{2610.28315v2}). Existing. Disclosure Begets Disclosure, source 76.
\item M5 (\nolinkurl{2604.02522v1}). Existing. Opal, source 9.
\item M6 (\nolinkurl{2609.32401v1}). Exclude. Fictional character beliefs and plot-state consistency, outside personal privacy.
\item M7 (\nolinkurl{2608.03130v1}). Existing. DP-MemView, source 8.
\end{itemize}
\noindent\textbf{Query K}\par
\begin{itemize}\setlength\itemsep{0pt}
\item K1 (\nolinkurl{2608.03700v1}). Existing. AntiSkillBench, source 59.
\item K2 (\nolinkurl{2605.05413v2}). Exclude. Moves procedural state into weights; benchmarks task success and token use without a privacy result.
\item K3 (\nolinkurl{2606.03143v1}). Existing. FederatedSkill, source 62.
\item K4 (\nolinkurl{2609.07131v1}). Exclude. Capability cloning risk without a personal-state protection result.
\item K5 (\nolinkurl{2610.21348v1}). Existing. FedSkillGit, source 17.
\item K6 (\nolinkurl{2603.06902v1}). Exclude. Entity-agnostic skills from simulated clinical tasks; evaluation concerns execution utility.
\end{itemize}

\paragraph{Additional discovery and primary-source checks.}
A separate pass used the same endpoint and relevance order, with \texttt{sortOrder=descending} and \texttt{submittedDate:[202301010000 TO 202611102359]}. We inspected titles and abstracts for leads and verified the selected works against the full text. These are discovery windows. The decisions below refer to the followed leads. An initial A2 request returned 25 of 28 hits and was rerun with a cap of 35.

\begin{description}
\item[A1 (6/6, cap 25).] {\footnotesize\ttfamily\raggedright (ti:personalization OR ti:personalized)\newline AND (ti:\char34{}language model\char34{} OR ti:\char34{}language models\char34{})\newline AND (abs:\char34{}on-device\char34{} OR abs:\char34{}user history\char34{})\newline AND abs:privacy\par}
\item[A2 (28/28, cap 35).] {\footnotesize\ttfamily\raggedright (ti:\char34{}agent skills\char34{} OR ti:\char34{}skill library\char34{})\newline AND (abs:privacy OR abs:permission OR abs:provenance)\par}
\item[A3 (27/27, cap 50).] {\footnotesize\ttfamily\raggedright (ti:personalized OR ti:personalization OR ti:continual)\newline AND (ti:prompt OR ti:prompts OR ti:adapter OR ti:preference)\newline AND (abs:privacy OR abs:\char34{}differential privacy\char34{} OR abs:\char34{}on-device\char34{})\newline AND (abs:\char34{}language model\char34{} OR abs:\char34{}language models\char34{} OR abs:agent OR abs:agents)\par}
\item[A4 (48/48, cap 50).] {\footnotesize\ttfamily\raggedright (ti:privacy OR ti:private OR ti:confidential)\newline AND (ti:\char34{}fine-tuning\char34{} OR ti:LoRA OR ti:adapter)\newline AND (abs:personalized OR abs:\char34{}single user\char34{} OR abs:\char34{}user data\char34{} OR abs:\char34{}user history\char34{})\par}
\end{description}
\begin{itemize}\setlength\itemsep{0pt}
\item A1.4, HomeLLaMA (\nolinkurl{2507.08878v1}). Include 131. Use IEEE TMC 10.1109/TMC.2026.3654976. Profile updates stay local in the controlled lab setting; PrivShield tests current-command exposure separately.
\item R1, Private personalized recommendation (\nolinkurl{10.1287/msom.2023.0271}). Include 132. Exact-title check. Formal comparator for a recipient inferring preferences from an observed action.
\item A1.2, OPPU (\nolinkurl{2402.04401v3}). Include 133. Use ACL 10.18653/v1/2024.emnlp-main.372. Static per-user adaptation is design context; privacy outcomes are outside the benchmark.
\item A2.3, SkillGuard (\nolinkurl{2606.03024v2}). Include 134. Runtime permission mediation evaluated on generic benchmark skills.
\item A2.14, SkillLifeBench (\nolinkurl{2610.35215v1}). Include 135. Paired authorized-use and recovery experiments in a skill benchmark.
\item A4.12, Separable Expert Architecture (\nolinkurl{2604.21571v1}). Include 136. Evaluated proxy omission on synthetic users; software deletion is described separately.
\item A3.12 / A4.7, PRISP (\nolinkurl{2601.06471v2}). Include 137. Use ACL 10.18653/v1/2026.acl-long.1146. Static few-shot isolation of the per-user adaptation.
\item R2, Zep Agent Skills (\nolinkurl{2611.01386v1}). Include 138. Keyword referral. Privacy-related admission and retirement are design context; the benchmark measures task outcomes.
\end{itemize}
The two sets of windows and the referrals added 14 sources. Twelve support a credited operation, including a new direct-memory result. The other two provide design context. The corpus at the end of Stage~12 contained 138 sources. The original Stage~2 exclusion record is incomplete, and the new query results bound the coverage inference to the recorded windows.
\subsection{Full-window screening by carrier}
\label{sec:supp-stage13}

On 19 November 2026 we repeated the five fielded expressions P, W, R, M, and K from Section~\ref{sec:supp-stage12} with the same submission dates. We set \texttt{max\_results=100}, \texttt{sortBy=relevance}, and \texttt{sortOrder=descending}. The parameter query required a second page starting at 100. We retrieved every hit returned for each expression. The arXiv date criterion applies to first submission, and the returned versions used for this round were all available by the cutoff. We used archived versions when available.

For the archival check, we ran five title and abstract filters over the official ACL Anthology BibTeX export with abstracts (build dated 17 November 2026). The eligible window was 24,844 main-conference and Findings records from 2024 through the cutoff. The archive contains the venue patterns, exact regular expressions, deterministic ID ordering, and the export's SHA-256. P and W query prompt and adaptation terms with a privacy term in the title and a language-agent or personalization term in the title or abstract. R requires a preference, alignment, reward or recommendation title and differential-privacy wording. M requires a memory or unlearning title and a privacy or lifecycle-governance term. K requires a skill title and a privacy, permission, or provenance term. Stable identifiers, abstracts, and primary-source checks support the candidate dispositions in the archived ledger.

\noindent\textbf{Hits and dispositions by query.} Entries give all retrieved hits and their disposition. A paper returned by two queries appears in both query rows. Unique totals deduplicate by stable identifier within each channel.

{\small\centering
\begin{tabular}{@{}lrrrrrrrr@{}}
\toprule
 & \multicolumn{4}{c}{\textbf{arXiv}} & \multicolumn{4}{c}{\textbf{ACL Anthology}} \\
\cmidrule(lr){2-5}\cmidrule(lr){6-9}
\textbf{Carrier} & Hits & Existing & Include & Other & Hits & Existing & Include & Other \\
\midrule
Prompt & 16 & 3 & 1 & 12 & 13 & 2 & 0 & 11 \\
Parameter & 149 & 15 & 13 & 121 & 17 & 2 & 1 & 14 \\
Preference & 38 & 11 & 11 & 16 & 1 & 0 & 0 & 1 \\
Memory & 7 & 5 & 0 & 2 & 68 & 3 & 4 & 61 \\
Skill & 6 & 3 & 0 & 3 & 3 & 0 & 1 & 2 \\
\midrule
All hits & 216 & 37 & 25 & 154 & 102 & 7 & 6 & 89 \\
Unique within channel & 191 & 28 & 21 & 142 & 98 & 6 & 5 & 87 \\
\bottomrule
\end{tabular}\par
}

\paragraph{Screening and linkage.}
We considered title and abstract for each record, checked duplicates against the starting corpus and within the round, and inspected primary texts for new inclusions. Previously represented works retained their indices and canonical citations. Transfer comparators contribute when they establish an additional observer, unit, operator, or directly useful comparative result. Static selection or architecture alone can enter as design context. An evaluated study of current-query rewriting, for example, can supply a transferable control without counting as evidence for release from retained personal memory. The candidate files record exclusions under the existing E codes, reasons, and original version or anthology identifiers. Existing and newly accepted works link to source indices or canonical proceedings IDs. The 21 arXiv additions and five ACL additions represent distinct new works after cross-channel matching.

\paragraph{Archival checks.}
MemGov, flagged in the review and recovered by the M query, supplies surface evidence for scoped context assembly~\cite{christou2026memgov}. COLLABMEM, PRISM, and SkillLadder add transfer evidence for team memory, client-private tuning, and skill sharing, respectively. TimlMemBench supplies risk and evaluation evidence~\cite{wu2026collabmem,wang2026prism,chen2026skillladder,zhou2026timlmembench}. A separate exact-title check in the NeurIPS proceedings adds Hanke et al.'s matched prompt and parameter adaptation comparisons~\cite{hanke2024openprivate}. The exact-title result is additional to the table's channel denominators.

\paragraph{Coding sensitivity.}
A separate AI recoder, given the rubric and primary source identifiers before seeing the new decisions, reviewed nine selected arXiv cases spanning prompt tuning, personal adapters, memory release, and preference learning (P10, P13, W9, W65, W79, W116, R19, R25, R28). Eight of nine initial distance calls agreed, counting the absence of a mechanism code as a distinct outcome. The difference concerned R25. The second reading classified its current-query rewrite under a fixed privacy profile as \texttt{T}; the initial reading treated the fixed rule as sufficient for \texttt{S}. We retained \texttt{T} because the test protects current-query content while applying a fixed rule, and does not exercise private facts read from retained memory~\cite{ramirez2026peep}. The alternative would move one source and one operation from \texttt{T} to \texttt{S}, leaving the direct-evidence finding intact.

This round added 27 sources, yielding 165 unique sources and 168 credited operation rows. Among the 128 credited sources, closest-source counts are 9 direct, 14 surface, and 105 transferable. The added operations comprise two surface and 23 transfer rows. The 11 direct rows from nine controlled memory sources remain the direct evidence in the screened corpus. The recorded lexical and venue windows define the coverage of this audit.

The anonymized archive's \texttt{literature\_screening/} directory contains replay scripts, exact arXiv request URLs and page counts, and both complete per-candidate TSV files. Those ledgers preserve the observed hits when the public index versions change.

\subsection{Observer and Withdrawal Follow-up}
\label{sec:supp-stage14}

On 19 November 2026 we ran a targeted follow-up with the same 10 November cutoff. We searched public scholarly and proceedings pages for combinations of language agents, joint or multi-channel observation, tool-result inference, unlearning, derived state, and activation after deletion. We traced the withdrawal references of Lin et al.'s expanded memory-governance survey (v5, \S6 and Appendix~D) and checked proceedings and version histories against primary texts~\cite{lin2026memorysecuritysurvey}. This was a targeted search for comparisons that could change the interpretation of Main Section~7. The archive records the query strings and targeted decisions. The shortlist is defined by those logged checks.

\paragraph{Known-source recovery check.}
We matched identifiers, DOIs, and normalized titles for the nine known direct sources in Table~\ref{tab:supp-75-ledger} against the complete Stage~13 arXiv and anthology ledgers. Five of nine were recovered (MemPrivacy, Opal, SP-Mem, Disclosure Begets Disclosure, and HomeLLaMA). HealthJev, Reglos, AfterDeleteBench, and Audience-Bound Persistent Memory appear in the corpus through earlier stages and had no match in either final carrier window. HomeLLaMA matched two arXiv queries and counts once. This check tests recovery of a known positive set, with each study counted once. Differences in retrieval coverage limit interpretation of zero cells across carriers.

\paragraph{Joint observation and withdrawal.}
K-Bench (source~\#166, v1) follows a referral from Lin et al. Its single-source injection test compares weight, prompt, and retrieval interventions using six observation channels. Its tool-path evidence changes how we interpret reduced answer disclosure after withdrawal~\cite{yu2026kbench}. TOP-Bench (source~\#167, EMNLP Findings) tests combined inference over tool returns and evaluates response mitigations in fixed tasks. The archive links its preprint to the proceedings record. The proceedings article was absent from the Stage~13 candidate hits even though that venue was in the archival window~\cite{qiao2026topbench}. CIPL (source~\#168, arXiv:2609.21686v1) evaluates recoverable leakage with channel-specific observations and compares audited and unaudited paths. The earlier public CIPL manuscript (arXiv:2603.22751v2) shares its evaluation framework and overlapping experimental cases. We retain the later expanded report as one source for this comparison~\cite{huang2026cipl}. We also rechecked the matched control and separate reconstruction, commitment, and activation outcomes in AfterDeleteBench. We recoded MemCut (existing source~\#43) to credit its fixed-budget cleanup comparison as \texttt{T}/B. It measures recovery after the intervention at a fixed checkpoint. The direct code for withdrawal requires carryover to later decisions or interactions~\cite{saleh2026afterdeletebench,zhang2026memcut}.

\paragraph{Independent evidence for the same operation.}
Stage~13 candidate W50 was excluded on operation overlap. Its full text (source~\#169) compares DP-SGD, LoRA, and full fine-tuning on common tasks, with separate empirical memorization, utility, and compute measures. That comparison is eligible even though the operators already occur in the corpus. We code the benchmark as transferable evidence for parameter updates~\cite{das2026finetuning}.

The follow-up adds four risk and evaluation sources and credits the existing MemCut comparator. At the end of Stage~14, the corpus had 169 sources and 173 credited operation rows. Closest-source totals for the 133 studies with a credited intervention were nine direct, 14 surface, and 110 transferable. The direct rows are unchanged by these decisions. They describe the screened corpus.

\subsection{Complete Observer and Withdrawal Windows and E6 Rescreening}
\label{sec:supp-stage15}

On 19 November 2026 we searched two bibliographic windows under the unchanged cutoff. The joint-observation query combines agent, privacy and terms for internal channels or tool returns. The withdrawal query combines agent and memory with terms for deletion, derived state or selective forgetting. We reran these title and abstract filters on the same 24,844-record ACL Anthology snapshot as Stage~13. We screened the one joint and three withdrawal matches in canonical identifier order. In Crossref we screened the first 12 results in API relevance order for each query using the 2024--2026 publication window through 10 November. The anonymized archive's protocol and ledgers record the exact expressions, date boundaries, source versions and reasons for all 28 candidate occurrences. The arXiv Atom search was rate-limited or timed out. The two completed search channels and the separate exact checks supply the screening evidence.

\paragraph{Linked-observer records.}
The ACL window recovers TOP-Bench~\cite{qiao2026topbench}. An exact DOI check adds AgentLeak as a risk and evaluation source with a transferable release result~\cite{elyagoubi2026agentleak}. The paper compares final-output and internal-channel controls in scripted multi-agent tasks (Sections~VI-E/F and VII-C). In its main traces, an apparently clean final answer can co-occur with internal exposure. The observed channels and tested controls inform the linked-recipient comparison in Main Section~7.3. The withdrawal window recovers TimlMemBench~\cite{zhou2026timlmembench}. The cutoff and targeted windows continue to limit generalization from the direct-evidence counts.

\paragraph{Reconsideration of operation-overlap exclusions.}
We obtained all 40 versioned primary abstracts for Stage~13 arXiv rows previously excluded under E6. We reassessed them for a distinct observer interface or an independent matched privacy, utility or resource result. Full-text verification retained ten additional works, including matched memorization and resource comparisons, post-training suppression under equal pruning budgets, and studies that vary what training recipients can see. W50 was already restored in Stage~14. The remaining 29 decisions are recorded individually. Supplementary Section~S6.4 gives the comparisons and evidence boundaries.

The corpus now contains 180 studies. The operation matrix contains 184 credited rows from 144 studies, with closest-source counts of nine direct, 14 surface and 121 transferable. The Stage~13 known-source recovery result remains five of nine. That count describes the original search windows.

\section{Deduplicated Evidence Role and Distance Ledger}
\label{sec:supp-ledger75}

\subsection{Source Roles and Evidence Distance}
\label{sec:supp-ledger75-rules}

Each source available by 10 November 2026 receives one primary evidence role. The source-level distance code reports its closest supported operation in Section~\ref{sec:supp-operation-evidence}, using \texttt{D}, then \texttt{S}, then \texttt{T}. A dash denotes a source without a coded protection operation. Section~\ref{sec:supp-operation-evidence} retains the individual operation codes and their distinct validation settings.

The representative, extended, and primitive strata include 121 sources, of which 120 support a counted operation. 24 risk and evaluation studies also support a protection operation. Together they yield 144 distinct sources with credited evidence.

\subsection{Coding Ledger and Reconciliation Summary}
\label{sec:supp-ledger75-table}

Table~\ref{tab:supp-75-ledger} preserves source indices across screening passes. The indices link the source ledger to the operation ledger. Bracketed citations follow the alphabetical ACM bibliography.

\begingroup
\footnotesize
\setlength{\tabcolsep}{2.6pt}
\renewcommand{\arraystretch}{1.18}
\begin{longtable}{>{\raggedleft\arraybackslash}p{0.035\linewidth}>{\raggedright\arraybackslash}p{0.230\linewidth}>{\raggedright\arraybackslash}p{0.185\linewidth}>{\raggedright\arraybackslash}p{0.105\linewidth}>{\centering\arraybackslash}p{0.045\linewidth}>{\raggedright\arraybackslash}p{0.285\linewidth}}
\caption{Source Roles, Manuscript Placement, and Closest Evidence Distance for 180 Sources}
\label{tab:supp-75-ledger}\\
\toprule
\textbf{\#} & \textbf{Source and Citation} & \textbf{Primary Role (Sub-role)} & \textbf{Main Text} & \textbf{Code} & \textbf{Presentation Mode or Analytical Scope / Boundary} \\
\midrule
\endfirsthead
\caption[]{Source Roles, Manuscript Placement, and Closest Evidence Distance (Continued)}\\
\toprule
\textbf{\#} & \textbf{Source and Citation} & \textbf{Primary Role (Sub-role)} & \textbf{Main Text} & \textbf{Code} & \textbf{Presentation Mode or Analytical Scope / Boundary} \\
\midrule
\endhead
\midrule
\multicolumn{6}{r}{\footnotesize\emph{Continued on next page}} \\
\endfoot
\bottomrule
\endlastfoot
1 & \texttt{DP-ES}~\cite{liu2026dpes} & Representative mechanism & \S3, \S4.1 & \texttt{T} & Main Table~3: Gaussian-noised prompt population scoring and RDP accounting \\
2 & \texttt{DP-OPT}~\cite{hong2024dpopt} & Representative mechanism & \S3, \S4.1, \S7.2 & \texttt{T} & Main Table~3: Local DP ensemble prompt generation and disjoint-set exponential selection \\
3 & \texttt{Mind the Privacy Unit}~\cite{chua2024privacyunit} & Representative mechanism & \S3, \S4.2 & \texttt{T} & Main Table~3: User-level clipped and noised LoRA parameter fine-tuning \\
4 & \texttt{Decoupled DP Reward}~\cite{cho2026decoupledreward} & Representative mechanism & \S3, \S4.3, \S7.4 & \texttt{T} & Main Table~3: Full preference-tuple DP reward head with post-processing Best-of-$N$ \\
5 & \texttt{DP Reward Estimation}~\cite{raychowdhury2024privatereward} & Representative mechanism & \S3, \S4.3, \S7.4 & \texttt{T} & Main Table~3: Central and local label-DP preference reward estimation \\
6 & \texttt{SecP-Tuning}~\cite{luo2026secptuning} & Representative mechanism & \S3, \S4.4 & \texttt{T} & Main Table~3: Two-server MPC forward execution with client-side soft-prompt updates \\
7 & \texttt{MemPrivacy}~\cite{chen2026memprivacy} & Representative mechanism & \S3, \S5.1 & \texttt{D} & Main Table~3: Typed pre-write entity substitution before cloud memory storage \\
8 & \texttt{DP-MemView}~\cite{kim2026dpmemview} & Representative mechanism & \S2.3, \S3, \S5.2, \S\S7.2--7.3 & \texttt{S} & Main Table~3: Attribute-level pure-DP multi-round memory read and transcript release \\
9 & \texttt{Opal}~\cite{kaviani2026opal} & Representative mechanism & \S3, \S5.2, \S7.3 & \texttt{D} & Main Table~3: Confidential TEE/ORAM memory ingestion, indexing, expiry, and summarization \\
10 & \texttt{MaskClaw}~\cite{zhao2026maskclaw} & Representative mechanism & \S3, \S6.1, \S7.4, \S8.4 & \texttt{T} & E3 tests candidate filtering in five scenarios without a committed-version reuse trace. E1/E2 cover static screenshot arbitration. \\
11 & \texttt{Safety in Self-Evolving Agents}~\cite{chen2026safetysurvey} & Surveys \& state bg.\ (Survey) & \S1 (Table~1) & --- & Adjacent safety survey (\texttt{SAVER} carrier$\times$adaptation taxonomy); not an original mechanism \\
12 & \texttt{Agents That Know Too Much}~\cite{lahjouji2026agentsknowtoomuch} & Surveys \& state bg.\ (Survey) & \S1 (Table~1) & --- & Data-centric LLM-agent privacy survey across data surfaces and governance controls \\
13 & \texttt{Prompt Privacy Survey}~\cite{edemacu2025promptsurvey} & Surveys \& state bg.\ (Survey) & \S1 (Table~1) & --- & ACM CSUR survey on privacy-preserving prompt engineering models and mechanisms \\
14 & \texttt{Preference Alignment Survey}~\cite{chen2026preferencesurvey} & Surveys \& state bg.\ (Survey) & \S1 & --- & Survey of privacy-preserving LLM preference alignment (\texttt{TechRxiv v1}) \\
15 & \texttt{PrivMeSA}~\cite{wang2026privmesa} & Representative mechanism & \S3, \S4.2 & \texttt{T} & Main Table~3: Offline shared clinical LoRA, separate cross-patient lesson writes, and outbound disclosure and linkage evaluation \\
16 & \texttt{HealthJev}~\cite{li2026healthjev} & Representative mechanism & \S3, \S5.1 & \texttt{D} & Main Table~3: Selective memory writeback governance for privacy-sensitive health interactions \\
17 & \texttt{FedSkillGit}~\cite{lan2026fedskillgit} & Representative mechanism & \S2.3, \S3, \S6.1 & \texttt{T} & Main Table~3: Pre-commit sanitization and auditing for federated shared skills (not single-user \texttt{D}) \\
18 & \texttt{Private Selection}~\cite{liu2019privateselection} & Representative mechanism & \S3, \S4.1 & \texttt{T} & Main Table~3: Randomized-stopping meta-selection over already differentially private candidates \\
19 & \texttt{KL-Regularization DP}~\cite{zhang2025klregularization} & Representative mechanism & \S3, \S4.3 & \texttt{T} & Main Table~3: Single-shot action-sampling label-DP induced by KL-regularized policies \\
20 & \texttt{Lifelong DP}~\cite{lai2022lifelongdp} & Representative mechanism & \S3, \S4.4 & \texttt{T} & Main Table~3: Task partitioning, objective perturbation, and continual model-version release \\
21 & \texttt{Concurrent Composition}~\cite{henzinger2026concurrentcomposition} & Representative mechanism & \S3, \S4.4 & \texttt{T} & Main Table~3: Concurrent composition across interleaved queries, updates, and joint transcripts \\
22 & \texttt{Self-Evolving Agents Survey}~\cite{gao2026selfevolvingsurvey} & Surveys \& state bg.\ (Survey) & \S1 (Table~1) & --- & Operational definition of feedback-driven persistent state evolution across carriers \\
23 & \texttt{Reglos}~\cite{yang2026reglos} & Representative mechanism & \S3, \S5.3, \S7.5, \S8.2, \S8.4 & \texttt{D} & Main Table~3: tombstone update, tested jointly with reader non-use; Table S22 separates mask, reader, and UI deletion \\
24 & \texttt{SBU}~\cite{wang2026agenticunlearning} & Representative mechanism & \S2.3, \S3, \S5.3, \S7.5, \S8.2 & \texttt{T} & Main Table~3: Dependency-closure memory cleanup and synchronized parameter suppression \\
25 & \texttt{Galaxy}~\cite{bao2025galaxy} & Representative mechanism & \S2.3, \S3, \S5.2 & \texttt{S} & Main Table~3: Tiered pre-cloud context masking and selective demasking in a personal agent \\
26 & \texttt{Flocks (PromptDPSGD / PromptPATE)}~\cite{duan2023stochasticparrots} & Extended/adjacent mech.\ (Extended) & \S4.1 & \texttt{T} & Separately proposes \texttt{PromptDPSGD} (DP-SGD on continuous soft prompts) and \texttt{PromptPATE} (PATE class-label voting over unlabeled public data) \\
27 & \texttt{DP Few-Shot Generation}~\cite{tang2024dpfewshot} & Extended/adjacent mech.\ (Extended) & \S4.1 & \texttt{T} & Generates reusable differentially private few-shot demonstrations from private corpora \\
28 & \texttt{DP Zeroth-Order Tuning}~\cite{tang2025dpzo} & Extended/adjacent mech.\ (Extended) & \S4.2 & \texttt{T} & Low-memory sample-level DP parameter fine-tuning via privatized zeroth-order loss differences \\
29 & \texttt{DP Gboard FL}~\cite{xu2023gboard} & Extended/adjacent mech.\ (Deployment) & \S4.2 & \texttt{T} & Production device-level \texttt{DP-FTRL} federated deployment; population rather than single-user loop \\
30 & \texttt{Unintended Memorization}~\cite{szep2026memorization} & Risk \& evaluation study & \S4.2 & --- & Benchmarks extraction of input-side PII memorized during fine-tuning across defenses \\
31 & \texttt{PrivDPO}~\cite{jiang2026privdpo} & Extended/adjacent mech.\ (Extended) & \S4.3 & \texttt{T} & Extends label-DP directly to DPO policy updates (pinned \texttt{arXiv:2608.05040v1}; CCS~2026 accepted) \\
32 & \texttt{DP Sparse Reward}~\cite{ding2026sparsereward} & Extended/adjacent mech.\ (Extended) & \S4.3 & \texttt{T} & Theoretical comparison of sample-DP vs.\ randomized-response label-DP in sparse BTL models \\
33 & \texttt{DP-FPL}~\cite{tran2025dpfpl} & Extended/adjacent mech.\ (Extended) & \S4.4 & \texttt{T} & Combines local and central DP for personalized federated multimodal soft-prompt learning \\
34 & \texttt{Secure Aggregation}~\cite{bonawitz2017secureaggregation} & Theoretical \& systems prim.\ (System) & \S4.4 & \texttt{T} & Cryptographic aggregation primitive hiding individual client uploads without constraining aggregate output \\
35 & \texttt{DP-FTRL}~\cite{kairouz2021dpftrl} & Theoretical \& systems prim.\ (System) & \S4.4 & \texttt{T} & Tree-based continual private prefix-sum aggregation without subsampling or shuffling amplification \\
36 & \texttt{DP CL Label Space}~\cite{tobaben2026labelspace} & Risk \& evaluation study (Dual-coded) & \S2.3 & \texttt{T} & Primary role: dynamic output label-space metadata leakage; also proposes private label-set release \\
37 & \texttt{Fully-Adaptive Composition}~\cite{whitehouse2023fullyadaptive} & Theoretical \& systems prim.\ (Theory) & \S4.4, \S8.3 & \texttt{T} & Formal privacy filters and odometers for adaptively chosen privacy parameters across rounds \\
38 & \texttt{Continual Mean Estimation}~\cite{george2024continualmean} & Theoretical \& systems prim.\ (Theory) & \S4.4, \S8.3 & \texttt{T} & User-level continual statistical release under repeated per-user stream contributions \\
39 & \texttt{SP-Mem}~\cite{wang2026spmem} & Extended/adjacent mech.\ (Extended) & \S2.3, \S5.1 & \texttt{D} & Separates sanitized conversational memory from isolated exact secrets with authorized recovery \\
40 & \texttt{PrivAgentFlow}~\cite{tang2026privagentflow} & Extended/adjacent mech.\ (Extended) & \S6.2, \S8.4 & \texttt{T} & Adapts workflow-node privacy policies from failure history; updates workflow governance, not user skills \\
41 & \texttt{MEXTRA}~\cite{wang2025mextra} & Risk \& evaluation study & \S1, \S5, \S8.5 & --- & Black-box prompt extraction attack against long-term LLM-agent memory stores \\
42 & \texttt{SPORE}~\cite{gao2026spore} & Risk \& evaluation study & \S1, \S5, \S8.5 & --- & Cross-session persistence-based exfiltration of isolated memory via malicious tool parameters \\
43 & \texttt{MemCut}~\cite{zhang2026memcut} & Risk \& evaluation study & \S5.3, \S7.5, \S8.2, \S8.5 & \texttt{T} & V1, Sec.~6.1 and App.~B.3: budget-matched candidate-pool sweep at a fixed recovery checkpoint. \\
44 & \texttt{Deployment Memorization}~\cite{chen2026deploymentmemorization} & Risk \& evaluation study & \S5.3, \S8.2 & --- & Evaluates forgetting residue across summary and derived memory layers in foundation-model agents \\
45 & \texttt{RemoteRAG}~\cite{cheng2025remoterag} & Extended/adjacent mech.\ (Adjacent) & \S5.2 & \texttt{T} & Query-side \texttt{DistanceDP} perturbation and candidate-range bounding for cloud RAG retrieval \\
46 & \texttt{PRAG}~\cite{li2026prag} & Extended/adjacent mech.\ (Adjacent) & \S5.2 & \texttt{T} & Homomorphically encrypted RAG retrieval primitive clarifying client vs.\ server visibility \\
47 & \texttt{SAGE}~\cite{zeng2025sage} & Extended/adjacent mech.\ (Adjacent) & \S5.1 & \texttt{T} & Replaces private RAG corpora with synthetic data; static corpus synthesis outside personal loops \\
48 & \texttt{Oblix}~\cite{mishra2018oblix} & Theoretical \& systems prim.\ (System) & \S5.2 & \texttt{T} & Dynamic oblivious search index hiding access patterns across insertions and deletions \\
49 & \texttt{Compass}~\cite{zhu2025compass} & Theoretical \& systems prim.\ (System) & \S5.2 & \texttt{T} & Encrypted high-accuracy semantic vector search primitive combining ORAM and oblivious traversal \\
50 & \texttt{RAG Unlearning}~\cite{wang2026ragunlearning} & Extended/adjacent mech.\ (Adjacent) & \S5.3 & \texttt{T} & Knowledge-base-driven behavioral unlearning in RAG without modifying backbone weights \\
51 & \texttt{Certified Data Removal}~\cite{guo2020certifiedremoval} & Theoretical \& systems prim.\ (Theory) & \S5.3, \S7.5, \S8.2 & \texttt{T} & Formal distributional indistinguishability reference between post-removal and retrained models \\
52 & \texttt{Verifiable Unlearning}~\cite{eisenhofer2025verifiableunlearning} & Theoretical \& systems prim.\ (Theory) & \S5.3, \S7.5, \S8.2 & \texttt{T} & Cryptographic execution proofs (SNARKs and hash chains) verifying unlearning computation \\
53 & \texttt{Deletion-as-Control}~\cite{cohen2023deletionascontrol} & Theoretical \& systems prim.\ (Theory) & \S5.3, \S7.5, \S8.2 & \texttt{T} & Formal control-based deletion semantics for ongoing interactive systems and derived state \\
54 & \texttt{FLOWSEAL}~\cite{shim2026controltheflow} & Extended/adjacent mech.\ (Adjacent) & \S5.2 & \texttt{T} & External information-flow lattice enforcement and controlled declassification for tool calls \\
55 & \texttt{PrivacyFlow}~\cite{staab2026privacyflow} & Extended/adjacent mech.\ (Adjacent) & \S5.2 & \texttt{T} & Pre-execution purpose, recipient inference, and consequence-aware information-flow risk control \\
56 & \texttt{Skill-V}~\cite{ma2026skillv} & Surveys \& state bg.\ (State bg.) & \S6.2, \S8.4 & --- & Versioned falsifiable skill contracts and historical replay; state non-regression background \\
57 & \texttt{Unfaithful Self-Evolvers}~\cite{zhao2026faithfulness} & Surveys \& state bg.\ (State bg.) & \S6.2, \S8.4 & --- & Causal evaluation testing whether distilled experiences actually govern downstream agent decisions \\
58 & \texttt{ExpeL}~\cite{zhao2024expel} & Surveys \& state bg.\ (State bg.) & \S6 & --- & Architectural precedent establishing distilled trajectory experiences as persistent agent state \\
59 & \texttt{AntiSkillBench}~\cite{xiang2026personaskills} & Risk \& evaluation study (Dual-coded) & \S1, \S6, \S8.5 & \texttt{T} & Benchmarks persona-skill leakage and tests sanitizing fixed dialogue traces before distillation (\S\S4.4, 5.3). \\
60 & \texttt{Memory Security Survey}~\cite{lin2026memorysecuritysurvey} & Surveys \& state bg.\ (Survey) & \S1, \S7.1, \S8.1 & --- & Six-stage LLM-agent long-term memory lifecycle security and governance survey \\
61 & \texttt{PrivacyLens}~\cite{shao2024privacylens} & Risk \& evaluation study & \S1, \S8.5 & --- & Evaluates gap between passive privacy-norm awareness and leakage in executable agent trajectories \\
62 & \texttt{FederatedSkill}~\cite{yang2026federatedskill} & Extended/adjacent mech.\ (Adjacent) & \S6.1 & \texttt{T} & Federated trajectory distillation, semantic diffing, and PII canary audit; not single-user loop \\
63 & \texttt{PANOLA}~\cite{ulusoy2022panola} & Representative mechanism & \S1, \S2.3, \S3, \S5.2 & \texttt{S} & Main Table~3: Learns user privacy preferences and negotiates outbound release; update path unprotected \\
64 & \texttt{Uncertainty-Aware PA}~\cite{ayci2023uncertaintypa} & Extended/adjacent mech.\ (Extended) & \S1, \S2.3, \S5.2 & \texttt{S} & Personalizes privacy labels and defers high-uncertainty disclosure choices to the user \\
65 & \texttt{AirGapAgent}~\cite{bagdasarian2024airgapagent} & Representative mechanism & \S2.3, \S3, \S5.2 & \texttt{S} & Main Table~3: Task-scoped context minimization prior to third-party conversational interaction \\
66 & \texttt{Privacy-Preserving BERT}~\cite{qu2021privatebert} & Extended/adjacent mech.\ (Extended) & \S4.2 & \texttt{T} & Local token/representation $d_{\chi}$-privacy with adaptive pretraining; non-longitudinal adjacency \\
67 & \texttt{Unlearning Leakage}~\cite{chen2021unlearningprivacy} & Risk \& evaluation study (Dual-coded) & \S5.3, \S7.5 & \texttt{T} & Primary role: version-differential membership leakage; includes label-only, scaling, and DP defenses \\
68 & \texttt{TAPE}~\cite{wang2025tape} & Theoretical \& systems prim.\ (System) & \S5.3, \S7.5 & \texttt{T} & Tailored posterior-difference auditing for multi-sample unlearning; empirical audit, not certificate \\
69 & \texttt{EFU}~\cite{mohammadi2025efu} & Extended/adjacent mech.\ (Extended) & \S5.3, \S7.5 & \texttt{T} & Functional encryption enforces authorized federated unlearning aggregation and hides request intent \\
70 & \texttt{Tracked Capabilities}~\cite{odersky2026trackedcapabilities} & Extended/adjacent mech.\ (Extended) & \S5.2, \S6.2 & \texttt{T} & Static capability typing, local purity, and derived labels constraining agent tool information flow \\
71 & \texttt{Backward-Secure SSE}~\cite{sun2018backwardsearch} & Extended/adjacent mech.\ (Extended) & \S5.2 & \texttt{T} & Puncturable-encryption backward security for dynamic encrypted indexes after entry deletion \\
72 & \texttt{Machine Unlearning Survey}~\cite{xu2024unlearnsurvey} & Surveys \& state bg.\ (Survey) & \S1 (Table~1), \S5.3 & --- & ACM CSUR taxonomy of exact/approximate unlearning, streaming deletion requests, and verification \\
73 & \texttt{SKIPPER}~\cite{xie2022skipper} & Risk \& evaluation study & \S6.2 & --- & Black-box compliance audit detecting policy-to-behavior discrepancies in virtual assistant skills \\
74 & \texttt{Privacy Norms for SPA}~\cite{abdi2021privacynorms} & Surveys \& state bg.\ (Norms) & \S5.2 & --- & Contextual-integrity characterization of smart-assistant recipients, data types, and norms \\
75 & \texttt{Collaborative Bandits DP}~\cite{wang2020collaborativebandits} & Extended/adjacent mech.\ (Extended) & \S4.2 & \texttt{T} & Central and local DP for online personalized collaborative bandits; single-reward adjacency \\
76 & Disclosure Begets Disclosure~\cite{kaneko2026disclosureloop} & Extended/adjacent mech. & \S5.1 & \texttt{D} & Write-time summary anonymization changes persistent personal memory in multi-session simulations. Response-side instruction is a separate empirical comparator. \\
77 & Algorithmic Self-Portrait~\cite{dash2026selfportrait} & Risk \& evaluation study & \S5.1, \S8.5 & \texttt{T} & Donated memories measure deployment risk. Attribution Shield is a query-reformulation prototype tested outside a longitudinal writeback loop. \\
78 & Agent-Memory Protocol~\cite{wu2026amp} & Surveys \& state bg. (Design) & \S5.1 & --- & Worked examples illustrate redaction, purpose-scoped context, and local restoration. They provide design context for storage and release. \\
79 & AfterDeleteBench~\cite{saleh2026afterdeletebench} & Risk \& evaluation study & \S5.3, \S7.5, \S8.2 & \texttt{D} & Behavioral audit with paired exposure controls. Tested closure retires or rebuilds derived memory under complete recorded lineage in synthetic single-user sessions. \\
80 & Bemula~\cite{xie2026bemula} & Risk \& evaluation study & \S5.3, \S8.5 & --- & Paired no-request controls separate failure to remember from forgetting and explicitly measure retention of unrelated facts. \\
81 & AUBench~\cite{lin2026aubenchcontext} & Risk \& evaluation study & \S5.3, \S8.5 & --- & Evaluates persistence across weights, external memory, and tools. Existing unlearning baselines serve as comparators. \\
82 & Pufferfish~\cite{kifer2014pufferfish} & Theoretical \& systems prim. & \S3, \S8.3 & \texttt{T} & Models protected secret pairs and plausible data-generating distributions for observable releases of correlated data. \\
83 & Time-Series Inferential Privacy~\cite{song2017timecomposition} & Theoretical \& systems prim. & \S4.4, \S8.3 & \texttt{T} & Separates general composition limitations from conditional Markov Quilt composition for related time series. \\
84 & Pufferfish Composition~\cite{bai2026pufferfishcomposition} & Theoretical \& systems prim. & \S4.4, \S8.3 & \texttt{T} & Characterizes when linear adaptive composition holds and constructs composable mechanisms under explicit influence conditions. \\
85 & Adaptive Machine Unlearning~\cite{gupta2021adaptiveunlearning} & Theoretical \& systems prim. & \S5.3, \S8.2 & \texttt{T} & Uses a DP-based reduction to extend deletion guarantees to requests that depend on previously released models. \\
86 & Forget Unlearning~\cite{chourasia2023trueprivacy} & Theoretical \& systems prim. & \S5.3, \S8.2 & \texttt{T} & Distinguishes deletion privacy over post-deletion model sequences from similarity to retraining under adaptive requests. \\
87 & \texttt{MemJoin}~\cite{yang2026memjoin} & Risk \& evaluation study (Dual-coded) & \S2.3, \S5.1, \S8.5 & \texttt{S} & Cross-session relation disclosure and empirical pre-reader cue removal on existing memory views. arXiv v1, \S\S3, 4.3--4.4, Tables~2--3 and App.~F.1. \\
88 & \texttt{CIMemories}~\cite{mireshghallah2026cimemories} & Risk \& evaluation study & \S5.2, \S8.5 & --- & ICLR 2026, \S\S4--5. Measures inappropriate attribute disclosure and task completeness across recipient--task contexts and repeated releases. \\
89 & \texttt{HUSH-Bench}~\cite{xu2026hushbench} & Risk \& evaluation study & \S5.2, \S8.5 & --- & arXiv v2, \S\S3.3--3.4, 4.4--4.5. Separates exact-target retrieval, unsolicited use, and use after invitation. \\
90 & \texttt{Memora}~\cite{uddin2026memora} & Risk \& evaluation study & \S5.3, \S8.2 & --- & Findings of ACL 2026. Forgetting-aware accuracy penalizes reliance on obsolete memory without testing erasure of its copies. \\
91 & \texttt{AIM}~\cite{johnson2026aim} & Extended/adjacent mech. & \S5.1 & \texttt{T} & arXiv v1, \S\S3.4, 3.7, 5. Visibility tagging and index-level access for a multi-user memory service. \\
92 & \texttt{PAHF}~\cite{liang2026pahf} & Surveys \& state bg. (State) & \S1 & --- & arXiv v1, \S\S3.2, 4.2. Clarification and feedback update per-user natural-language memory. Evaluation targets preference adaptation. \\
93 & \texttt{PeCL}~\cite{zhan2026pecl} & Extended/adjacent mech. & \S4.2 & \texttt{T} & EMNLP 2026, \S\S3.2--3.3 and App.~G. Token-embedding perturbation and retention control in continual task learning. \\
94 & MaRS/FiFA~\cite{alqithami2025mars} & Extended/adjacent mech. & \S5.1, \S8.1 & \texttt{T} & \S\S3.2--3.4 and 6.6.4: sensitivity-aware retention with provenance and simulated leakage scores. Optional DP tie-break has a local scope. \\
95 & Secure Forgetting~\cite{ye2026secureforgetting} & Extended/adjacent mech. & \S5.3 & \texttt{T} & Secs.~III--VI and VIII: state, trajectory and environment unlearning in public tasks with behavioral and trained-without comparisons. \\
96 & \texttt{CAMS}~\cite{dhivyasree2026cams} & Extended/adjacent mech. & \S5.1 & \texttt{T} & \S\S3.5.4 and 3.6, and experiments: trusted write screen and re-audit, plus extraction interception. Joint pipeline benchmark measures extraction defense. \\
97 & \texttt{P2Skill}~\cite{ryu2026p2skill} & Extended/adjacent mech. & \S6.1 & \texttt{T} & \S\S3.2--3.3 and 4: reusable redaction/routing skills learned offline. A cloud evaluator sees originals during refinement. \\
98 & \texttt{CounterPersona}~\cite{wang2026counterpersona} & Extended/adjacent mech. & \S6.1 & \texttt{T} & \S\S3--5: new controlled records reduce fidelity of unauthorized distillation from a fixed retained archive. \\
99 & \texttt{ConMem}~\cite{zhang2026conmem} & Extended/adjacent mech. & \S5.1 & \texttt{T} & \S\S3--5: counterfactual memory admission over individual reconstructed histories. Counts storage decisions, without carried-over writes. \\
100 & \texttt{FSFM}~\cite{gu2026fsfm} & Extended/adjacent mech. & \S5.1 & \texttt{T} & \S\S3--4: sensitive-record filtering and pruning in a capacity benchmark. Tests retained records, not post-deletion disclosure. \\
101 & \texttt{Profile-DP}~\cite{li2026profiledp} & Extended/adjacent mech. & \S5.2 & \texttt{T} & \S4.4 and App.~A: withheld-coordinate DP over a fixed profile with public release supports. Fresh releases incur additional privacy cost. \\
102 & \texttt{UMPeek}~\cite{li2026umpeek} & Risk \& evaluation study (Dual-coded) & \S5.2 & \texttt{S} & \S VII, Tables IV--V: stateful counterfactual release control evaluated on personalized task benchmarks. The deployment analysis separately evaluates inference risk. \\
103 & Online private KL-RLHF~\cite{wu2025onlineprivaterlhf} & Extended/adjacent mech. & \S4.3 & \texttt{T} & \S5 and Alg.~2: local preference-label randomization enters online contextual-bandit updates. The empirical runs in \S6 cover the offline algorithm. \\
104 & RAG Privacy Accounting~\cite{zhang2026ragaccounting} & Surveys \& state bg. (Survey) & \S1 (Table~1) & --- & \S4: compares protected granularity, accounting scope, and release boundaries under repeated retrieval. \\
105 & Agent Security and Privacy~\cite{he2026agentsecuritysurvey} & Surveys \& state bg. (Survey) & \S1 (Table~1) & --- & CSUR, \S\S3--5: organizes agent-workflow threats, defenses, and case studies. \\
106 & Federated Recommendations~\cite{slokom2026federatedrecommendations} & Extended/adjacent mech. & \S4.2 & --- & \S\S2.1 and 3: field deployment with clipping and noise; measured outcomes concern recommendation and control, not privacy. \\
107 & \texttt{MemGate}~\cite{zhang2026memgate} & Extended/adjacent mech. & \S5.2 & \texttt{S} & \S\S IV--V: query-conditioned gate on retrieved memories evaluated in scripted tasks across memory runtimes. \\
108 & Revoked but Still Authoritative~\cite{shen2026revokedmemory} & Risk \& evaluation study (Dual-coded) & \S5.3 & \texttt{T} & \S\S4--5 and App.~C.1: retrieval guard for superseded policies. Writeback experiments measure recurrence from derived entries. \\
109 & Persistent Memory Admission~\cite{liu2026memoryadmission} & Extended/adjacent mech. & \S5.2 & \texttt{S} & \S\S3--5: tests which retained entries enter working context and compares how the admitted content is presented. \\
110 & Audience-Bound Memory~\cite{liu2026audiencememory} & Extended/adjacent mech. & \S5.1, Supp. S6.2 & \texttt{D} & \S\S3.1--3.5 and 6.4: binds persisted items to audiences and constrains derivation in controlled multi-party histories for one owner. Evaluated context admission has its own \texttt{S} row. \\
111 & \texttt{ARIEL}~\cite{flemings2026ariel} & Extended/adjacent mech. & \S5.2 & \texttt{S} & \S\S3--6: predicts disclosure appropriateness from past user judgments in non-adversarial benchmark requests. The measured outcome is judgment agreement. \\
112 & \texttt{P2B}~\cite{malekzadeh2020p2b} & Extended/adjacent mech. & \S4.2 & \texttt{T} & \S\S2.1 and 3--5: samples and shares encoded bandit feedback through a trusted shuffler for a shared warm-start model. The analysis uses context-vector adjacency. \\
113 & Privacy Collapse~\cite{goel2026privacycollapse} & Risk \& evaluation study & \S4.2, Supp. S6.2 & \texttt{T} & Offline fine-tuning degrades contextual disclosure behavior. \S5.4 and App.~I test partial mitigations, not contribution privacy. \\
114 & Execution-State Unlearning~\cite{yao2026executionunlearning} & Extended/adjacent mech. & \S7.5 & \texttt{T} & Theorem~1 and \S5 cover replayable runtime state in controlled harnesses, not a complete single-user withdrawal lifecycle. \\
115 & Always-On Agents~\cite{ding2026alwayson} & Surveys \& state bg. & \S1, \S7.1, \S8.1 & --- & \S2.3, \S8 and \S9 organize durable state, governance, and evaluation of recovery. \\
116 & Task-Graph CI~\cite{wu2026tgci} & Extended/adjacent mech. & \S5.2 & \texttt{T} & \S\S4--5 test propagation-scoped release policies in inter-agent task graphs. \\
117 & AgentCIScope~\cite{guo2026agentciscope} & Risk \& evaluation study & \S8.5 & --- & \S\S3--4 measure executable cross-application disclosures; repeated runs do not test persistent adaptation. \\
118 & PiSAs~\cite{gupta2026pisas} & Risk \& evaluation study & \S5.2, \S8.5 & \texttt{T} & \S5.1 and Apps.~16 and 19.1 compare instructions and architectures in shared multi-user workflows. \\
119 & AgentCIBench~\cite{goel2026agentcibench} & Risk \& evaluation study & \S8.5 & --- & Archival EMNLP benchmark tests disclosures in executable personal applications, not protected state updates. \\
120 & Field-Tier Minimization~\cite{oswal2026fieldtier} & Extended/adjacent mech. & \S5.2 & \texttt{T} & \S\S3, 8--9 analyze and test projection of action parameters to the mediator and audit view. \\
121 & Claws in Plain Sight~\cite{dong2026claws} & Risk \& evaluation study & \S2.3 & \texttt{T} & \S7 tests a pre-execution monitor in deterministic synthetic replay against an inert sink. \\
122 & SkillJack~\cite{ying2026skilljack} & Risk \& evaluation study & \S6.2 & \texttt{T} & \S4.9 and Table~3 compare admission review and runtime blocking in controlled skill banks; outcomes are policy-violation proxies. \\
123 & SovereignPA-Bench~\cite{liu2026sovereignpa} & Risk \& evaluation study & \S8.5 & \texttt{S} & \S\S3 and~5 evaluate a bundled consent and privacy policy on fixed personal scenarios, not protected state transitions. \\
124 & PrivacyChecker~\cite{wang2025privacychecker} & Extended/adjacent mech. & \S5.2 & \texttt{S} & \S\S3--4 test contextual release mediation on PrivacyLens and executable tool workflows, not persistent updates. \\
125 & SecFPP~\cite{hou2025secfpp} & Extended/adjacent mech. & \S4.4 & \texttt{T} & V1, Secs. IV--V: protects prompts in federated clustering subject to disclosed distance leakage. \\
126 & FedSEPT~\cite{wang2026fedsept} & Extended/adjacent mech. & \S4.4 & \texttt{T} & V1, Theorem 4.1 and Sec. V: DP accounting for client releases over federated prompt rounds. \\
127 & Private fine-tuning audit~\cite{du2025privatefinetune} & Extended/adjacent mech. & \S4.2 & \texttt{T} & DBSec 2025: compares extraction and membership risk across DP fine-tuning settings. \\
128 & Rank-aware preference release~\cite{chen2026rankaware} & Extended/adjacent mech. & \S4.3 & \texttt{T} & V2, Secs. 4 and 6: privatizes one item relocation per ranking before offline alignment. \\
129 & TIPO~\cite{lin2026tipo} & Extended/adjacent mech. & \S4.2 & \texttt{T} & V1, Secs. 4--6: trajectory tuning and compliance for fixed privacy personas. \\
130 & Square preference optimization~\cite{zhou2025squarepo} & Extended/adjacent mech. & \S4.3 & \texttt{T} & ICML 2025: offline central full-sample and local label privacy under different assumptions. \\
131 & HomeLLaMA~\cite{huang2026homellama} & Extended/adjacent mech. & \S5.1 & \texttt{D} & IEEE TMC, Secs. IV-D, VI-B, VII-B: local profile writes and reuse in a lab study. \\
132 & Private personalized recommendation~\cite{fu2026privaterecs} & Extended/adjacent mech. & \S7.3 & \texttt{T} & M\&SOM: policies bound preference inference from the observed recommended item. \\
133 & OPPU~\cite{tan2024oppu} & Surveys \& state bg. & \S7.4 & --- & EMNLP 2024, Sec. 2.2 and Ethics: local per-user adapters; benchmarks evaluate personalization. \\
134 & SkillGuard~\cite{pan2026skillguard} & Extended/adjacent mech. & \S6.2, \S7.4 & \texttt{T} & V2, Secs. IV and VII--VIII: manifest and runtime operation checks on benchmark skills. \\
135 & SkillLifeBench~\cite{zhang2026skilllife} & Risk \& evaluation study & \S6.2, \S7.4 & \texttt{T} & V1, Secs. 3.6 and 4.4: compares recovery with execution provenance and checks useful outcomes. \\
136 & Separable Expert Architecture~\cite{schneider2026sea} & Extended/adjacent mech. & Supp. S6.3 & \texttt{T} & V1, Secs. 2.4 and 4.2: inference-time bypass of static synthetic personal adapters. \\
137 & PRISP~\cite{park2026prisp} & Extended/adjacent mech. & \S4.2 & \texttt{T} & ACL 2026, Secs. 3.2 and 4.2: few-shot per-user adapter isolation on fixed tasks. \\
138 & Zep Agent Skills~\cite{chalef2026zep} & Surveys \& state bg. & \S6.1 & --- & V1, Secs. 5.3--5.6 and 6: documented scanning, version admission and deletion; evaluation measures task outcomes. \\
139 & MemGov (ACL)~\cite{christou2026memgov} & Extended/adjacent mechanism & \S5.2, \S7.5 & \texttt{S} & AACL-IJCNLP, Secs.~3--4, Tables~2--3, Fig.~6: governed context selection and working-set repair. \\
140 & Open vs. closed private adaptation~\cite{hanke2024openprivate} & Extended/adjacent mechanism & \S4.1 & \texttt{T} & NeurIPS, Secs.~3--4: adaptation and inference provider views, utility and cost. \\
141 & RAG-CT~\cite{lyu2026ragct} & Extended/adjacent mechanism & \S5.2 & \texttt{T} & V1, Secs.~3.2 and 4: empirical read gate for PII in two generic RAG corpora. \\
142 & LaMP retrieval vs. PEFT~\cite{salemi2025ragpeft} & Surveys \& state bg. & \S7.6 & --- & ICTIR, Secs.~3--5: controlled per-user retrieval and parameter comparisons on LaMP; utility endpoints. \\
143 & HE LoRA~\cite{frery2025helora} & Extended/adjacent mechanism & Supp. S6.4 & \texttt{T} & V1, Secs.~4--6: encrypted delegated adapter training; convergence and system cost. \\
144 & GuardedTuning~\cite{shi2025guardedtuning} & Extended/adjacent mechanism & Supp. S6.4 & \texttt{T} & V3, Secs.~2--4: client/provider trade-offs for split and offsite tuning. \\
145 & Zero-knowledge personal advice~\cite{watanabe2025zkadvice} & Extended/adjacent mechanism & Supp. S6.4 & \texttt{T} & WWW Companion, Secs.~3--4: proof-backed trait disclosure and advice behavior in fixed scenarios. \\
146 & Quantized privacy-unit audit~\cite{zhang2026routingdp} & Risk \& evaluation & Supp. S6.4 & \texttt{T} & V1, Secs.~3--6: cross-record sensitivity inflation and record-isolating controls. \\
147 & Fixed-compute user DP~\cite{charles2025fixedcompute} & Extended/adjacent mechanism & \S4.2 & \texttt{T} & SaTML, cohort-user privacy accounting for example or user sampling and fixed-budget benchmarks. \\
148 & PriME~\cite{kim2025prime} & Extended/adjacent mechanism & \S4.2 & \texttt{T} & EMNLP, Secs.~3--4: adapter merging with explicit donor-risk objective and membership tests. \\
149 & CSLR~\cite{shihab2026cslr} & Extended/adjacent mechanism & \S4.4 & \texttt{T} & V1, Secs.~3, 5--6: client-level replay-release DP, theorem and cross-task evaluation. \\
150 & DP-Forward~\cite{du2023dpforward} & Extended/adjacent mechanism & Supp. S6.4 & \texttt{T} & CCS, Secs.~3--4: forward representation perturbation in fine-tuning and inference. \\
151 & Just Fine-tune Twice~\cite{shi2022justfinetune} & Extended/adjacent mechanism & Supp. S6.4 & \texttt{T} & EMNLP, Secs.~3--5: selective token privacy and an imperfect policy function. \\
152 & Continual participation privacy~\cite{chan2026participation} & Theoretical/systems primitive & Supp. S6.4 & \texttt{T} & V2, Secs.~3--5: certified buffering and adaptation under single-edit stream adjacency. \\
153 & TriShield~\cite{wei2026trishield} & Extended/adjacent mechanism & Supp. S6.4 & \texttt{T} & V2, threat model and results: malicious-server privacy backdoor mitigation in federated PEFT. \\
154 & Client-side personalized drafts~\cite{salemi2026p3} & Extended/adjacent mechanism & \S5.2 & \texttt{S} & V1, Secs.~3.2 and 4.2, Fig.~5: fixed profiles and inference about user attributes or identity. \\
155 & Private ordinal alignment~\cite{qi2026surveyalignment} & Extended/adjacent mechanism & Supp. S6.4 & \texttt{T} & V1, Secs.~3--5: ordinal questionnaire labels and three privacy mechanisms. \\
156 & DPPrefSyn~\cite{gao2026dpprefsyn} & Extended/adjacent mechanism & \S4.3 & \texttt{T} & ICML, Secs.~4--5: private reward estimation and portable synthetic comparisons. \\
157 & CIDER~\cite{guo2026cider} & Risk \& evaluation & Supp. S6.4 & --- & V2, Secs.~2.2 and 4.1--4.2: errors in predicting individualized disclosure judgments. \\
158 & PREMIA~\cite{feng2025premia} & Risk \& evaluation & Supp. S6.4 & --- & AISTATS, Secs.~3--4: preference membership inference after offline policy alignment. \\
159 & PEEP / PROFIT~\cite{ramirez2026peep} & Extended/adjacent mechanism & Supp. S6.4 & \texttt{T} & ACL Findings, Secs.~3--6: local rewriting of current queries under fixed disclosure profiles. \\
160 & Privacy-preference mixing~\cite{yang2026p3m} & Extended/adjacent mechanism & Supp. S6.4 & \texttt{T} & V1, Secs.~III--IV: data composition, memorization, membership and task metrics. \\
161 & PrivAct~\cite{cheng2026privact} & Extended/adjacent mechanism & \S4.3 & \texttt{T} & ICML, Secs.~3--4: shared contextual preference training and action disclosure. \\
162 & COLLABMEM~\cite{wu2026collabmem} & Extended/adjacent mechanism & Supp. S6.4 & \texttt{T} & EMNLP, Secs.~4 and 6: role-scoped graph memory in team workflows. \\
163 & PRISM~\cite{wang2026prism} & Extended/adjacent mechanism & \S4.4 & \texttt{T} & EMNLP, method and experiments: local private updates in federated continual tuning. \\
164 & TimlMemBench~\cite{zhou2026timlmembench} & Risk \& evaluation & \S5.3 & --- & EMNLP Findings, Secs.~4--5 and App.~B: native ingestion followed by diagnostic memory governance probes. \\
165 & SkillLadder~\cite{chen2026skillladder} & Extended/adjacent mechanism & \S6.1 & \texttt{T} & EMNLP Findings, Secs.~2.4, 2.6 and 3.3: compliance and utility of synchronized skills. \\
166 & K-Bench~\cite{yu2026kbench} & Risk \& evaluation & \S7.5 & \texttt{T} & V1, Secs.~3--5: tests forgetting across six observed channels with one source carrier per condition. \\
167 & TOP-Bench~\cite{qiao2026topbench} & Risk \& evaluation & \S7.3 & \texttt{T} & EMNLP Findings, Secs.~3--6: joint inference from tool returns and evaluated output interventions. \\
168 & CIPL~\cite{huang2026cipl} & Risk \& evaluation & \S7.3 & \texttt{T} & V1, Secs.~3--4 and App.~I: compares monitored tool paths and unmonitored output paths. \\
169 & Fine-tuning trade-offs~\cite{das2026finetuning} & Risk \& evaluation & \S7.6 & \texttt{T} & V2, Secs.~4--5: measures memorization, utility and efficiency under full tuning, LoRA and DP-SGD. \\
170 & AgentLeak~\cite{elyagoubi2026agentleak} & Risk \& evaluation & \S7.3 & \texttt{T} & IEEE Access, Secs.~VI-E/F and VII-C: final-output and internal-channel controls. \\
171 & VeilTune~\cite{xie2026veiltune} & Extended/adjacent mechanism & Supp. S6.4 & \texttt{T} & V1, Secs.~3.2 and 5.6: semi-honest split of data-owner and model-owner training views. \\
172 & PrivPair~\cite{jin2026privpair} & Extended/adjacent mechanism & Supp. S6.4 & \texttt{T} & V1, Secs.~III-B and VII-C--F: empirical leakage tests at a federated split interface. \\
173 & Private PEFT comparisons~\cite{ma2024privatepeft} & Risk \& evaluation & \S7.6, Supp. S6.4 & \texttt{T} & V1, Sec.~4: common-task comparison of resource costs and empirical membership inference. \\
174 & PrivTune~\cite{liu2026privtune} & Extended/adjacent mechanism & Supp. S6.4 & \texttt{T} & INFOCOM and v3, Secs.~IV--VI: privacy and utility of cloud-visible token representations. \\
175 & PACZero~\cite{ertan2026paczero} & Extended/adjacent mechanism & Supp. S6.4 & \texttt{T} & V3, Secs.~3--4: hidden-subset mutual information and tuning utility. \\
176 & JEPA-Prune~\cite{shaaban2026jepaprune} & Risk \& evaluation & Supp. S6.4 & \texttt{T} & V1, Secs.~5.1 and 7: targeted versus random pruning and remaining weight-side disclosure. \\
177 & pAirZero~\cite{cai2026pairzero} & Extended/adjacent mechanism & Supp. S6.4 & \texttt{T} & V1, Secs.~IV-C and VII-C: privacy under a wireless aggregate model and estimated device cost. \\
178 & HE-OFT~\cite{kanpak2026heoft} & Extended/adjacent mechanism & Supp. S6.4 & \texttt{T} & V1, Secs.~IV-F and V-E/F: encrypted fine-tuning and a measured label-only service interface. \\
179 & DP-DyLoRA~\cite{xu2026dpdylora} & Extended/adjacent mechanism & Supp. S6.4 & \texttt{T} & ICC and v4, Secs.~V--VII: synchronized public rank and DP aggregate with simulated noise. \\
180 & AEGIS~\cite{tao2026aegis} & Risk \& evaluation & Supp. S6.4 & \texttt{T} & V1, Secs.~III-A and VI-C/F: measured protection across components of a shared gradient. \\
\end{longtable}
\endgroup

\begin{table}[htbp]
\centering
\small
\setlength{\tabcolsep}{4.5pt}
\renewcommand{\arraystretch}{1.16}
\caption{Primary Roles and Closest Evidence Distance Across 180 Unique Sources}
\label{tab:supp-75-summary}
\begin{tabular}{@{}>{\raggedright\arraybackslash}p{0.42\linewidth}>{\raggedleft\arraybackslash}p{0.07\linewidth}>{\raggedleft\arraybackslash}p{0.07\linewidth}>{\raggedleft\arraybackslash}p{0.07\linewidth}>{\raggedleft\arraybackslash}p{0.12\linewidth}>{\raggedleft\arraybackslash}p{0.11\linewidth}@{}}
\toprule
\textbf{Primary Evidence Role Stratum (Main Table~2)} & \textbf{\texttt{D}} & \textbf{\texttt{S}} & \textbf{\texttt{T}} & \textbf{Non-mech.\ (\texttt{---})} & \textbf{Row Total} \\
\midrule
Representative protection mechanisms (Main Table~3) & 4 & 4 & 14 & 0 & \textbf{22} \\
Extended / adjacent mechanisms \& deployment evidence & 4 & 7 & 71 & 1 & \textbf{83} \\
Theoretical \& systems primitives & 0 & 0 & 16 & 0 & \textbf{16} \\
Risk \& evaluation studies & 1 & 3 & 20 & 16 & \textbf{40} \\
Surveys \& state background & 0 & 0 & 0 & 19 & \textbf{19} \\
\midrule
\textbf{Mechanism-Eligibility / Corpus Column Total} & \textbf{9} & \textbf{14} & \textbf{121} & \textbf{36} & \textbf{180} \\
\bottomrule
\end{tabular}
\end{table}

% Operator-Level Verification Ledger for the 22 Representative Mechanisms
% Self-contained section for CSUR Supplementary Material / Appendix (acmart compatible).
% Literature cutoff: 2026-11-10. This section audits the 22 representative mechanisms.

\section{Operator-Level Verification Ledger for the 22 Representative Mechanisms}
\label{sec:supp-22-mechanisms-ledger}

This section provides the detailed primary-source verification ledger for the 22 representative privacy mechanisms synthesized in the main text. The codes in Table~\ref{tab:supp-22-mechanisms-ledger} identify the focal operation for each exemplar. Section~\ref{sec:supp-operation-evidence} separates additional operations within those exemplars and applies the same coding across the mechanism-eligible corpus. Two orthogonal dimensions are recorded throughout:
\begin{enumerate}
  \item \textbf{Evidence distance (\texttt{D}/\texttt{S}/\texttt{T}).} \texttt{D} denotes a protected update to durable single-user state carried across decisions or interactions. \texttt{S} denotes a tested read or release of retained state in a personal-agent task. \texttt{T} denotes results transferable from another setting. Each code refers to the focal operation.
  \item \textbf{Guarantee class:} Differential privacy (DP), cryptographic or hardware-enforced confidentiality (MPC, TEE/ORAM), and empirical governance or behavioral gating are tracked separately so that scenario proximity (\texttt{D}/\texttt{S}/\texttt{T}) is never conflated with formal guarantee strength.
\end{enumerate}

Table~\ref{tab:supp-22-mechanisms-ledger} records, for each of the 22 representative mechanisms, its state carrier and coding, exact protection unit, private-input access boundary and trusted endpoints, constrained observable interface together with multi-round and withdrawal conditions, and verified primary-source locators (sections, definitions, theorems, propositions, algorithms, figures, and tables in the archival publication or pinned preprint). Where an archival paper or ACL/Nature-style manuscript uses unnumbered top-level headings (e.g.,~\cite{kim2026dpmemview,li2026healthjev}), locators cite the exact formal statements, named headings, figures, tables, and numbered appendices rather than imputed section numbers.

\paragraph{Representative selection and boundary checks.}
For prompt and parameter updates, DP-OPT, DP-ES, Private Selection, and user-level DP fine-tuning distinguish candidate generation, private scoring, selection over private candidates, and contribution clipping. PromptDPSGD and zeroth-order tuning remain in the extended set as alternative implementation routes. PrivMeSA provides an empirical cohort-trained risk-reward contrast. Decoupled reward modeling, DP Reward Estimation, and KL-Regularization DP distinguish tuple protection, label protection, and single-output sampling. PrivDPO and sparse reward estimation extend the label case. SecP-Tuning represents confidential execution, and Lifelong DP and Concurrent Composition separate a version-release protocol from composition across mechanisms.

For memory, MemPrivacy and HealthJev contrast representation substitution with writeback decisions, and SP-Mem is discussed as a variant. Opal represents dynamic confidential maintenance. Oblix and Compass provide alternative indexing trust models, and Compass supplies batched ORAM proof results used by Opal~\cite{mishra2018oblix,zhu2025compass,kaviani2026opal}. DP-MemView, Galaxy, PANOLA, and AirGapAgent distinguish attribute-limited reading, cloud masking, learned disclosure policies, and disclosure to a third party. The uncertainty-aware assistant extends the learned policy case. Reglos and SBU distinguish invalidation in a personal workflow from joint parameter and memory cleanup in a public-task evaluation. MaskClaw and FedSkillGit distinguish scenario-level skill filtering from federated publication. FederatedSkill extends the latter. These groupings explain the main table choices and leave the other studies available in Table~\ref{tab:supp-75-ledger} and the relevant sections.

Section~\ref{sec:supp-boundary-review} records the earlier boundary recheck and the Stage~4 decisions. AUBench~\cite{lin2026aubenchcontext} contributes to risk and evaluation counts. Xu et al.~\cite{xu2026dynamicgraphcontext} remains a contextual graph-model reference outside the corpus.

{\footnotesize
\setlength{\tabcolsep}{3.5pt}
\setlength{\emergencystretch}{1.2em}
\renewcommand{\arraystretch}{1.18}
\begin{longtable}{%
  >{\raggedright\arraybackslash}p{\dimexpr0.18\textwidth-2\tabcolsep\relax}%
  >{\raggedright\arraybackslash}p{\dimexpr0.16\textwidth-2\tabcolsep\relax}%
  >{\raggedright\arraybackslash}p{\dimexpr0.22\textwidth-2\tabcolsep\relax}%
  >{\raggedright\arraybackslash}p{\dimexpr0.26\textwidth-2\tabcolsep\relax}%
  >{\raggedright\arraybackslash}p{\dimexpr0.18\textwidth-2\tabcolsep\relax}%
}
\caption{Operator-level primary-source verification ledger for the 22 representative privacy mechanisms (\texttt{D} $=4$, \texttt{S} $=4$, \texttt{T} $=14$).}
\label{tab:supp-22-mechanisms-ledger}\\
\toprule
\textbf{Mechanism, Carrier \& Code} &
\textbf{Protection Unit} &
\textbf{Private Input Access \& Trusted Endpoint} &
\textbf{Constrained Interface, Multi-Round \& Withdrawal Semantics} &
\textbf{Primary-Source Locators} \\
\midrule
\endfirsthead

\caption[]{Operator-level primary-source verification ledger for the 22 representative privacy mechanisms (continued).}\\
\toprule
\textbf{Mechanism, Carrier \& Code} &
\textbf{Protection Unit} &
\textbf{Private Input Access \& Trusted Endpoint} &
\textbf{Constrained Interface, Multi-Round \& Withdrawal Semantics} &
\textbf{Primary-Source Locators} \\
\midrule
\endhead

\midrule
\multicolumn{5}{r}{\footnotesize\emph{Continued on next page}}\\
\endfoot

\bottomrule
\endlastfoot

DP-ES~\cite{liu2026dpes}\newline
\emph{Prompt policy}\newline
Code: \texttt{T}\newline
(\emph{Approx.\ DP}) &
Single replace-one (bounded) example record $(x,y)$ under uniform without-replacement minibatch subsampling ($q = b/n$); not user-level history. &
Trusted optimizer and scoring LLM endpoint read raw private $(x,y)$ pairs; initialization and mutation models are data-independent and do not access private records. &
Gaussian noise privatizes minibatch evaluation scores across the $K$-prompt candidate population; survivor selection, mutation, and released discrete prompt follow by post-processing against prompt observers. Bounded $T \times K$ scoring calls compose via RDP (forfeiting subsampling amplification $q=b/n$ if scored deterministically on a small recent-turn buffer); new private feedback requires fresh accounting; no withdrawal operator. &
\S3 (Def.~1, Alg.~1, Props.~1--2); \S6 (Thm.~3); \S5 Table~4; App.~B.1, B.4; App.~C.1 Tables~6--7; App.~D.3. \\
\addlinespace[3pt]

DP-OPT~\cite{hong2024dpopt}\newline
\emph{Prompt policy}\newline
Code: \texttt{T}\newline
(\emph{Approx.\ DP}) &
Single add/remove record in disjoint private training and validation splits. &
Trusted local ensemble generator reads private training records; trusted local validator reads disjoint private validation records; cloud LLM is untrusted. &
Local ensemble voting (RDP/approx.-RDP) and \texttt{LimitedDomain} token selection (per-token truncation failure $\delta_0$, accumulating $m\delta_0$ across $m$ tokens) generate candidates, and exponential mechanism selects one discrete prompt released to the cloud LLM. Parallel composition relies on disjoint splits at the target intra-user protection unit (recurring intra-user facts across record-disjoint splits revert to sequential composition); cross-session reuse requires re-accounting; no withdrawal operator. &
\S3 (Def.~3.1); \S4.2 (Algs.~1--2); \S5 Table~3; App.~A.1 (Alg.~3), A.2--A.2.2 (Defs.~A.1--A.4, Alg.~4, Thms.~A.1--A.4), A.3. \\
\addlinespace[3pt]

Private Selection~\cite{liu2019privateselection}\newline
\emph{Candidate selection}\newline
Code: \texttt{T}\newline
(\emph{Pure/Approx.\ DP}) &
Single-person dataset entry shared across base DP candidate mechanisms. &
Central meta-selector and candidate oracles access private dataset; each base oracle must already satisfy $(\varepsilon_1,0)$- or $(\varepsilon_1,\delta_1)$-DP under identical adjacency. &
Random-stopping meta-selection releases one winning candidate/score to output observers at a constant multiple ($2\varepsilon_1$ or $3\varepsilon_1$) of pure $(\varepsilon_1,0)$-DP base cost without requiring low-sensitivity quality scores; for approximate $(\varepsilon_1,\delta_1)$-DP base oracles ($\delta_1 > 0$, Thm.~3.1), failure probability $\delta$ scales with the expected stopping horizon ($\propto (\delta_1/\gamma)\log(1/\delta_2)$) alongside additive $\varepsilon$ overhead. Covers adaptive oracle draws within one run; repeated cross-session selections compose; no withdrawal operator. &
\S1 (Def.~1.1, Thms.~1.2--1.5, Applications); \S3 (Alg.~1, Thm.~3.1). \\
\addlinespace[3pt]

User-Level DP Fine-Tuning~\cite{chua2024privacyunit}\newline
\emph{Parameters (LoRA)}\newline
Code: \texttt{T}\newline
(\emph{User-level Approx.\ DP}) &
All records contributed by one user in a multi-user training corpus. &
Trusted central trainer reads raw multi-record user histories and either enforces a $k$-record user cap under example-level DP-SGD with Group Privacy conversion (\texttt{Alg.~1}) or aggregates per-user updates, clips at the user level, and injects Gaussian noise (\texttt{User-wise DP-SGD}, \texttt{Algs.~2--3}). &
Bounds single-user participation leakage in released LoRA weights (\texttt{Alg.~1} Group Privacy incurs $(\varepsilon,\delta)$ group-conversion $\delta$ inflation unless accounted directly in RDP/zCDP or clipped at the user level via \texttt{Algs.~2--3}). Training steps compose under user-level accounting (authors note implementation uses shuffled epochs rather than Poisson subsampling assumed by accountant); does not define intra-user fact adjacency or unlearning. &
\S2 (Group Privacy \texttt{Alg.~1}; User-wise DP-SGD \texttt{Algs.~2--3}); \S4--\S6 (Table~8); App.~B.2--B.3; App.~C.4 Table~12. \\
\addlinespace[3pt]

PrivMeSA~\cite{wang2026privmesa}\newline
\emph{Shared LoRA and lesson memory}\newline
Code: \texttt{T}\newline
(\emph{Empirical risk reward}) &
Outbound consultation details and quasi-identifier linkage across visits (non-DP). &
The trusted local clinical system holds raw records and the shared policy and store. A fixed distiller sees completed exchanges. Remote specialists observe messages the local gate sends. &
Offline GRPO trains one shared policy across 1{,}700 patients. A separate distiller writes general lessons. At test time the fixed policy adds lessons across 897 encounters from 200 held-out patients. Disclosure and registry-based linkage measure the outbound view. No per-user adapter or per-user lesson update is tested. &
\S\S3.1--3.4; \S4.1 (patient splits and test-time learning); App.~B, C.1, and E.1 (Fig.~4). \\
\addlinespace[3pt]

Decoupled DP Reward~\cite{cho2026decoupledreward}\newline
\emph{Evaluator / preference}\newline
Code: \texttt{T}\newline
(\emph{Tuple Approx.\ DP}) &
One full preference tuple $(x, y_1, y_2,\allowbreak \text{pref})$. &
Trusted central trainer reads raw preference tuples to train a lightweight DP reward head; both the fixed public candidate generator $\pi_0$ and the reward head's frozen representation backbone $\phi(x,a)$ are strictly independent of the private preference dataset. &
Releases DP reward head; downstream candidate generation and Best-of-$N$ selection incur zero extra privacy cost by post-processing provided $\pi_0$, $\phi$, and inference never re-access raw preferences. Does not protect unseen private deployment prompts, continual retraining without re-accounting, or tuple withdrawal. &
\S3.2 (Alg.~1, Prop.~6); \S4 (Thms.~9--10); \S5.2 Tables~1--2; App.~B.1, B.3--B.4, B.7 (Table~4, Thm.~11). \\
\addlinespace[3pt]

DP Reward Estimation~\cite{raychowdhury2024privatereward}\newline
\emph{Evaluator / preference}\newline
Code: \texttt{T}\newline
(\emph{Central\slash local label-DP}) &
Single pairwise preference label only; prompts\slash contexts and action pairs are assumed public. &
Central mode trusts the estimator with raw labels and applies regularized objective perturbation for $(\varepsilon,\delta)$-central label-DP; local mode applies randomized response at the user endpoint before label transmission for pure $\varepsilon$-local label-DP. &
Bounds single-label influence on released linear BTL reward parameters against output observers (and against the server in local mode). Leaves sensitive prompt/response text unprotected; repeated label releases compose across rounds; no withdrawal operator. &
\S2 (Defs.~2.2--2.3, Rem.~2.4); \S3 (Thms.~3.1--3.2); \S4 (Thms.~4.1--4.2, 4.5); \S5 (Thms.~5.1--5.3, Cor.~5.4); \S6. \\
\addlinespace[3pt]

KL-Regularization DP~\cite{zhang2025klregularization}\newline
\emph{Preference output}\newline
Code: \texttt{T}\newline
(\emph{Single-action label-DP}) &
Main theorem: single preference label (contexts/actions public; appendix also analyzes full-tuple add/remove). &
Trusted trainer computes pessimistic BTL MLE and KL-regularized stochastic policy from raw feedback under linear-reward, boundedness, and feature-coverage assumptions. &
Intrinsic policy stochasticity yields pure $(\varepsilon,0)$-label-DP (Thm.~3) or approximate $(\varepsilon,\delta)$-label-DP under high-probability coverage bounds (Cor.~3) for \emph{one} sampled action on a fixed prompt without explicit noise injection, degrading under imperfect policy optimization via bidirectional $D_\infty$ error (Prop.~3). Policy parameters themselves are not privatized; $T$ action samplings compose linearly/adaptively; sequential online RLHF and withdrawal remain open. &
\S3.2--\S3.3 (Defs.~1--3); \S4.3 (Thm.~3, Cor.~3); \S4.4 (Props.~2--3); \S5. \\
\addlinespace[3pt]

SecP-Tuning~\cite{luo2026secptuning}\newline
\emph{Soft prompt}\newline
Code: \texttt{T}\newline
(\emph{2PC/MPC confidentiality}) &
Cryptographic confidentiality of input tokens and soft-prompt shares (no DP adjacency unit). &
Two semi-honest, non-colluding online servers execute secret-shared forward passes; a trusted third-party dealer $T$ supplies offline correlated randomness (collusion of $T$ with either server breaks online input secrecy); data owner locally holds labels, plaintext loss, and CMA-ES prompt updates. &
Hides inputs and intermediate activations from servers during forward-only tuning relative to an approximated fixed-point Random Feature Attention ($\Pi_{\text{RFA}}$) circuit rather than exact Softmax attention; reconstructs up to $8{,}000$ plaintext forward logit outputs to the data owner during CMA-ES tuning. Does not bound statistical leakage from allowed outputs/answers; composes per protocol execution under non-collusion; no withdrawal operator. &
\S4.1--\S4.3; \S5 Tables~1--3; App.~\S6.4--\S6.5 (Algs.~1--2, Security Analysis, Table~5). \\
\addlinespace[3pt]

Lifelong DP~\cite{lai2022lifelongdp}\newline
\emph{Continual parameters}\newline
Code: \texttt{T}\newline
(\emph{Tuple-level Lifelong DP}) &
Tuple-level neighboring task and episodic-memory datasets (Def.~2). The protocol uses fixed, disjoint task datasets or batches for episodic reuse. &
Trusted central learner and episodic-memory manager access raw task streams and enforce strict partition non-overlap and prescribed memory reuse. &
Objective/gradient perturbation releases a sequence of model checkpoints under a constant tuple-level pure $\varepsilon$-Lifelong-DP ($\delta = 0$, Thm.~2) bound against observers of all versions. Correlated tuples remain within the guarantee when the same tuple adjacency and protocol apply. Multi-tuple fact changes and extra observation paths require separate analysis. &
\S3 (Defs.~2--3, Alg.~1); \S4 (Thm.~2); \S5 (Alg.~2); \S6. \\
\addlinespace[3pt]

Concurrent Composition~\cite{henzinger2026concurrentcomposition}\newline
\emph{Multi-mechanism composition}\newline
Code: \texttt{T}\newline
(\emph{DP composition framework}) &
Event-level or user-level stream adjacency specified via an explicit verification function. &
Meta-framework over stateful continual DP mechanisms; requires each component mechanism to satisfy DP independently under declared verification semantics. &
Bounds privacy of the full interactive transcript under adaptive interleaving of dataset updates and queries. Proves naive parallel composition fails for approximate $(\varepsilon,\delta)$-DP ($\delta>0$) under adaptive updates while giving tight replacement bounds and privacy filters (and recovering parallel composition for pure DP, Cor.~5.5); does not privatize non-DP modules. &
\S3 (Def.~3.7); \S4 (Thm.~4.12); \S5 (Thms.~5.2, 5.4, Cor.~5.5); \S7 (Thm.~7.6); \S8. \\
\addlinespace[3pt]

MemPrivacy~\cite{chen2026memprivacy}\newline
\emph{Memory (writeback)}\newline
Code: \texttt{D}\newline
(\emph{Conditional system isolation}) &
Policy-classified sensitive spans (PL2--PL4 raw values); non-DP. &
Trusted edge device runs an offline-trained span detector, replaces sensitive spans with typed placeholders, stores persistent value--placeholder mappings locally, and restores values locally after cloud generation. &
Cloud memory services (LangMem, Mem0, Memobase) store and retrieve only placeholder-sanitized text across sessions. Unrecognized spans, allowed typed structure, and cross-placeholder inferences remain visible to cloud; no formal DP bound or downstream derived-state deletion. &
\S3; \S4.1 (Alg.~1); \S4.4; \S5.1--\S5.3 (Tables~2--4); App.~D.3--D.4 (Tables~6--7). \\
\addlinespace[3pt]

DP-MemView~\cite{kim2026dpmemview}\newline
\emph{Memory (read/release)}\newline
Code: \texttt{S}\newline
(\emph{Per-attribute pure DP}) &
Attribute-group replacement adjacency across fixed personal memories implying attribute $a$. &
Trusted memory interface holds raw fixed memories, private selection scores, and per-attribute ledgers; downstream response LLM and querying app receive only selected public views. &
The exponential mechanism selects from a fixed public view vocabulary, charges every intersecting attribute ledger under complete policy indexing, and emits a fixed fallback before exceeding $B_a$. Theorem~1 gives pure $B_a$-DP for full adaptive transcripts when all changed slots belong to $I_a$ and the interface obeys \texttt{C1--C6}. If other charged groups cover all changed slots, Corollary~1 gives the minimum-cover bound $\kappa(M,M')$. Dynamic writeback, slot mutation, and unlearning lie outside the fixed-layout guarantee. &
Problem Definition \& DP-MemView (Def.~1 [C1--C6], Thm.~1, Cor.~1); Experiments Tables~2--4; App.~A (Alg.~1); App.~B.1--B.5 (Lemma~1, Prop.~1); App.~C.5. \\
\addlinespace[3pt]

Galaxy~\cite{bao2025galaxy}\newline
\emph{Memory / context}\newline
Code: \texttt{S}\newline
(\emph{Empirical release gating}) &
Context spans graded L1--L4 prior to external cloud LLM transmission (non-DP). &
Trusted local runtime hosts KoRa agent, Persona cognition tree, Kernel meta-agent, and LLM-based Privacy Gate; external cloud LLM is untrusted. &
Privacy Gate applies L1--L4 graded masking before cloud calls and selective demasking on return (reducing PrivacyLens leakage from $50.5\%$ to $18.5\%$ in Kernel ablation). Paper does not prove all internal Persona/Kernel evolution calls traverse Gate or isolate Gate from full Kernel ablation; no unlearning operator. &
\S3 (Persona, Kernel, Privacy Gate, Fig.~4); \S4 Table~1 (PrivacyLens evaluation \& ablation). \\
\addlinespace[3pt]

Opal~\cite{kaviani2026opal}\newline
\emph{Memory (maintenance)}\newline
Code: \texttt{D}\newline
(\emph{TEE/ORAM confidentiality}) &
Request sequences sharing identical declared leakage traces over equal-sized memory worlds (non-DP). &
The trusted logical client sends root key $k$ and durable counter $\mathsf{ctr}$ with each \texttt{Query}/\texttt{Ingest} over an authenticated encrypted channel. The attested controller checks ORAM roots and the counter-bound MAC and coordinates trusted embedding and LLM enclaves. &
Fixed-budget storage traces cover ingestion and piggybacked ``oblivious dreaming'' up to declared leakage (operation type, trace shape, checkpoint boundaries). The prototype uses disk-backed Ring ORAM and IVF-PQ. The proof imports Compass's batched-access trace privacy and authenticated-storage integrity. Sealed state freshness ends at the latest checkpoint. User-requested derived-summary erasure is a separate operation. &
\S2.1--2.2 (Thm.~2.1), \S3.3--3.4, \S5, \S6.1 and 6.7, App.~A. Compass~\cite{zhu2025compass}, \S2.2 and App.~C. \\
\addlinespace[3pt]

HealthJev~\cite{li2026healthjev}\newline
\emph{Memory (writeback)}\newline
Code: \texttt{D}\newline
(\emph{Empirical governance}) &
Semantically identified privacy-sensitive health interactions and layered memory items (non-DP). &
Trusted agent runtime, layered memory store (L2 profile, L3 procedures, L4 episodic traces), and induction/governance modules read raw user interactions. &
Selectively retains, revises, or excludes privacy-sensitive interactions prior to persistent writeback and supports layer-wise withholding during retrieval across 90-day simulated longitudinal profiles. Empirical probe compliance only (JevGate filters faulty lessons rather than acting as a formal privacy mechanism); no formal deletion proof. &
Main text (Selective Memory Updates, Fig.~1; Longitudinal Benchmark, Fig.~2; JevGate Results; Discussion on writeback exclusion). \\
\addlinespace[3pt]

Reglos~\cite{yang2026reglos}\newline
\emph{Memory \& harness policy}\newline
Code: \texttt{D}\newline
(\emph{Behavioral forgetting \& control}) &
Rule-matched sensitive identifiers (keys, card/ID numbers), individual facts, and user forget requests (non-DP). &
Application runtime, task/extraction LLMs, append-only fact store, and outer-loop coding-agent proposer/critics/evaluator all reside inside the trusted system boundary. &
The credited result pairs a tombstone update with a reader policy and tests behavioral non-use on PersonaMem-v2. The append-only log retains invalidated facts by default. Pre-write masking and manual deletion are documented separately. &
\S3.1, \S4.1 (Fig.~5), \S4.3 (forgetting case), Apps.~A and C (Table~8). \\
\addlinespace[3pt]

SBU~\cite{wang2026agenticunlearning}\newline
\emph{Parameters \& memory}\newline
Code: \texttt{T}\newline
(\emph{Empirical joint unlearning}) &
Targeted forget set $\mathcal{D}_F$, exclusive derived memory artifacts, and coupled parametric traces (non-DP). &
Trusted unlearning executor holds provenance/dependency graph, reference counts, retrieval blocklist, and model weights. &
Blocks retrieval via immediate blocklist, prunes exclusive provenance closure ($\mathcal{C}\leftarrow\text{Dep}(\mathcal{D}_F)$) while logically invalidating shared multi-source artifacts, and suppresses parametric backflow via KL alignment to a high-entropy reference prior ($100\%\to 0\%$ forget hit rate in Store--Query--Delete--Probe). Validated on public medical QA rather than longitudinal single-user histories; empirical MIA/QA metrics, not certified removal. &
\S3.2 (Alg.~1); \S4 (Table~1, Fig.~6: Store--Query--Delete--Probe agent loop). \\
\addlinespace[3pt]

MaskClaw~\cite{zhao2026maskclaw}\newline
\emph{Candidate filtering}\newline
Code: \texttt{T}\newline
(\emph{Empirical behavior tests}) &
Candidate text skills and downstream actions in designed privacy scenarios (non-DP). &
A local controller uses fixed scenario corrections. The sandbox uses an external multimodal judge (App.~K, Table~28), adding a separate observation boundary. &
E3 reports optimization iterations, held-out behavior and sandbox screening in five fixed scenarios. It does not trace a stored skill across a later user interaction. The separate 832-case static evaluation does not test updates. &
v2, \S4.5 (design), \S5.5 and \S\S6.3--6.4 (E3), Apps.~E--F (Tables~18--20), and App.~K (Table~28). \\
\addlinespace[3pt]

FedSkillGit~\cite{lan2026fedskillgit}\newline
\emph{Shared skill}\newline
Code: \texttt{T}\newline
(\emph{Empirical payload filtering}) &
Fixed private-value set entries and free-text spans inside candidate federated skill commits (non-DP). &
Trusted client runs local typed-placeholder sanitizer and private-value auditor before upload; central version-control server is untrusted for raw private values (no secure aggregation). &
Blocks skill patches/summaries failing local private-value audit and retries sanitization prior to federated version merge ($33\%\to 81\%$ first-pass audit rate). Evolves a shared multi-client skill rather than a single-user private library; server observes committer identity, timing, and accepted-revision sequence; no formal miss-rate bound or commit revocation proof. &
\S III-A--D; \S III-C.2 (Eqs.~5--7); \S IV Tables~I, III; App.~A-D, A-E, B. \\
\addlinespace[3pt]

PANOLA~\cite{ulusoy2022panola}\newline
\emph{Privacy policy}\newline
Code: \texttt{S}\newline
(\emph{Empirical release governance}) &
Co-owned OSN content items subject to multi-party sharing conflicts (non-DP). &
Trusted personal assistant observes raw user sharing feedback, maintains internal RL confidence/effectiveness tables, and submits integer bids to the PANO auction mechanism. &
Pre-release OSN sharing negotiation via feedback-learned bidding policy in Clarke-Tax auctions (\emph{privacy-aware evolution}). Protects outbound sharing decisions in multi-agent OSN simulations, but does not privatize the RL preference table, user feedback stream, or auction bid transcript (\emph{not privacy-protected evolution}); no withdrawal operator. &
\S2 (PANO Clarke-Tax auction); \S3 (Alg.~1: RL confidence update \& bid selection); \S4 (OSN persona simulations). \\
\addlinespace[3pt]

AirGapAgent~\cite{bagdasarian2024airgapagent}\newline
\emph{Context release}\newline
Code: \texttt{S}\newline
(\emph{Empirical context minimization}) &
Task-irrelevant personal profile fields under contextual-integrity norms (non-DP). &
Trusted local context-guided data minimizer reads full user profile and stated task; third-party-facing conversational agent and external counterparties are untrusted against context hijacking. &
Restricts execution agent's context window to minimizer-approved fields and escalates scope shifts to explicit user confirmation ($97\%$ protection vs.\ $45\%$ unmitigated baseline under single-query context hijacking). Protects session release over static synthetic profiles without persistent self-evolving minimizer updates, formal multi-round bounds, or state unlearning. &
\S3 (Contextual Integrity threat model); \S4 (Minimizer architecture, Fig.~1); \S5 (Fig.~2, Table~1). \\

\end{longtable}
}

\subsection{Targeted Recheck of Screening and Evidence-Distance Decisions}
\label{sec:supp-boundary-review}

\paragraph{Scope and record.}
On 18 November 2026 we rechecked seven records from Table~\ref{tab:supp-75-ledger} and one excluded candidate from Table~S4. We selected cases that could plausibly change category if a reader coded the entire architecture, treated any long-running loop as personal, or conflated the paper's primary role with its eligibility. We used the archival PANOLA text, the revision-pinned versions indicated below, and the full-text revision of the Tobaben et al.\ journal paper. Each entry gives the decision in the frozen ledger or screening log, a plausible competing reading, the primary-source locator, and the ruling. The existing classifications were visible during this single-pass recheck, which used AI assistance to locate and read the cited source passages. Competing readings are decision checks. They are not votes from separate coders. The earlier FedSkillGit classification in BR6 is an exception in that the draft history records an actual change of code.

\paragraph{Adjudication rule.}
We first checked whether the paper specifies its own protection step, distinguishing benchmark-only work from studies that also contribute a mitigation. We then located the protected operation and the evaluation setting independently of the guarantee type. The direct code \texttt{D} requires evidence that protection covers a persistent update in a workflow serving one user over time. Prototype and simulated histories qualify with the evaluation setting noted. The surface code \texttt{S} applies when the personal workflow's reported protection concerns reading or disclosure. The transfer code \texttt{T} applies to eligible operators evaluated outside those target settings. When an architecture supports multiple operations, an operation can support its own code only where the paper describes and evaluates that protection. Where a stronger reading lacks direct evidence, we retain the code supported by the documented operator and setting.

\begin{description}
\item[BR1. PANOLA (source~\#63).] Baseline: representative mechanism, \texttt{S}. Candidate alternative: \texttt{D}, since the bidding policy persists and learns. In the archival paper, \S3.1 describes learning from auction outcomes and \S4.1--4.3 evaluate the sharing decisions of simulated personas. The reported protection concerns release decisions for jointly owned content. Those evaluations provide no protection evidence for policy updates or their feedback. \emph{Ruling:} retain \texttt{S} for empirical release governance.~\cite{ulusoy2022panola}

\item[BR2. Galaxy (S4:R02, source~\#25).] Baseline: representative mechanism, \texttt{S}. Candidate alternative: \texttt{D}, since the Persona and Kernel evolve. In arXiv:2508.03991v1, \S3 and Fig.~4 place masking before cloud transmission. The comparison in \S4, Table~1 removes the entire Kernel, so it cannot isolate protection of the internal update calls. \emph{Ruling:} retain \texttt{S} for context release in a personal agent and limit the empirical claim to the evaluated configuration.~\cite{bao2025galaxy}

\item[BR3. Opal (source~\#9).] Baseline: representative mechanism, \texttt{D}. Candidate alternatives: \texttt{S} if treated as private retrieval, or \texttt{T} if only live-user deployment qualified. In arXiv:2604.02522v1, \S2.1--2.2 and \S3.3--3.4 cover ingestion and maintenance under a stated host leakage model. Sections~4 and~6 describe a synthetic longitudinal personal workload and prototype evaluation. \emph{Ruling:} retain \texttt{D} for confidential maintenance. Record synthetic-workload evidence separately from the conditional TEE/ORAM guarantee.~\cite{kaviani2026opal}

\item[BR4. SP-Mem (source~\#39).] Baseline: extended mechanism, \texttt{D}. Candidate alternative: \texttt{S} if only the consent gate is considered. In arXiv:2608.16551v1, \S2.1--2.2 and Appendix~D.1, Algorithm~3 specify sanitization before writing searchable memory and a separate store for exact values. Section~4.2 evaluates the writing step on constructed histories. Section~2.3 and Algorithm~4 cover consent at retrieval. \emph{Ruling:} retain \texttt{D} on the basis of write-time protection. The study remains an extended example of representation and storage separation, with consent-based retrieval discussed separately.~\cite{wang2026spmem}

\item[BR5. SBU (S4:R01, source~\#24).] Baseline: representative mechanism, \texttt{T}. Candidate alternative: \texttt{D}, since the withdrawal operator modifies both memory and parameters. In arXiv:2602.17692v2, \S3.2 and Algorithm~1 specify dual-path cleanup. Section~4.1 and the Ethical Statement identify public medical question-answering data. The reported setting does not follow the same user's personal history across sessions. \emph{Ruling:} retain \texttt{T} for transferable empirical unlearning. Its closed loop does not establish target-setting evidence.~\cite{wang2026agenticunlearning}

\item[BR6. FedSkillGit (source~\#17).] Baseline: representative mechanism, \texttt{T}. An earlier draft considered \texttt{D} because clients modify persistent skills across rounds. In arXiv:2610.21348v1, \S III-A--C locates privacy filtering on transmitted client commits. The server evolves a shared skill, and \S IV-C.2 and Table~III measure first-pass payload audit performance. \emph{Ruling:} confirm the earlier correction to \texttt{T}. The protected operation serves a federated shared skill, and the ablation supports a payload-filtering claim.~\cite{lan2026fedskillgit}

\item[BR7. Tobaben et al.\ (source~\#36).] Baseline: risk and evaluation primary role, with mechanism code \texttt{T}. Candidate alternative: risk only, since the paper exposes an output-label side channel. In arXiv:2411.04680v6, \S4.1 describes that side channel, \S4.2--4.3 specify private label release and a data-independent label space, and \S6 evaluates continual classifiers. \emph{Ruling:} retain the risk role for the side-channel finding and \texttt{T} for the transferable mitigation. The source contributes once to the corpus total.~\cite{tobaben2026labelspace}

\item[BR8. AUBench (S4:R03).] Initial decision: \texttt{E4} for mechanism coding, with a contextual citation. Candidate alternative: \texttt{T} because the evaluation covers coordinated memory and parameter forgetting. In arXiv:2610.17659v1, \S4.1--4.2 define a benchmark with controlled facts across weights, memory, and tools. The experiments and Appendix~A.1 compare unlearning methods within that benchmark. \emph{Ruling:} retain \texttt{E4} because the contribution used here is the evaluation protocol and its tests. Stage~4 retains \texttt{E4} for mechanisms and includes AUBench in the risk and evaluation stratum (\#81 in Table~\ref{tab:supp-75-ledger}).~\cite{lin2026aubenchcontext}
\end{description}

\paragraph{Outcome.}
The eight decisions retain the original operator-distance rulings. The Stage~4 recheck changes AUBench's corpus inclusion while preserving its mechanism disposition. It adds two dual-coded risk studies, because they contribute identifiable interventions in addition to evaluation evidence. AfterDeleteBench modifies persistent state in a controlled, synthetic loop and receives \texttt{D}. Attribution Shield is evaluated on query reformulation and a memory-extraction proxy and receives \texttt{T}. The Stage~5 follow-up adds MemJoin as a dual-coded risk study with \texttt{S} for its tested read-view intervention. The other additions and their boundaries appear in Section~\ref{sec:supp-targeted-search} and Table~\ref{tab:supp-75-ledger}. These decision checks establish source-level traceability. They do not estimate independent coding agreement.

\paragraph{Stage 7 operator rulings.}
PrivMeSA moves from \texttt{D} to \texttt{T}. The parameter intervention trains shared LoRA offline (\S3.3 and \S4.1); persistent lesson writes use a separate frozen distiller (\S3.4 and App.~C.1); and the release evaluation measures outgoing messages and cross-visit linkage (\S4.1 and App.~B). The test-time setting holds weights fixed and adds reusable lessons across patients (App.~E.1). Each is \texttt{T} in this review's single-user scope. The primary role remains representative, counted once~\cite{wang2026privmesa}.

Stage~7 provisionally credited MaskClaw's controlled admission as \texttt{D}. The Stage~10 source check changes that ruling. ConMem remains \texttt{T}: it evaluates candidate decisions from reconstructed histories without carrying the resulting memory across successive writes~\cite{zhang2026conmem}.

\paragraph{Stage 10 adjudication: MaskClaw, arXiv v2.}
Section~4.5 and Eq.~(9) specify \texttt{AuditMerge} into local state. Section~5.5 and App.~E (Tables~18--19) document curated corrections, optimizer iterations, and held-out skill-use tests in five scenarios; the arbitration-rule store is fixed. Section~6.4 and App.~F (Table~20) measure candidate-level pass rates under the text and sandbox gates. App.~K (Table~28) identifies the sandbox judge as an external multimodal endpoint. These are positive results for candidate behavior and screening. They do not identify a committed version loaded and used in a subsequent interaction by the same user. An optimization iteration is not such an interaction. \emph{Ruling:} change E3 from \texttt{D}/L to \texttt{T}/B. Keep the separate E1/E2 screen-disclosure tests at \texttt{T}/B: they test static GUI observations, not the release of retained personal state after adaptation. No \texttt{D} credit is assigned to either evaluation~\cite{zhao2026maskclaw}.

\subsection{Direct Evidence and Representative-Set Sensitivity}
\label{sec:supp-representative-sensitivity}

The revised corpus has nine \texttt{D} sources. Table~\ref{tab:supp-operation-evidence} gives the operation-level coding, including separately tested interfaces for the same source.
\begin{description}
\item[Write, storage, and derivation, five sources.] MemPrivacy tests typed writeback masking on memory benchmarks including PersonaMem-v2 (v3, \S\S4.1 and~5.3). HealthJev tests selective writeback over simulated health interactions (v2, memory policy and longitudinal evaluation). SP-Mem tests pre-write sanitization and separate exact-value storage on constructed histories (v1, \S\S2.1--2.3 and~4.2). Kaneko et al.\ test summary anonymization in multi-session support simulations (v2, \S\S2.5 and~3.5, Fig.~4). Audience-Bound Persistent Memory tests ingest and derived audience constraints in controlled histories (v1, \S\S3.1--3.5 and~6.4)~\cite{chen2026memprivacy,li2026healthjev,wang2026spmem,kaneko2026disclosureloop,liu2026audiencememory}.
\item[Local profile maintenance, one source.] HomeLLaMA specifies a local profile-update and retrieval path and evaluates carryover with ten participants (\S\S IV-D, VI-B.1, VII-B and Fig.~16). Its privacy scope is direct profile transmission to the cloud. Command obfuscation and cloud fallback frequency are evaluated separately~\cite{huang2026homellama}.
\item[Maintenance, one source.] Opal evaluates an enclave and ORAM prototype on synthetic personal workloads, with ingest and background updates inside the stipulated host leakage model (v1, \S\S2--4 and~6)~\cite{kaviani2026opal}.
\item[Withdrawal, two sources.] Reglos maintains per-user facts and invalidation in its persistent memory system and reports benchmark behavior (v1, \S\S3.1 and~4). AfterDeleteBench retires or rebuilds recorded descendants in a synthetic loop with fixed weights and complete annotated lineage (v1, \S\S3--4)~\cite{yang2026reglos,saleh2026afterdeletebench}.
\end{description}

We substituted scope-challenging sources from Stage~7 into the 22-row comparison at neighboring operations and checked whether the replacement changed the carrier and guarantee conclusions:
\begin{description}
\item[HealthJev to ConMem.] Direct selective admission becomes transferable candidate screening. The revised set has $(D,S,T)=(4,4,14)$ and the substitution gives $(3,4,15)$. MemPrivacy still supplies direct writeback evidence~\cite{li2026healthjev,zhang2026conmem,chen2026memprivacy}.
\item[MaskClaw to P2Skill.] The count stays $(4,4,14)$. The emphasis shifts from scenario-level candidate screening to offline refinement observed by a cloud evaluator. Neither supplies direct evidence for protected persistent skill reuse~\cite{zhao2026maskclaw,ryu2026p2skill}.
\item[PrivMeSA to PeCL.] The count remains $(4,4,14)$. The replacement shifts the unit from disclosure-risk optimization on clinical cohorts to sensitive token embeddings in a task-incremental benchmark. Both leave per-user parameter adaptation untested~\cite{wang2026privmesa,zhan2026pecl}.
\item[SBU to Secure Forgetting.] The count remains $(4,4,14)$. Both use public tasks; the supported claim changes from joint memory and parameter mitigation to specified state, trajectory, and environment behavior relative to a reference. Neither substitution supplies a longitudinal personal erasure guarantee~\cite{wang2026agenticunlearning,ye2026secureforgetting}.
\end{description}
With all four substitutions, the count is $(3,4,15)$. These substitutions test exemplar choice within the screened corpus, not search recall or independent agreement.

\subsection{Independent recoding of selected operations}
\label{sec:supp-independent-recoding}

A separate AI reviewer received the evidence-distance rubric, named operations, and public source identifiers. Its context excluded the manuscript, evidence ledger, and previous ratings. It reviewed the nine prespecified operations below without seeing the author codes. Two new sources were then checked under the same separation. For SEA the reviewer distinguished bypass from physical deletion, yielding 12 calls in total. The table reports the independent calls and the adjudications. Agreement is descriptive for these selected cases.

{\small
\noindent\begin{tabular}{@{}>{\raggedright\arraybackslash}p{0.38\linewidth}>{\raggedright\arraybackslash}p{0.16\linewidth}>{\raggedright\arraybackslash}p{0.38\linewidth}@{}}
\toprule
\textbf{Operation} & \textbf{Second call $\rightarrow$ final} & \textbf{Controlling source observation} \\
\midrule
MaskClaw safety gate & T/B $\to$ T/B & V2, \S6.4 and App.~F, Table~20 count screened candidates. App.~E behavioral tests do not link a committed sandbox survivor to later use~\cite{zhao2026maskclaw}. \\
PrivMeSA LoRA training & T/B $\to$ T/B & V1, \S\S3.3, 4.1--4.2, App.~E.2 use a shared policy and measure outbound disclosure~\cite{wang2026privmesa}. \\
PrivMeSA lesson write & T/B $\to$ T/B & V1, \S3.4, App.~E.1, Fig.~4 report a shared store across 200 held-out patients. Its outcome combines writing and retrieval~\cite{wang2026privmesa}. \\
Reglos pre-ingest mask & ---/I $\to$ ---/I & V1, \S3.1 and App.~A describe masking. The privacy results concern reader behavior~\cite{yang2026reglos}. \\
Reglos tombstone with reader policy & D/L $\to$ D/L & V1, \S3.1 and forgetting results in \S4.3 support the joint intervention. The read operation has a separate S row for the same result~\cite{yang2026reglos}. \\
Reglos manual UI deletion & ---/I $\to$ ---/I & V1, \S\S3.1, 5 and App.~C, Table~8 document the control and events~\cite{yang2026reglos}. \\
Opal maintenance & D/F+L $\to$ D/F+L & V1, \S3.3, \S6.7 and App.~A.1 support a host-trace result and synthetic personal-state carryover~\cite{kaviani2026opal}. \\
AfterDeleteBench lineage closure & D/L $\to$ D/L & V1, \S5.3 and App.~D.3, Table~6 test fixed-weight reconstruction, activation, and commitment~\cite{saleh2026afterdeletebench}. \\
DP-MemView adaptive release & S/F+B $\to$ S/F+B & V1, Theorem~1 and RQ1 use fixed adjacent stores and adaptive release accounting~\cite{kim2026dpmemview}. \\
\bottomrule
\end{tabular}\par}

\paragraph{New borderline operations.}
HomeLLaMA local profile writes received \texttt{D}/L in both passes. Sections~IV-D and VI-B.1 specify the local-only path for persistent profiles. Section~VII-B and Fig.~16 examine carried-over state and count optional fallback in a lab study. The judgment is limited to direct transmission. The separate current-command exposure test receives \texttt{T}/B~\cite{huang2026homellama}.

For SEA proxy omission, the second reviewer assigned \texttt{S}/B because bypass changes whether a retained user proxy can influence the response. We retain \texttt{T}/B because the measured outcomes are static language-model style and domain prompts, without a personal-agent task or evolving history. The test uses held-out prompts for four synthetic profiles and measures divergence after omitting the proxy. Both passes marked filesystem deletion ---/I because the reported verification uses the bypass path (\S2.4 and \S\S3--4.2)~\cite{schneider2026sea}.

\paragraph{Effect on the result.}
Eleven of 12 distance calls match the final coding. Applying the alternate SEA call changes the closest-source tally from $(9,14,121)$ to $(9,15,120)$ and leaves the direct-evidence finding intact. Requiring a quantitative disclosure test specifically on the stored HomeLLaMA profile would remove its direct credit and retain its current-command transfer row. The direct count would be eight sources and ten rows, still all in memory. These alternatives test the chosen scope boundaries on audited operations.

\section{Extended Protocols and Tabular Results for Mechanism Diagnostics}\label{sec:supp-diagnostics}

This section documents the complete protocols, execution-quality ledgers, multi-layer observation contingencies, full-grid parameter trajectories, state-scale closure topologies, offline trace-omission configurations, and anonymized reproducibility specifications for the bounded mechanism diagnostics summarized in the main text. Across all diagnostic batches, protocols were frozen prior to execution, evaluated exclusively with synthetic markers or public offline preference features, and analyzed strictly within each batch's designated experimental unit without cross-batch pooling or population-level statistical inference.

\subsection{Protocol Specifications, Execution Quality, and Computational Cost}\label{sec:supp-diag-protocols}

Four diagnostic suites operationalize the update-operator, protection-unit, and withdrawal distinctions developed in the review:
\begin{enumerate}
\item \textbf{Fixed 12-Episode Synthetic-Marker Withdrawal Diagnostic (\texttt{formal\_\allowbreak protocol\_v1} and \texttt{revision\_\allowbreak v2\_\allowbreak formal\_12x2}).} Twelve synthetic episodes (each containing five initial records and one or two target source records $R$) are instantiated under two fixed build seeds (\texttt{b1}: 42; \texttt{b2}: 43), yielding 24 paired episode-builds that represent repeated builds of the same synthetic cases rather than 24 independent users. For each episode and build seed, full-state retention (\texttt{K}), direct source-root deletion (\texttt{D}), and observed read-to-write successor-closure deletion (\texttt{L}) restore the same pre-intervention snapshot. The executor verifies their digests before deletion, and \texttt{D} and \texttt{L} use the same deletion operation. The reference arm (\texttt{N}) builds a fresh store from the target-free source history for that episode and seed. Each branch is probed immediately after intervention ($t_0$) and after five pre-frozen, target-free exogenous follow-up updates ($t_1$), using two target probes (\texttt{target\_direct}, \texttt{target\_\allowbreak rephrased}) and one non-target exact-field control probe (\texttt{utility}), totaling $24 \times 4 \times 2 \times 3 = 576$ logical probes per run. Primary exact-marker recoverability $Y \in \{0,1\}$ equals $1$ if either target probe's raw output contains the complete synthetic marker, $0$ if both target probes are technically valid and negative, and missing ($\text{bounds } [0,1]$) otherwise. Run \texttt{v1} and prospectively revised Run \texttt{v2} use separate manifests and ledgers; \texttt{v2} raised the native evolution output cap from $2{,}048$ to $4{,}096$ tokens (retaining $256$ for the evidence reader within a $32{,}768$-token context window) and recorded pure native \texttt{invalid\_link} events as audited degradations rather than group-fatal aborts. The two runs are never pooled.
\item \textbf{Label-Private Logistic Parameter Update Grid (Stage~A: \texttt{run\_\allowbreak synthetic\_v2} and \texttt{run\_\allowbreak movielens\_v1}).} A $d=16$ projected logistic preference model ($\ell_2$ gradient clipping threshold $C=0.25$, step size $\eta=0.5$, parameter projection radius $4.0$) is trained over $N=64$ fixed training feature slots under label-replacement adjacency of $k \in \{1,4,16,64\}$ private label slots ($\Delta_k = 2kC/N$, $\delta = 10^{-6}$). Gaussian noise is calibrated analytically via R\'enyi differential privacy (RDP) sequential composition ($\rho = T \Delta_k^2 / (2\sigma^2)$) and conversion (optimizing $(\alpha,\alpha\rho)$-RDP over $\alpha > 1$ at $\alpha^* = 1 + \sqrt{\log(1/\delta)/\rho} > 1$ to yield the upper bound $\varepsilon \le \rho + 2\sqrt{\rho\log(1/\delta)}$, which is conservative relative to exact Gaussian hockey-stick divergence); IEEE-754 closed-form inversion $\rho = (\sqrt{\log(1/\delta)+\varepsilon} - \sqrt{\log(1/\delta)})^2$ is verified against numerical back-checks with maximum absolute error $7.11 \times 10^{-15}$. The synthetic suite evaluates $24$ fixed preference profiles across $T \in \{1,2,4,8,16,32\}$, $\varepsilon \in \{0.5,1,2,3,4,6,8,16\}$, and five noise seeds ($\{701,907,1213,1601,2027\}$), totaling $25{,}488$ non-zero update trajectories ($23{,}040$ fixed-budget, $2{,}160$ fixed-$\sigma$, and $288$ deterministic no-noise baselines) plus $24$ zero-update baselines ($267{,}624$ optimizer steps). The public anchor evaluates $24$ MovieLens~100K users ($\ge 84$ ratings each; ratings $\ge 4$ mapped to $+1$ and $\le 3$ to $-1$ over fixed projections of genre metadata, $20$ evaluation pairs per user) at $T=16$ across the same $(k,\varepsilon)$ grid and three noise seeds ($\{701,907,1213\}$), totaling $2{,}352$ non-zero update trajectories ($2{,}304$ fixed-budget and $48$ deterministic baselines) plus $24$ zero-update baselines ($37{,}632$ optimizer steps), yielding $27{,}840$ non-zero training trajectories overall. The MovieLens label guarantee conditions on the selected user, the timestamp-ordered item slots, and their projected features. The held-out item--label pairs are fixed for evaluation. The condition excludes user selection, timing and item history from the protected changes. The proof analyzes noise draws private to the randomized mechanism, with the released head as its output. Benchmark runners fix seeds for reproducibility, and the archive reports accuracy aggregates without a separate DP guarantee for those releases. An accompanying analytical calibration suite (\texttt{survey\_\allowbreak accounting}) independently verifies post-processing invariance under repeated display ($P_0$), linear RDP accumulation ($P_1$), Monte Carlo mean-squared error agreement ($100$ repeats), and a deterministic adjacency counterexample for noiseless clipped sample means ($P_2$).
\item \textbf{Four-History Memory Cleanup and Source-Reintroduction Diagnostic (Stage~B: \texttt{run\_\allowbreak dev\_v3}).} Four fixed 12-record synthetic histories with inductive synthesis updates (\texttt{dev\_h0}--\texttt{dev\_h3}) are built five times (\texttt{dev\_h2} built under both \texttt{b1} and \texttt{b2}, averaged within history prior to equal-weight aggregation across the four histories). Primary arms \texttt{K}, \texttt{D}, \texttt{L}, \texttt{M1} (cardinality-matched random deletion, which removes $12/12$ records here because conservative closures saturate all five builds), and \texttt{N} are evaluated at $t_0$ and after five continued updates ($t_5$, matching the five-update horizon labeled $t_1$ in Formal~\texttt{v2}) across five probes per checkpoint (\texttt{target\_direct}, \texttt{target\_\allowbreak rephrased}, \texttt{near\_field}, \texttt{far\_field}, \texttt{structured\_task}). Single-unit calibration controls on \texttt{dev\_h2/b1} additionally inspect fractional deletion ($\alpha=0.5$, removing $6/12$ records across \texttt{L} and matched random arms \texttt{M1}--\texttt{M3}) and partial trace retention ($p=0.5$, retaining $81/163$ edges). Longitudinal trajectory \texttt{E2} tracks \texttt{dev\_h2/b1} across update steps $t_6, t_{10}, t_{15}, t_{20}$ under a normal target-free path and an explicit source-reintroduction path (re-submitting withdrawn source \texttt{source:\allowbreak dev\_h2:t1} at $t_{10}$), adding arm \texttt{G} (\texttt{L} plus an explicit withdrawn-source registry gate). Total Stage~B evaluation comprises $480$ isolated-process probes ($192$ target probes and $288$ utility/task probes) and $275$ follow-up events.
\item \textbf{State-Scale Dependency Closure and Offline Trace-Omission Diagnostic (Stage~C: \texttt{run\_\allowbreak scale\_v1}).} Eight single-seed (\texttt{b1}: 42) synthetic A-MEM states cross four sizes ($|S| \in \{12,24,48,96\}$) with two write orderings (\texttt{blocked} vs.\ \texttt{interleaved} over identical record sets, $360$ total ingested records, zero answer probes). Subsequently, an offline structural ablation evaluates $336$ trace-retention configurations across the eight frozen read-to-write graphs at seven edge-retention fractions $p \in \{0, 0.125, 0.25, 0.5, 0.75, 0.875, 1\}$ ($\lfloor p |E| \rfloor$ retained edges) under nested random edge omission (five permutation seeds $\{11,23,37,53,71\}$ per state) and centered contiguous log gaps.
\end{enumerate}

Table~\ref{tab:supp-run-quality} reports execution completeness, token volume, and wall-clock cost across all batches without pooling denominators. In Formal~\texttt{v1}, six of 24 episode-builds failed to complete: three hit the $2{,}048$-token generation cap with \texttt{done\_reason=\allowbreak length} during neighbor evolution (\texttt{pilot\_01/b2}, \texttt{pilot\_06/b2}, \texttt{pilot\_11/b1}), and three produced out-of-range native link indices (\texttt{CONNECTIONS: 0, 2} against two shown records) that the strict \texttt{v1} adapter escalated into fatal state errors (\texttt{pilot\_09/b1}, \texttt{pilot\_10/b2}, \texttt{pilot\_12/b1}), leaving $108$ missing probes and wide all-assignment missing-data bounds for $Y_{\mathrm{D}} - Y_{\mathrm{L}}$ at $t_1$ ($[-0.2083, 0.2917]$; $1/18 = 0.0556$ among complete pairs). Under the prospective \texttt{v2} protocol, all $24/24$ episode-builds, $96/96$ cold-restart checks, and $576/576$ probes completed with zero length truncations (maximum output $1{,}979$ tokens) and zero naturally occurring invalid links ($749$ accepted links in \texttt{b1}, $768$ in \texttt{b2}), yielding point-identified $t_1$ bounds $[0.0417, 0.0417]$ ($1/24$). In Stage~B (\texttt{run\_dev\_v3}), all $480$ probes and $1{,}909$ local backend calls succeeded within the $2{,}400$-call hard cap, but applying the pre-specified $1.30\times$ safety multiplier yielded $2{,}481.7 > 2{,}400$ buffered calls, so technical acceptance passed while the formal 24-history expansion gate remained closed (\texttt{formal\_gate\_\allowbreak passed=false}). The archive's Stage~A acceptance record (\texttt{stage\_a\_\allowbreak acceptance.md}) specifies the gate, and the Stage~B development acceptance record (\texttt{development\_\allowbreak acceptance.json}) records the observed inputs and decision. Table~\ref{tab:supp-memory-outcomes} details the complete arm-by-checkpoint outcomes across Formal~\texttt{v2} and Stage~B.

\begin{table}[htbp]
\centering
\caption{Execution quality, completion coverage, and computational cost across diagnostic batches (unpooled). For LLM-backed memory suites, cumulative backend inference time and end-to-end wall-clock runtime (including isolated-process probe restarts and digest verification) are reported separately; paid API cost is zero across all runs.}\label{tab:supp-run-quality}
\scriptsize
\setlength{\tabcolsep}{2.6pt}
\begin{tabular}{@{}l l c r r r r r@{}}
\toprule
Batch / Component & Units (Done/Plan) & Probes (V/I/M) & Calls & Input Tok. & Output Tok. & Call (s) & Wall (s) \\
\midrule
Formal \texttt{v1} (\texttt{12x2}) & $18/24$ builds & $452/16/108$ & $2{,}418$ & $2{,}925{,}400$ & $475{,}222$ & $3{,}356.76$ & $7{,}380.35$ \\
Formal \texttt{v2} (\texttt{12x2}) & $24/24$ builds & $565/11/0$ & $2{,}941$ & $3{,}554{,}512$ & $567{,}243$ & $3{,}977.95$ & $8{,}988.39$ \\
\quad \texttt{v2} build \texttt{b1} & $12/12$ builds & $282/6/0$ & $1{,}468$ & $1{,}755{,}702$ & $280{,}426$ & $1{,}968.30$ & --- \\
\quad \texttt{v2} build \texttt{b2} & $12/12$ builds & $283/5/0$ & $1{,}473$ & $1{,}798{,}810$ & $286{,}817$ & $2{,}009.65$ & --- \\
\midrule
Stage~B \texttt{run\_dev\_v3}$^{\dagger}$ & $5/5$ builds ($4$h) & $476/4/0$ & $1{,}909$ & $3{,}910{,}093$ & $486{,}321$ & $3{,}578.26$ & $7{,}266.44$ \\
\quad \texttt{native\_ingest} & $275/275$ events & --- & $1{,}429$ & $1{,}587{,}060$ & $462{,}152$ & $3{,}052.86$ & --- \\
\quad \texttt{evidence\_reader} & $480/480$ probes & $476/4/0$ & $480$ & $2{,}323{,}033$ & $24{,}169$ & $525.40$ & --- \\
\midrule
Stage~C \texttt{run\_scale\_v1}$^{\ddagger}$ & $8/8$ states ($360$r) & $0$ (struct.) & $1{,}294$ & $1{,}394{,}595$ & $362{,}367$ & $2{,}386.97$ & $2{,}440.09$ \\
\midrule
Stage~A \texttt{synthetic\_v2} & $25{,}488$ (+$24$ $T{=}0$) & $267{,}624$ steps & $0$ & $0$ & $0$ & --- & $7.05$ \\
Stage~A \texttt{movielens\_v1} & $2{,}352$ (+$24$ $T{=}0$) & $37{,}632$ steps & $0$ & $0$ & $0$ & --- & $0.99$ \\
Aux.\ RDP accounting & $100$ MC repeats & $6$ invariants & $0$ & $0$ & $0$ & --- & $1.26$ \\
\bottomrule
\end{tabular}
\par\smallskip\noindent\parbox{\linewidth}{\raggedright\scriptsize
Probes (V/I/M): technically valid / guard-rejected invalid / missing. $^{\dagger}$Stage~B median/max per-call latency: $0.98$\,s / $17.09$\,s overall ($0.89$\,s / $17.09$\,s native ingest; $1.12$\,s / $2.00$\,s evidence reader). Hard call limit: $2{,}400$; $1.30\times$ buffered requirement: $2{,}481.7$ (\texttt{formal\_gate\_passed=false}). $^{\ddagger}$Stage~C call budget utilization: $1{,}294 / 1{,}800$ ($71.89\%$; $3.594$ calls/record; min/max call latency $0.21$\,s / $15.25$\,s; $0$ backend errors; $2$ filler-turn native \texttt{invalid\_link} degradations $= 0.556\%$ per record).}
\end{table}

\begingroup
\scriptsize
\setlength{\tabcolsep}{3.5pt}
\renewcommand{\arraystretch}{1.08}
\begin{longtable}{@{}l l l l c l l r@{}}
\caption{Complete memory-intervention outcomes across Formal~\texttt{v2} (12 episodes $\times$ 2 builds; counts as Yes/No/Missing and missing-data identification bounds), Stage~B primary 4-history aggregation (history-weighted mean $\pm$ SD across 4 fixed histories), and Stage~B \texttt{E2} single-history trajectories (\texttt{dev\_h2/b1}).}\label{tab:supp-memory-outcomes}\\
\toprule
Stratum & Ckpt. & Arm(s) & Raw Target Hit & Guarded & Util.\ / Near, Far & Task / Bounds & $|S|$ \\
\midrule
\endfirsthead
\caption[]{Complete memory-intervention outcomes across Formal~\texttt{v2}, Stage~B primary, and Stage~B \texttt{E2} (Continued)}\\
\toprule
Stratum & Ckpt. & Arm(s) & Raw Target Hit & Guarded & Util.\ / Near, Far & Task / Bounds & $|S|$ \\
\midrule
\endhead
\midrule
\multicolumn{8}{r}{\emph{Continued on next page}} \\
\endfoot
\bottomrule
\endlastfoot
\multicolumn{8}{@{}l}{\textbf{Panel A: Formal \texttt{v2} Fixed 12-Episode Batch} (\texttt{figure\_data\_t0\_t1.csv}; Yes/No/Missing per arm)} \\
\texttt{b1} ($12$) & $t_0$ & \texttt{K} & \texttt{12/0/0} ($100.0\%$) & --- & \texttt{12/0/0} ($100.0\%$) & exact pt. & $5$ \\
\texttt{b1} ($12$) & $t_0$ & \texttt{D} & \texttt{2/10/0} ($16.7\%$) & --- & \texttt{8/4/0} ($66.7\%$) & exact pt. & $3\text{--}4$ \\
\texttt{b1} ($12$) & $t_0$ & \texttt{L} & \texttt{0/12/0} ($0.0\%$) & --- & \texttt{0/12/0} ($0.0\%$) & exact pt. & $0$ \\
\texttt{b1} ($12$) & $t_0$ & \texttt{N} & \texttt{0/12/0} ($0.0\%$) & --- & \texttt{12/0/0} ($100.0\%$) & exact pt. & $5$ \\
\texttt{b2} ($12$) & $t_0$ & \texttt{K} & \texttt{12/0/0} ($100.0\%$) & --- & \texttt{12/0/0} ($100.0\%$) & exact pt. & $5$ \\
\texttt{b2} ($12$) & $t_0$ & \texttt{D} & \texttt{1/11/0} ($8.3\%$) & --- & \texttt{8/4/0} ($66.7\%$) & exact pt. & $3\text{--}4$ \\
\texttt{b2} ($12$) & $t_0$ & \texttt{L} & \texttt{0/12/0} ($0.0\%$) & --- & \texttt{0/12/0} ($0.0\%$) & exact pt. & $0$ \\
\texttt{b2} ($12$) & $t_0$ & \texttt{N} & \texttt{0/12/0} ($0.0\%$) & --- & \texttt{12/0/0} ($100.0\%$) & exact pt. & $5$ \\
Tot.\ ($24$) & $t_0$ & \texttt{K} & \texttt{24/0/0} ($100.0\%$) & --- & \texttt{24/0/0} ($100.0\%$) & exact pt. & $5$ \\
Tot.\ ($24$) & $t_0$ & \texttt{D} & \texttt{3/21/0} ($12.5\%$) & --- & \texttt{16/8/0} ($66.7\%$) & exact pt. & $3\text{--}4$ \\
Tot.\ ($24$) & $t_0$ & \texttt{L} & \texttt{0/24/0} ($0.0\%$) & --- & \texttt{0/24/0} ($0.0\%$) & exact pt. & $0$ \\
Tot.\ ($24$) & $t_0$ & \texttt{N} & \texttt{0/24/0} ($0.0\%$) & --- & \texttt{24/0/0} ($100.0\%$) & exact pt. & $5$ \\
\addlinespace[2pt]
\texttt{b1} ($12$) & $t_1$ & \texttt{K} & \texttt{12/0/0} ($100.0\%$) & --- & \texttt{10/0/2} ($83.3\%$) & util.\ $[83.3,100]\%$ & $10$ \\
\texttt{b1} ($12$) & $t_1$ & \texttt{D} & \texttt{0/12/0} ($0.0\%$) & --- & \texttt{8/4/0} ($66.7\%$) & exact pt. & $8\text{--}9$ \\
\texttt{b1} ($12$) & $t_1$ & \texttt{L} & \texttt{0/12/0} ($0.0\%$) & --- & \texttt{0/12/0} ($0.0\%$) & exact pt. & $5$ \\
\texttt{b1} ($12$) & $t_1$ & \texttt{N} & \texttt{0/12/0} ($0.0\%$) & --- & \texttt{11/1/0} ($91.7\%$) & exact pt. & $10$ \\
\texttt{b2} ($12$) & $t_1$ & \texttt{K} & \texttt{12/0/0} ($100.0\%$) & --- & \texttt{11/0/1} ($91.7\%$) & util.\ $[91.7,100]\%$ & $10$ \\
\texttt{b2} ($12$) & $t_1$ & \texttt{D} & \texttt{1/11/0} ($8.3\%$) & --- & \texttt{8/4/0} ($66.7\%$) & exact pt. & $8\text{--}9$ \\
\texttt{b2} ($12$) & $t_1$ & \texttt{L} & \texttt{0/12/0} ($0.0\%$) & --- & \texttt{0/12/0} ($0.0\%$) & exact pt. & $5$ \\
\texttt{b2} ($12$) & $t_1$ & \texttt{N} & \texttt{0/12/0} ($0.0\%$) & --- & \texttt{11/1/0} ($91.7\%$) & exact pt. & $10$ \\
Tot.\ ($24$) & $t_1$ & \texttt{K} & \texttt{24/0/0} ($100.0\%$) & --- & \texttt{21/0/3} ($87.5\%$) & util.\ $[87.5,100]\%$ & $10$ \\
Tot.\ ($24$) & $t_1$ & \texttt{D} & \texttt{1/23/0} ($4.2\%$) & --- & \texttt{16/8/0} ($66.7\%$) & $Y_{\mathrm{D}}{-}Y_{\mathrm{L}}{=}4.2\%$ & $8\text{--}9$ \\
Tot.\ ($24$) & $t_1$ & \texttt{L} & \texttt{0/24/0} ($0.0\%$) & --- & \texttt{0/24/0} ($0.0\%$) & exact pt. & $5$ \\
Tot.\ ($24$) & $t_1$ & \texttt{N} & \texttt{0/24/0} ($0.0\%$) & --- & \texttt{22/2/0} ($91.7\%$) & exact pt. & $10$ \\
\midrule
\multicolumn{8}{@{}l}{\textbf{Panel B: Stage~B Primary 4-History Batch} (\texttt{primary\_memory\_aggregate.csv}; \texttt{dev\_h2} averaged over \texttt{b1,b2} first)} \\
4 hist. & $t_0$ & \texttt{K} & $100.0 \pm 0.0\%$ & $100.0\%$ & $100.0\%,\; 100.0\%$ & $100.0 \pm 0.0\%$ & $12.0$ \\
4 hist. & $t_0$ & \texttt{D} & $100.0 \pm 0.0\%$ & $75.0 \pm 50.0\%$ & $100.0\%,\; 100.0\%$ & $100.0 \pm 0.0\%$ & $10.5$ \\
4 hist. & $t_0$ & \texttt{L, M1} & $0.0 \pm 0.0\%$ & $0.0\%$ & $0.0\%,\; 0.0\%$ & $0.0 \pm 0.0\%$ & $0.0$ \\
4 hist. & $t_0$ & \texttt{N} & $0.0 \pm 0.0\%$ & $0.0\%$ & $100.0\%,\; 100.0\%$ & $100.0 \pm 0.0\%$ & $12.0$ \\
4 hist. & $t_5$ & \texttt{K} & $100.0 \pm 0.0\%$ & $100.0\%$ & $100.0\%,\; 100.0\%$ & $100.0 \pm 0.0\%$ & $17.0$ \\
4 hist. & $t_5$ & \texttt{D} & $100.0 \pm 0.0\%$ & $75.0 \pm 50.0\%$ & $100.0\%,\; 100.0\%$ & $100.0 \pm 0.0\%$ & $15.5$ \\
4 hist. & $t_5$ & \texttt{L, M1} & $0.0 \pm 0.0\%$ & $0.0\%$ & $0.0\%,\; 0.0\%$ & $0.0 \pm 0.0\%$ & $5.0$ \\
4 hist. & $t_5$ & \texttt{N} & $0.0 \pm 0.0\%$ & $0.0\%$ & $100.0\%,\; 100.0\%$ & $100.0 \pm 0.0\%$ & $17.0$ \\
\midrule
\multicolumn{8}{@{}l}{\textbf{Panel C: Stage~B \texttt{E2} Longitudinal Trajectories on \texttt{dev\_h2/b1}} (\texttt{e2\_trajectory.csv}; at $t_6, t_{10}, t_{15}, t_{20}$)} \\
Normal & $t_6\text{--}t_{20}$ & \texttt{K, D} & $1, 1, 1, 1$ & $1, 1, 1, 1$ & $1, 1$ (all ckpts) & $1$ (all ckpts) & $17\text{--}32$ \\
Normal & $t_6\text{--}t_{20}$ & \texttt{L, G} & $0, 0, 0, 0$ & $0, 0, 0, 0$ & $0, 0$ (all ckpts) & $0$ (\texttt{L} $t_{20}$ miss.) & $6\text{--}20$ \\
Normal & $t_6\text{--}t_{20}$ & \texttt{N} & $0, 0, 0, 0$ & $0, 0, 0, 0$ & $1, 1$ (all ckpts) & $1$ (all ckpts) & $18\text{--}32$ \\
Reintro. & $t_6\text{--}t_{20}$ & \texttt{L} & $0, 1, 1, 1$ & $0, 1, 1, 1$ & $0, 0$ (all ckpts) & $0$ (all ckpts) & $6\text{--}20$ \\
Reintro. & $t_6\text{--}t_{20}$ & \texttt{G} & $0, 0, 0, 0$ & $0, 0, 0, 0$ & $0, 0$ (all ckpts) & $0$ (all ckpts) & $6\text{--}19$ \\
Reintro. & $t_6\text{--}t_{20}$ & \texttt{N} & $0, 1, 1, 1$ & $0, 1, 1, 1$ & $1, 1$ (all ckpts) & $1$ (all ckpts) & $18\text{--}32$ \\
\end{longtable}
\noindent\parbox{\linewidth}{\raggedright\scriptsize
Note: On \texttt{dev\_h2/b1} at $t_0$, single-unit dose calibration controls (\texttt{dose\_controls.csv}) removing $6/12$ records ($\alpha=0.5$ under \texttt{L} and random arms \texttt{M1}--\texttt{M3}) or retaining $81/163$ trace edges ($p=0.5$ random omission and contiguous gap, yielding $0$ non-root reachable descendants and removing $1/12$ root record) all produced raw target recovery $=1$ and near/far/task retention $=1$. In Stage~B Panel~B, \texttt{D} state record count SD across the 4 histories is $0.58$ ($10\text{--}11$ at $t_0$; $15\text{--}16$ at $t_5$). In \texttt{E2} reintroduction, \texttt{K} ($18,22,27,32$) and \texttt{D} ($17,21,26,31$) remain at $1$ across $t_6\text{--}t_{20}$; \texttt{G} blocks the withdrawn-source write at $t_{10}$ ($|S|=6,9,14,19$ vs.\ $6,10,15,20$ in \texttt{L}).}
\endgroup

\subsection{Multi-Layer Observation Surface and Reader-Guard Audit}\label{sec:supp-diag-layers}

\begingroup\emergencystretch=1.5em
Because suppression at an output verifier does not imply erasure from persistent state or non-retrieval during context assembly, Stage~B records four binary indicators per target probe: persistent state presence (\texttt{state\_\allowbreak present}, whether the exact marker remains in any active record or evolved context field), retrieval exposure (\texttt{retrieval\_\allowbreak hit}, whether a marker-bearing record enters the retrieved prompt context), raw generation recovery (\texttt{raw\_hit}, whether the unconstrained model output contains the exact marker), and post-guard target delivery (\texttt{guarded\_hit}, alongside binary \texttt{guard\_\allowbreak accepted}). Crucially, \texttt{guarded\_hit=0} and \texttt{guard\_\allowbreak accepted=0} are distinct events: when no target is generated (\texttt{raw\_hit=0}) and the model emits a schema-valid \texttt{null} abstinence with consistent evidence metadata, the reader guard accepts the negative response (\texttt{guard\_\allowbreak accepted=1}, \texttt{verified\_\allowbreak missing}).

Table~\ref{tab:supp-evidence-contingency} presents the joint contingency distribution across all $192$ Stage~B target probes (\texttt{evidence\_\allowbreak joint\_\allowbreak patterns.csv}). Exactly half ($96/192$) are clean negatives across all four layers (\texttt{L}, \texttt{M1}, \texttt{N}, and blocked \texttt{G}/\texttt{E2} checkpoints), and $91/192$ ($47.40\%$) are positive across all four layers. Five target probes exhibit divergence across layers---all within source-only deletion arm \texttt{D}: two \texttt{target\_\allowbreak direct} probes (\texttt{dev\_h0/b1} at $t_0$ and $t_5$) retrieve marker-bearing evolved context into the prompt (\texttt{state=1}, \texttt{retrieval=1}) without emitting the marker in raw output (\texttt{raw=0}, \texttt{guard\_\allowbreak accepted=1}), whereas three \texttt{target\_\allowbreak rephrased} probes emit the exact marker in raw output (\texttt{raw=1}) but are rejected by the citation/\allowbreak consistency guard (\texttt{guard\_\allowbreak accepted=0}, \texttt{guarded\_\allowbreak hit=0}). In \texttt{dev\_h0/b1} at $t_5$, counting only guard-accepted answers would misclassify \texttt{D} as negative on that history ($75.0\%$ four-history mean), whereas retaining raw hits captures the $100.0\%$ persistent survival of the marker across all four histories. Table~\ref{tab:supp-guard-rejections} audits every reader-guard rejection across both Stage~B ($4/480$ probes) and Formal~\texttt{v2} ($11/576$ probes across seven episode-build/\allowbreak arm/\allowbreak checkpoint cells; all $576$ and $480$ probes had \texttt{schema\_\allowbreak valid=true} and unchanged SHA-256 state digests).\par
\endgroup

\begin{table}[htbp]
\centering
\caption{Four-layer observation-surface contingency table across all $192$ Stage~B target probes (\texttt{evidence\_joint\_patterns.csv} and \texttt{evidence\_layers\_target\_probes.csv}).}\label{tab:supp-evidence-contingency}
\scriptsize
\setlength{\tabcolsep}{2.6pt}
\begin{tabular}{@{}c c c c c l r r l@{}}
\toprule
\texttt{state} & \texttt{retr.} & \texttt{raw} & \texttt{guard} & \texttt{acc.} & Guard Disposition Reason & $n$ & \% & Observing Arms / Cells \\
\midrule
$0$ & $0$ & $0$ & $0$ & $1$ & \texttt{verified\_missing} & $96$ & $50.00$ & \texttt{L, M1, N, G} (clean / blocked) \\
$1$ & $1$ & $0$ & $0$ & $1$ & \texttt{verified\_missing} & $2$ & $1.04$ & \texttt{dev\_h0/b1 D} ($t_0, t_5$ \texttt{direct}) \\
$1$ & $1$ & $1$ & $0$ & $0$ & \texttt{missing/conflicting\_evid.} & $2$ & $1.04$ & \texttt{dev\_h0/b1 D} ($t_0, t_5$ \texttt{rephr.}) \\
$1$ & $1$ & $1$ & $0$ & $0$ & \texttt{quote\_not\_verbatim} & $1$ & $0.52$ & \texttt{dev\_h3/b1 D} ($t_0$ \texttt{rephr.}) \\
$1$ & $1$ & $1$ & $1$ & $1$ & \texttt{verified\_verbatim} & $91$ & $47.40$ & \texttt{K, D}, $\alpha{=}0.5$, $p{=}0.5$, reintro.\ \texttt{L, N} \\
\midrule
\multicolumn{6}{@{}l}{\textbf{Total target probes} ($189/192$ \texttt{guard\_accepted=1}; $3/192$ rejected)} & $\mathbf{192}$ & $\mathbf{100.00}$ & +$288$ utility/task ($287$ acc.) \\
\bottomrule
\end{tabular}
\end{table}

\begin{table}[htbp]
\centering
\caption{Complete item-level audit of reader-guard rejections in Stage~B ($4/480$ probes; \texttt{guard\_rejections.csv}) and Formal~\texttt{v2} ($11/576$ probes: $6$ in \texttt{b1}, $5$ in \texttt{b2}; \texttt{formal\_v2\_results\_report.md}). Every row satisfied JSON \texttt{schema\_valid=true} and preserved identical pre/post-probe state digests.}\label{tab:supp-guard-rejections}
\scriptsize
\setlength{\tabcolsep}{2.3pt}
\begin{tabular}{@{}l l c c l l l@{}}
\toprule
Batch & Unit & Arm & Ckpt. & Probe(s) ($n$) & Guard Rejection Code & Retained Raw Observation \& Scoring \\
\midrule
Stage~B & \texttt{dev\_h0/b1} & \texttt{D} & $t_0$ & \texttt{rephr.} ($1$) & \texttt{missing\_or\_conflicting} & Exact marker in raw output; raw hit $=1$ \\
Stage~B & \texttt{dev\_h0/b1} & \texttt{D} & $t_5$ & \texttt{rephr.} ($1$) & \texttt{missing\_or\_conflicting} & Exact marker in raw output; raw hit $=1$ \\
Stage~B & \texttt{dev\_h3/b1} & \texttt{D} & $t_0$ & \texttt{rephr.} ($1$) & \texttt{quote\_not\_verbatim} & Exact marker in raw output; raw hit $=1$ \\
Stage~B & \texttt{dev\_h2/b1} & \texttt{E2\_L} & $t_{20}$ & \texttt{task} ($1$) & \texttt{task\_evidence\_invalid} & Expected task hit ($1$), target ($0$); missing \\
\midrule
Formal \texttt{v2} & \texttt{pilot\_01/b1} & \texttt{K} & $t_1$ & \texttt{both tgt.} ($2$) & \texttt{missing\_or\_conflicting} & Exact marker \& value in both; $Y=1$ \\
Formal \texttt{v2} & \texttt{pilot\_05/b1} & \texttt{D} & $t_0$ & \texttt{both tgt.} ($2$) & \texttt{missing\_or\_conflicting} & Exact marker \& value in both; $Y=1$ \\
Formal \texttt{v2} & \texttt{pilot\_05/b2} & \texttt{K} & $t_1$ & \texttt{utility} ($1$) & \texttt{quote\_unsupported\_val} & Control value correct in raw; missing $[0,1]$ \\
Formal \texttt{v2} & \texttt{pilot\_06/b1} & \texttt{K} & $t_1$ & \texttt{utility} ($1$) & \texttt{quote\_not\_verbatim} & Control value correct in raw; missing $[0,1]$ \\
Formal \texttt{v2} & \texttt{pilot\_06/b2} & \texttt{D} & $t_0$ & \texttt{both tgt.} ($2$) & \texttt{missing\_or\_conflicting} & Exact marker in both (direct val.\ false); $Y=1$ \\
Formal \texttt{v2} & \texttt{pilot\_06/b2} & \texttt{D} & $t_1$ & \texttt{both tgt.} ($2$) & \texttt{missing\_or\_conflicting} & Sole \texttt{D} $t_1$ hit: exact marker in both; $Y=1$ \\
Formal \texttt{v2} & \texttt{pilot\_07/b1} & \texttt{K} & $t_1$ & \texttt{utility} ($1$) & \texttt{quote\_unsupported\_val} & Control value correct in raw; missing $[0,1]$ \\
\bottomrule
\end{tabular}
\par\smallskip\noindent\parbox{\linewidth}{\raggedright\scriptsize
Full machine codes in ledgers: \texttt{missing\_or\_conflicting\_evidence}, \texttt{quote\_does\_not\_support\_field\_value}, \texttt{quote\_not\_verbatim}, and \texttt{task\_evidence\_invalid}; \texttt{both tgt.} denotes \texttt{target\_direct} and \texttt{target\_rephrased}.}
\end{table}

\subsection{Label-Private Parameter Update Grid and Baselines}\label{sec:supp-diag-param}

Tables~\ref{tab:supp-param-baselines} and~\ref{tab:supp-param-grid} report the complete baseline and differentially private parameter-update grids from \texttt{parameter\_\allowbreak baselines.csv} ($16$ baseline conditions) and \texttt{parameter\_\allowbreak grid\_\allowbreak summary.csv} ($224$ $(dataset, T, k, \varepsilon)$ cells). Noise seeds are averaged within each synthetic profile or MovieLens~100K user first; tables report the equal-weight mean and standard deviation across the $24$ fixed profiles or users (reflecting between-profile heterogeneity rather than sampling error). Across the synthetic grid, deterministic $\ell_2$ clipping at $C=0.25$ without noise yields an invariant $80.8\% \pm 5.0\%$ balanced accuracy across all $T \in \{1,2,4,8,16,32\}$ (compared with $80.8\%\text{--}81.4\%$ unclipped), confirming that utility degradation in Table~\ref{tab:supp-param-grid} is driven by group sensitivity $\Delta_k = 2kC/N$ rather than clipping bias; mean gradient clipping fractions range from $0.858$ to $1.000$ and mean test log-loss increases from $0.632$ ($T=32, k=1, \varepsilon=16$) to $0.837$ ($T=32, k=64, \varepsilon=4$). On MovieLens~100K, class imbalance elevates zero-update raw accuracy to $62.7\%$ while balanced accuracy sits at $50.0\%$, and the $T=16$ clipped no-noise baseline reaches $56.3\% \pm 12.6\%$ balanced accuracy, bounding all noisy curves near the public genre-only linear ceiling.

\begin{table}[htbp]
\centering
\caption{Zero-update, clipped no-noise ($C=0.25$), and unclipped no-noise logistic preference baselines across update horizons $T$ (\texttt{parameter\_baselines.csv}; $24$ profiles/users per row).}\label{tab:supp-param-baselines}
\scriptsize
\setlength{\tabcolsep}{4.0pt}
\begin{tabular}{@{}l l l r c r@{}}
\toprule
Dataset & Baseline Condition & Horizon $T$ & Units & Balanced Accuracy Mean $\pm$ SD (\%) & Raw Accuracy (\%) \\
\midrule
Synthetic & \texttt{zero\_update} & $0$ & $24$ & $50.00 \pm 0.00$ & $48.97$ \\
Synthetic & \texttt{clipped\_no\_noise} & $1, 2, 4, 8, 16, 32$$^{\dagger}$ & $24$ & $80.77 \pm 5.03$ & $80.79$ \\
Synthetic & \texttt{unclipped\_no\_noise} & $1$ & $24$ & $80.77 \pm 5.03$ & $80.79$ \\
Synthetic & \texttt{unclipped\_no\_noise} & $2$ & $24$ & $80.78 \pm 5.05$ & $80.81$ \\
Synthetic & \texttt{unclipped\_no\_noise} & $4$ & $24$ & $80.78 \pm 5.05$ & $80.81$ \\
Synthetic & \texttt{unclipped\_no\_noise} & $8$ & $24$ & $80.85 \pm 5.06$ & $80.88$ \\
Synthetic & \texttt{unclipped\_no\_noise} & $16$ & $24$ & $80.98 \pm 5.22$ & $81.01$ \\
Synthetic & \texttt{unclipped\_no\_noise} & $32$ & $24$ & $81.36 \pm 5.27$ & $81.40$ \\
\midrule
MovieLens 100K & \texttt{zero\_update} & $0$ & $24$ & $50.00 \pm 0.00$ & $62.71$ \\
MovieLens 100K & \texttt{clipped\_no\_noise} & $16$ & $24$ & $56.33 \pm 12.62$ & $62.71$ \\
MovieLens 100K & \texttt{unclipped\_no\_noise} & $16$ & $24$ & $55.21 \pm 13.40$ & $61.88$ \\
\bottomrule
\end{tabular}
\par\smallskip\noindent\parbox{\linewidth}{\raggedright\scriptsize
$^{\dagger}$All six \texttt{clipped\_no\_noise} rows on synthetic preferences ($T \in \{1,2,4,8,16,32\}$) yield identical full-batch clipped trajectories ($0.807681 \pm 0.050342$ balanced accuracy; $0.807943$ raw accuracy) because full-batch unit-norm direction updates saturate the $C=0.25$ clip threshold uniformly across step counts.}
\end{table}

\begingroup
\scriptsize
\setlength{\tabcolsep}{2.7pt}
\renewcommand{\arraystretch}{1.08}
\begin{longtable}{@{}l r r r r r r r r r r@{}}
\caption{Complete 224-cell label-private parameter update grid (\texttt{parameter\_grid\_summary.csv}). Entries report balanced accuracy mean (SD) in percent across $24$ fixed profiles ($5$ noise seeds each) for Synthetic preferences and $24$ fixed users ($3$ noise seeds each) for MovieLens~100K under $(\varepsilon,\delta)$-DP ($\delta=10^{-6}$) for replacing $k$ labels among $64$ fixed public slots.}\label{tab:supp-param-grid}\\
\toprule
Dataset & $T$ & $k$ & $\varepsilon{=}0.5$ & $\varepsilon{=}1$ & $\varepsilon{=}2$ & $\varepsilon{=}3$ & $\varepsilon{=}4$ & $\varepsilon{=}6$ & $\varepsilon{=}8$ & $\varepsilon{=}16$ \\
\midrule
\endfirsthead
\caption[]{Complete 224-cell label-private parameter update grid (Continued)}\\
\toprule
Dataset & $T$ & $k$ & $\varepsilon{=}0.5$ & $\varepsilon{=}1$ & $\varepsilon{=}2$ & $\varepsilon{=}3$ & $\varepsilon{=}4$ & $\varepsilon{=}6$ & $\varepsilon{=}8$ & $\varepsilon{=}16$ \\
\midrule
\endhead
\midrule
\multicolumn{11}{r}{\emph{Continued on next page}} \\
\endfoot
\bottomrule
\endlastfoot
Synthetic & 1 & 1 & 53.1 (4.8) & 59.1 (3.4) & 64.2 (3.8) & 67.8 (3.6) & 71.3 (4.0) & 75.0 (4.7) & 76.6 (4.0) & 79.5 (4.8) \\
Synthetic & 1 & 4 & 50.6 (3.4) & 52.2 (3.5) & 54.2 (3.9) & 55.6 (3.3) & 56.5 (4.2) & 60.3 (4.3) & 63.0 (2.9) & 69.8 (3.9) \\
Synthetic & 1 & 16 & 49.6 (3.0) & 50.7 (3.9) & 50.1 (3.4) & 50.7 (4.4) & 51.5 (3.8) & 51.1 (4.1) & 52.8 (3.5) & 57.0 (4.3) \\
Synthetic & 1 & 64 & 50.3 (2.3) & 49.2 (3.9) & 49.6 (4.0) & 50.9 (3.2) & 51.6 (4.0) & 50.4 (3.7) & 52.3 (3.9) & 52.2 (4.4) \\
\midrule
Synthetic & 2 & 1 & 54.7 (3.5) & 57.3 (3.3) & 63.2 (4.4) & 67.7 (3.5) & 70.7 (3.8) & 75.4 (3.8) & 77.1 (4.1) & 79.2 (4.8) \\
Synthetic & 2 & 4 & 50.3 (4.6) & 51.0 (3.7) & 54.4 (4.0) & 55.9 (3.2) & 57.7 (3.8) & 61.1 (2.7) & 62.3 (3.8) & 71.2 (3.6) \\
Synthetic & 2 & 16 & 49.9 (4.5) & 50.1 (4.7) & 51.3 (4.7) & 50.4 (3.7) & 51.7 (5.3) & 53.9 (2.7) & 55.3 (3.4) & 54.3 (3.4) \\
Synthetic & 2 & 64 & 49.3 (3.7) & 48.9 (3.2) & 50.7 (3.2) & 50.2 (3.3) & 49.8 (4.1) & 51.0 (2.6) & 50.9 (3.7) & 51.9 (3.8) \\
\midrule
Synthetic & 4 & 1 & 54.2 (4.7) & 58.2 (3.9) & 63.6 (2.9) & 68.0 (3.0) & 71.2 (3.8) & 74.6 (4.0) & 76.9 (4.0) & 79.2 (4.5) \\
Synthetic & 4 & 4 & 49.4 (3.3) & 51.5 (3.2) & 53.4 (3.5) & 53.8 (5.1) & 58.2 (3.4) & 60.0 (4.0) & 62.8 (3.5) & 69.2 (3.9) \\
Synthetic & 4 & 16 & 49.7 (3.2) & 50.7 (4.6) & 50.6 (3.8) & 50.0 (3.9) & 51.7 (4.0) & 52.2 (3.7) & 54.2 (3.5) & 56.6 (2.9) \\
Synthetic & 4 & 64 & 49.1 (4.1) & 50.9 (2.6) & 49.4 (4.5) & 50.5 (4.2) & 49.0 (4.2) & 50.5 (2.7) & 50.8 (3.4) & 53.7 (3.7) \\
\midrule
Synthetic & 8 & 1 & 53.1 (3.1) & 58.1 (3.3) & 64.6 (2.9) & 67.9 (3.7) & 70.9 (3.7) & 75.3 (4.7) & 76.5 (4.3) & 79.2 (4.7) \\
Synthetic & 8 & 4 & 50.5 (3.9) & 51.3 (4.2) & 52.9 (4.8) & 55.3 (3.1) & 56.6 (3.8) & 60.0 (3.3) & 63.2 (3.1) & 69.6 (3.7) \\
Synthetic & 8 & 16 & 50.7 (4.1) & 50.0 (3.6) & 49.9 (3.3) & 49.9 (4.7) & 50.8 (3.6) & 52.6 (3.9) & 53.3 (4.2) & 55.8 (4.0) \\
Synthetic & 8 & 64 & 50.4 (3.6) & 48.6 (4.1) & 50.5 (5.1) & 50.1 (3.4) & 50.8 (3.6) & 51.3 (4.1) & 49.7 (3.9) & 51.0 (3.7) \\
\midrule
Synthetic & 16 & 1 & 55.1 (3.6) & 57.5 (3.1) & 64.6 (3.5) & 68.7 (3.9) & 71.1 (3.7) & 74.7 (4.1) & 76.8 (4.5) & 79.1 (4.6) \\
Synthetic & 16 & 4 & 50.6 (3.3) & 51.5 (3.7) & 55.5 (3.1) & 55.9 (3.8) & 57.4 (4.3) & 60.2 (4.6) & 63.9 (4.1) & 69.8 (3.5) \\
Synthetic & 16 & 16 & 50.5 (4.3) & 50.2 (4.1) & 51.5 (4.5) & 51.4 (4.1) & 51.1 (3.6) & 53.3 (3.1) & 54.2 (3.0) & 57.2 (4.0) \\
Synthetic & 16 & 64 & 48.7 (3.4) & 50.2 (2.7) & 50.1 (4.4) & 49.5 (3.7) & 48.4 (3.6) & 51.0 (4.2) & 50.4 (4.0) & 52.0 (4.6) \\
\midrule
Synthetic & 32 & 1 & 52.9 (3.9) & 58.6 (4.9) & 63.6 (3.7) & 68.1 (3.0) & 71.7 (3.8) & 74.8 (4.5) & 76.2 (4.4) & 79.3 (4.7) \\
Synthetic & 32 & 4 & 50.7 (3.6) & 52.1 (3.8) & 51.7 (2.9) & 55.1 (2.9) & 57.4 (4.2) & 60.4 (3.8) & 63.2 (3.4) & 69.5 (3.6) \\
Synthetic & 32 & 16 & 49.7 (3.4) & 50.9 (3.2) & 50.7 (3.2) & 51.0 (4.2) & 50.3 (4.1) & 53.8 (4.4) & 53.3 (3.4) & 56.0 (3.6) \\
Synthetic & 32 & 64 & 50.7 (3.5) & 50.0 (5.1) & 50.5 (3.8) & 49.7 (3.9) & 47.9 (4.9) & 51.3 (3.5) & 49.9 (4.3) & 49.0 (4.5) \\
\midrule
ML-100K & 16 & 1 & 50.7 (4.0) & 48.1 (6.9) & 52.2 (7.0) & 54.5 (11.5) & 52.6 (11.3) & 54.4 (12.4) & 53.6 (11.8) & 55.6 (12.9) \\
ML-100K & 16 & 4 & 48.7 (8.9) & 52.1 (8.9) & 49.7 (8.8) & 52.0 (9.8) & 53.8 (7.9) & 52.2 (9.0) & 56.1 (10.2) & 54.9 (11.9) \\
ML-100K & 16 & 16 & 50.6 (6.2) & 49.6 (6.8) & 51.7 (7.9) & 53.9 (8.4) & 51.3 (7.1) & 47.1 (7.3) & 50.4 (9.2) & 50.2 (7.4) \\
ML-100K & 16 & 64 & 46.7 (7.7) & 48.0 (9.1) & 51.1 (9.0) & 51.9 (6.1) & 49.9 (7.7) & 51.2 (7.1) & 51.0 (10.4) & 50.7 (5.1) \\
\end{longtable}
\endgroup

\subsection{State-Scale Dependency Closure and Offline Trace-Edge Omission}\label{sec:supp-diag-scale}

Table~\ref{tab:supp-state-scale} reports the full structural topology of the eight single-seed Stage~C A-MEM states (\texttt{figure5\_\allowbreak state\_\allowbreak scale.csv}), supplementing the main text with maximum source-to-descendant hop distance, total recorded read-to-write trace edges $|E|$, non-root reachable descendant denominator $|C \setminus R|$, and exact filler-record identifiers for the two native \texttt{invalid\_link} degradations (\texttt{t\_fill\_039} in \texttt{48-blocked} and \texttt{t\_fill\_083} in \texttt{96-interleaved}). Note that total trace edges $|E|$ increase with state size and interleaved evolution ($45$ to $827$ edges), whereas conservative reachable closure in \texttt{96-interleaved} drops to $5/96$ ($5.21\%$, maximum hop distance $3$, $|C \setminus R|=4$) when target ingestion occurs at write position $14$. Because write ordering and target position vary jointly under a single build seed, these values establish structural non-monotonicity without isolating causal main effects.

Table~\ref{tab:supp-trace-ablation} presents all $336$ offline trace-omission configurations (\texttt{trace\_ablation.csv}) across the eight frozen graphs and seven retention levels $p \in \{0, 0.125, 0.25, 0.5, 0.75, 0.875, 1\}$. By listing the five individual random-omission seed missed counts ($s \in \{11,23,37,53,71\}$) alongside their within-state mean (SD) and the deterministic centered contiguous-gap count, the table losslessly encodes every configuration ($56 \times 5 + 56 \times 1 = 336$ rows). At intermediate retention levels ($p=0.5$ and $p=0.75$), contiguous logging outages sever entire temporal evolution chains and therefore miss substantially more reachable descendants than dispersed random omissions of identical edge count ($79.0\%$ vs.\ $36.9\%$ at $p=0.5$, $\Delta = +42.1$\,pp; $51.5\%$ vs.\ $13.7\%$ at $p=0.75$, $\Delta = +37.8$\,pp).

\begin{table}[htbp]
\centering
\caption{Extended structural topology of the eight single-seed Stage~C A-MEM states (\texttt{figure5\_state\_scale.csv}; \texttt{ministral-3:8b}, temperature $0$, seed $42$, zero answer probes).}\label{tab:supp-state-scale}
\scriptsize
\setlength{\tabcolsep}{3.0pt}
\begin{tabular}{@{}r l r r r r r r r l@{}}
\toprule
$|S|$ & Ordering & Pos. & $|C|$ & Closure (\%) & $|C{\setminus}R|$ & Holders & Max Hop & Edges $|E|$ & Degraded Links \\
\midrule
$12$ & \texttt{blocked} & $1$ & $12$ & $100.00$ & $11$ & $1$ & $9$ & $223$ & $0$ \\
$12$ & \texttt{interleaved} & $2$ & $11$ & $91.67$ & $10$ & $1$ & $2$ & $45$ & $0$ \\
$24$ & \texttt{blocked} & $1$ & $24$ & $100.00$ & $23$ & $1$ & $8$ & $277$ & $0$ \\
$24$ & \texttt{interleaved} & $4$ & $24$ & $100.00$ & $23$ & $1$ & $5$ & $279$ & $0$ \\
$48$ & \texttt{blocked}$^*$ & $1$ & $48$ & $100.00$ & $47$ & $1$ & $9$ & $373$ & $1$ (\texttt{t\_fill\_039}) \\
$48$ & \texttt{interleaved} & $7$ & $48$ & $100.00$ & $47$ & $1$ & $4$ & $629$ & $0$ \\
$96$ & \texttt{blocked} & $1$ & $96$ & $100.00$ & $95$ & $1$ & $8$ & $565$ & $0$ \\
$96$ & \texttt{interleaved}$^*$ & $14$ & $5$ & $5.21$ & $4$ & $1$ & $3$ & $827$ & $1$ (\texttt{t\_fill\_083}) \\
\bottomrule
\end{tabular}
\par\smallskip\noindent\parbox{\linewidth}{\raggedright\scriptsize
$^*$Single pure native \texttt{invalid\_link} degradation occurred during an unrelated filler-record write; adapter-internal closure and independent graph-traversal verification matched identically across all $8/8$ states.}
\end{table}

\begingroup
\scriptsize
\setlength{\tabcolsep}{3.2pt}
\renewcommand{\arraystretch}{1.06}
\begin{longtable}{@{}l r r r r l r@{}}
\caption{Complete 336-configuration offline trace-edge omission ledger across eight frozen Stage~C read-to-write graphs and seven retention fractions $p$ (\texttt{trace\_ablation.csv}). Denominator $|C \setminus R|$ counts current non-root records reachable from the target source in the complete graph ($p=1$). Random omission reports within-state mean (SD) missed percentage across five fixed permutation seeds ($s \in \{11,23,37,53,71\}$) and exact per-seed missed counts; contiguous gap removes the centered $(1-\lfloor p|E|\rfloor/|E|)$ segment of the chronological edge log. Bottom rows report equal-weight means (SD across states) over the eight states.}\label{tab:supp-trace-ablation}\\
\toprule
State ($|S|$, Org.) & $p$ & Kept/$|E|$ & $|C{\setminus}R|$ & Rand.\ Mean (SD) \% & Rand.\ Seed Counts ($11,23,37,53,71$) & Contig.\ Gap ($m/|C{\setminus}R|$, \%) \\
\midrule
\endfirsthead
\caption[]{Complete 336-configuration offline trace-edge omission ledger (Continued)}\\
\toprule
State ($|S|$, Org.) & $p$ & Kept/$|E|$ & $|C{\setminus}R|$ & Rand.\ Mean (SD) \% & Rand.\ Seed Counts ($11,23,37,53,71$) & Contig.\ Gap ($m/|C{\setminus}R|$, \%) \\
\midrule
\endhead
\midrule
\multicolumn{7}{r}{\emph{Continued on next page}} \\
\endfoot
\bottomrule
\endlastfoot
12, blocked & 0 & 0/223 & 11 & 100.0 ( 0.0) & \texttt{11,11,11,11,11} & 11/11 (100.0\%) \\
12, blocked & 0.125 & 27/223 & 11 & 100.0 ( 0.0) & \texttt{11,11,11,11,11} & 11/11 (100.0\%) \\
12, blocked & 0.25 & 55/223 & 11 &  92.7 (16.3) & \texttt{11,11,11,7,11} & 11/11 (100.0\%) \\
12, blocked & 0.5 & 111/223 & 11 &  61.8 (52.4) & \texttt{11,11,11,1,0} & 9/11 ( 81.8\%) \\
12, blocked & 0.75 & 167/223 & 11 &   0.0 ( 0.0) & \texttt{0,0,0,0,0} & 9/11 ( 81.8\%) \\
12, blocked & 0.875 & 195/223 & 11 &   0.0 ( 0.0) & \texttt{0,0,0,0,0} & 0/11 (  0.0\%) \\
12, blocked & 1 & 223/223 & 11 &   0.0 ( 0.0) & \texttt{0,0,0,0,0} & 0/11 (  0.0\%) \\
\midrule
12, interlv. & 0 & 0/45 & 10 & 100.0 ( 0.0) & \texttt{10,10,10,10,10} & 10/10 (100.0\%) \\
12, interlv. & 0.125 & 5/45 & 10 &  90.0 (10.0) & \texttt{10,10,8,9,8} & 8/10 ( 80.0\%) \\
12, interlv. & 0.25 & 11/45 & 10 &  68.0 (16.4) & \texttt{7,7,4,8,8} & 7/10 ( 70.0\%) \\
12, interlv. & 0.5 & 22/45 & 10 &  40.0 (12.2) & \texttt{4,6,3,3,4} & 3/10 ( 30.0\%) \\
12, interlv. & 0.75 & 33/45 & 10 &   4.0 ( 5.5) & \texttt{0,1,0,1,0} & 1/10 ( 10.0\%) \\
12, interlv. & 0.875 & 39/45 & 10 &   2.0 ( 4.5) & \texttt{0,1,0,0,0} & 1/10 ( 10.0\%) \\
12, interlv. & 1 & 45/45 & 10 &   0.0 ( 0.0) & \texttt{0,0,0,0,0} & 0/10 (  0.0\%) \\
\midrule
24, blocked & 0 & 0/277 & 23 & 100.0 ( 0.0) & \texttt{23,23,23,23,23} & 23/23 (100.0\%) \\
24, blocked & 0.125 & 34/277 & 23 & 100.0 ( 0.0) & \texttt{23,23,23,23,23} & 23/23 (100.0\%) \\
24, blocked & 0.25 & 69/277 & 23 &  99.1 ( 1.9) & \texttt{23,23,23,22,23} & 23/23 (100.0\%) \\
24, blocked & 0.5 & 138/277 & 23 &  61.7 (52.5) & \texttt{23,2,23,0,23} & 22/23 ( 95.7\%) \\
24, blocked & 0.75 & 207/277 & 23 &  20.0 (44.7) & \texttt{23,0,0,0,0} & 1/23 (  4.3\%) \\
24, blocked & 0.875 & 242/277 & 23 &   0.0 ( 0.0) & \texttt{0,0,0,0,0} & 0/23 (  0.0\%) \\
24, blocked & 1 & 277/277 & 23 &   0.0 ( 0.0) & \texttt{0,0,0,0,0} & 0/23 (  0.0\%) \\
\midrule
24, interlv. & 0 & 0/279 & 23 & 100.0 ( 0.0) & \texttt{23,23,23,23,23} & 23/23 (100.0\%) \\
24, interlv. & 0.125 & 34/279 & 23 &  97.4 ( 2.4) & \texttt{22,23,23,22,22} & 23/23 (100.0\%) \\
24, interlv. & 0.25 & 69/279 & 23 &  94.8 ( 4.8) & \texttt{22,22,23,22,20} & 22/23 ( 95.7\%) \\
24, interlv. & 0.5 & 139/279 & 23 &  20.0 (37.6) & \texttt{0,0,20,2,1} & 18/23 ( 78.3\%) \\
24, interlv. & 0.75 & 209/279 & 23 &  15.7 (35.0) & \texttt{0,0,18,0,0} & 4/23 ( 17.4\%) \\
24, interlv. & 0.875 & 244/279 & 23 &   0.0 ( 0.0) & \texttt{0,0,0,0,0} & 2/23 (  8.7\%) \\
24, interlv. & 1 & 279/279 & 23 &   0.0 ( 0.0) & \texttt{0,0,0,0,0} & 0/23 (  0.0\%) \\
\midrule
48, blocked$^*$ & 0 & 0/373 & 47 & 100.0 ( 0.0) & \texttt{47,47,47,47,47} & 47/47 (100.0\%) \\
48, blocked$^*$ & 0.125 & 46/373 & 47 & 100.0 ( 0.0) & \texttt{47,47,47,47,47} & 47/47 (100.0\%) \\
48, blocked$^*$ & 0.25 & 93/373 & 47 &  86.4 (30.4) & \texttt{47,47,47,15,47} & 47/47 (100.0\%) \\
48, blocked$^*$ & 0.5 & 186/373 & 47 &   3.0 ( 3.2) & \texttt{0,3,0,1,3} & 44/47 ( 93.6\%) \\
48, blocked$^*$ & 0.75 & 279/373 & 47 &   0.0 ( 0.0) & \texttt{0,0,0,0,0} & 44/47 ( 93.6\%) \\
48, blocked$^*$ & 0.875 & 326/373 & 47 &   0.0 ( 0.0) & \texttt{0,0,0,0,0} & 6/47 ( 12.8\%) \\
48, blocked$^*$ & 1 & 373/373 & 47 &   0.0 ( 0.0) & \texttt{0,0,0,0,0} & 0/47 (  0.0\%) \\
\midrule
48, interlv. & 0 & 0/629 & 47 & 100.0 ( 0.0) & \texttt{47,47,47,47,47} & 47/47 (100.0\%) \\
48, interlv. & 0.125 & 78/629 & 47 &  98.3 ( 2.3) & \texttt{47,45,47,45,47} & 47/47 (100.0\%) \\
48, interlv. & 0.25 & 157/629 & 47 &  54.9 (24.2) & \texttt{10,42,24,27,26} & 47/47 (100.0\%) \\
48, interlv. & 0.5 & 314/629 & 47 &   5.1 ( 4.1) & \texttt{1,1,5,4,1} & 38/47 ( 80.9\%) \\
48, interlv. & 0.75 & 471/629 & 47 &   0.0 ( 0.0) & \texttt{0,0,0,0,0} & 36/47 ( 76.6\%) \\
48, interlv. & 0.875 & 550/629 & 47 &   0.0 ( 0.0) & \texttt{0,0,0,0,0} & 3/47 (  6.4\%) \\
48, interlv. & 1 & 629/629 & 47 &   0.0 ( 0.0) & \texttt{0,0,0,0,0} & 0/47 (  0.0\%) \\
\midrule
96, blocked & 0 & 0/565 & 95 & 100.0 ( 0.0) & \texttt{95,95,95,95,95} & 95/95 (100.0\%) \\
96, blocked & 0.125 & 70/565 & 95 & 100.0 ( 0.0) & \texttt{95,95,95,95,95} & 95/95 (100.0\%) \\
96, blocked & 0.25 & 141/565 & 95 &  88.8 (24.9) & \texttt{95,42,95,95,95} & 92/95 ( 96.8\%) \\
96, blocked & 0.5 & 282/565 & 95 &  23.4 (42.9) & \texttt{3,7,0,6,95} & 68/95 ( 71.6\%) \\
96, blocked & 0.75 & 423/565 & 95 &   0.0 ( 0.0) & \texttt{0,0,0,0,0} & 27/95 ( 28.4\%) \\
96, blocked & 0.875 & 494/565 & 95 &   0.0 ( 0.0) & \texttt{0,0,0,0,0} & 13/95 ( 13.7\%) \\
96, blocked & 1 & 565/565 & 95 &   0.0 ( 0.0) & \texttt{0,0,0,0,0} & 0/95 (  0.0\%) \\
\midrule
96, interlv.$^*$ & 0 & 0/827 & 4 & 100.0 ( 0.0) & \texttt{4,4,4,4,4} & 4/4 (100.0\%) \\
96, interlv.$^*$ & 0.125 & 103/827 & 4 & 100.0 ( 0.0) & \texttt{4,4,4,4,4} & 4/4 (100.0\%) \\
96, interlv.$^*$ & 0.25 & 206/827 & 4 & 100.0 ( 0.0) & \texttt{4,4,4,4,4} & 4/4 (100.0\%) \\
96, interlv.$^*$ & 0.5 & 413/827 & 4 &  80.0 (27.4) & \texttt{4,2,4,4,2} & 4/4 (100.0\%) \\
96, interlv.$^*$ & 0.75 & 620/827 & 4 &  70.0 (27.4) & \texttt{4,2,2,4,2} & 4/4 (100.0\%) \\
96, interlv.$^*$ & 0.875 & 723/827 & 4 &  20.0 (44.7) & \texttt{4,0,0,0,0} & 0/4 (  0.0\%) \\
96, interlv.$^*$ & 1 & 827/827 & 4 &   0.0 ( 0.0) & \texttt{0,0,0,0,0} & 0/4 (  0.0\%) \\
\midrule
\textbf{8-state mean} & 0 & --- & 8 states & \textbf{100.0} ( 0.0) & $\Delta= +0.0$\,pp & \textbf{100.0\%} ( 0.0) \\
\textbf{8-state mean} & 0.125 & --- & 8 states & \textbf{ 98.2} ( 3.5) & $\Delta= -0.7$\,pp & \textbf{ 97.5\%} ( 7.1) \\
\textbf{8-state mean} & 0.25 & --- & 8 states & \textbf{ 85.6} (16.0) & $\Delta= +9.7$\,pp & \textbf{ 95.3\%} (10.4) \\
\textbf{8-state mean} & 0.5 & --- & 8 states & \textbf{ 36.9} (28.6) & $\Delta=+42.1$\,pp & \textbf{ 79.0\%} (22.0) \\
\textbf{8-state mean} & 0.75 & --- & 8 states & \textbf{ 13.7} (24.1) & $\Delta=+37.8$\,pp & \textbf{ 51.5\%} (40.2) \\
\textbf{8-state mean} & 0.875 & --- & 8 states & \textbf{  2.8} ( 7.0) & $\Delta= +3.7$\,pp & \textbf{  6.4\%} ( 5.8) \\
\textbf{8-state mean} & 1 & --- & 8 states & \textbf{  0.0} ( 0.0) & $\Delta= +0.0$\,pp & \textbf{  0.0\%} ( 0.0) \\
\end{longtable}
\endgroup

\subsection{Double-Blind Reproducibility and Artifact Verification Notes}\label{sec:supp-diag-repro}

\begingroup\emergencystretch=4em
To support independent verification while preserving double-blind review anonymity and strict data-protection boundaries, the companion reproducibility packager (\texttt{package\_\allowbreak anonymized\_\allowbreak artifact.py}) constructs and verifies \texttt{CSUR\_\allowbreak Reproducibility\_\allowbreak Artifact\_\allowbreak Anonymous.zip} under five engineering and governance rules:
\begin{enumerate}
\item \textbf{Repository-Relative Anonymized Layout and Dual-Digest Manifest.} The packager maps source workspace directories (\texttt{survey\_\allowbreak runtime\_\allowbreak 2026-\allowbreak 11-\allowbreak 11/}, \texttt{survey\_\allowbreak extension\_\allowbreak v5\_\allowbreak 2026-\allowbreak 11-\allowbreak 17/}, and \texttt{survey\_\allowbreak accounting\_\allowbreak 2026-\allowbreak 11-\allowbreak 11/}) into three self-contained archive modules---\texttt{survey\_\allowbreak runtime/} (Formal \texttt{v1}/\texttt{v2} 12-episode synthetic-marker adapters, frozen protocols, and \texttt{figure\_\allowbreak ready\_\allowbreak results/}), \texttt{survey\_\allowbreak extension\_v5/} (Stage~A parameter grids, Stage~B 4-history memory cleanup and \texttt{E2} reintroduction, Stage~C 8-state scale pilot, and \texttt{organized\_\allowbreak results\_\allowbreak 2026-\allowbreak 11-\allowbreak 17/}), and \texttt{survey\_\allowbreak accounting/} (analytical Gaussian RDP composition and Monte Carlo calibration). The archive README reports the packaged file count. The archive root contains \texttt{README.md}, \texttt{verify\_\allowbreak offline.py}, and \texttt{sanitized\_\allowbreak manifest.json}. Across the $10$ sanitized configuration, protocol, and analysis-manifest files (\texttt{runtime\_\allowbreak config*.json}, \texttt{formal\_\allowbreak protocol.json}, \texttt{prepare\_\allowbreak protocol.py}, and \texttt{analysis\_\allowbreak manifest.json}), workstation-specific filesystem paths are rewritten to \texttt{<LOCAL\_\allowbreak EMBEDDING\_\allowbreak DIR>/\allowbreak all-MiniLM-\allowbreak L6-v2} and repository-relative \texttt{survey\_\allowbreak extension\_v5/} prefixes. To preserve auditability against the five historical \texttt{freeze\_\allowbreak manifest.json} records, \texttt{sanitized\_\allowbreak manifest.json} pairs each pre-sanitization SHA-256 digest with its post-sanitization archive digest and verifies unchanged digests for all remaining frozen files.
\item \textbf{Pinned Upstream Baseline and Deterministic Local Inference.} Memory diagnostics pin the open-source MIT-licensed A-MEM baseline to upstream commit \texttt{f9a173de\allowbreak c325d84f\allowbreak cc97e782\allowbreak 8775c7b2\allowbreak 41c33466}, with Formal~\texttt{v2}, Stage~B, and Stage~C applying the prospective patch \texttt{revision\_v2/\allowbreak amem\_\allowbreak upstream\_\allowbreak v2.patch} to log and isolate pure \texttt{invalid\_\allowbreak link} events without clamping or remapping indices. Third-party upstream clones (\texttt{amem/\allowbreak upstream/}) are excluded from archive redistribution and cloned separately per \texttt{survey\_\allowbreak runtime/\allowbreak README.md} prior to running \texttt{test\_\allowbreak structure.py}. All LLM-backed runs bind exclusively to a project-isolated loopback Ollama endpoint (\texttt{http://\allowbreak 127.0.0.1:\allowbreak 11435}) serving \texttt{ministral-\allowbreak 3:8b} (pinned SHA-256 digest \texttt{1922accd\allowbreak 5827ebe6\allowbreak 829e5363\allowbreak 69195db2\allowbreak 5eaf6645\allowbreak 28dc6620\allowbreak 6d646ea3\allowbreak bb386b71}) with \texttt{temperature=\allowbreak 0}, \texttt{num\_ctx=\allowbreak 32768}, single-concurrency GPU execution, fixed build seeds (\texttt{b1=42}, \texttt{b2=43}), zero external network egress, and zero paid API calls. Stage~B development ran on an idle identical-capacity workstation GPU (\texttt{GPU1}, $32\,\text{GiB}$ cap) documented as a pre-submission resource allocation note while keeping formal hardware freezes separate.
\item \textbf{Process-Isolated Non-Mutating Probe Execution.} Every memory read probe executes in a fresh worker subprocess from a serialized state snapshot and verifies exact SHA-256 state-digest equality before and after probe execution ($576/576$ in Formal~\texttt{v2}; $480/480$ in Stage~B), ensuring that evaluation queries cannot mutate persistent memory, neighbor links, or subsequent checkpoints. Independent cold-restart workers ($96/96$ in Formal~\texttt{v2}; $25/25$ in Stage~B) further confirm deterministic snapshot restoration.
\item \textbf{Synthetic-Marker Hygiene and Excluded Raw Artifacts.} Memory histories, target values, and control probes use purely synthetic markers containing zero real user records or personally identifiable information. The packager strictly excludes model weights and local embedding checkpoints, raw multi-megabyte LLM backend call transcripts (\texttt{backend\_\allowbreak calls.jsonl}, \texttt{probes.jsonl}, \texttt{ollama.log}) and serialized runtime memory states (retaining compact Stage~C \texttt{trace.json} lineage edge graphs and hash-locked structured CSV/JSON ledgers), raw MovieLens~100K rating files and sweep intermediates, and macOS AppleDouble (\texttt{.\_*}) sidecar files. For Stage~A MovieLens~100K experiments, the archive distributes deterministic feature-projection and evaluation scripts (seed \texttt{20261117}) alongside aggregate CSV summaries; reproducers obtain raw MovieLens~100K ratings directly from the official GroupLens release.
\item \textbf{Offline Verification Inside a Temporary Unpack Directory.} Invoking \texttt{python3} \texttt{package\_\allowbreak anonymized\_\allowbreak artifact.py} \texttt{--output} \texttt{build/\allowbreak CSUR\_\allowbreak Reproducibility\_\allowbreak Artifact\_\allowbreak Anonymous.zip} \texttt{--verify} (or running \texttt{python3} \texttt{verify\_\allowbreak offline.py} after extracting the archive) executes deterministic CPU verification using packaged data and the dependencies listed in the archive README: (i) zero host-path and sidecar hygiene scans alongside the complete \texttt{sanitized\_\allowbreak manifest.json} and five-manifest freeze-hash verification; (ii) structural, denominator, and claim-boundary checks matching \texttt{validate\_\allowbreak organized\_\allowbreak results.py} together with all $17/17$ SHA-256 digests in \texttt{organized\_\allowbreak results\_\allowbreak 2026-\allowbreak 11-\allowbreak 17/\allowbreak artifact-\allowbreak manifest.json} (invoking \texttt{validate\_\allowbreak organized\_\allowbreak results.py} directly on its raw-source re-hash step requires the omitted raw \texttt{probes.jsonl}/\allowbreak \texttt{backend\_\allowbreak calls.jsonl} logs and sweep files); (iii) Stage~C eight-state \texttt{trace.json} closure recomputation and recomputation of the 336 packaged trace-omission rows; and (iv) all $11$ offline \texttt{pytest}/\allowbreak \texttt{unittest}/\allowbreak \texttt{validate\_\allowbreak protocol.py} suites (\texttt{test\_\allowbreak reader} across four adapter snapshots, \texttt{test\_\allowbreak metrics}, \texttt{validate\_\allowbreak protocol.py}, \texttt{test\_\allowbreak synthetic\_\allowbreak parameter\_\allowbreak update}, \texttt{test\_\allowbreak movielens\_\allowbreak parameter\_\allowbreak update}, \texttt{test\_\allowbreak structural\_\allowbreak controls}, \texttt{test\_\allowbreak dev\_\allowbreak data}, \texttt{test\_\allowbreak reader\_\allowbreak v5}, \texttt{test\_\allowbreak runtime\_\allowbreak controls}, \texttt{test\_\allowbreak scale\_\allowbreak data}, and \texttt{test\_\allowbreak accounting}).
\end{enumerate}

\paragraph{Offline verification and model reruns.}
The archive is ready for offline verification after extraction. The historical model runner checks the original freeze digests and rejects the sanitized \texttt{runtime\_\allowbreak config.json}. To prepare a local rerun, first verify the extracted archive against its sanitized manifest. Then use a separate working copy, restore the pinned upstream revision and patch, provision the local model and embeddings, and configure the local paths. Record those inputs in a separate rerun manifest and use a runner that validates it. The archive retains the original freeze and sanitized digests for provenance. The supplied historical runner does not implement the local rerun path.
\endgroup

\subsection{Fixed Synthetic-Marker Diagnostic: Deletion Scope versus Utility Retention}\label{sec:diag-12ep}

To illustrate the withdrawal claims in Main Table~5, we run a fixed synthetic-marker diagnostic on a versioned A-MEM store across 12 synthetic episodes (5 initial records each) $\times$ 2 deterministic build seeds, yielding 24 paired episode-builds (Tables~\ref{tab:supp-run-quality}--\ref{tab:supp-memory-outcomes}). Within each episode-build, keep (\texttt{K}), root-only deletion (\texttt{D}), and conservative observed read-to-write closure deletion (\texttt{L}) start from the same saved snapshot. We verify their state digests before intervention. The reference arm (\texttt{N}) uses the same episode and build seed to construct a fresh store from a target-free source history. All four arms are probed at $t_0$ and after five target-free updates ($t_1$, called $t_5$ in Stage~B). The 576 probes cover both checkpoints.

\begin{figure}[htbp]
\centering
\includegraphics[width=\textwidth,height=0.40\textheight,keepaspectratio]{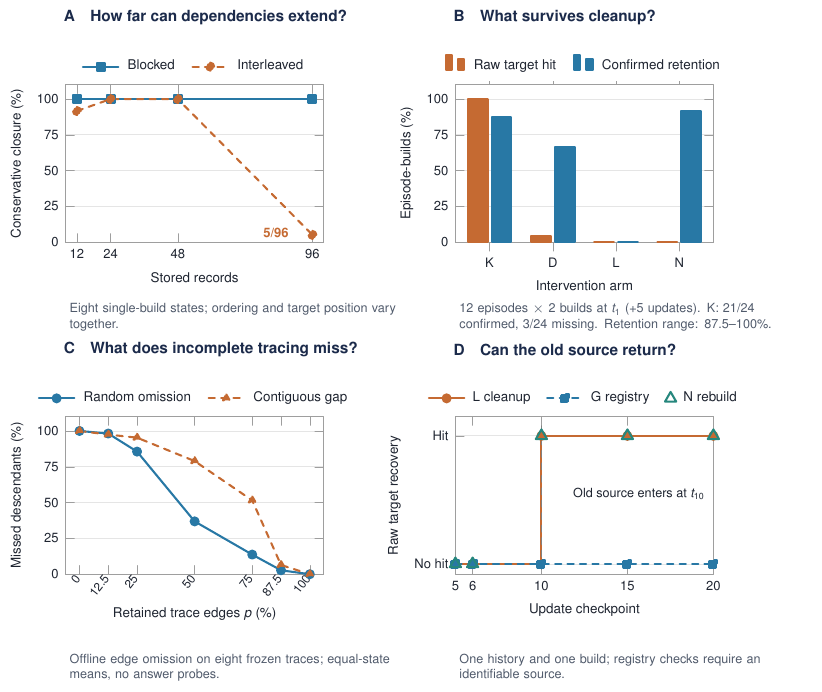}
\caption{Four unpooled memory-withdrawal diagnostics (Tables~\ref{tab:supp-run-quality}--\ref{tab:supp-guard-rejections}, \ref{tab:supp-state-scale}--\ref{tab:supp-trace-ablation}). \textbf{A}: Read-to-write closure share across eight single-seed states (ordering and target position co-vary). \textbf{B}: Target recovery and non-target retention at $t_1$ ($+5$ updates, 12 episodes $\times$ 2 builds). \texttt{K} has 21 confirmed retentions and three missing readings, which yields a retention range of 87.5--100\%. \textbf{C}: Missed descendants across 336 offline trace-omission configs (graph reachability, not answer leakage). \textbf{D}: Single-history reintroduction at $t_{10}$ (\texttt{L} closure, \texttt{G} closure plus registry, \texttt{N} rebuild).}
\label{fig:supp-withdrawal}
\Description{Four-panel vector diagnostic figure. Panel A plots closure percentages across 12--96 records for blocked (100 percent) and interleaved (91.7, 100, 100, and 5.2 percent [5/96]) episode-builds. Panel B shows t1 (+5 updates) bars for K, D, L, N (raw hits: 100, 4.2, 0, 0 percent; confirmed retention: 87.5, 66.7, 0, 91.7 percent, with a range of 87.5 to 100 percent for K due to missing readings). Panel C plots missed descendants versus edge retention p (79.0 vs 36.9 percent at p=0.5; 51.5 vs 13.7 percent at p=0.75). Panel D shows L and N jumping to 1 at t10 reintroduction while G stays 0.}
\end{figure}

All 24 episode-builds completed. All 576 probes returned schema-valid JSON and preserved the state digest. The evidence guard accepted 565 probes and rejected 11. Raw-output scoring retains the rejected responses. Raw target hits are 24/24 (\texttt{K}), 3/24 (\texttt{D}), 0/24 (\texttt{L}), and 0/24 (\texttt{N}) at $t_0$, and 24/24 (\texttt{K}), 1/24 (\texttt{D}), 0/24 (\texttt{L}), and 0/24 (\texttt{N}) at $t_1$ (Figure~\ref{fig:supp-withdrawal}B). Non-target exact-field retention at $t_1$ is 21/24 (\texttt{K}, 3 missing), 16/24 (\texttt{D}, losses concentrated where target and control fields co-occurred in the root record), 0/24 (\texttt{L}), and 22/24 (\texttt{N}). Because conservative closures saturated all five initial records, \texttt{L} eliminated tested initial personalization alongside target recovery, and \texttt{D} retained 16/24 tested fields with 1/24 post-update raw recovery. With three unresolved utility readings in \texttt{K}, its retention is identified only within 87.5--100\%. The comparison measures the cost of saturation. AfterDeleteBench isolates a tracked descendant under a complete harness, and Bemula conditions preservation on paired controls~\cite{saleh2026afterdeletebench,xie2026bemula}. A comparison of selective cleanup in this store would need both partial deletion and retained utility measured on the same states.

\subsection{Extended Diagnostics: Protection Units, Cleanup Scope, and Source Reintroduction}\label{sec:diag-extended}

\begin{figure}[htbp]
\centering
\includegraphics[width=\textwidth,height=0.25\textheight,keepaspectratio]{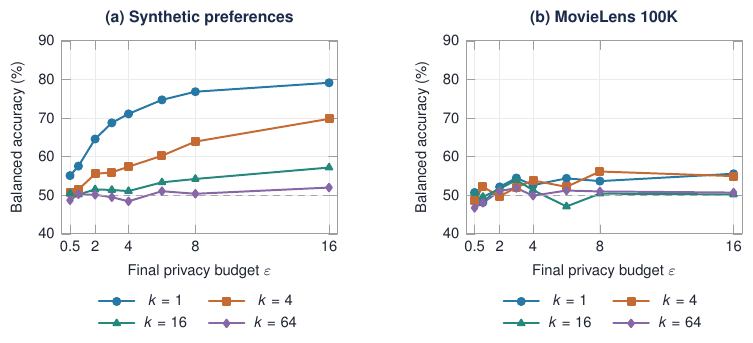}
\caption{Protection-unit granularity versus balanced accuracy in Stage~A ($T=16$) under approximate $(\varepsilon,\delta)$-DP ($\delta=10^{-6}$, $N=64$ fixed training slots, $C=0.25$, $\Delta_k=2kC/N$) across $\varepsilon\in\{0.5,\dots,16\}$ and $k\in\{1,4,16,64\}$ for (a) synthetic preferences and (b) MovieLens~100K. MovieLens label adjacency holds the selected user, ordered items, and projected features fixed. The test pairs are fixed for evaluation. Reported means come from reproducibly seeded benchmark runs (Tables~\ref{tab:supp-param-baselines}--\ref{tab:supp-param-grid}).}
\label{fig:supp-param}
\Description{Two vector line charts plotting balanced accuracy against varepsilon at T=16 under delta=1e-6 and N=64 slots for k=1, 4, 16, 64. Synthetic curves (a) reach 79.1, 69.8, 57.2, and 52.0 percent at varepsilon=16; MovieLens 100K curves (b) cluster near 56.3 percent, reaching 55.6, 54.9, 50.2, and 50.7 percent.}
\end{figure}

Stage~A tests a logistic preference head on 24 synthetic profiles and 24 MovieLens~100K users. Across settings, baselines, and repeated noise seeds, there are 27,840 non-zero training trajectories (Tables~\ref{tab:supp-param-baselines}--\ref{tab:supp-param-grid}). The randomized mechanism protects changes to up to $k$ binary labels in $N=64$ fixed training slots under approximate $(\varepsilon,\delta)$-DP ($\delta=10^{-6}$, $\ell_2$ clip $C=0.25$, sensitivity $\Delta_k = 2kC/N$) for $k \in \{1, 4, 16, 64\}$ protected slots ($k=64$ permits replacing all 64 labels). For MovieLens, the guarantee conditions on the selected user, ordered item slots, and projected genre features. The held-out item--label pairs are fixed in the comparison. User selection and item history fall outside this label adjacency. The guarantee concerns the randomized head under private noise draws. We use fixed seeds for the benchmark and report aggregate accuracy. These reported aggregates have no separate release guarantee. At $T=16$ and $\varepsilon=16$ (Figure~\ref{fig:supp-param}), widening $k$ from $1$ to $4, 16,$ and $64$ drops synthetic balanced accuracy from $79.1\%$ to $69.8\%$, $57.2\%$, and $52.0\%$. On MovieLens~100K (clipped baseline $56.3\%$ using genre metadata), $\varepsilon=16$ accuracies are $55.6\%$, $54.9\%$, $50.2\%$, and $50.7\%$. These are descriptive comparisons at one budget and update count. The synthetic means decline across this slice, while the MovieLens means lie below the clipped baseline and vary nonmonotonically. Equal nominal budgets apply to different protected units.

\begin{figure}[htbp]
\centering
\includegraphics[width=\textwidth,height=0.20\textheight,keepaspectratio]{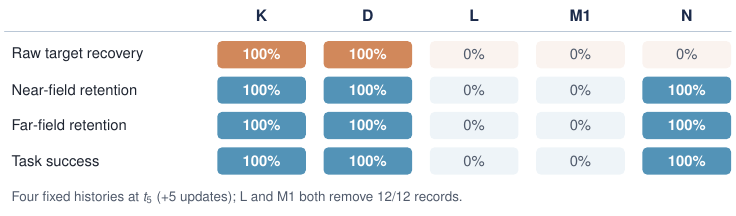}
\caption{Stage~B memory-cleanup outcomes at $t_5$ ($+5$ updates, aligned with $t_1$ in Section~\ref{sec:diag-12ep}; 4 histories, Tables~\ref{tab:supp-memory-outcomes}--\ref{tab:supp-evidence-contingency}). \texttt{L} and \texttt{M1} both delete all 12 initial records. Their agreement tests saturation. Raw hits reflect unguarded generation.}
\label{fig:supp-cleanup}
\Description{Matrix heatmap at t5 across K, D, L, M1, N: raw hits are 100 percent for K, D and 0 percent for L, M1, N; near-field, far-field, and task retention are 100 percent for K, D, N and 0 percent for L, M1.}
\end{figure}

Stage~B tests four 12-record synthetic histories across five builds (\texttt{dev\_h2} built twice, averaged first), adding cardinality-matched random deletion \texttt{M1}. This development-stage batch did not advance to formal 24-history expansion because buffered calls exceeded the pre-specified budget gate ($1{,}909 \times 1.30 = 2{,}481.7 > 2{,}400$). Section~S4.1 identifies the protocol and acceptance records. At five updates ($t_5$, aligned with $t_1$ in Section~\ref{sec:diag-12ep}; Figure~\ref{fig:supp-cleanup}), raw target recovery is $100\%$ for \texttt{K}, \texttt{D} versus $0\%$ for \texttt{L}, \texttt{M1}, \texttt{N}; near-field, far-field, and task utility reach $100\%$ for \texttt{K}, \texttt{D}, \texttt{N} versus $0\%$ for \texttt{L}, \texttt{M1} under $12/12$ saturation. At that dose both arms remove the same entire initial store. A comparison that isolates provenance targeting would need non-saturated states and deletion arms that remove the same number of records. Across 480 probes ($476$ guard-accepted), three rejected \texttt{D} probes emitted raw target markers. At $t_5$, history-weighted recovery for \texttt{D} is $75\%$ at the guarded output and $100\%$ at the raw output. The two rates describe different observation interfaces. On \texttt{dev\_h2/b1} ($t_6$--$t_{20}$; Figure~\ref{fig:supp-withdrawal}D), re-supplying the retracted source at $t_{10}$ restores target hits ($0 \to 1$) for \texttt{L} and \texttt{N}, and the registry in \texttt{G} (\texttt{L} plus source registry) rejected that exact source identifier in the tested replay.

\subsection{Memory Scale, Write Ordering, and Conservative Dependency Closure}\label{sec:diag-scale}

Stage~C tests closure saturation across eight single-seed (\texttt{seed=42}, temperature~$0$) A-MEM states of $12, 24, 48,$ and $96$ records under \texttt{blocked} versus \texttt{interleaved} write orderings ($360$ writes, zero probes; telemetry in Tables~\ref{tab:supp-run-quality}, \ref{tab:supp-state-scale}--\ref{tab:supp-trace-ablation}). Every state keeps one exact \texttt{content} hit, whereas closures span 5--96 records (Table~\ref{tab:supp-scale-summary}, Figure~\ref{fig:supp-withdrawal}A): all \texttt{blocked} builds saturate ($100.0\%$), while \texttt{96-interleaved} drops to $5/96$ ($5.2\%$ after $91.7\%, 100.0\%, 100.0\%$), leaving non-saturated deletion utility untested. Because write ordering and target arrival index co-vary (\texttt{blocked} pos.~1 vs.\ \texttt{interleaved} pos.~$2,4,7,14$) under one seed, causal ordering effects remain confounded.

\begin{table}[htbp]
\centering
\caption{Stage~C State-Scale and Conservative Dependency-Closure Diagnostic}\label{tab:supp-scale-summary}
\footnotesize
\setlength{\tabcolsep}{5pt}
\renewcommand{\arraystretch}{1.06}
\begin{tabular}{r l r r r r}
\toprule
\textbf{Records} & \textbf{Ordering} & \textbf{Target Pos.} & \textbf{Closure Size} & \textbf{Closure Share} & \textbf{\texttt{content} Hits} \\
\midrule
12 & Blocked            &  1 & 12 & 100.0\% & 1 \\
12 & Interleaved        &  2 & 11 &  91.7\% & 1 \\
24 & Blocked            &  1 & 24 & 100.0\% & 1 \\
24 & Interleaved        &  4 & 24 & 100.0\% & 1 \\
48 & Blocked$^*$        &  1 & 48 & 100.0\% & 1 \\
48 & Interleaved        &  7 & 48 & 100.0\% & 1 \\
96 & Blocked            &  1 & 96 & 100.0\% & 1 \\
96 & Interleaved$^*$    & 14 &  5 &   5.2\% & 1 \\
\bottomrule
\end{tabular}
\par\smallskip\noindent\parbox{\linewidth}{\raggedright\footnotesize
\emph{Note:} Single deterministic build (\texttt{seed=42}) per row; Target Pos.\ is 1-based ingestion index; asterisks ($^*$) denote single native \texttt{invalid\_link} filler fallbacks. Extended topology and telemetry appear in Supplementary Section~S4 (Tables~\ref{tab:supp-run-quality} and~\ref{tab:supp-state-scale}).}
\end{table}

Across 336 offline trace-omission configurations over these eight read-to-write graphs (Figure~\ref{fig:supp-withdrawal}C; Table~\ref{tab:supp-trace-ablation}), contiguous chronological log gaps have higher mean missed-descendant rates than random omissions at the two reported retention levels ($79.0\%$ vs.\ $36.9\%$ at $p=0.5$; $51.5\%$ vs.\ $13.7\%$ at $p=0.75$). We weight states equally and average the five random seeds within each state. These structural rates measure reachability loss on the recorded graphs.

\section{Operation-Level Evidence Matrix}
\label{sec:supp-operation-evidence}

\subsection{Unit and decision rules}

The matrix indexes a source, a specific protective operation, and its reported evaluation or formal setting. Source identifiers refer to the pinned references in Table~\ref{tab:supp-75-ledger}. We split a study when the protected view, intervention, or setting changes. For example, Reglos's tombstone update and reader policy share a joint behavioral result, while its manual deletion control has a narrower documented scope. If a test reports a pipeline outcome, the corresponding rows identify that shared outcome and leave the component effect unassigned. Benchmark controls without an identifiable protection step remain in Table~\ref{tab:supp-75-ledger}. Table~\ref{tab:supp-22-mechanisms-ledger} supplies fuller trust assumptions and primary-text locators for the 22 exemplars.

For an individual row, \texttt{D} requires evidence that a protected update affects durable single-user state carried across decisions or interactions. A candidate revision within an optimization run is not evidence of persistence. \texttt{S} denotes protection evaluated at a read or release interface over retained personal state, including fixed profiles instantiated in personal-agent tasks. \texttt{T} denotes an adjacent setting or a transferable primitive. The same source can contribute more than one code. A dash marks a feature whose scope is described but whose protection claim has no separately creditable outcome or applicable formal bound. The source-level tally in Table~\ref{tab:supp-75-summary} uses the closest credited code (\texttt{D} before \texttt{S} before \texttt{T}) once per source.

Setting tags describe the evidence independently of distance. \textbf{F} denotes a formal result within its stated assumptions. \textbf{B} is a fixed-dataset, isolated-component, or replay benchmark. \textbf{L} is a controlled longitudinal personal history or simulated workflow with evidence of state carryover. \textbf{E} is a field study or production deployment. \textbf{I} marks design or implementation evidence without a reported protection result. A dash in the distance column excludes the row from protection counts; for example, \texttt{---} / E records a field setting without a supported privacy outcome.

\subsection{Coverage and deduplication}

Table~\ref{tab:supp-operation-family-summary} counts coded rows and gives distinct source counts in parentheses. A study can appear in multiple families and distance columns. The \texttt{D} evidence covers memory write or maintenance (seven sources) and memory withdrawal (two), all in controlled settings. HomeLLaMA includes real participants in a lab protocol. The skill operations were evaluated in transfer settings. The matrix additionally retains uncredited design or deployment rows.

\begin{table}[htbp]
\centering
\footnotesize
\setlength{\tabcolsep}{3pt}
\renewcommand{\arraystretch}{1.08}
\caption{Operation Rows and Distinct Sources by Family and Evidence Distance}
\label{tab:supp-operation-family-summary}
\begin{tabular}{@{}>{\raggedright\arraybackslash}p{0.48\linewidth}rrrr@{}}
\toprule
\textbf{Operation Family} & \textbf{D} & \textbf{S} & \textbf{T} & \textbf{Total} \\
\midrule
Prompt generation, tuning and selection & 0 (0) & 0 (0) & 13 (10) & 13 (10) \\
Parameter and preference learning & 0 (0) & 0 (0) & 51 (46) & 51 (46) \\
Memory write and maintenance & 9 (7) & 0 (0) & 11 (10) & 20 (17) \\
Skill and harness adaptation or sharing & 0 (0) & 0 (0) & 12 (11) & 12 (11) \\
Read, release and execution & 0 (0) & 23 (21) & 35 (33) & 58 (53) \\
Withdrawal and audit & 2 (2) & 0 (0) & 18 (17) & 20 (19) \\
Cross-operation accounting and aggregation & 0 (0) & 0 (0) & 10 (10) & 10 (10) \\
\midrule
\textbf{All families, deduplicated source totals} & 11 (9) & 23 (21) & 150 (124) & 184 (144) \\
\bottomrule
\end{tabular}
\end{table}

\begingroup
\footnotesize
\setlength{\tabcolsep}{2.5pt}
\renewcommand{\arraystretch}{1.14}
\begin{longtable}{@{}>{\raggedleft\arraybackslash}p{0.048\linewidth}>{\raggedright\arraybackslash}p{0.108\linewidth}>{\raggedright\arraybackslash}p{0.27\linewidth}>{\centering\arraybackslash}p{0.082\linewidth}>{\raggedright\arraybackslash}p{0.435\linewidth}@{}}
\caption{Source, Protective Operation, and Evidence Setting}
\label{tab:supp-operation-evidence}\\
\toprule
\textbf{\#} & \textbf{Family} & \textbf{Carrier and Operation} & \textbf{Code / Setting} & \textbf{Protected View, Result, or Scope} \\
\midrule
\endfirsthead
\caption[]{Source, Protective Operation, and Evidence Setting (Continued)}\\
\toprule
\textbf{\#} & \textbf{Family} & \textbf{Carrier and Operation} & \textbf{Code / Setting} & \textbf{Protected View, Result, or Scope} \\
\midrule
\endhead
\midrule
\multicolumn{5}{r}{\footnotesize\emph{Continued on next page}} \\
\endfoot
\bottomrule
\endlastfoot
1 & Prompt & Text prompt / private candidate scoring & \texttt{T} / F+B & Noised fitness scores under example adjacency. Scorer holds raw data. Noisy scores support downstream selection. \\
2 & Prompt & Text prompt / private candidate generation & \texttt{T} / F+B & Token-level mechanism composes over released text for record changes in the training split. \\
2 & Prompt & Text prompt / private candidate selection & \texttt{T} / F+B & Exponential selection over validation scores. Parallel accounting requires disjoint affected units across the two splits. \\
3 & Learning & LoRA / group accounting for example clipping & \texttt{T} / F+B & Contribution cap and group privacy bound one participant's records in a cohort. \\
3 & Learning & LoRA / user-wise gradient clipping and noise & \texttt{T} / F+B & User-wise update protects released adapter under cohort adjacency and the sampling accountant. \\
4 & Learning & Reward head / private tuple-based fitting & \texttt{T} / F+B & Trusted trainer protects full tuples in the released head. Public generator and fixed features support downstream selection. \\
5 & Learning & Reward head / central objective perturbation & \texttt{T} / F+B & Choice-label DP of reward parameters, with prompt and response pairs held public. \\
5 & Learning & Reward head / local randomized response & \texttt{T} / F+B & Randomized response privatizes the label at the feedback endpoint. The learner receives perturbed labels. \\
6 & Prompt & Soft prompt / secure tuning forward pass & \texttt{T} / F+B & MPC protects tokens and shares from non-colluding compute servers. The data owner receives logits. \\
7 & Memory & Personal memory / sanitize before cloud write & \texttt{D} / L & Persistent cloud store and retrieval see typed placeholders, conditional on the edge detector and local exact-value map. \\
7 & Release & Personal memory / local output restoration & \texttt{S} / L & Cloud inference uses placeholders. The trusted edge restores exact values. Personalization and masking are tested in the pipeline. \\
8 & Release & Personal memory / capped adaptive view selection & \texttt{S} / F+B & Pure DP for a fixed-layout attribute group, with a trusted reader; Thm.~1 and Cor.~1 bound covered transcript changes. \\
9 & Memory & Personal memory / oblivious ingestion & \texttt{D} / F+L & Attested ingest protects raw record and padded access trace against the host under the declared leakage model. \\
9 & Memory & Personal memory / oblivious maintenance & \texttt{D} / F+L & Expiry, indexing and compression run in enclave. The host sees a fixed trace shape. The claim concerns access traces. \\
9 & Release & Personal memory / oblivious recall & \texttt{S} / F+L & Indexed search gives host confidentiality under declared leakage. The client remains an authorized recipient. \\
10 & Skill & Skill candidates / screening and held-out tests & \texttt{T} / B & E3 tests revisions in five fixed scenarios and sandbox filtering (Apps.~E--F). No committed version is traced through a later interaction; the external judge sees candidates. \\
10 & Release & GUI observation / static screen disclosure gate & \texttt{T} / B & Static arbitration and screenshot checks (E1/E2) evaluate screen exposure. They do not test protected persistent skill updates. \\
15 & Learning & Shared clinical LoRA / risk-reward update & \texttt{T} / B & Offline cohort optimization uses empirical disclosure risk. The reported privacy outcome concerns outbound disclosures. \\
15 & Memory & Shared lessons / distilled writeback & \texttt{T} / B & Test-time writes use fixed weights and share lessons across patients. The evaluation measures the combined system. \\
15 & Release & Clinical consultation / outbound disclosure & \texttt{T} / B & Evaluates disclosed content and linkage across visits under a shared policy. Report covers recipient-facing disclosure. \\
16 & Memory & Personal memory / selective health writeback & \texttt{D} / L & Simulated patient history and privacy policies exercise writeback and scope rules. \\
16 & Release & Personal memory / layer-specific retrieval and disclosure & \texttt{S} / L & Privacy probes use the same 90-day simulated patient histories. Withholding sensitive profile, procedure and episode traces is scoped to tested policy. \\
17 & Skill & Shared skill patch / pre-commit audit & \texttt{T} / B & Client sanitizer and private-value tests filter payloads before version merge. The server still sees commit metadata. \\
18 & Prompt & Candidate artifact / private meta-selection & \texttt{T} / F & Randomized stopping protects selected output given DP candidate oracles and hidden rejected draws. \\
19 & Release & Policy / single private action sampling & \texttt{T} / F+B & Bounds a single action's dependence on a preference label. Fitted policy and multiple actions require separate treatment. \\
20 & Learning & Continual parameters / objective perturbation & \texttt{T} / F+B & Tuple-level DP for model versions under specified task partitions and prescribed reuse. \\
21 & Accounting & Interleaved updates / concurrent accounting & \texttt{T} / F & Composition results for interactive transcripts when each component satisfies stated privacy semantics. \\
23 & Memory & Personal memory / pre-ingest masking & --- / I & Implemented in \S3.1 and App.~A. Reported sensitive-answer results do not isolate write-side secret retention or model-input exposure. \\
23 & Withdrawal & Personal memory / tombstone update & \texttt{D} / L & Append-only retraction changes eligibility for reuse. Pairing it with the harness policy improves benchmark non-use (Fig.~5 and \S4.3). \\
23 & Release & Personal memory / harness non-reuse policy & \texttt{S} / L & The benchmark tests the harness policy together with tombstones and measures response behavior. \\
23 & Withdrawal & Personal memory / manual UI deletion & --- / I & Scope of user controls is documented. No all-copy deletion, residue probe, or withdrawal proof is reported for the UI action. \\
23 & Skill & Shared harness / candidate patch screening & \texttt{T} / B & Offline patch evaluation tests reuse rules and task behavior on benchmark questions, rather than an adaptive per-user harness. \\
24 & Withdrawal & Memory / blocklist and dependency cleanup & \texttt{T} / B & Graph-tracked descendants are pruned or invalidated. Ablations assess forgetting in a public medical QA loop. \\
24 & Withdrawal & Parameters / synchronized suppression & \texttt{T} / B & KL-based update suppresses parametric residue. Joint and ablation results cover QA, retrieval and empirical membership tests. \\
25 & Release & Personal context / tiered cloud masking & \texttt{S} / B & Masks contextual values ahead of model calls; PrivacyLens compares a full Kernel ablation, without isolating this gate. \\
26 & Prompt & Soft prompt / private gradient update & \texttt{T} / F+B & DP-SGD clips and noises record-level updates to released continuous prompts. \\
26 & Prompt & Demonstrations / noisy teacher votes & \texttt{T} / F+B & Disjoint teacher prompts label public inputs via PATE. Release contains discrete demonstrations. \\
27 & Prompt & Demonstrations / DP synthesis & \texttt{T} / F+B & The mechanism releases exemplars from a fixed corpus using DP generation accounting. \\
28 & Learning & Parameters / private zeroth-order tuning & \texttt{T} / F+B & Privatizes per-example loss differences used to choose scalar steps under example adjacency. \\
29 & Learning & Shared language model / device-level FL training & \texttt{T} / F+E & Production Gboard DP-FTRL protects device contribution under stated participation constraints. \\
31 & Learning & DPO policy / private label update & \texttt{T} / F+B & Bounds preference-label influence. Prompt and response text and reference model require their own protection assumptions. \\
32 & Learning & Sparse reward model / sample-private estimator & \texttt{T} / F & Sample-DP estimation bound for sparse Bradley--Terry--Luce comparisons. \\
32 & Learning & Sparse reward model / label-private estimator & \texttt{T} / F & Randomized response bounds labels with fixed public pairs and distinct estimation-error guarantees. \\
33 & Prompt & Soft prompt factors / local DP upload & \texttt{T} / F+B & Uploaded low-rank factors are locally privatized. Personalized residuals remain on the client. \\
33 & Prompt & Shared soft prompt / central aggregation & \texttt{T} / F+B & Distributed prompt receives central DP protection under the federation assumptions. \\
34 & Accounting & Shared parameters / secure update summation & \texttt{T} / F+B & Crypto protocol hides individual updates before aggregate release under threshold and protocol assumptions. \\
35 & Accounting & Continual training / noisy prefix sums & \texttt{T} / F+B & Tree-based private aggregation across continual updates. Repeated participation requires the stated accountant. \\
36 & Release & Continual classifier / private label-set release & \texttt{T} / F+B & Protects dynamic label-space metadata under the studied output mechanism. \\
36 & Release & Continual classifier / fixed public label space & \texttt{T} / B & Avoids data-dependent label-index release by fixing support. The learned scores require separate control. \\
37 & Accounting & Adaptive transcript / filtering and accounting & \texttt{T} / F & Filters enforce a cap. Odometers report bounds under adaptively chosen mechanisms and budgets. \\
38 & Accounting & Streaming statistics / continual mean release & \texttt{T} / F & User-level guarantee under limits on stream contributions. The stream contains multiple contributors. \\
39 & Memory & Personal memory / sanitize and partition writes & \texttt{D} / L & Constructed persona histories test sanitized records and isolated exact-value store (\S2.1--2.2 and \S4.2). \\
39 & Release & Personal memory / retrieve or request consent & \texttt{S} / L & Tests privacy behavior of sanitized retrieval and conditional exact-value recovery under simulated consent (\S2.3, \S4.4). \\
40 & Skill & Shared workflow policies / feedback-driven update & \texttt{T} / B & Benchmarks evolving workflow-level privacy rules shared across users against workflow failures. \\
43 & Withdrawal & Memory turns / vary the pool for selective cleanup & \texttt{T} / B & Sec.~6.1 and Table~9 vary candidate reach and deletion budget. Native and residual recovery use fixed post-deletion probes. \\
45 & Release & RAG query / metric-private perturbation & \texttt{T} / F+B & Cloud observers see privatized request embeddings with a distance-indexed bound. Retrieved outputs are a separate interface. \\
46 & Release & RAG corpus / encrypted retrieval & \texttt{T} / F+B & Homomorphic scoring protects search inputs at the server within protocol assumptions. Downstream generation remains visible where allowed. \\
47 & Memory & RAG corpus / synthetic index construction & \texttt{T} / B & Compares leakage and utility when a fixed private retrieval corpus is replaced with synthetic text. \\
48 & Memory & Oblivious index / dynamic maintenance & \texttt{T} / F+B & Encrypted insertions and deletes conceal access patterns under the index leakage model. \\
48 & Release & Oblivious index / confidential search & \texttt{T} / F+B & Index hides read access patterns from its host, with a specified trusted enclave. \\
49 & Release & Vector index / oblivious traversal & \texttt{T} / F+B & Server observes encrypted ORAM accesses within declared leakage. Vector similarity search has no personal update loop. \\
50 & Withdrawal & RAG corpus / knowledge-base-driven forgetting & \texttt{T} / B & Retrieval change reduces targeted behavior on benchmark questions. The method holds backbone weights fixed. \\
51 & Withdrawal & Parameters / certified data removal & \texttt{T} / F+B & Formal distributional comparison of removed and retrained models under the stated learner assumptions. \\
52 & Withdrawal & Parameters / cryptographic execution verification & \texttt{T} / F+B & Proof verifies declared unlearning computation and state transition. Certification depends on underlying deletion rule. \\
53 & Withdrawal & Interactive state / deletion-as-control & \texttt{T} / F & Defines post-deletion control against a counterfactual trajectory under stated interaction semantics. \\
54 & Release & Agent tools / lattice enforcement and declassification & \texttt{T} / F+B & Tracks provenance before writes to external tools. Authorized release and the enforced endpoints delimit the scope. \\
55 & Release & Agent tools / purpose and risk checks & \texttt{T} / B & Inspects inferred consequences and permits, blocks, or requests confirmation for outgoing tool information. \\
59 & Skill & Persona skill / sanitize distillation input & \texttt{T} / B & Rewrites fixed traces before authorized distillation and measures attribute exposure and behavioral fidelity (\S\S4.4, 5.3). \\
59 & Skill & Persona skill / perturb distilled skill & \texttt{T} / B & Empirically tests perturbation of synthesized skills under the evaluated distillers. \\
62 & Skill & Shared skills / sanitize trajectory-derived patches & \texttt{T} / B & Local compression and PII-canary audits reduce content exposure before sharing. Model use remains outside the payload test. \\
63 & Release & Sharing policy / consent-aware release decisions & \texttt{S} / B & Learned bidding governs proposed OSN disclosure on simulated personas. Policy-learning feedback and bids remain visible where allowed. \\
64 & Release & Sharing policy / uncertainty-aware release decisions & \texttt{S} / B & A personalized classifier defers uncertain sensitive-content decisions to a user. The test covers the release decision. \\
65 & Release & Personal profile / task-scoped minimization & \texttt{S} / B & Isolated minimizer supplies a synthetic profile subset to the task agent. Protection is measured under tested recipient tasks. \\
66 & Learning & Text representations / local substitution before training & \texttt{T} / F+B & Metric privacy limits inference from noisy tokens or vectors. Adaptive pretraining addresses utility recovery. \\
67 & Release & Released versions / label-only disclosure & \texttt{T} / B & Limits precision of post-unlearning outputs and measures differential version leakage empirically. \\
67 & Release & Released versions / posterior scaling & \texttt{T} / B & Measures version-difference leakage after scaling permitted posterior outputs. \\
67 & Learning & Parameters / private pretraining mitigation & \texttt{T} / F+B & DP learner is compared as a defense against version-difference leakage. Joint release needs its own accounting. \\
68 & Withdrawal & Parameters / posterior-difference audit & \texttt{T} / B & Tests evidence of deletion with version access and shadow models. The outcome is scoped to the audit probes. \\
69 & Accounting & Federated unlearning / encrypted update aggregation & \texttt{T} / F+B & Controlled function evaluation protects submitted uploads from server visibility under threshold and setup assumptions. \\
69 & Withdrawal & Federated unlearning / withdrawal intent masking & \texttt{T} / B & Shared update seeks to conceal whether a request was an unlearning request. Inference from returned aggregates remains a separate question. \\
70 & Release & Agent tools / typed capabilities and release constraints & \texttt{T} / F+B & Type and semantic rules restrict effectful paths from tagged observations to tool sinks under trusted policy and runtime. \\
71 & Withdrawal & Encrypted index / backward-private deletion updates & \texttt{T} / F+B & Puncturing bounds server leakage after deleting indexed entries under the construction's leakage profile. \\
75 & Learning & Collaborative bandit / central private updates & \texttt{T} / F+B & Central mechanism protects a single reward observation across online shared policy updates under its assumptions. \\
75 & Learning & Collaborative bandit / local reward privatization & \texttt{T} / F+B & Client-side reward protection in online setting. Permitted action transcript remains in scope. \\
76 & Memory & Personal memory / anonymize summaries at write time & \texttt{D} / L & Matched multi-session simulations measure inference from transformed summaries and response/helpfulness effects (\S\S2.5, 3.5). \\
76 & Release & Memory-conditioned replies / disclosure instruction & \texttt{S} / L & Instructs the agent to limit elicitation and repetition, then measures inference and coherence in simulation. \\
76 & Release & Current messages / sender-side anonymization & \texttt{T} / L & Redacts each user turn before the support agent first sees it. This fresh-input control changes conversation and retained summaries indirectly. \\
77 & Memory & Memory input / Attribution Shield query reformulation & \texttt{T} / B & Evaluates proxy memory extraction after reformulation. Donated memories characterize exposure and do not validate a longitudinal intervention. \\
78 & Memory & Memory exchange / redaction before storage & --- / I & Worked examples describe sanitized storage and local retention of recovery values, conditional on trusted mapping. \\
78 & Release & Memory exchange / purpose-bound packing and restoration & --- / I & Worked examples select context by purpose and resolve values at the authorized local endpoint. \\
79 & Withdrawal & Personal memory / retire or rebuild tracked descendants & \texttt{D} / L & Synthetic repeated sessions compare closure cleanup and clean references with complete recorded lineage and fixed model weights. \\
82 & Accounting & Correlated data / secret-pair release accounting & \texttt{T} / F & Defines inferential privacy against a declared distribution class and pairs of alternate secrets. \\
83 & Accounting & Time series / successive inferential releases & \texttt{T} / F & Composition results under explicit dependency and Markov Quilt assumptions. \\
84 & Accounting & Correlated data / adaptive inferential composition & \texttt{T} / F & Characterizes when sequential bounds can sum under stated influence conditions. \\
85 & Withdrawal & Parameters / unlearning under adaptive requests & \texttt{T} / F+B & DP-based reduction controls effects of requests selected from previous released models, conditional on its base learner. \\
86 & Withdrawal & Parameters / privacy over post-deletion models & \texttt{T} / F & Formal deletion privacy targets the release sequence when training and deletion decisions are adaptive. \\
87 & Release & Personal memory / remove shared linking cues before generation & \texttt{S} / L & Empirical relation-disclosure results test matched cases and preserve some visible values (\S\S4.3--4.4, Tables~2--3). \\
87 & Release & Personal memory / mask retrieved values & \texttt{S} / L & Separate comparison changes retrieved values, measuring disclosure and task completion after the common reader (Table~3). \\
91 & Memory & Multi-user memory / assign visibility on writes & \texttt{T} / B & The prototype labels entries with scopes that determine later index access. The evaluation covers the combined pipeline. \\
91 & Release & Multi-user memory / enforce index retrieval scope & \texttt{T} / B & The trusted index excludes unauthorized cross-user entries. No single-user fact-inference guarantee. \\
93 & Learning & Continual model / perturb token embeddings & \texttt{T} / F+B & Appendix~G calibrates a fixed-scale Gaussian step. Changing invocation or repeating access changes the accounting problem. \\
93 & Memory & Continual training memory / sensitivity-guided retention & \texttt{T} / B & Task-incremental evaluation measures retention and adaptation. \\
94 & Memory & Multi-agent memory / sensitivity-aware eviction & \texttt{T} / B & Simulated policies compare retention, modeled leakage and utility using tracked provenance (\S\S3.2--3.4, 6.6.4). \\
94 & Memory & Multi-agent memory / DP eviction tie-break & --- / I & Optional selection rule has local scope. Scarce firings are insufficient for a system-level privacy outcome. \\
95 & Withdrawal & Agent state and trajectories / unlearning interventions & \texttt{T} / B & Compares state, trajectory and environment behavior against trained-without controls on public tasks (Secs.~III--VI, VIII). \\
96 & Memory & Agent memory / write screening and re-audit & \texttt{T} / B & Trusted checks handle new entries and already-stored content. Outcome evaluates the joint pipeline, without a separate write-side bound. \\
96 & Release & Agent memory / intercept extraction requests & \texttt{T} / B & Extraction interception is measured as part of the shared test harness alongside combined interventions. \\
97 & Skill & Reusable skills / learn redaction and routing & \texttt{T} / B & Offline refinement creates reusable controls from benchmark prompts. An external evaluator sees originals during refinement. \\
97 & Release & Cloud subtask / release redacted or locally routed views & \texttt{T} / B & Leakage measurement concerns cloud-bound runtime subtasks under the learned skill policies. \\
98 & Skill & Personal archives / impair unauthorized distillation & \texttt{T} / B & Appended record interventions reduce behavioral imitation fidelity for a fixed archive. Retained original text remains available. \\
99 & Memory & Personal memory / candidate admission decision & \texttt{T} / B & Compares downstream over-retention and privacy risk of proposed writes on reconstructed histories, without successive writes. \\
100 & Memory & Agent memory / sensitive-record filtering & \texttt{T} / B & Measures which records are retained in a capacity benchmark. Does not measure later disclosure. \\
100 & Memory & Agent memory / sensitivity-aware pruning & \texttt{T} / B & Capacity prunes stored records. Benchmark provides no descendant withdrawal or recovery evidence. \\
101 & Release & Static profile / private slot and block release & \texttt{T} / F+B & Theorem~1 and App.~A compare individual and joint attributes, fixed public domains, cached reuse and fresh composition. \\
102 & Release & Personalized choices / counterfactual exposure control & \texttt{S} / B & Adaptive release checks and cumulative exposure scoring are evaluated on personalized benchmarks (\S VII, Tables~IV--V). \\
103 & Learning & Online reward and policy / privatize feedback labels & \texttt{T} / F & User-side randomized response precedes update; Theorem and Algorithm~2 cover online contextual bandit learning (\S5). \\
103 & Learning & Offline reward and policy / privatize feedback labels & \texttt{T} / F+B & Fixed-data estimation and policy selection under label LDP. Offline implementation is tested in \S6. \\
106 & Learning & Shared recommender / perturb uploaded BPR updates & --- / E & Clipping and Gaussian perturbation in a 53-day field deployment; measured outcomes and nominal budget do not substantiate a privacy claim. \\
107 & Release & Personal memory / context-qualified pre-reader gate & \texttt{S} / B & Evaluates context-aware retrieval filtering on constructed personal histories in multiple memory backends and OpenClaw. \\
108 & Withdrawal & Stored policies / enforce metadata-based revocation & \texttt{T} / B & Guard is tested for retrieval and writeback behavior in controlled policy supersession tasks. Scope is stale-policy non-use. \\
109 & Release & Personal memory / classify and admit recalled candidates & \texttt{S} / B & Tests whole-entry and unit-level admission from fixed pools under leakage and utility outcomes. Judge sees raw candidates. \\
109 & Release & Personal memory / render admitted candidates & \texttt{S} / B & Matched-verdict experiment reports inconclusive differences between channel rendering and permission notes. \\
110 & Memory & Personal memory / bind origin audience at ingest & \texttt{D} / L & Persisted origin labels and trusted identifiers are tested in controlled histories with one owner and multiple audiences. \\
110 & Memory & Personal memory / intersect derived audiences & \texttt{D} / L & Mutation and transformation tests exercise provenance-bound derivation and suppression when admissibility fails (\S6.4). \\
110 & Release & Personal memory / authorize context admission & \texttt{S} / F+L & Exact-context exclusion under trusted identity, provenance and mediation. Tests cover per-attempt retrieval, not answer noninterference. \\
110 & Withdrawal & Personal memory / grant revocation and narrowing & --- / I & App.~E.2 specifies next-attempt authorization semantics for these unevaluated operations. \\
111 & Release & Sharing requests / entailment-based decision & \texttt{S} / B & Prior personal judgments inform a non-adversarial release decision. The benchmark measures judgment agreement and abstention, with disclosure and history protection outside that outcome. \\
112 & Learning & Contextual bandits / share encoded feedback for warm starts & \texttt{T} / F+B & A trusted shuffler filters sampled feedback for server-side shared learning. The analysis uses one-context-vector adjacency, and benchmarks measure recommendation utility. \\
113 & Learning & Parameters / norm-aware training-data selection & \texttt{T} / B & Filtering or mixing partially restores contextual disclosure behavior in offline fine-tuning tests (\S5.4, App.~I). No contribution-privacy bound. \\
114 & Withdrawal & Runtime prompts, summaries, plans, and cache / selective replay & \texttt{T} / F+B & Theorem~1 and \S5 compare replay with a never-exposed execution under fixed weights, deterministic transitions or fixed randomness, and counterfactual-equivalent suffix inputs. Committed effects are outside the replay contract. \\
116 & Release & Multi-agent memory / propagation-scoped gate & \texttt{T} / B & Full-Scope and Auto-Region restrict downstream disclosure across task-graph edges. MAPLE-Bench reports violations and utility on controlled workflows (\S\S4--5). \\
118 & Release & Shared-agent context / constrain inter-agent sharing & \texttt{T} / B & Privacy prompts and topology are compared on contextual violations in shared multi-user tasks (\S5.1, Apps.~16 and 19.1). The prompt contrast does not isolate the write-side effect. \\
120 & Release & Action parameters / minimize mediator and audit views & \texttt{T} / B & Reference-policy tests and projection analysis cover field suppression and residual leaks (\S\S3, 8--9), not recipient behavior or personal-state updates. \\
121 & Release & Proposed tool arguments / pre-execution monitoring & \texttt{T} / B & Deterministic replay to a local inert sink tests bounded provenance and block/permit semantics (\S7.2, Table~5), not independently sampled calls. \\
122 & Skill & Candidate skill / extraction-time review & \texttt{T} / B & Exploratory review blocks flagged entries in benchmark skill banks (\S4.9, Table~3). Trigger rates are proxies for policy violations. \\
122 & Release & Skill execution / runtime action monitor & \texttt{T} / B & Runtime blocking is compared with extraction review and their combination in the same exploratory study; no validated personal disclosure guarantee. \\
123 & Release & Personal tool decision / bundled privacy and consent policy & \texttt{S} / B & \S\S3 and~5 compare frozen policy prompts on personal scenarios with memory and consent context. Table~2 measures privacy and consent risk without isolating the component effect. \\
124 & Release & Personal context / contextual flow filtering & \texttt{S} / B & PrivacyChecker evaluates scoped removal or abstraction of information flows before output in PrivacyLens and tool workflows (\S\S3--4). Residual leakage is measured. \\
125 & Prompt & Federated prompt / protect during clustering & \texttt{T} / F+B & Private clustering limits prompt visibility to the aggregator. The server sees distances, and Theorem~1 characterizes leakage under a Gaussian model. \\
126 & Prompt & Federated prompt / privatize client updates & \texttt{T} / F+B & Theorem~4.1 accounts for client update releases across communication rounds under the stated sampler and clipping. Section~V evaluates heterogeneous image datasets. \\
127 & Learning & Weights / DP fine-tuning and empirical audit & \texttt{T} / B & Compares extraction, membership risk, and utility across DP fine-tuning methods on fixed datasets. \\
128 & Learning & Listwise feedback / private top subset release & \texttt{T} / F+B & Definition~4.1 and Theorem~B.2 cover item relocation within a fixed ranking domain. Offline experiments evaluate training from sanitized preferences. \\
129 & Learning & Policy / trajectory preference alignment & \texttt{T} / B & Protocol varies privacy persona and measures GUI compliance and task success on fixed tasks. The result concerns behavior after training. \\
130 & Learning & Offline policy / private preference optimization & \texttt{T} / F & The central mechanism protects sample changes including prompts and responses. The local mechanism protects choice labels. Both assume offline feedback. \\
131 & Memory & Personal profiles / local write and merge & \texttt{D} / L & Secs.~IV-D and VI-B.1 specify local retention against direct cloud transmission. Sec.~VII-B and Fig.~16 demonstrate carryover with ten lab participants. The study measures pipeline use and fallback frequency. \\
131 & Release & Current commands / optional cloud obfuscation & \texttt{T} / B & Sec.~VI-B.2 tests command exposure in DevFinder. The cloud receives transformed current requests and decoys. The test targets command identification. \\
132 & Release & Recommender / observable choice policy & \texttt{T} / F & The interceptor observes a recommendation and updates beliefs about a preference ranking. The policy is analyzed under the specified preference and observation model. \\
134 & Release & Skill runtime / enforce contextual permissions & \texttt{T} / B & Secs.~IV, VII and VIII evaluate permission checks on benchmark skills and report violations and retained task utility. \\
135 & Skill & Skill runtime / evidence-guided recovery & \texttt{T} / B & Secs.~3.6 and 4.4 compare matched recovery with and without execution provenance and test authorized task completion. \\
136 & Withdrawal & Personal adapter / omit retained proxy & \texttt{T} / B & Secs.~2.4 and 4.2 test bypass for fixed synthetic profiles. The software deletion path is specified as a separate operation. \\
136 & Withdrawal & Personal adapter / filesystem deletion & --- / I & The deletion protocol names the proxy directory. The verification measurements use inference-time bypass. \\
137 & Learning & Isolated LoRA / adapt on per-user history & \texttt{T} / B & Secs.~3.2 and 4.2 compare privacy interfaces and adaptation on fixed personal datasets. User artifacts remain separate from the shared anchor. \\
138 & Skill & Skills / scan candidates and retire descendants & --- / I & Secs.~5.3--5.6 specify scanner and lineage controls. Secs.~6--7 report benchmark task outcomes. \\
139 & Release & Memory context / eligibility and pairwise repair & \texttt{S} / B & Filters annotated owners, purposes and lifecycle states before context assembly. Repair certificates verify stated constraints over the working set. \\
140 & Prompt & Prompts and demonstrations / local DP generation & \texttt{T} / F+B & A cloud host receives privatized prompt artifacts under the reported adaptation conditions and can still see raw inference queries. \\
140 & Learning & Local weights / private tuning benchmark & \texttt{T} / B & Matched tasks compare local private fine-tuning and closed-model alternatives; benchmark uses fixed tasks. \\
141 & Release & RAG / score-based query gate & \texttt{T} / B & Entropy and score margins drive empirical blocking; fixed Healthcare and Enron experiments measure PII exposure after retrieval. \\
143 & Learning & LoRA / encrypted delegated fine-tuning & \texttt{T} / B & Remote workers handle encrypted data-derived computations. Owner-side steps and outputs have separate trust requirements. \\
144 & Learning & Split or offsite weights / client-provider isolation & \texttt{T} / B & Matched tuning architectures compare reconstruction resistance, task performance and cost. Client and provider observe different internal representations. \\
145 & Release & Profile traits / proof-based disclosure & \texttt{T} / B & A verifier receives a proof about an authorized rule outcome. Evaluates proof overhead and advice grounding in static settings. \\
146 & Learning & Adapter training / isolate precision dispatch & \texttt{T} / F+B & Batch-dependent outlier routing can change other clipped gradients. Record-owned dispatch restores the stated single-record sensitivity condition. \\
147 & Learning & Weights / user-level accounting under sampling & \texttt{T} / F+B & Compares example- and user-sampled private updates under user adjacency and fixed compute budgets. \\
148 & Learning & Personal adapter / empirical donor-risk merging & \texttt{T} / B & LaMP experiments measure target-user utility and membership inference for users supplying shared adapters. \\
149 & Learning & Replay summaries / private list release & \texttt{T} / F+B & Public anchors align client-private replay lists. Theorem~4 covers replay release under client-level adjacency; released trained heads require separate accounting. \\
150 & Learning & Model representations / perturb forward pass for tuning & \texttt{T} / F+B & Matrix Gaussian noise privatizes the stated example-dependent training representations. \\
150 & Release & Query representations / privatize forward embedding & \texttt{T} / F+B & Local forward-pass perturbation addresses embedding inversion and attribute inference in fixed-task evaluations. \\
151 & Learning & Weights / redacted first stage and private second stage & \texttt{T} / F+B & The sensitive-token policy defines adjacency; experiments test missed sensitive tokens and utility of the two-stage design. \\
152 & Accounting & Streaming releases / buffer single-edit inputs & \texttt{T} / F & The buffering wrapper and one-round privacy profile yield a trajectory guarantee with a delay cost in the stated adaptive model. \\
153 & Learning & Federated PEFT / update-subspace and optimizer defense & \texttt{T} / B & Reports resistance to a specified malicious-server reconstruction mechanism and measures task utility under the tested conditions. \\
154 & Release & Personal history / client-side draft modification & \texttt{S} / B & Client conditions on retained history. The audit measures linkability and attribute inference from query and response; full iterative traces require another assessment. \\
155 & Learning & Policy / private ordinal preference training & \texttt{T} / F+B & Fixed survey inputs supply label-private rating supervision. The benchmark compares response perturbation, private selection and private gradient updates. \\
156 & Learning & Preference data / DP synthesis from public prompts & \texttt{T} / F+B & Preference-model estimation consumes private comparisons; later examples are generated using protected models and public prompts. \\
159 & Release & Current queries / local privacy-profile rewriting & \texttt{T} / B & Measures disclosure and utility when the provider receives a rewritten current query. Transfer to release of facts read from memory requires another test. \\
160 & Learning & DPO / vary synthetic privacy-preference mix & \texttt{T} / B & Offline benchmark reports canary and membership effects across data sources alongside utility. No formal bound is assigned. \\
161 & Learning & Agent policy / contextual privacy preference training & \texttt{T} / B & Offline training and held-out contexts test disclosure behavior, with separate task-utility outcomes. \\
162 & Release & Shared graph memory / role-based access & \texttt{T} / B & Tests allowed and denied access under task roles at the collaborative retrieval interface. \\
163 & Learning & Federated adapters / local perturbation and consolidation & \texttt{T} / F+B & Protects a local sample change in client updates under its LDP analysis. Consolidation uses released aggregates in a multi-client setting. \\
165 & Skill & Shared skills / pre-sync sanitization and lineage & \texttt{T} / B & Separate simulated users test covered entity retention and source attribution. Transfer concerns skill publication; source attribution remains above chance. \\
166 & Withdrawal & One secret source / multi-channel unlearning test & \texttt{T} / B & Compares suppression and joint recovery over six channels. Each run injects the fact into exactly one carrier. \\
167 & Release & Tool returns / limit sensitive inference in responses & \texttt{T} / B & Tests output mitigations under combinations of tool returns in fixed tasks. Reasoning traces have separate measurements. \\
168 & Release & Tool calls / audit external transmission & \texttt{T} / B & App.~I contrasts a monitored tool path with an unmonitored output path, with separate exact and semantic measurements. \\
169 & Learning & Weights / DP-SGD and LoRA comparison & \texttt{T} / B & Tests empirical exposure, utility and compute on fixed tasks. This row credits the comparison of training controls. \\
170 & Release & Multi-agent output and internal-channel controls & \texttt{T} / B & Compares final-view and internal exposure on fixed cooperative workflows, with channel-specific defense outcomes. \\
171 & Learning & Split ownership / confidential adapter tuning & \texttt{T} / F+B & The simulation result covers protocol views under a semi-honest model. The model owner receives the final adapter. \\
172 & Learning & Cut-point activations and gradients / obfuscation & \texttt{T} / B & Compares the training server's split-interface exposure and performance. Federated upload is another view. \\
173 & Learning & Full tuning and PEFT / clipping and noise & \texttt{T} / B & Compares membership inference, utility and resources under matched tasks. Attack AUC is an empirical outcome. \\
174 & Learning & Split token representations / perturb before upload & \texttt{T} / B & Measures inversion and attribute inference. Input-dependent calibration requires separate release accounting. \\
175 & Learning & Sign updates / bound hidden-subset information & \texttt{T} / F+B & Protects a sampled subset under a declared candidate universe and distribution. The output is a training transcript and its post-processed model. \\
176 & Withdrawal & Weights / reference-guided neuron pruning & \texttt{T} / B & Compares extraction and utility against equal-budget pruning baselines at a fixed post-training checkpoint. \\
177 & Learning & Wireless aggregation / controlled noisy transmission & \texttt{T} / F+B & Bounds an adjacent record at the specified receiver under channel assumptions. Resource comparisons use simulation and estimates. \\
178 & Learning & Federated head / encrypted aggregate and restricted inference & \texttt{T} / F+B & Separates a conditional cryptographic result from measured exposure and cost for score and label-only output. \\
179 & Learning & Federated LoRA / synchronize the round rank & \texttt{T} / F+B & Protects user contributions under the stated aggregation and accountant. Large-cohort results use simulated noise. \\
180 & Learning & Shared update / joint gradient-component masking & \texttt{T} / B & Compares recovery when jointly visible components are protected together in a single-step server view. \\
\end{longtable}
\endgroup

\section{Primary Evidence for the Deployment Comparisons}\label{sec:supp-deployment-cases}

Main Section~7 compares histories with and without a private evening-booking restriction while holding the public task, slots, and permissions fixed. The sources below evaluate different settings. The booking comparisons identify their observation interfaces and the assumptions needed to transfer each result.

\subsection{Booking: Host Trace, Recipient Transcript, and Visible Action}

\paragraph{State and recipient comparators.}
The Always-On Evaluation Protocol (v1, \S\S9.2--9.4, Table~18) reconstructs state from an event stream and checks authority epochs, deletion propagation, scope, and approval of external actions~\cite{ding2026alwayson}. VMG (EMNLP archival, \S5, Eqs.~3 and 5, Appendix~D) uses principal-scoped retrieval and a query distribution over which to probe reconstruction after deletion~\cite{lin2026memorysecuritysurvey}. Its expanded version (arXiv:2604.16548v5, \S5 and Appendix~D) also specifies authorized utility and paired deletion comparisons. In the booking case, use these checks at the state and retrieval interfaces. At the service interface, compare request text, selected slot, refusals, and timing across both histories before withdrawal and under the same permitted follow-up queries afterward. Score distinguishing the presence of the restriction, booking completion, and target recovery separately. For example, returning a valid evening booking only with the unrestricted history is sufficient for a distinguishing event even when both bookings satisfy authorization and the restriction is absent from message text.

\paragraph{Prompt and preference.}
DP-OPT (ICLR 2024, \S3 and Algorithms~1--2) uses local data for generation and exponential selection over a disjoint validation split~\cite{hong2024dpopt}. Ray Chowdhury et al.\ (Definitions~2.2--2.3, \S\S4--5) protect the comparison label, while Cho and Sun (\S2.2, \S3.2, Algorithm~1) privatize a reward learner operating on comparison tuples~\cite{raychowdhury2024privatereward,cho2026decoupledreward}. For transfer, bound how often one fact occurs in each split and how many entire tuples it changes. A change in a candidate response is not label adjacency. Keep the reference sampler and online context independent of the private history except through already accounted releases.

\paragraph{Memory and recipient.}
Opal v1 \S2.2, Figure~3, and Theorem~2.1 define a host security game with public operation types, fixed trace shapes, and an exogenous checkpoint schedule. The attestation-hybrid model uses a durable monotonic client counter, and freshness of sealed client state is relative to the last provider checkpoint (\S3.4)~\cite{kaviani2026opal}. DP-MemView v1 Definition~1 (C1--C6), Theorem~1, and Corollary~1 bound adaptive application transcripts for covered same-scaffold store differences~\cite{kim2026dpmemview}. A public ingest triggered by a private fact changes the first interface. A new public slot or raw read changes the second. The booking recipient's joint view requires coverage of every additional observable.

\paragraph{Matched release experiment.}
DP-MemView v1 \S4, Table~2 holds the query trajectory and response backbone fixed. In the synthetic PairedMem pairs, the scaffold and non-target content match and 4--6 slots for the target attribute vary. With Qwen2.5-7B-Instruct, $T=16$, and online DP-MemView $B_a=2$, $K=3$, the (AUC, tRec) pairs are $(0.832,0.524)$ for TaskMin, $(0.500,0.386)$ for TypedMask, and $(0.490,0.506)$ for online DP-MemView. Here AUC measures the frozen auditor's ability to distinguish histories from held-out responses and tRec scores target-required behavior via the blinded utility judge. Appendix~C.4, Table~8 details the adapted baselines: TaskMin sends a fixed task-family subset and TypedMask sends slot-typed placeholders for protected items. The latter yields identical conditioning and chance AUC by construction. The same table reports public-corpus transfer AUCs of $0.535$--$0.563$ across the three backbones for online DP-MemView. Compare each trace under its own corpus and observer. Changed public layouts, autonomous writes, and real booking actions need matched tests~\cite{kim2026dpmemview}.

\paragraph{Mapping the history change.}
Let $H_0$ have no evening restriction and let $H_1$ contain one. The latter adds a labeled example to the training set and a separate labeled example to the validation set. It also replaces a neutral value in one preallocated slot in attribute group $I_a$. Hold slot metadata, the booking request, authority and the three publicly available slots fixed. Let $G$ denote the complete list of candidate prompts, $P$ the selected prompt, $V$ the augmented DP-MemView transcript and $A$ the booking exchange including the request, response, timing and errors. The linked recipient view is $(P,V,A)$. Raw examples, memory values, intermediate scores and $G$ stay at the trusted local endpoint.

\paragraph{Candidate generation with stable voting groups.}
Use the ensemble generation and LimitedDomain selection structure of DP-OPT~\cite{hong2024dpopt}. In this variant, set the public sampling rate to $q=1$. Partition by a fixed public map from stable record identifiers to a fixed number of groups. The assignment and order of other records are independent of the added record. Use one fixed-model vote per group with a public default vote for an empty group. At each fixed generation prefix, condition on independent per-group model randomness for that call. Only the group containing the new example can change its vote. The histogram thus changes by at most one in any coordinate and at most two coordinates change. Calibration for vote replacement can be reduced to the single-vote insertion and removal case in DP-OPT Appendix~A.2.1, Theorem~A.2. Apply top-one LimitedDomain with a fixed public vocabulary, deterministic tie breaking, a public domain limit and single-vote sensitivity parameters $\Delta_0=\Delta_\infty=1$. If its single-vote bound is $(\eta,\gamma)$, connect the two histograms by removing the old ballot and inserting the new one. The same mechanism is defined on the intermediate histogram. The two DP inequalities give the bound $(2\eta,(1+e^\eta)\gamma)$ in each direction. Choose $\eta=\varepsilon_0/2$ and $\gamma=\delta_0/(1+e^\eta)$. The resulting per-call bound is $(\varepsilon_0,\delta_0)$ for the history pair.

Precommit a public candidate count and a total budget of $K$ private calls across the list. Every call counts, including end-of-sequence outputs, $\bot$ and any retry. Termination, retry and candidate assembly use the call count and the privatized tokens only. At the cap, stop and fill any unused positions with a public default prompt. For the privacy analysis, pad the private call transcript to $K$ with data-independent symbols. Each step is private for every fixed prefix, including padded steps at zero cost. Basic adaptive composition gives $(K\varepsilon_0,K\delta_0)$ for that transcript and hence for $G$.

\paragraph{Selection, reading and the recipient transcript.}
For each fixed $G$, use the count of correct predictions on the validation set as the score and apply the exponential mechanism with exponent $\varepsilon_v$ times the score divided by two. The score changes by at most one for either history. For each fixed $P$, apply DP-MemView Theorem~1 with C1--C6 on the same-scaffold stores. Future queries can depend on $P$ and earlier accounted outputs. For example, allocate $B_a=0.50$ to two reads at $0.25$ each. Each read can select from the same two public views recommending an early or a late daytime appointment, with a score of one if the view matches the slot's recommendation rule and zero otherwise. The known sensitivity is one, and each read charges attribute $a$. The booking controller chooses the daytime slot from these views and public information. At this interface, request timing has a fixed public schedule, and responses and error handling use only accounted outputs and public inputs. Thus $(P,V,A)$ is post-processing of $(G,P,V)$, whose sequential bound is $(K\varepsilon_0+\varepsilon_v+B_a,K\delta_0)$~\cite{hong2024dpopt,kim2026dpmemview,whitehouse2023fullyadaptive}.

As one budget example, schedule four candidate prompts with at most eight private attempts each, set $\varepsilon_0=0.025$, $\delta_0=2\times10^{-8}$ and $\varepsilon_v=0.30$, and use the two-read budget above. Generation contributes $(0.80,6.4\times10^{-7})$ and the recipient transcript satisfies $(1.60,6.4\times10^{-7})$-DP for this pair. These are analytical budgets. The calculation uses the fixed-group generation variant, the declared recipient view and the common release schedule.

\paragraph{Failure with the same recipients.}
Hold $G$, $P$, $V$, the input histories and the recipient boundary fixed as mechanisms and inputs. Change the booking rule to inspect the raw restriction and choose the evening slot in $H_0$ and a daytime slot in $H_1$. If $E$ is the evening request at the service, $\Pr[E\mid H_0]=1$ and $\Pr[E\mid H_1]=0$. The DP inequality forces $\delta\geq 1$. The protected prompt and read kernels still run as specified. The booking request is the output that leaves the accounted path~\cite{fu2026privaterecs,kim2026dpmemview}.

\paragraph{Failure after adding the host view.}
Restore the protected booking rule and extend the observer to host logs. Emit an extra host-visible ingest to an auxiliary store exactly in $H_1$, leaving the recipient path as specified. The ingest event now has probabilities zero and one under the two histories. The extended view requires $\delta\geq 1$ even though the original recipients receive the same releases as in the composable case. Opal \S2.2 and Theorem~2.1 compare histories with the same public leakage. The extra ingest changes that leakage. Adding a public slot instead changes DP-MemView's scaffold. A public schedule with the same host-visible operation counts addresses the ingest event. Bringing the full host view into the positive bound additionally requires a conditional simulation from the accounted recipient transcript and public inputs, or compatible separate accounting~\cite{kaviani2026opal,kim2026dpmemview}.

\paragraph{Evidence on linked observations.}
TOP-Bench (EMNLP Findings, \S\S3--6) constructs 1,000 cases under a two-stage tool protocol. Individual returns are insufficient for the designated inference by construction. The study measures inference after combining returns, and evaluates prompt and training interventions for the agent's output behavior. CIPL (v1, \S\S3--4, Apps.~D.4 and I) distinguishes internal selection from external recovery, evaluates tools and answers separately, and contrasts a monitored tool path with an unmonitored output path. These fixed-task studies inform the observer and probe choices. The longitudinal booking comparison additionally requires a new write and reuse of the resulting state~\cite{qiao2026topbench,huang2026cipl}.

\subsection{Adaptive Parameters and Skills: Separate Policy from Enforcement}

In the ACL archival \emph{Privacy Collapse} study, \S4.1 and Figure~2 compare proactive and conservative training conditions on PrivacyLens and CIMemories. The former evaluates privacy decisions in tool-use scenarios, not actual tool executions in production. Section~5.4 and Appendix~I isolate partial mitigation via data filtering and mixing; there is no DP guarantee for the training data~\cite{goel2026privacycollapse}. In MaskClaw v2, E3 (\S5.5, \S\S6.3--6.4, Apps.~E--F) compares text scoring, held-out skill use, and sandbox filtering. Appendix~F Table~20 gives the calendar denominator (189 candidates) and rates (43.39\% strict text and 16.93\% sandbox). Appendix~K Table~28 identifies the external judge; Section~\ref{sec:supp-boundary-review} gives the persistence ruling~\cite{zhao2026maskclaw}. Audience-Bound Persistent Memory (v1, \S\S3.1--3.5 and 6.4) constrains context admission under its own authorization and mediation assumptions~\cite{liu2026audiencememory}. An external tool gate after generation and repeated regression checks across admitted versions are design implications, not properties established by any of these results in combination.

\subsection{Withdrawal: Match Verification to the Carrier}

Reglos v1 \S3.1 and \S4 (Figure~5 and forget-case analysis) support a scoped non-use claim for a tombstone combined with reader enforcement~\cite{yang2026reglos}. MemGov (archival, \S\S3.3--3.4, 4.3, Table~3, and Appendix~D) certifies eligibility and pairwise consistency decisions on a bounded working set~\cite{christou2026memgov}. In the 202-case eligibility profile ($k=5$), both the metadata-filter baseline (metadata/vector retrieval) and MemGov (BM25 plus governance) return zero forbidden objects. Answer hit, measured by inclusion of the expected memory ID, is 26.2\% versus 70.3\%. In separate pairwise profiles, unsafe merges are 22.3\% versus zero observed. The certificate is local to the supplied metadata and detected constraints, and the retrieval difference prevents attributing the answer-hit gap to the repair layer alone.

AfterDeleteBench v1 \S5.3 and Appendix~D.3, Table~6 distinguish existing descendants from future writes. For the fixed-weight Qwen IID condition (200 profiles), exact deletion gives reconstruction, commitment, and later activation of $0.735/0.735/0.510$. The sealed literal gate gives $0.735/0/0.680$. Recorded-lineage closure yields $0/0/0$, and the corresponding clean-world rates are zero. The gate acts at candidate write time, so later activation can draw on surviving state even when commitment is blocked. The measurements depend on the audit interface and complete recorded lineage in the closure arm~\cite{saleh2026afterdeletebench}.

K-Bench v1 \S\S3--5 isolates a single source at a time (weights, prompt context, free-text retrieval, or structured retrieval) and probes six observable channels. In the structured-retrieval comparison (Section~5.3 and Figure~6b), StaR reduces recovery in the answer channel while tool observations still expose the target and the aggregate recovery rate is unchanged. This paired comparison isolates a missed observation channel under its scaffold. Combine its channel checks with a never-exposed reference and multi-carrier persistence when testing withdrawal from a personal agent~\cite{yu2026kbench}.

The A-MEM diagnostic in Section~\ref{sec:diag-12ep} (Tables~\ref{tab:supp-run-quality}--\ref{tab:supp-memory-outcomes} and Figure~\ref{fig:supp-withdrawal}B) compares 24 paired episode-builds at $t_1$ after five target-free writes. The root-only, conservative closure, and never-exposed arms have raw exact-marker hit counts of 1, 0, and 0, and confirmed non-target exact-field recovery counts of 16, 0, and 22. The closure removes all five initial records in each build. The raw target endpoint keeps guard-rejected responses, while utility requires confirmed exact-field recovery. The equal-cardinality random deletion arm in the development run also deletes the entire initial store (Section~S4.7). To isolate selective cleanup, match the deletion count in non-saturated stores and measure semantic recovery, later activation, and task utility alongside exact-marker hits. MemCut v1 Section~6.1 and Appendix~B.3, Table~9 hold the budget at ten turns on a 144-fact subset spanning three systems and two write models. Widening the candidate pool from 20 to 60 reduces reported residual recovery by 12.5--23.9 percentage points for model-guided cleanup. Appendix~B.4, Table~11 also compares off-target deletions on Turn RAG. These controls assess candidate reach and deletion cost at a fixed evaluation checkpoint. AfterDeleteBench measures subsequent lifecycle outcomes~\cite{zhang2026memcut,saleh2026afterdeletebench}.

Yao et al.\ v1 Theorem~1 and \S5 require sanitized suffix observations that match the counterfactual execution, compatible runtime, fixed weights and deterministic decoding or fixed randomness, and the absence of committed external side effects after the target enters~\cite{yao2026executionunlearning}. SBU v2 \S4.3 Table~6 tests memory-only, parameter-only, and joint interventions in public medical QA~\cite{wang2026agenticunlearning}. SEA v1 \S2.4 and \S4.2 compare outputs after bypass of a preference proxy fitted on synthetic data~\cite{schneider2026sea}. Its deletion procedure is separate. For the booking case, probe the reloaded store, plans, and active proxy on the same tasks after recovery, and track any external booking as a separate recipient-side state.

\subsection{Additional transfer and risk comparisons}
\label{sec:supp-transfer-additions}

\paragraph{Training and replay.}
Selective DP makes sensitive token flags part of the privacy unit, and DP-Forward moves perturbation to the forward representation~\cite{shi2022justfinetune,du2023dpforward}. In an adapter stack with batch-dependent precision dispatch, a changed record can affect other clipped gradients. Record-owned dispatch restores the sensitivity condition used by per-record accounting~\cite{zhang2026routingdp}. Delegated HE-LoRA keeps encrypted computations at remote workers. In GuardedTuning, split and offsite variants expose different intermediate representations to the client and provider~\cite{frery2025helora,shi2025guardedtuning}. TriShield tests a separate malicious-coordinator threat in federated PEFT~\cite{wei2026trishield}. For single-edit streams, buffering can provide a trajectory bound with an explicit delay cost~\cite{chan2026participation}. These results motivate separate checks for the unit of a released update, the compute party's view, and the schedule of fresh private reads.

\paragraph{Preference and read release.}
Private ordinal supervision protects a rating under fixed questionnaire inputs~\cite{qi2026surveyalignment}. PREMIA and privacy-preference mixing measure membership and memorization in offline training~\cite{feng2025premia,yang2026p3m}. These endpoints complement behavioral disclosure tests after adaptation. For read release, PROFIT measures cloud-visible rewrites of current queries under fixed policies, proof-backed advice reveals a designated rule outcome, and COLLABMEM tests scope enforcement at a shared graph interface~\cite{ramirez2026peep,watanabe2025zkadvice,wu2026collabmem}. CIDER supplies user-labeled disclosure judgments for evaluating personalized policies~\cite{guo2026cider}. Each study fixes something that can change inside a personal loop, such as the policy, predicate, role, or history. Tests after revision should record whether such a change affects the same release observation.

\paragraph{Matched tuning and observer comparisons from the rescreening.}
Ma et al.\ compare full tuning and parameter-efficient tuning on the same tasks~\cite{ma2024privatepeft}. On IMDb with DistilBERT, non-private full tuning and LoRA have observed membership-inference AUCs of $0.59$ and $0.52$ (Figure~2). For the private variants on the same P5000, full tuning and LoRA use $8.50$ and $6.10$ GB, respectively (Table~4). These are task-specific observations. PACZero protects which subset is drawn from a stated public candidate family, with the private update sequence and its post-processed model as outputs (v3, Section~3.4, Theorem~5)~\cite{ertan2026paczero}. Mapping that guarantee to the booking fact would require a distribution over histories and an explicit correspondence between the fact and the hidden subset.

\paragraph{What crosses a training boundary.}
VeilTune separates the data owner's training view from the model owner's output adapter under its semi-honest simulation assumptions (v1, Sections~3.2 and 5.6)~\cite{xie2026veiltune}. The output adapter has its own disclosure boundary. PrivPair's matched split-interface evaluation measures the server-visible activations and gradients (v1, Sections~III-B and VII-C--F)~\cite{jin2026privpair}. PrivTune measures inversion and attribute inference on perturbed token representations, with input-dependent calibration requiring a separate transcript analysis~\cite{liu2026privtune}. AEGIS isolates the effect of protecting several gradient components together in its single-step comparison (v1, Sections~III-A and VI-C/F)~\cite{tao2026aegis}. These results tell the reviewer which messages enter the mechanism-specific view. Later release of model weights or linked rounds requires its own assessment.

\paragraph{Conditional communication and deployment costs.}
pAirZero bounds a record change at the receiver of a wireless aggregate under stated channel and power assumptions (v1, Lemma~1). Its device-memory and upload figures are estimates for the modeled system (Table~II)~\cite{cai2026pairzero}. DP-DyLoRA uses a common public rank within a round and accounts for user contributions through the aggregate. Large-cohort costs and accuracy are evaluated under simulated noise conditions (arXiv:2405.06368v4, Sections~V-A, VI-F and VII)~\cite{xu2026dpdylora}. HE-OFT compares score and label-only inference from the same trained head, with measured latency and model recovery under the tested configurations (v1, Section~V-E/F, Tables~7 and 10). Section~IV-F distinguishes those measurements from the smudging parameters required by its proof~\cite{kanpak2026heoft}. These network, training and inference costs attach to different observers and should be reported at their respective interfaces.

\paragraph{Weight-side suppression and persistent withdrawal.}
In JEPA-Prune, an equal-budget random and targeted pruning comparison on the same fine-tuned model separates pruning quantity from selection (v1, Section~5.1, Tables~1--2). Its evaluation measures extraction and utility immediately after the intervention. The effect of additional adaptation and recovery remains a separate question (Section~7)~\cite{shaaban2026jepaprune}. A personal agent may also recover the same fact from a retained store. Detecting that path requires the later-activation and recipient probes in Supplementary Section~S6.3.

\subsection{Bounded Rescreening and Sensitivity of Main Table 4}\label{sec:supp-table4-audit}

\paragraph{Candidate set.}
On 19 November 2026 we rechecked 12 distinct records against the 10 November literature cutoff. We took three anchors per scenario: prompt, read, and host mechanisms; label, tuple, and skill-admission mechanisms; and non-use, lineage, and model-removal mechanisms. We paired each trio with one previously excluded record covering a similar operation: attributed retrieval (Agent Zero Memory, Stage~4 scope exclusions), feedback-gated skill learning (Stage~1 R07), and learned memory deletion (Stage~15 Crossref W4). We checked the cited archival or pinned primary texts at the locators below. This purposive rescreen tests the boundary decisions supporting Main Table~4. It is neither a new literature search nor a reconstruction of the historical Stage~2 exclusion trail.

\paragraph{Evidence rules.}
\textbf{C (component):} Keep a proof or measurement that addresses the mapped input, operation, and observer or behavioral endpoint; distinguish a proof, a scoped outcome, and a proxy. Exclude ordinary utility, attribution, or state management without a protection result.
\textbf{D* (direct outcome):} Require a measured outcome at the claimed interface after the affected state carries into a later interaction. A proof, fixed-store release test, candidate-only test, or workload performance result alone does not qualify. Non-use qualifies only for the reader's behavior. To support the \textbf{linked-view} claim, evidence must cover the same fact across booking, committed skill reuse, and withdrawal for an observer retaining earlier outputs. D* is a sensitivity filter, not a recoding of \texttt{D}/\texttt{S}/\texttt{T}.

\begin{longtable}{@{}>{\raggedright\arraybackslash}p{0.20\linewidth}>{\raggedright\arraybackslash}p{0.36\linewidth}>{\raggedright\arraybackslash}p{0.37\linewidth}@{}}
\caption{Primary-source rescreen of 12 records supporting or bordering Main Table 4.}\label{tab:supp-table4-audit}\\
\toprule
\textbf{Case and record} & \textbf{Primary locator and result checked} & \textbf{Disposition and sensitivity} \\
\midrule
\endfirsthead
\toprule
\textbf{Case and record} & \textbf{Primary locator and result checked} & \textbf{Disposition and sensitivity} \\
\midrule
\endhead
\bottomrule
\endfoot
Book: DP-OPT~\cite{hong2024dpopt} &
\S3, Algorithms~1--2, App.~A.2.1: private token votes and selection on a separate split. &
Keep C: record-level primitives for the bounded variant. Fails D*: no linked booking after an evolving write. A fact affecting both splits cannot use parallel composition. \\
\addlinespace
Book: DP-MemView~\cite{kim2026dpmemview} &
V1, Definition~1, Theorem~1, Table~2: fixed-scaffold attribute adjacency and fixed-store release experiments. &
Keep C for the bounded read transcript. Fails D*: repeated queries do not test a new write followed by reuse. A different public layout falls outside the theorem. \\
\addlinespace
Book: Opal~\cite{kaviani2026opal} &
V1, \S2.2, Theorem~2.1 and \S6: host security at matched public leakage; dynamic workload measures cost and utility. &
Keep C: conditional host proof. Fails D* for recipient privacy: workload results do not measure that observer. An extra visible ingest changes declared leakage. \\
\addlinespace
Book boundary: Agent Zero Memory &
\href{https://arxiv.org/html/2608.29606v1}{V1}, \S\S3.2--3.3, 4.1, 5: provenance-aware search, accuracy, cost, and latency. &
Keep excluded under C and D*. Attribution does not test exposure of a changed fact at the service or host. \\
\midrule
Learn: DP Reward Estimation~\cite{raychowdhury2024privatereward} &
PMLR, Definitions~2.2--2.3 and \S\S4--5: private comparison label with fixed contexts and action pairs. &
Keep C for label changes; altered candidate content is out of scope. Fails D*: no test of later skill execution. \\
\addlinespace
Learn: Decoupled DP Reward~\cite{cho2026decoupledreward} &
V1, \S\S2.2, 3.2 and Algorithm~1: one-entry replacement, private reward learner, and downstream post-processing. &
Keep C for tuple changes after accounting for all replacements; raw candidate generation and provider logs are outside the bound. Fails D*: no test after committing a skill. \\
\addlinespace
Learn: MaskClaw~\cite{zhao2026maskclaw} &
V2, \S\S5.5, 6.3--6.4 and App.~F Table~20: text and sandbox gates over 189 calendar candidates. &
Keep C as admission evidence only. Fails D*: gate scores do not measure disclosure when the accepted version is committed and later invoked. \\
\addlinespace
Learn boundary: Future-feedback skill evolution &
\href{https://arxiv.org/html/2607.18973v1}{V1}, \S\S4.3, 5.2--5.3, 7: offline prediction of feedback; no distilled answer-skill deployment test. &
Keep excluded under C and D*. Data preparation and predictive accuracy do not test confidentiality of skill updates or releases. \\
\midrule
Withdraw: Reglos~\cite{yang2026reglos} &
V1, \S\S3.1, 4, Figure~5 and forget-case analysis: tombstone plus reader policy, with scored later answers. &
Keep C and D* for reader non-use only. No physical cleanup, version-view guarantee, or isolated tombstone effect follows. \\
\addlinespace
Withdraw: AfterDeleteBench~\cite{saleh2026afterdeletebench} &
V1, \S\S3, 5.3, App.~D.3 Table~6: paired never-exposed histories and carried memory after deletion and candidate write. &
Keep C and D* for the tested memory endpoints. Complete lineage and fixed weights preclude extrapolation to all prompt, weight, and skill recovery. \\
\addlinespace
Withdraw: Certified Data Removal~\cite{guo2020certifiedremoval} &
PMLR, \S2 and \S4: compare the removal output with a model trained without the target record. &
Keep C as a formal marginal comparator, not as a bound for the before/after pair. Fails D*: no linked agent-recovery outcome. \\
\addlinespace
Withdraw boundary: AgeMem &
\href{https://aclanthology.org/2026.acl-long.981/}{ACL archival}, \S3.2 Table~1 and \S4: learned DELETE tool; task and memory-management evaluation. &
Keep excluded under C and D*. An available deletion action is not a test of private residue or authorized forgetting. \\
\end{longtable}

Under C, nine anchors remain eligible at their stated scope and the three boundary exclusions stand. Under D*, only the two withdrawal anchors provide direct evidence, for reader non-use or tracked-memory outcomes. Booking remains formal transfer plus component evidence; label and tuple results remain conditional, and skill admission remains a proxy. None of the 12 meets the linked-view criterion. The stricter rule changes which findings count as direct validation, not the mathematical scope exclusions or the original corpus counts and codes. These results apply to the checked set, not to the entire literature.

\bibliographystyle{ACM-Reference-Format}
\bibliography{references}